\documentclass[aps,prb,10pt,twocolumn,amsfonts,longbibliography,citeautoscript]{revtex4-1}

\usepackage{color}
\usepackage{bm}
\usepackage{bbold}
\usepackage{amsmath}
\usepackage{amssymb}
\usepackage{epsfig}
\usepackage{verbatim}
\usepackage[T1]{fontenc}
\usepackage{array}

\newcommand{\be}{\begin{equation}}
\newcommand{\ee}{\end{equation}}
\newcommand{\bw}{\begin{widetext}}
\newcommand{\ew}{\end{widetext}}
\newcommand{\bea}{\begin{eqnarray}}
\newcommand{\eea}{\end{eqnarray}}
\newcommand{\la}{\langle}
\newcommand{\ra}{\rangle}
\newcommand{\dg}{^\dagger}
\newcommand{\p}{\partial}
\newcommand{\rd}{{\rm d}}
\newcommand{\s}{\sigma}
\def\nn{\nonumber\\}

\allowdisplaybreaks[1]
\usepackage[colorlinks=true,bookmarks=false,citecolor=blue,linkcolor=magenta,hyperfootnotes=true,urlcolor=blue]{hyperref}

\begin{document}
\title{Signatures of rare states and thermalization in a theory with confinement}

\author{Neil J. Robinson}
\email{n.j.robinson@uva.nl}
\affiliation{Institute for Theoretical Physics, University of Amsterdam, Science Park 904, 1098 XH Amsterdam, The Netherlands}

\author{Andrew J. A. James}
\email{andrew.james@open.ac.uk}
\affiliation{School of Physical Sciences, The Open University, Milton Keynes, MK7 6AA, United Kingdom}
\affiliation{London Centre for Nanotechnology, University College London, Gordon Street, London WC1H 0AH, United Kingdom}

\author{Robert M. Konik}
\email{rmk@bnl.gov}
\affiliation{Condensed Matter Physics \& Materials Science Division, Brookhaven National Laboratory, Upton, NY 11973-5000, USA}

\date{\today}

\begin{abstract}
There is a dichotomy in the nonequilibrium dynamics of quantum many body systems. In the presence of integrability, expectation values of local operators equilibrate to values described by a generalized Gibbs ensemble, which retains extensive memory about the initial state of the system. On the other hand, in generic systems such expectation values relax to stationary values described by the thermal ensemble, fixed solely by the energy of the state. At the heart of understanding this dichotomy is the eigenstate thermalization hypothesis (ETH): individual eigenstates in nonintegrable systems are thermal, in the sense that expectation values agree with the thermal prediction at a temperature set by the energy of the eigenstate. In systems where ETH is violated, thermalization can be avoided. Thus establishing the range of validity of ETH is crucial in understanding whether a given quantum system thermalizes. Here we study a simple model with confinement, the quantum Ising chain with a longitudinal field, in which ETH is violated. Despite an absence of integrability, there exist rare (nonthermal) states that persist far into the spectrum. These arise as a direct consequence of confinement: pairs of particles are confined, forming new `meson' excitations whose energy can be extensive in the system size. We show that such states are nonthermal in both the continuum and in the low-energy spectrum of the corresponding lattice model. We highlight that the presence of such states within the spectrum has  important consequences, with certain quenches leading to an absence of thermalization and local observables evolving anomalously.  
\end{abstract}

\maketitle

\section{Introduction}
Understanding the nonequilibrium dynamics of quantum many body systems has become one of the central goals of physics in recent years~\cite{polkovnikov2011nonequilibrium,*gogolin2016equilibration,*dalessio2016quantum,*essler2016quench,*calabrese2016quantum,*cazalilla2016quantum,*bernard2016conformal,*caux2016quench,*vidmar2016generalized,*langen2016prethermalization,*ilievski2016quasilocal,*vasseur2016nonequilibrium,*deluca2016equilibration}. This has been motivated by groundbreaking progress in experiments on cold atomic gases~\cite{weiss2006quantum,*lewenstein2007ultracold}, which realized unprecedented control and isolation of quantum systems. Experiments have highlighted a lack of understanding of fundamental issues, such as thermalization: How and when does a quantum system thermalize? What does thermalization mean in an isolated quantum system undergoing unitary time evolution? 

The eigenstate thermalization hypothesis (ETH)~\cite{deutsch1991quantum,srednicki1994chaos,rigol2008thermalization,*rigol2009breakdown,rigol2012alternatives,reimann2015eigenstate,deutsch2018eigenstate} plays a central role in answering such questions. It states conditions that matrix elements of an operator $\hat A$ must satisfy in order for its expectation value on an eigenstate to agree with the microcanonical (thermal) prediction. These conditions can be summarized as 
\begin{align}
\hat A_{\alpha,\beta} = A(E) \delta_{\alpha,\beta} + e^{-S(E)/2}f_A(E,\omega) R_{\alpha,\beta},
\label{eth}
\end{align}
where $\hat A_{\alpha,\beta} = \la E_\beta | \hat A | E_\alpha\ra$ are matrix elements in the basis of eigenstates $|E_\alpha\ra$, $E = (E_\alpha + E_\beta)/2$ and $\omega = E_\alpha-E_\beta$. ETH tells us that the diagonal matrix elements $\hat A_{\alpha,\alpha}$ are controlled by a smooth function $A(E)$.~\footnote{For neighboring energy eigenvalues $E$ and $E'$ ($E'>E$) this function satisfies $A(E')-A(E) \propto e^{-R/\tilde R}$, with $R$ being the system size and $\tilde R$ being some (possibly $E$-dependent) constant.} The off-diagonal elements are suppressed by the thermodynamic entropy $S(E)$ and depend on both a smooth function $f_A(E,\omega)$ and the random variable $R_{\alpha,\beta}$, which has zero mean and unit variance. 

It is generally assumed that generic quantum many body systems satisfy ETH, in the sense that matrix elements of local observables obey Eq.~\eqref{eth} and hence expectation values are thermal (see, e.g., the brief argument of Ref.~\cite{mondaini2017eigenstate}). It is, however, known that generic finite size systems \textit{can} have rare (nonthermal) eigenstates which violate ETH~\cite{biroli2010effect,santos2010localization,santos2010onset}. Such states are often called nontypical, see Ref.~\cite{richter2018realtime} for a recent example. The presence of rare states in the finite volume leads to two interpretations of ETH. The so-called \textit{``weak ETH''} supposes that the fraction of rare to thermal states vanishes in the infinite volume limit. Weak ETH is not sufficient to imply thermalization: such a scenario applies to integrable models~\cite{biroli2010effect,ikeda2013finitesize,alba2015eigenstate} where thermalization is avoided due to the presence of an extensive number of local conservation laws. On the other hand, the \textit{``strong ETH''} postulates that all rare states must vanish in the infinite volume. This much stronger condition then implies thermalization, with expectation values of local operators coinciding with the microcanonical ensemble (MCE) average. In nonequilibrium scenarios, such as following a quantum quench, thermalization is signaled by the diagonal ensemble (DE) prediction agreeing with the MCE constructed at the appropriate energy density~\cite{rigol2008thermalization,rigol2009breakdown}. 

Distinguishing the weak and strong ETH scenarios is a challenging problem for state-of-the-art numerical methods. Evidence for the weak ETH has been seen in many nonintegrable systems, primarily through exact diagonalization of small systems~\cite{santos2010localization,santos2010onset,khlebnikov2013thermalization,ikeda2013finitesize,beugeling2014finitesize,sorg2014relaxation,khodja2015relevance,cosme2015relaxation,mondaini2016eigenstate,lan2017eigenstate,mondaini2017eigenstate,lan2017eigenstate}. Generally the rare states are observed at the very edges of the spectrum, which is not entirely surprising with many cases of low energy emergent integrability being known (see, e.g., Ref.~\cite{essler2005applications}). Evidence consistent with strong ETH, on the other hand, is less well established. Numerical studies of lattice hardcore bosons with next-neighbor and next-next-neighbor interactions, as well as the quantum Ising chain with certain values of transverse and longitudinal fields, suggest that it is valid in some contexts~\cite{kim2014testing}.

In the context of weak ETH, there are some cases where the nonthermal rare eigenstates are found away from the very edges of the spectrum. Firstly, in the case of AKLT-like models, certain eigenstates with finite energy density  can be represented exactly as matrix product states and shown to be nonthermal (through, e.g., the nonthermal entanglement entropies associated with such states)~\cite{moudgalya2018exact,moudgalya2018entanglement}. Another scenario, so-called many-body quantum scars, arises in models that are associated with one-dimensional Rydberg gases~\cite{bernien2017probing}. These models have been seen to have a polynomial-in-the-system-size number of nonthermal states which are distributed throughout the many-body spectrum of the model~\cite{turner2017quantum,khemani2018signatures,ho2018periodic,lin2018exact,turner2018quantum,choi2018emergent}. There are also examples of nonthermal behavior in nonintegrable models with constrained dynamics, which includes the aforementioned models of Rydberg atoms and quantum dimer models~\cite{ates2012thermalization,ji2013equilibration,vanhorssen2015dynamics,lan2018quantum}, and models with long-range interactions~\cite{liu2018confined,lerose2019quasilocalized}.

One significant issue in studying ETH is the lack of available techniques. Almost all studies are confined to the exact diagonalization of small systems ($L \lesssim 20$ sites), which results in large finite size effects. This makes it hard to extrapolate results to the infinite volume and hence make concrete statements about strong or weak ETH. Some progress has been made in the last few years using tools from typicality~\cite{steinigeweg2014pushing,bartsch2017necessity,richter2018realtime}, with which one can push to slightly larger system sizes ($L\lesssim 35$). Other techniques that can compute real-time dynamics (such as time evolving block decimation and related algorithms~\cite{kollath2007quench,manmana2007strongly,trotzky2012probing,sorg2014relaxation,james2015quantum,james2018nonperturbative}; numerical renormalization group~\cite{brandino2015glimmers}; equations of motion~\cite{moeckel2008interaction,moeckel2009realtime,eckstein2009new,essler2014quench,nessi2014equations,nessi2015glasslike,bertini2015prethermalization,bertini2016thermalization}; and Boltzmann equations~\cite{moeckel2008interaction,moeckel2009realtime,bertini2015prethermalization,bertini2016thermalization,furst2012matrixvalued,furst2013matrixvalued,biebl2017thermalization}) allow one to establish whether expectation values of operators approach their thermal values following a quench from a given state, but have little to say about ETH. With this in mind, it is desirable to develop techniques that can study system sizes beyond those accessible to exact diagonalization, or in scenarios that exact (full) diagonalization cannot study. Here, to investigate thermalization and the validity of ETH in a large non-integrable system---in both equilibrium and nonequilibrium scenarios---we focus on a \textit{continuum model} and use recently developed extensions of the truncated spectrum method~\cite{rakovsky2016hamiltonian,james2018nonperturbative}. We will see that these results are consistent with the lattice simulations away from the scaling limit.

Before introducing the model that we study herein, it would be remiss of us not to mention another example in which ETH is violated: many body localization. Many body localization can arise in interacting disordered models, as shown in the seminal work of Basko, Aleiner and Altshuler~\cite{basko2006metal} (see also the recent review articles~\cite{imbrie2017local,parameswaran2017eigenstate,agarwal2017rareregion,deng2017manybody,abanin2017recent,alet2018manybody}). Much like the scenarios discussed above, study of such models is almost entirely restricted to exact diagonalization (we note the exception of excited state DMRG~\cite{khemani2016obtaining,yu2017finding}), where violation of ETH has been established for numerous examples (see, e.g., Ref.~\cite{alet2018manybody} and references therein). The violation of ETH reflects an emergent integrability in the localized phase, with models possessing an extensive number of local conserved quantities~\cite{imbrie2016manybody,imbrie2017local}.

\subsection{Rare states in the perturbed Ising field theory}
In this paper we will argue that the perturbed Ising field theory possesses rare states.  This model, which arises as the continuum limit of the quantum Ising chain~\cite{mccoy1978twodimension,sachdev2011quantum}, is described by the Hamiltonian
\begin{align}
H(m,g) = \int \rd x \Big[ i \Big( \bar \psi \p_x \bar \psi - \psi \p_x \psi + m\bar \psi \psi\Big) + g \s \Big]. \label{FT}
\end{align}
Here $\psi$ ($\bar \psi$) is a left (right) moving Majorana fermion field, $m$ is the fermion mass, and $g$ is equivalent to a longitudinal field in the lattice model. In the ordered phase, the fermions can (loosely) be thought of as domain wall excitations in the ferromagnetic spin configuration. The longitudinal field $g$ acts as a \textit{nonlocal} confining potential for the domain walls~\cite{mccoy1978twodimension}. While the field theory is nonintegrable for generic values of the parameters $m$ and $g$, there exist two special lines in parameter space ($m=0$ and $g=0$) along which the model is integrable~\cite{onsager1944crystal,zamolodchikov1989integrals}. 

\begin{figure}[t]
\includegraphics[width=0.45\textwidth]{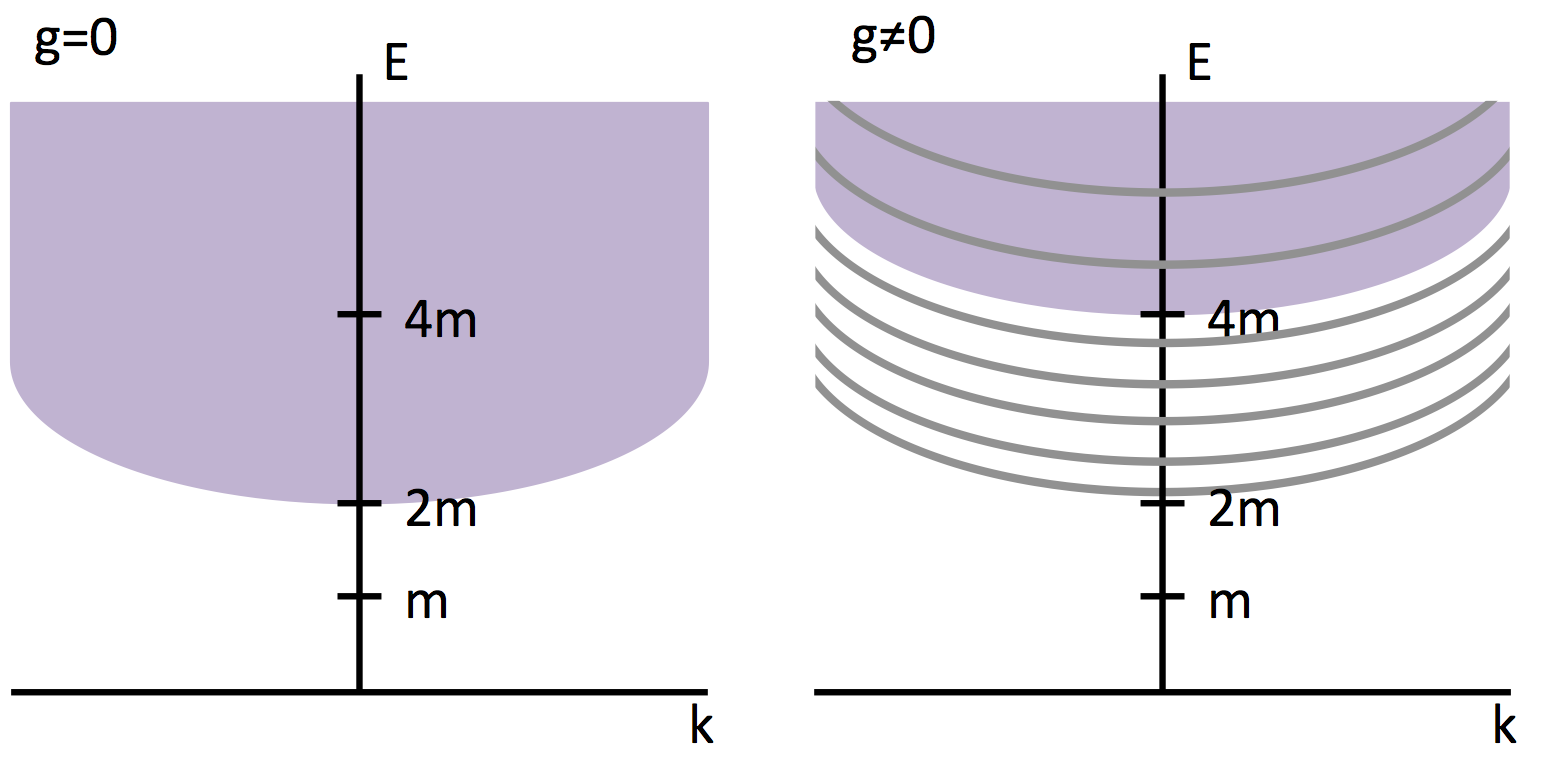} 
\caption{Schematic of the low energy spectrum of the Ising field theory~\eqref{FT}. The longitudinal field $g$ leads to a complete restructuring of the low energy continuum (shaded) and the appearance of confined `meson' states (solid lines).}
\label{Fig:Spectrum}
\end{figure}

Herein we will mostly focus on the ordered phase of the model, corresponding to $m>0$. Within this phase, the spectrum of the model depends intimately on the longitudinal field: when $g=0$ a flipped spin fractionalizes into two independent  domain walls, which can freely propagate though the system. Thus at low energies, when $g=0$, there is a two particle continuum of excitations, separated from the ground state by an energy gap of $2m$. On the other hand, when $g\neq 0$ there are profound changes in the spectrum. The presence of the longitudinal field, which is nonlocal in terms of the domain wall fermions, induces a linear potential between domain wall excitations, leading to confinement~\cite{mccoy1978twodimension,delfino1996nonintegrable,delfino1998nonintegrable}. This is very much reminiscent of the formation of mesons in quantum chromodynamics (analogies between magnetic systems and quantum chromodynamics have recently been emphasized in Ref.~\cite{sulejmanpasic2017confinement}). The low energy spectrum now completely restructures: the two domain wall continuum at $g=0$ collapses into well-defined meson excitations for $g\neq 0$, with a new multimeson continuum forming above energies $E\sim4m$. We sketch this schematically in Fig.~\ref{Fig:Spectrum}. We note that this restructuring of the continuum has been observed in the quasi-one-dimensional Ising ferromagnet CoNb$_2$O$_6$~\cite{coldea2010quantum,morris2014hierarchy} and the XXZ antiferromagnet SrCo$_{2}$V$_{2}$O$_{8}$~\cite{wang2015spinon,wang2016from}. 

Working directly with a continuum theory might, at first glance, seem more difficult than working on the lattice. However, we will see that low energy eigenstates of the Hamiltonian~\eqref{FT} can be constructed, and are representative of the behavior in the infinite volume limit. We will do so using truncated spectrum methods~\cite{james2018nonperturbative}, a well-established toolbox in the study of the Ising field theory~\eqref{FT}. One question that it is natural to ask before starting is: What is the status of ETH in a field theory? Is it expected to hold? Here one can turn to the original work of Deutsch~\cite{deutsch1991quantum}, which explicitly considered a continuum model and argued that it would be expected to thermalize. Srednicki's later paper~\cite{srednicki1994chaos} considered a continuum system with a bounded spectrum, i.e. the theory is equipped with a cutoff. We will see that expectation values of local operators in the vast majority of low energy eigenstates of~\eqref{FT} exhibit features consistent with ETH, while there are a subset of rare states that appear to violate it.  

With regards to ETH in~\eqref{FT}, there is one additional point worth noting. We can split the Hamiltonian~\eqref{FT} into two pieces: a noninteracting part $H(m,0)$ and an interaction term $g\int \rd x \s(x)$. The interaction term is strongly renormalization group relevant, with the scaling dimension of the operator $\s$ being $1/8$. Defining the theory with an explicit cutoff $\Lambda$ and performing a weak-coupling ($g \ll 1$) renormalization group analysis, the interaction strength $g(\Lambda)$ flows to zero in the ultraviolet $\Lambda \to \infty$. As a result, the high energy eigenstates of the theory~\eqref{FT} should correspond to noninteracting ($g=0$) eigenstates. Concomitantly the high energy sector of the theory~\eqref{FT} has an emergent integrability and hence we should not expect ETH to be valid there. Hence we will concentrate our efforts on the low energy sector, where the interaction term has a strong effect, and ask whether ETH is valid in this subspace. 

This work complements results presented by the authors in Ref.~\cite{james2018nonthermal}, in particular extending the one-dimensional aspect of that work.

\subsection{Layout}

This work proceeds as follows. In Sec.~\ref{Sec:RareStates}, we examine the matrix elements of the local spin operator between eigenstates constructed with truncated spectrum methods. We will see clear signatures of  rare states that exist in the low energy spectrum of the field theory~\eqref{FT}. We discuss ETH and behavior of these rare states with increasing system size. In Sec.~\ref{Sec:Nature} we establish the nature of the rare states, relating them to the two-fermion ``meson'' confined states of the model. We study the lifetime of the meson excitations to understand their stability and influence on ETH in the infinite volume limit. 

Following this, in Sec.~\ref{Sec:Dynamics} we turn our attention to the nonequilibrium dynamics following a quench of the longitudinal field strength $g$. The rare states established in the equilibrium spectrum will also be apparent in the DE, signaling a lack of thermalization after certain quenches. In Sec.~\ref{Sec:Lattice} we relate our work to the analogous lattice problem and discuss the presence of rare states there, comparing the finite size scaling analysis on the lattice with that in the continuum. We follow with our conclusions in Sec.~\ref{Sec:Conclusions} and cover a number of technical points in the appendices. 

\section{Testing ETH in a theory with confinement}
\label{Sec:RareStates}

In order to test ETH within the field theory~\eqref{FT}, we must be able to construct eigenstates. We do so using truncated spectrum methods, which we briefly summarize below. A detailed overview of the technique can be found in the recent review~\cite{james2018nonperturbative}.

\subsection{Truncated spectrum methods}

\subsubsection{General philosophy}

Truncated spectrum methods (TSMs) are a general approach to treating certain classes of field theory problem~\cite{james2018nonperturbative}. The best known is the truncated conformal space approach (TCSA)~\cite{yurov1990truncated,james2018nonperturbative}, where the Hamiltonian can be written in the following form: 
\begin{align}
H = H_{\rm CFT} + V. \label{tcsa}
\end{align}
Here the first part of the Hamiltonian, $H_{\rm CFT}$, describes a conformal field theory (CFT)~\cite{BigYellowBook}, while $V$ is a renormalization group relevant, but otherwise arbitrary, operator.
The central idea of TSMs is to use the eigenstates of a `known' theory (the CFT in Eq.~\eqref{tcsa}) as a computational basis for constructing the full Hamiltonian~\eqref{tcsa}. 

The presence of a relevant operator, $V$, leads to a strong mixing of the low-energy computational basis states. This ultimately leads to a failure in any perturbative treatment -- requiring instead a nonperturbative approach. The TSM is one such method: using the fact that $V$ does \textit{not} strongly couple basis states of largely differing energies, one motivates a truncation of the Hilbert space of basis states through the introduction of an energy cutoff $E_\Lambda$. For low energy eigenstates of $H$ this is a reasonable approximation. To obtain a finite spectrum, even with an energy cutoff $E_\Lambda$, one must then place the system in a finite volume $R$. Subsequently one diagonalizes the (finite) Hamiltonian to obtain approximate energies and eigenvectors of the full problem~\eqref{tcsa}. 

More generally, the TSM can be applied to problems of the form 
\begin{align}
H = H_\text{known} + V, \label{tsm}
\end{align}
where $H_\text{known}$ is a theory where one knows how to construct eigenstates. As in Eq.~\eqref{tcsa}, this may be a CFT, but more generally one can consider \textit{integrable} theories. Then, one needs to be able to compute matrix elements of $V$ in the basis of $H_\text{known}$, and one can follow the procedure above to obtain approximate eigenstates and eigenvalues. TSMs were first used by Yurov and Zamolodchikov~\cite{yurov1991truncated} to study the field theory~\eqref{FT} in what they called the truncated-free-fermionic-space approach. These methods can also be extended in order to mitigate truncation effects, for example using numerical renormalization group techniques~\cite{konik2007numerical,konik2009renormalization,brandino2010energy,james2018nonperturbative}, which can be a necessity in more complicated theories (we discuss convergence of the TSM for the problem at hand in Appendix~\ref{App:ConvTCSA}).

To summarize, TSMs allow us to construct well converged low-energy eigenstates of strongly correlated quantum field theories. These field theories, written in terms of a solvable part and a renormalization group relevant operator, are hard to tackle with alternative methods and the TSM can yield considerable insights into their properties. It is necessary to work in the finite volume (so the spectrum is discrete) and with a finite cutoff (to make the Hamiltonian matrix finite), so it is vital to systematically vary the cutoff and the system size to ensure one is describing the physics of the thermodynamic limit and not finite-size or truncation effects. More detailed discussions of this method can be found in the recent review~\cite{james2018nonperturbative}.

We proceed by focusing our attention on the zero momentum sector of the theory, which contains the system's ground state. Restricting ourselves to a particular momentum sector incurs no loss of generality [as momentum is conserved by our theory Eq.~\eqref{FT}]. Eigenstates obtained with TSMs (discussed in further detail in the next section) will be used to compute observables and, following recent works~\cite{rakovsky2016hamiltonian,hodsagi2018quench}, nonequilibrium dynamics. 

\subsubsection{As applied to the Ising field theory~\eqref{FT}}

Let us briefly recap some details of the TSM for the Ising field theory; a recent summary of known results and this method can be found in Ref.~\cite{james2018nonperturbative}. We work in the finite volume $R$ and begin by separating the Hamiltonian~\eqref{FT} into a `known' piece and a perturbation, cf. Eq.~\eqref{tsm}, 
\begin{align}
H(m,g) &= H_0(m) + g V, \\
H_0(m) &= \int_0^R \rd x\, i \Big( \bar \psi\p_x \bar\psi - \psi\p_x \psi + m \bar\psi\psi \Big), \\
V &= \int_0^R \rd x \,\s(x). 
\end{align}
Here our known piece, $H_0(m)$, describes a system of free fermions with mass $m$. 

Let us now consider the eigenstates of the known theory~\cite{mccoy1978twodimension}. The Hilbert space of the model is split into two sectors, known as Neveu-Schwartz (NS, antiperiodic boundary conditions) and Ramond (RM, periodic boundary conditions), in which the momenta of fermions is quantized in a different manner: $p_\text{NS} = 2\pi n/R$ with $n\in\mathbb{Z}+1/2$ and $p_\text{R} = 2\pi n/R$ with $n\in\mathbb{Z}$, respectively. Eigenstates are obtained by acting on the vacuum (within a given sector) with fermion creation operators
\begin{align}
|\{k\}_N \ra_\text{NS} &= a\dg_{k_1}\ldots a\dg_{k_N} |0\ra_\text{NS}, \quad k_i \in \mathbb{Z}+\frac12, \nonumber \\
|\{q\}_N\ra_\text{RM} &= a\dg_{q_1}\ldots a\dg_{q_N} |0\ra_\text{RM}, \quad ~~q_i \in \mathbb{Z}. 
\end{align}
Here $\{k\}_N = \{k_1,\ldots,k_N\}$ is a convenient shorthand notation, and henceforth we use $k_i$ ($q_i$) to signify momenta in the NS (RM) sector. The fermion creation operators obey the canonical anticommutation relations
\begin{align}
\{ a_k, a\dg_{k'} \} = \delta_{k,k'}, \quad \{ a_q, a\dg_{q'} \} = \delta_{q,q'}, \quad \{a_k, a\dg_q \} = 0.
\end{align}
Eigenstates of $H_0(m)$ with $N$ particles in the finite volume $R$ have energies
\begin{align}
E_N(R)_\nu = E_0(R)_\nu + \sum_{j=1}^N \omega_{p_{\nu,i}}(R),
\end{align}
where $\nu = \text{NS},\,\text{RM}$ and $p_{\nu,i} = k_i\,(q_i)$ for $\nu = \text{NS}\,\text{(RM)}$. The dispersion relation for the fermions is given by
\begin{align}
\omega_{p_{\nu}}(R) = \sqrt{m^2 + \bigg(\frac{2\pi}{R} p_\nu \bigg)^2}, \label{dispersion}
\end{align}
while the vacuum energy is 
\begin{align}
E_0(R)_\nu &= \frac{m^2 R}{8\pi} \log m^2 \nonumber\\
& - |m| \int_{-\infty}^\infty \frac{\rd\theta}{2\pi} \cosh\theta\log\Big(1 \pm e^{-|m|R\cosh\theta}\Big), 
\end{align}
with the upper (lower) sign applying for $\nu= \text{NS}\,\text{(RM)}$. 

With these expressions for the eigenstates and eigenvalues of $H_0(m)$ established, we turn our attention to the full problem~\eqref{FT}. For nonzero longitudinal field strength $g$, the spin operator $\s(x)$ is present in the Hamiltonian and has to obey periodic boundary conditions $\s(x) = \s(x+R)$. This imposes a restriction on the fermion states, which varies depending on the sign of $m$ ($m>0$ corresponds to the ordered phase)
\begin{align}
& m > 0: 
\left\{ \begin{array}{l}
|\{k\}_N\ra_\text{NS},~\text{with}~N\in2\mathbb{Z}, \\
|\{q\}_N\ra_\text{RM},~~~\text{with}~N\in2\mathbb{Z}, 
        \end{array} \right. \nonumber\\
  &   \label{compbasis} \\
& m < 0:
\left\{ \begin{array}{l} 
|\{k\}_N\ra_\text{NS},~\text{with}~N\in2\mathbb{Z},\\
|\{q\}_N\ra_\text{RM},~~~\text{with}~N\in2\mathbb{Z}+1. 
\end{array}
  \right. \nonumber
\end{align}
The spin operator $\s(x)$ has non-zero matrix elements between states in different sectors, thus coupling them. Detailed expressions for these matrix elements can be found in the literature~\cite{berg1979construction,fonseca2001ising,bugrij2000correlation,bugrij2001form,james2018nonperturbative}. 

With both a computational basis, Eqs.~\eqref{compbasis}, and matrix elements at hand, one proceeds by introducing an energy cutoff $E_\Lambda$ for the basis states, forming the (dense) Hamiltonian matrix, and then diagonalizing it. We illustrate how the number of computational basis states, $N_\text{states}$, varies as a function of the energy cutoff $E_\Lambda$ in Table~\ref{table:states}. Subsequently, the eigenstates obtained from the truncated spectrum procedure can be used to compute observables, as we discuss in the following sections. 

\begin{table}
\caption{The number of computational basis states $N_\text{states}$ in the zero momentum sector, as a function of energy cutoff $E_\Lambda$ for a system of size $R=35$.}
\vskip 10pt
\begin{tabular}{cc}
\hline
\hline
\hspace{0.75cm}$E_\Lambda$\hspace{0.75cm} & \hspace{0.75cm}$N_\text{states}$\hspace{0.75cm} \\
\hline
8 & 2305 \\
9 & 6269 \\
9.5 & 9809 \\
10 & 15309 \\
10.5 & 23498 \\
\hline
\hline
\end{tabular}
\label{table:states}
\end{table}

\subsection{Diagonal matrix elements}

\begin{figure}
\begin{tabular}{l}
(a) $R=25$ \\
\includegraphics[width=0.45\textwidth]{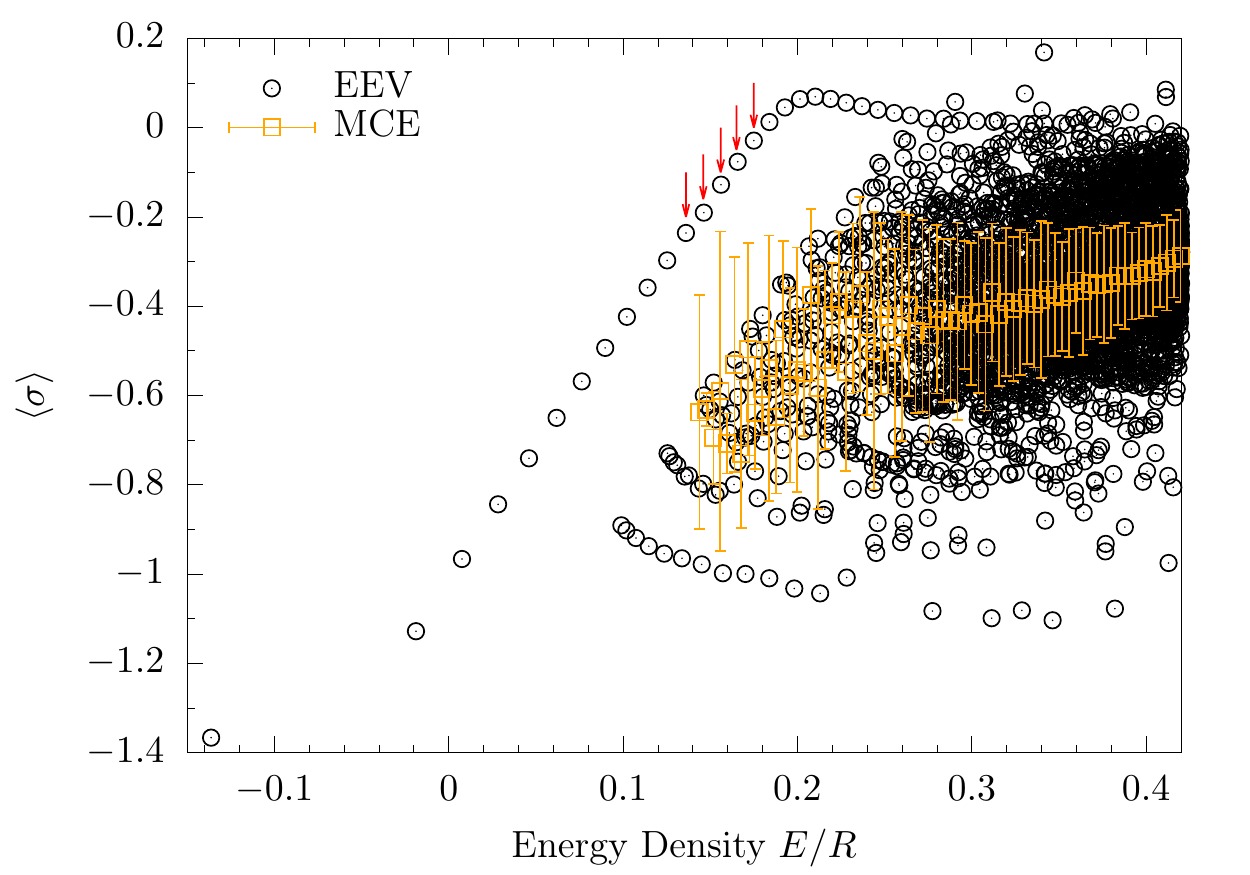} \\ 
(b) $R = 35$ \\
\includegraphics[width=0.45\textwidth]{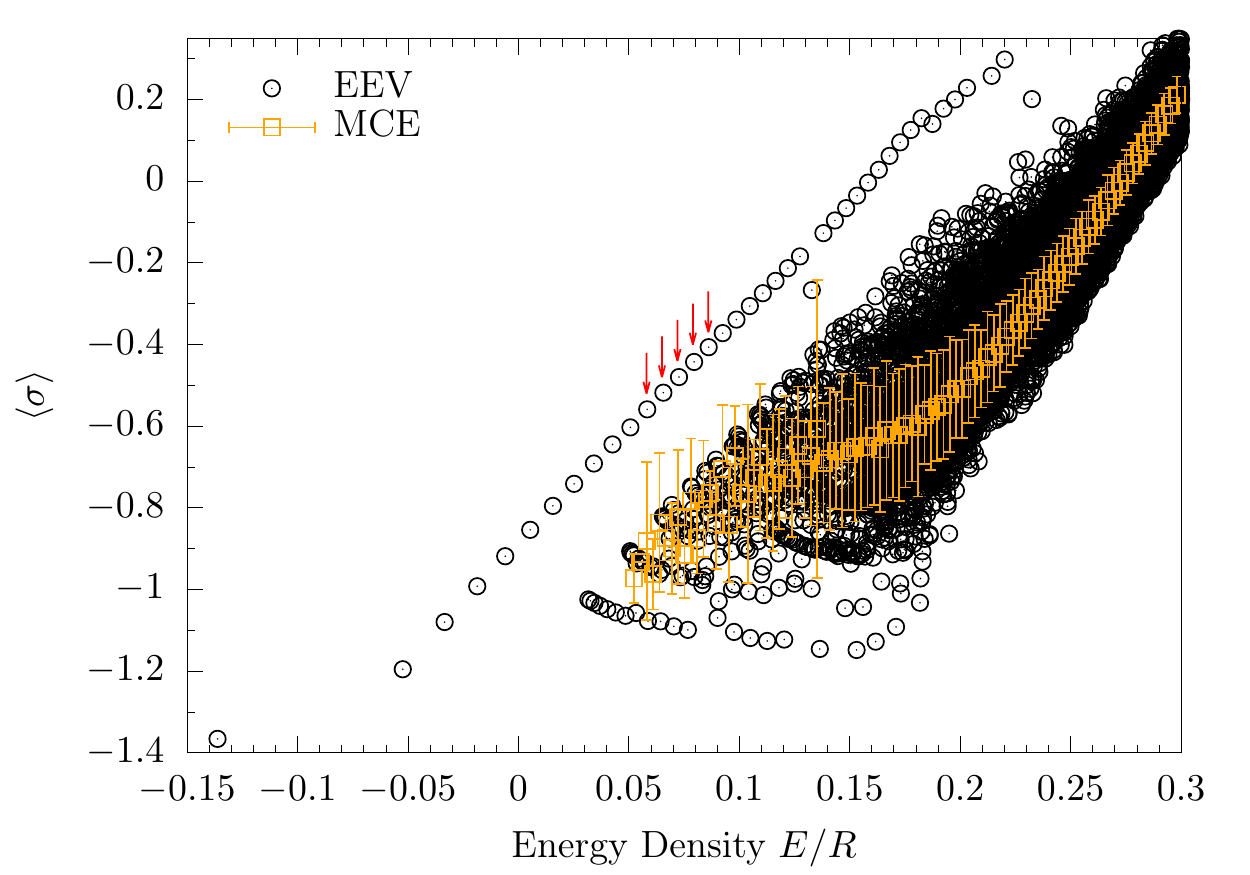} \\ 
(c) $R = 45$ \\
\includegraphics[width=0.45\textwidth]{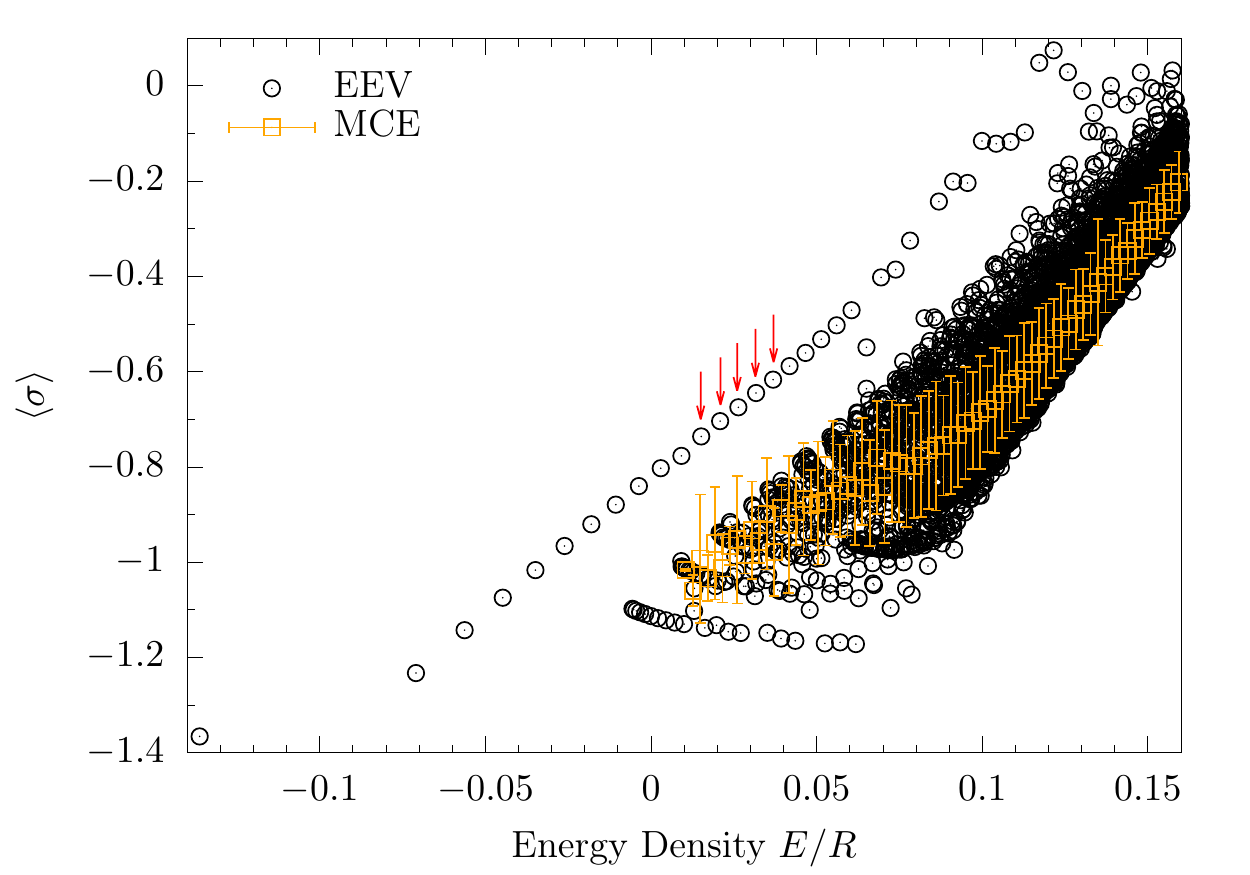}
\end{tabular}
\caption{Comparison of the eigenstate expectation values (EEV, black) of the local magnetization $\s$ with the microcanonical ensemble prediction (MCE, orange) for eigenstates at energy density $E/R$ in the Ising field theory~\eqref{FT} with $m=1$, $g=0.1$ on the ring of size (a) $R=25$; (b) $R=35$; (c) $R=45$. Eigenstates are constructed via the TSM with an energy cutoff of (a) $E_\Lambda = 13.5$; (b) $E_\Lambda = 10.5$; (c) $E_\Lambda = 9$. States containing up to ten particles are considered in the truncated Hilbert space. The MCE is constructed by uniformly averaging over an energy window of size $\Delta E = 0.1$, with error bars denoting the standard deviation. The red arrows highlight the nonthermal states considered further in Fig.~\ref{Fig:fss}.}
\label{Fig:EEVs}
\end{figure}

Let us begin by examining the behavior of diagonal matrix elements. As seen in Eq.~\eqref{eth}, under ETH these matrix elements should be smooth as a function of the energy of the eigenstate~\cite{deutsch1991quantum,srednicki1994chaos,rigol2008thermalization,rigol2009breakdown,reimann2015eigenstate,deutsch2018eigenstate}
\begin{align}
\hat A_{\alpha,\alpha} = A(E_\alpha), \quad \Big| A_{\alpha+1,\alpha+1} - A_{\alpha,\alpha} \Big| \propto e^{-R/R_0},
\label{ethdiag}
\end{align} 
where $R$ is the system size and $R_0 > 0$ is a dimensionful constant. To examine whether Eqs.~\eqref{ethdiag} hold in the perturbed Ising field theory, we compute the expectation value of the local spin operator $\s(0)$ within eigenstates [so-called eigenstate expectation values (EEVs)] in a finite volume, $R$. Results as a function of energy density $E/R$ with $m=1$, $g=0.1$ for $R=25,35,45$ are presented in Figs.~\ref{Fig:EEVs}. It is immediately apparent that in the finite volume EEVs are \textit{not} a smooth function of the energy: in particular there is a clear band of states lying above the majority with significantly different EEVs. 

The unusual nature of this band of states is further highlighted by a comparison to the microcanonical ensemble (MCE) thermal result
\begin{align}
  A_\text{MCE}(E) = \frac{1}{N_{\Delta E}} \sum_{E-\frac{\Delta E}{2} < E' < E + \frac{\Delta E}{2}} \la E' | \hat A | E' \ra,
\label{Eq:MCE}
\end{align}
which is constructed by averaging EEVs over an energy window of size $\Delta E = 0.1$ ($N_{\Delta E}$ is the number of states within the energy window). Error bars on the MCE reflect the standard deviation of the data averaged over.\footnote{We say that the EEVs and MCE agree when the EEVs fall within one standard deviation of the MCE. This spread of the EEVs reflects the second term in Eq.~\eqref{eth}, which features the random variable $R_{\alpha,\beta}$, which has mean zero and unit variance. We see that the standard deviation decreases with increasing system size $R$, reflecting the extensivity of the thermodynamic entropy $S(E)$.}  It is clear that numerous states fall outside the MCE prediction (plus one standard deviation), both close to the threshold of the multi-particle continuum ($E/R\sim0$), as well as at energies far within the continuum. The band of states above the continuum, whose EEVs differ significantly from the MCE result, are rare states by definition: their EEVs are nonthermal.

We note that when reading figures such as Fig.~\ref{Fig:EEVs}, one should always regard EEVs at larger energy densities as having truncation effects. This can be seen, for example, in the upturn of the MCE (increase in the gradient of the slope as a function of $E/R$) at higher energy densities in Figs.~\ref{Fig:EEVs}(b) and (c) (this also occurs in Fig.~\ref{Fig:EEVs}(a) at higher energy densities than plotted), and the decrease in spread of the continuum at larger energy densities. Convergence checks as a function of energy cutoff are presented in Appendix~\ref{App:ConvTCSA}; it should be noted that the separation between the rare states and multiparticle continuum is \textit{well converged} and not a truncation effect. We also note that the MCE converges rapidly in the energy cutoff. In the next section, we examine the issue of finite size scaling: does this difference between rare and thermal states persist to the infinite volume? 

Before moving on to the finite size scaling, it is worth briefly commenting on the energy cutoff $E_\Lambda$ as a function of system size $R$. As we see in the definition of the single particle dispersion relation, Eq.~\eqref{dispersion}, at fixed energy cutoff $E_\Lambda$, as $R$ is varied the maximum value of the integer $p_\nu$ is increased. Thus to keep a fixed energy cutoff, one needs to consider polynomially more states as $R$ is increased. This quickly becomes computationally problematic, so generally $E_\Lambda$ has to be decreased as $R$ is increased, as has been done in Fig.~\ref{Fig:EEVs}. 

\subsubsection{Finite size scaling analysis}
\label{Sec:EEVRdep}

It is interesting to consider the behavior of the system at different system sizes. One of the major advantages of TSMs is that one can construct thousands of low-energy eigenstates to high precision in rather large systems (up to $R\sim 50|m|^{-1}$, where $|m|^{-1}$ is the correlation length of the unperturbed model). Examples for a number of system sizes are presented in Fig.~\ref{Fig:EEVs}. At each value of the system size, focusing on the the low-energy part of the spectrum (where convergence is best) we see that the same essential features appear: there is a continuum of states where (roughly) the MCE and EEV results agree, and above this there is a separate band of rare states, whose EEVs are distinctly nonthermal. 

We see that the majority of EEVs are consistent with Eq.~\eqref{eth}: with increasing volume $R$ the thermal continuum narrows (reflecting the extensivity of the thermodynamic entropy $S(E)$) and the MCE encompasses the majority of results.
Similar sharpening of the distribution of EEVs, and improved agreement with the MCE, is observed in exact diagonalization of lattice models (see, e.g., Refs.~\cite{santos2010localization,ikeda2013finitesize,kim2014testing,steinigeweg2014pushing,beugeling2014finitesize,khodja2015relevance,mondaini2016eigenstate,mondaini2017eigenstate,lan2017eigenstate}). We note that here the agreement is particularly clear compared to lattice calculations, as TSMs allow us to access many states in the low-energy spectrum at relatively large system size. 

\begin{figure}
\begin{tabular}{lll}
(a) $n=11$ & & (b) $n=12$ \\
\includegraphics[width=0.23\textwidth]{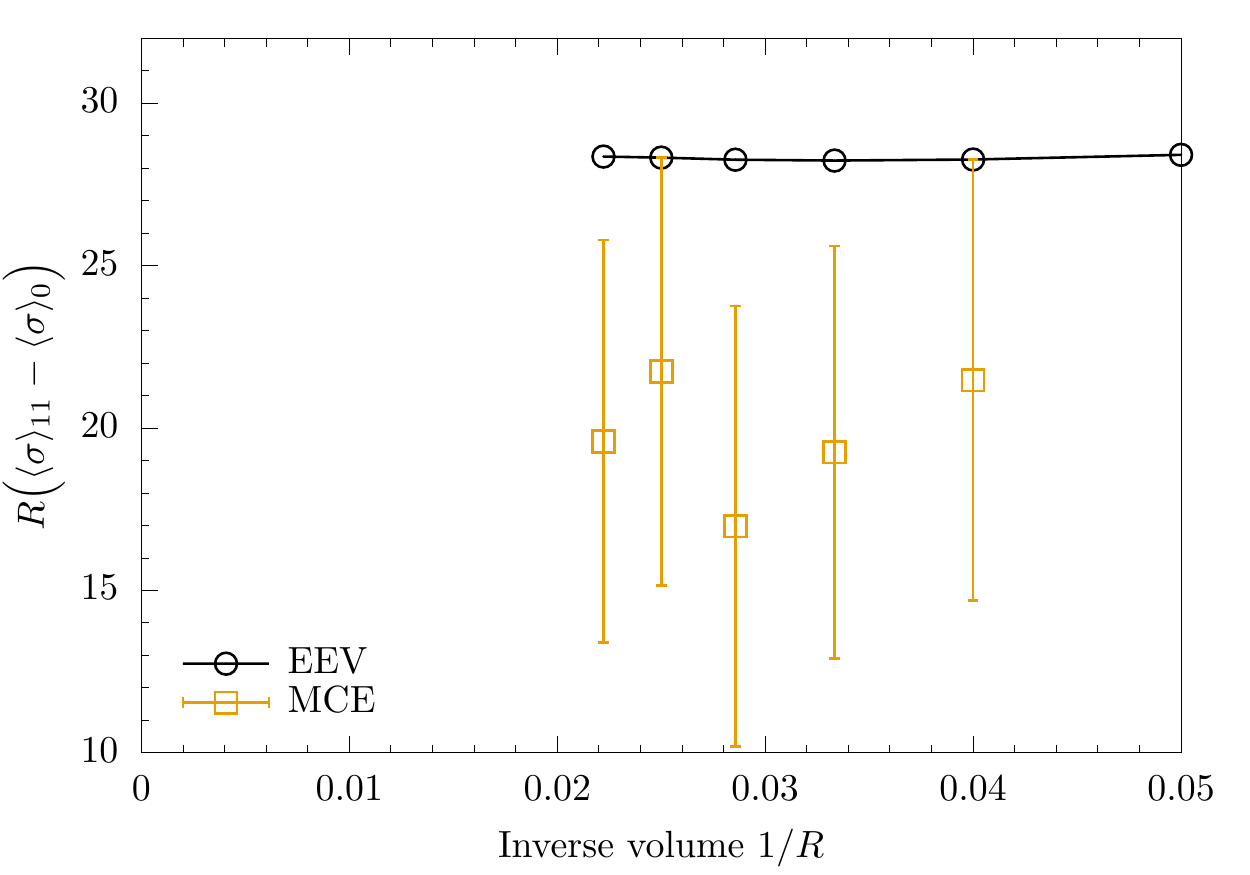} & & \includegraphics[width=0.23\textwidth]{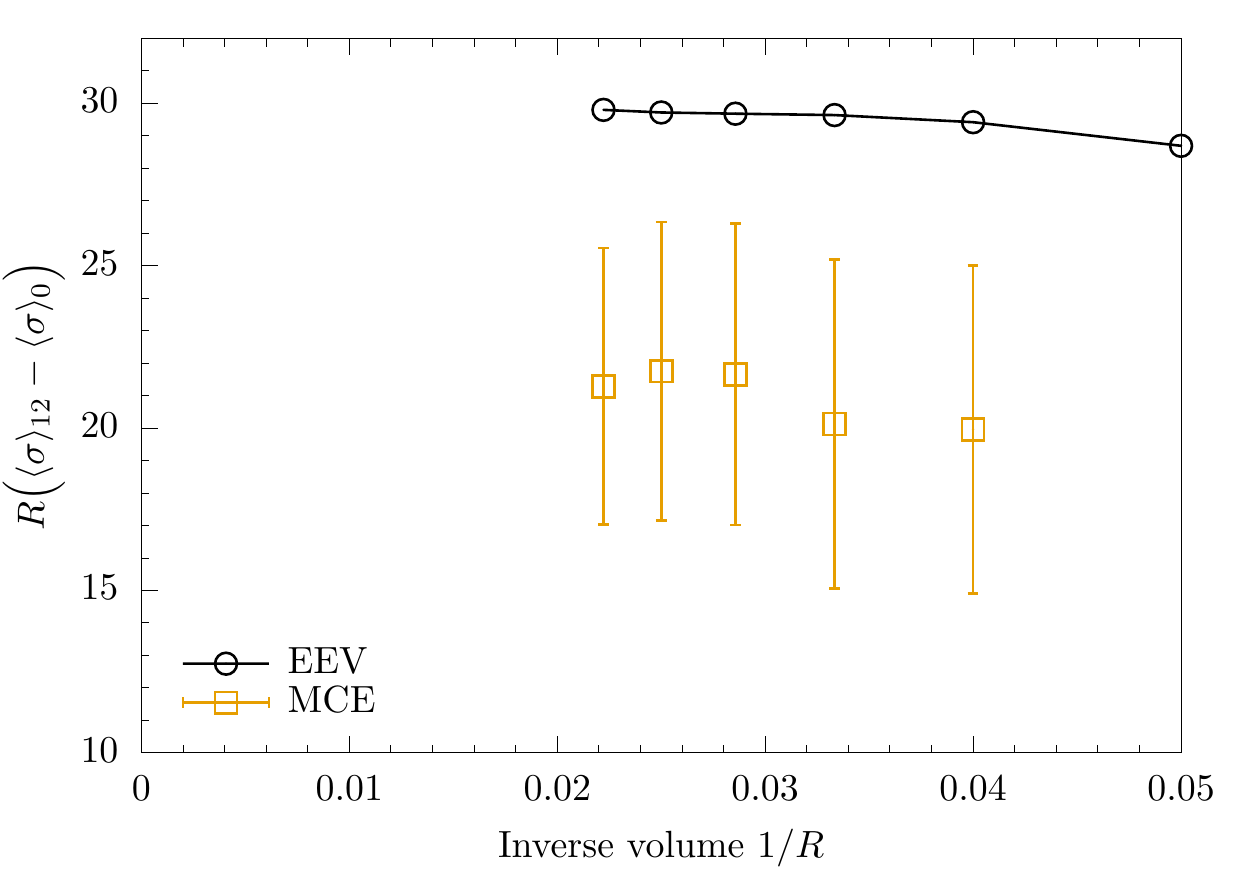}\\
(c) $n=13$ & & (d) $n=14$ \\
\includegraphics[width=0.23\textwidth]{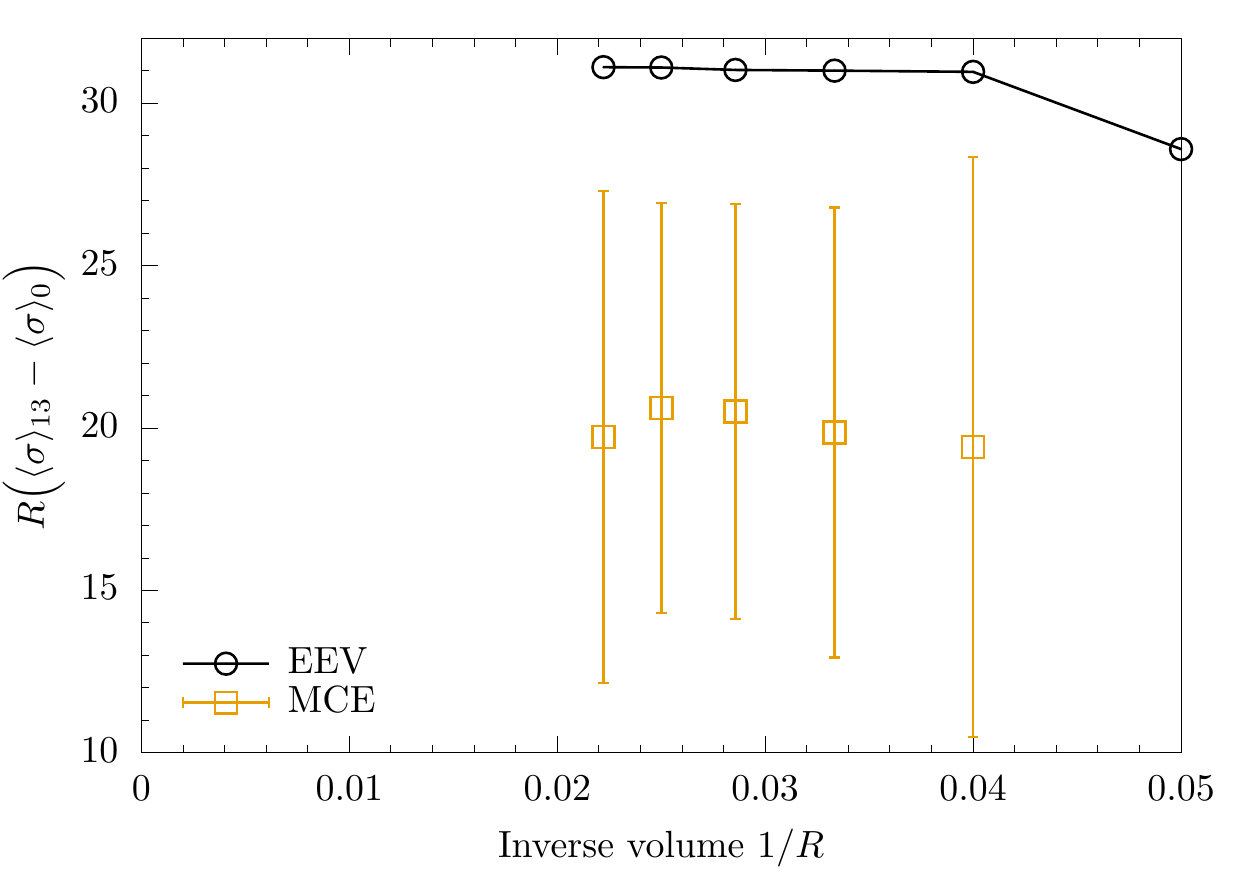} & & \includegraphics[width=0.23\textwidth]{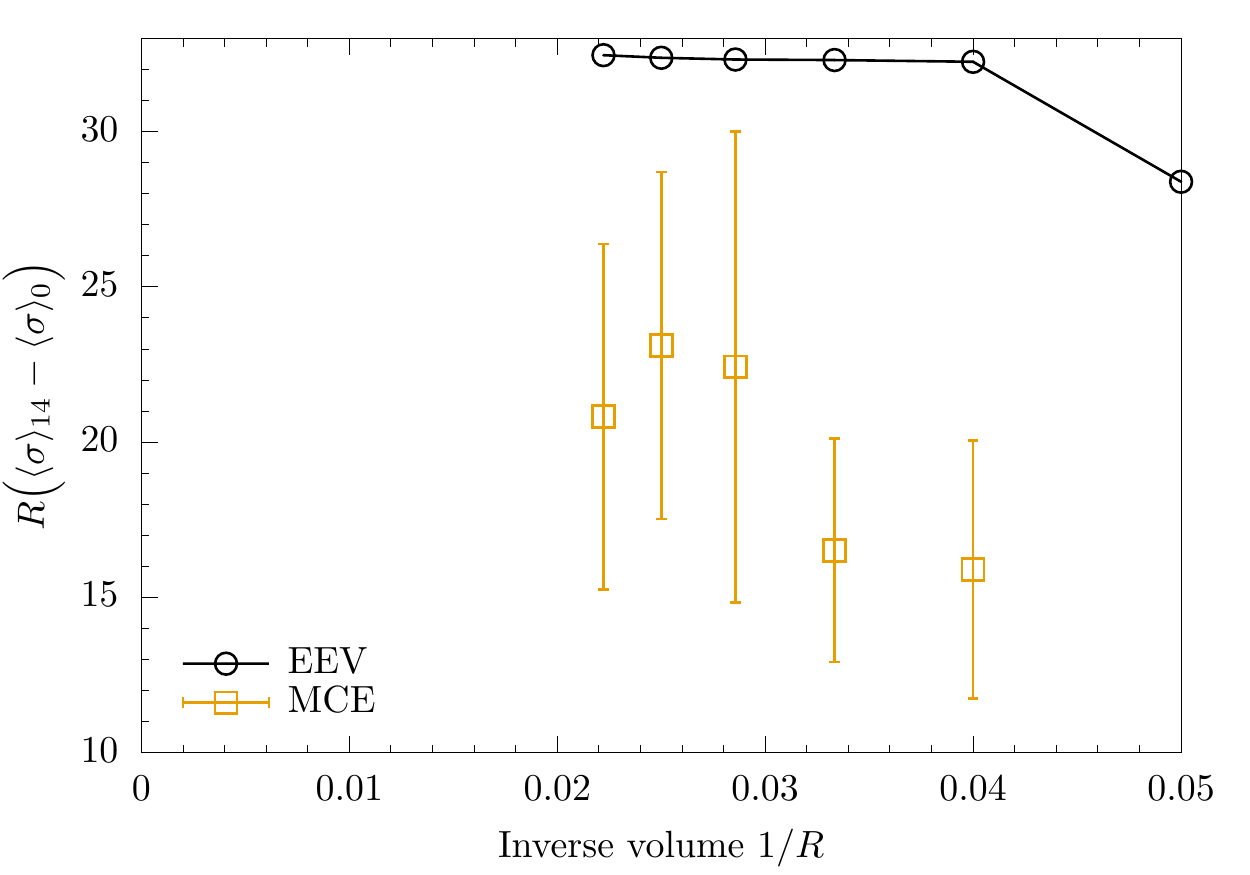}\\
(e) $n=15$ & & (f) averaged \\
\includegraphics[width=0.23\textwidth]{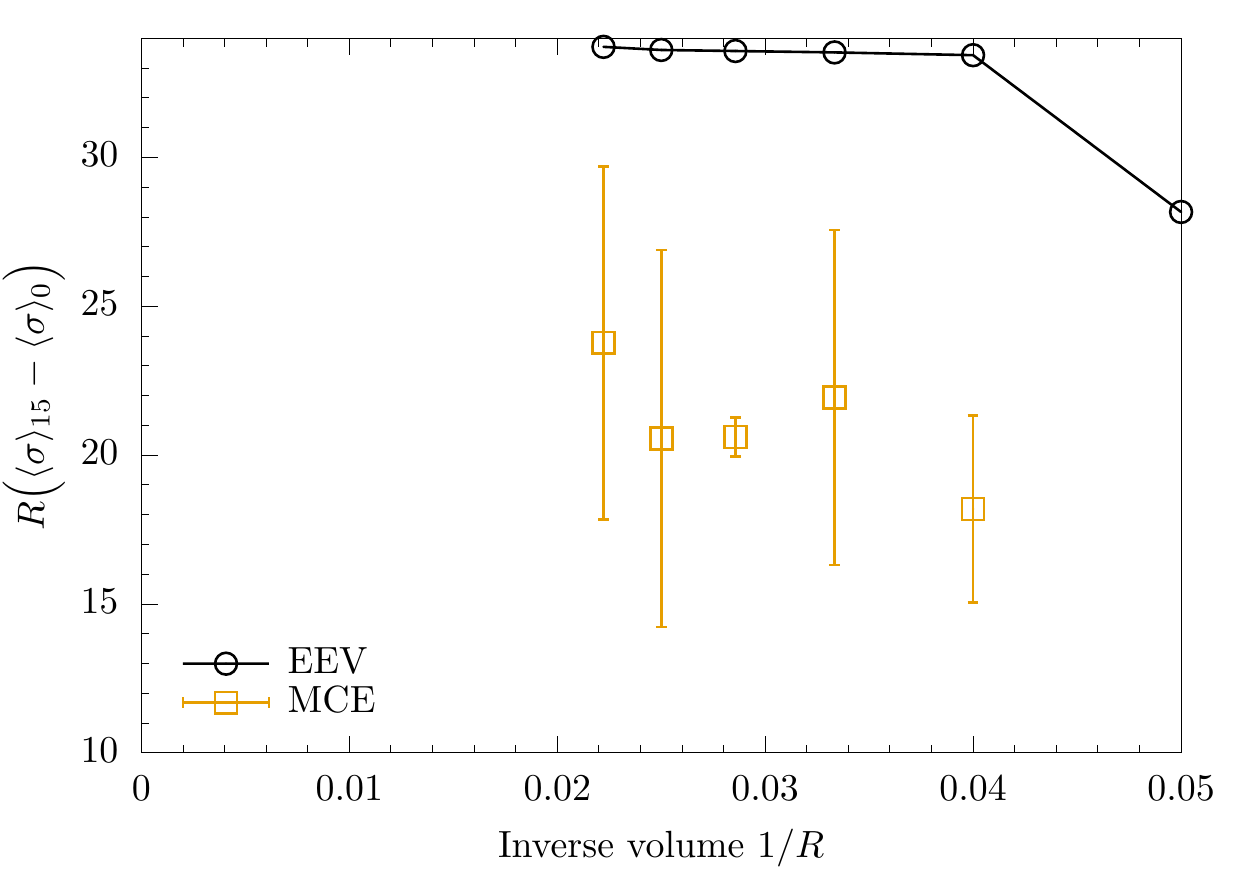} & & \includegraphics[width=0.23\textwidth]{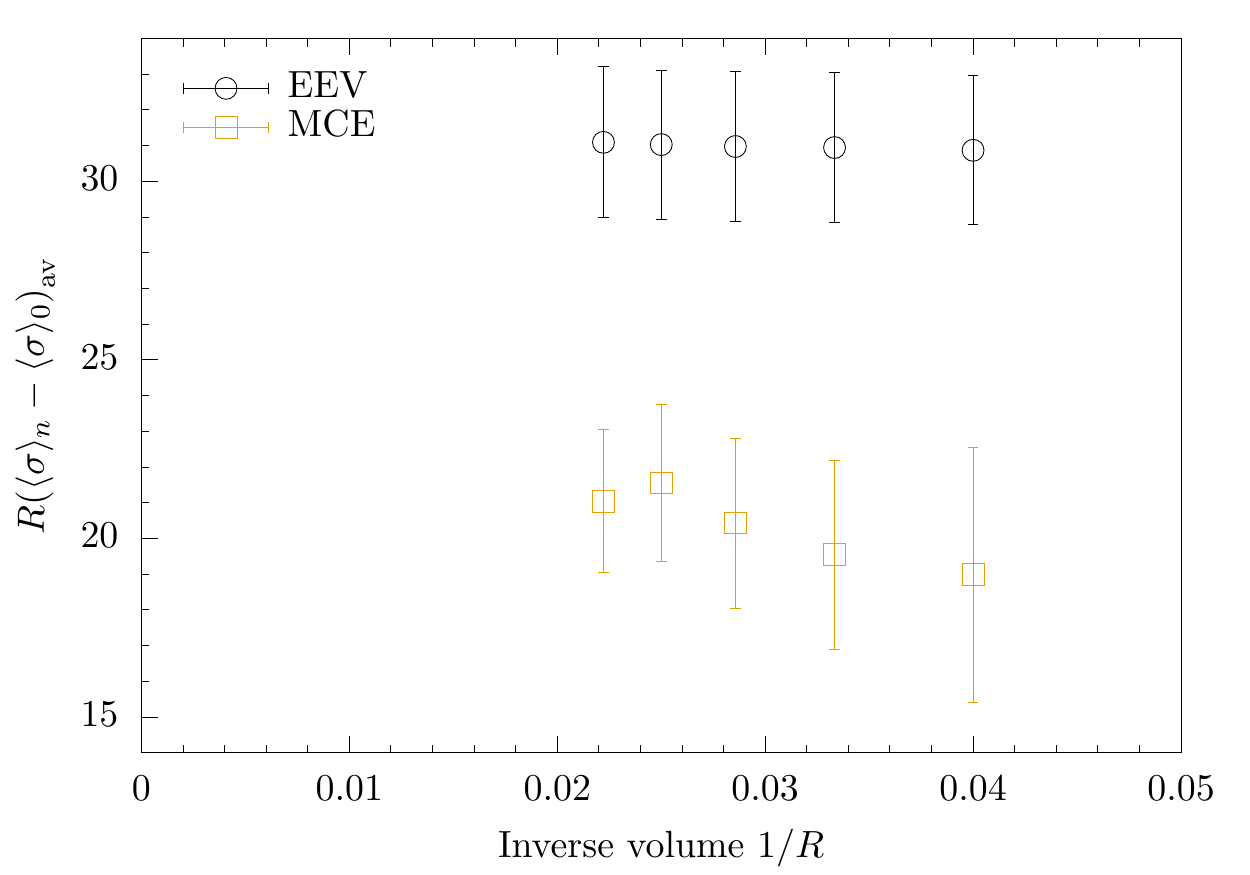} 
\end{tabular}
\caption{(a)--(e) The finite size scaling the total magnetization $R\la\s\ra$ of the $n=11-15$ rare states (EEV), compared to the MCE prediction. (f) The average magnetization of the $n=11-15$ rare states compared to that predicted by the average of the MCE results. Clearly the rare states and their proximate thermal states have different magnetization, and this persists to the $R\to\infty$ limit. In all cases, error bars show the standard deviation of the data averaged over. Note that $R(\la\s\ra_n-\la\s\ra_0)$ measures the number of overturned spins in the $n$th meson as compared to the (ordered) ground state.}
\label{Fig:fss}
\end{figure}

The band of rare states persists to the largest volumes we can access, and so it is natural to ask what becomes of these in the infinite volume limit, $R\to\infty$? As we cannot work directly in this limit (the spectrum of the theory becomes continuous and there are an infinite number of states below any non-zero energy cut-off), we infer results from a finite size scaling analysis. 

Ordering the rare states by energy, we focus on the five states numbered $n=11\to15$, which have energies well within the multiparticle continuum. These states are highlighted by the red arrows in Fig.~\ref{Fig:EEVs}. For all system sizes that we access, these states are well converged for the maximum value of the cutoff $E_\Lambda$ achieved with reasonable computational resources. We compare the finite size scaling of these EEVs to the MCE constructed at the same average energy density, as shown in Fig.~\ref{Fig:fss}. In each case, the MCE prediction for the magnetization of the state is different to the EEV of the rare state, and the finite size scaling analysis suggests that this persists to the infinite volume limit, $1/R \to 0$. This is particularly clear if one computes the average magnetization of the five rare states and compares to the average from the MCE, as shown in Fig.~\ref{Fig:fss}(e).

The convergence of the whole EEV spectrum as a function of system size $R$ is presented in Sec.~\ref{sec:EEVRdep}, where we compare to similar results on a corresponding lattice model. For the low-energy parts of the spectrum (including the rare states), the results are converged to those of the thermodynamic limit, $R\to\infty$.

\subsection{Off-diagonal matrix elements}
As we have already seen, ETH~\eqref{eth} also supposes that off-diagonal matrix elements of an operator have a certain structure,
\begin{align}
\hat A_{\alpha,\beta} = e^{-S(E)/2}f_A(E,\omega) R_{\alpha,\beta}, \qquad \alpha\neq\beta. 
\label{ethoffdiag}
\end{align}
Here $S(E)$ is the thermodynamic entropy, $f_A(E,\omega)$ is a smooth function of both the average $E$  and difference $\omega$ of the energies of the eigenstates $E_\alpha,E_\beta$. $R_{\alpha,\beta}$ is a random variable with zero average and unit variance. The off-diagonal matrix elements are exponentially suppressed by $S(E)$, an extensive quantity. We have already seen signatures of this in the previous section: the variance of the diagonal matrix elements decreases with increasing system size. From Eq.~\eqref{ethoffdiag}, we would expect average off-diagonal matrix elements to be much smaller than the diagonal ones (as these are not suppressed by $S(E)$)~\cite{deutsch1991quantum,srednicki1994chaos,rigol2008thermalization,rigol2009breakdown,reimann2015eigenstate,deutsch2018eigenstate}.

\begin{figure}
\includegraphics[width=0.45\textwidth]{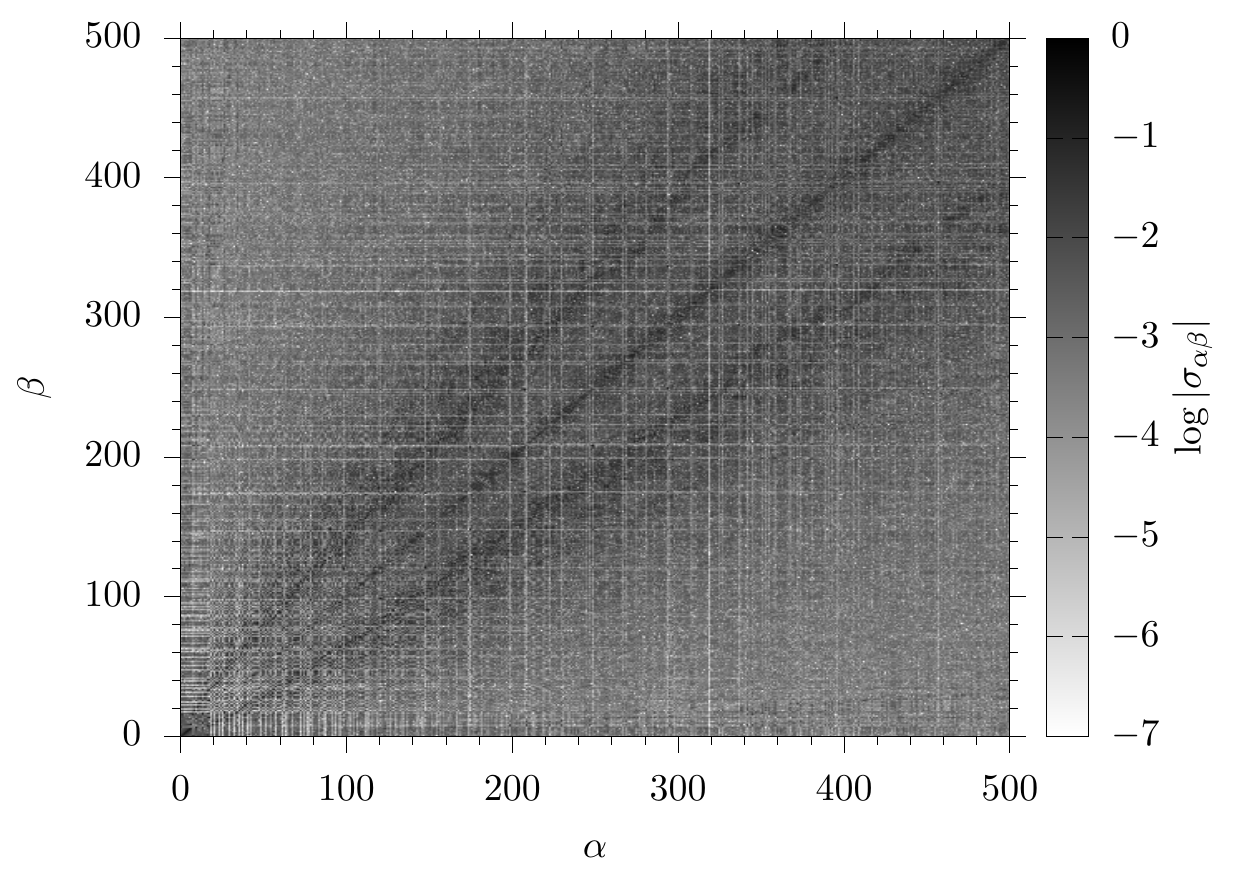}
\caption{Off-diagonal matrix elements of the local spin operator $\vert \la E_\beta | \s(0) | E_\alpha\ra \vert$ ($\alpha\neq\beta$) between the first 500 eigenstates constructed with the TSM for $m=1$, $g=0.1$ and $R=35$, cf. Fig.~\ref{Fig:EEVs}(b). Parameters are $m=1$, $g=0.1$. Matrix elements satisfying $|\s_{\alpha\beta}| < 10^{-7}$ are plotted as white. Note the visible vertical/horizontal lines are not remnants of the plotting, but show eigenstates that are only coupled very weakly to other states by the spin operator $\s(0)$. }
\label{Fig:Offdiag}
\end{figure}

To examine this, we compute the off-diagonal matrix elements of $\s(0)$ in the basis of eigenstates: 
\begin{align}
\s_{\alpha\beta} = \la E_\beta | \s(0) | E_\alpha \ra, \qquad \alpha\neq\beta. 
\end{align}
For $\alpha,\beta=1,\ldots,500$ we present these in Fig.~\ref{Fig:Offdiag}, where we consider the same set of parameters as in Fig.~\ref{Fig:EEVs}. From inspection of the numerical data, as well as Fig.~\ref{Fig:Offdiag}, we see that off-diagonal matrix elements are generally much smaller [$O(10^{-2})$ and smaller] than the diagonal elements (although we note that there are some off-diagonal elements that are comparable to the diagonal ones for $R=35$). It is also apparent that there is significant structure present within the off-diagonal elements (cf. Eq.~\eqref{ethoffdiag}), with clear lines of zeros in the $(\alpha,\beta)$ plane, as well as regions with larger matrix elements (on average). Many of the vertical/horizontal lines with small matrix elements coincide with the index of the rare states. Prominent examples include the $\alpha=175$, $\alpha=319$ and $\alpha=457$ lines (by symmetry, the same lines along the $\beta$ axis), as well as the many lines of suppressed matrix elements at small $\alpha$ and $\beta$.

\subsubsection{System size dependence}

\begin{figure}
\includegraphics[width=0.45\textwidth]{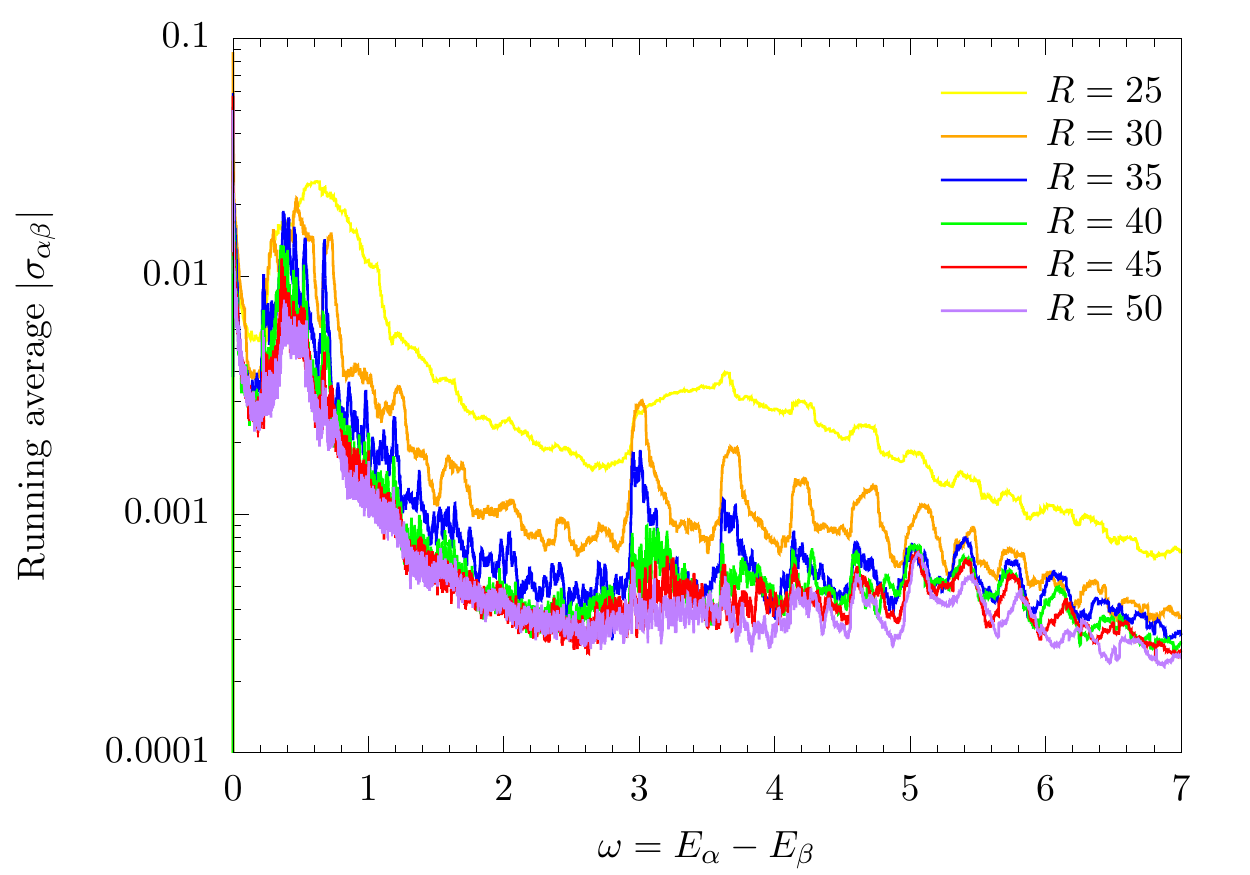}
\caption{The running average of the absolute value of off-diagonal matrix elements $|\s_{\alpha\beta}|$ between eigenstates with energies satisfying $(E_\alpha+E_\beta) < 8$. For the largest system, $R=50$, this condition restricts to the lowest (in energy) $\sim 1000$ eigenstates (and fewer at smaller volumes). The running average is computed over $300$ data points.}
\label{Fig:OffDiagFSS}
\end{figure}

To examine the system size dependence of the off-diagonal elements, we follow~~\textcite{mondaini2017eigenstate} and study the behavior of the absolute value of the off diagonal matrix elements. In particular, within the lowest 1000 eigenstates we consider matrix elements where the energy of the eigenstates $|E_\alpha\ra$, $|E_\beta\ra$ satisfies $(E_\alpha + E_\beta) <  8$ (we note that this corresponds to $|\omega| = |E_\alpha-E_\beta| \lesssim 11$ for all system sizes considered). We then consider the behavior of these elements as a function of the energy differences, $\omega = E_\alpha-E_\beta$, and how this varies with system size. To aid in extracting this behavior, we take a running average and this is shown in Fig.~\ref{Fig:OffDiagFSS}. 

Figure~\ref{Fig:OffDiagFSS} highlights a number of features of the off-diagonal matrix elements between low-energy eigenstates. Firstly, as mentioned in the previous section for $R=35$, we see that the average off-diagonal matrix element is small $<10^{-2}$. Secondly, this average smallness decreases with increasing system size $R$, consistent with general expectations from ETH. However, due to limitations of our methodology (being restricted to low energy eigenstates), we cannot ascertain whether generic off-diagonal elements are suppressed exponentially in the volume (via the extensivity of the thermodynamic entropy for generic, finite energy density, eigenstates).

\section{The nature of the rare states}
\label{Sec:Nature}

In the previous section, we have seen that rare states with nonthermal EEVs are present within the model~\eqref{FT}. A natural question is then: do these states share common characteristics? Looking at Fig.~\ref{Fig:EEVs} there is an obvious first guess as to their physical characteristics: the rare states extend in a band from the lowest-energy excitations in the system, which are the well-known `meson' confined states, with wave functions of the form  
\begin{align}
|\psi_n\ra = \sum_{\nu=\text{NS},\text{RM}} \sum_{p_\nu} \Psi_{n,\nu}(p_\nu) a\dg_{p_\nu}a\dg_{-p_\nu} |\nu\ra.
\label{mesonState}
\end{align}
These states consist of linearly-confined pairs of domain walls (described by the fermions $a\dg_{p_\nu}$ in the $\nu=\mathrm{RM},\mathrm{NS}$ sectors of the Hilbert space, with vacuum $|\nu\ra$ in each sector, see e.g. Ref.~\cite{james2018nonperturbative}). On the basis of Figs.~\ref{Fig:EEVs} it is easy to suggest that the rare states are simply higher energy meson states. 

\begin{figure}
\includegraphics[width=0.45\textwidth]{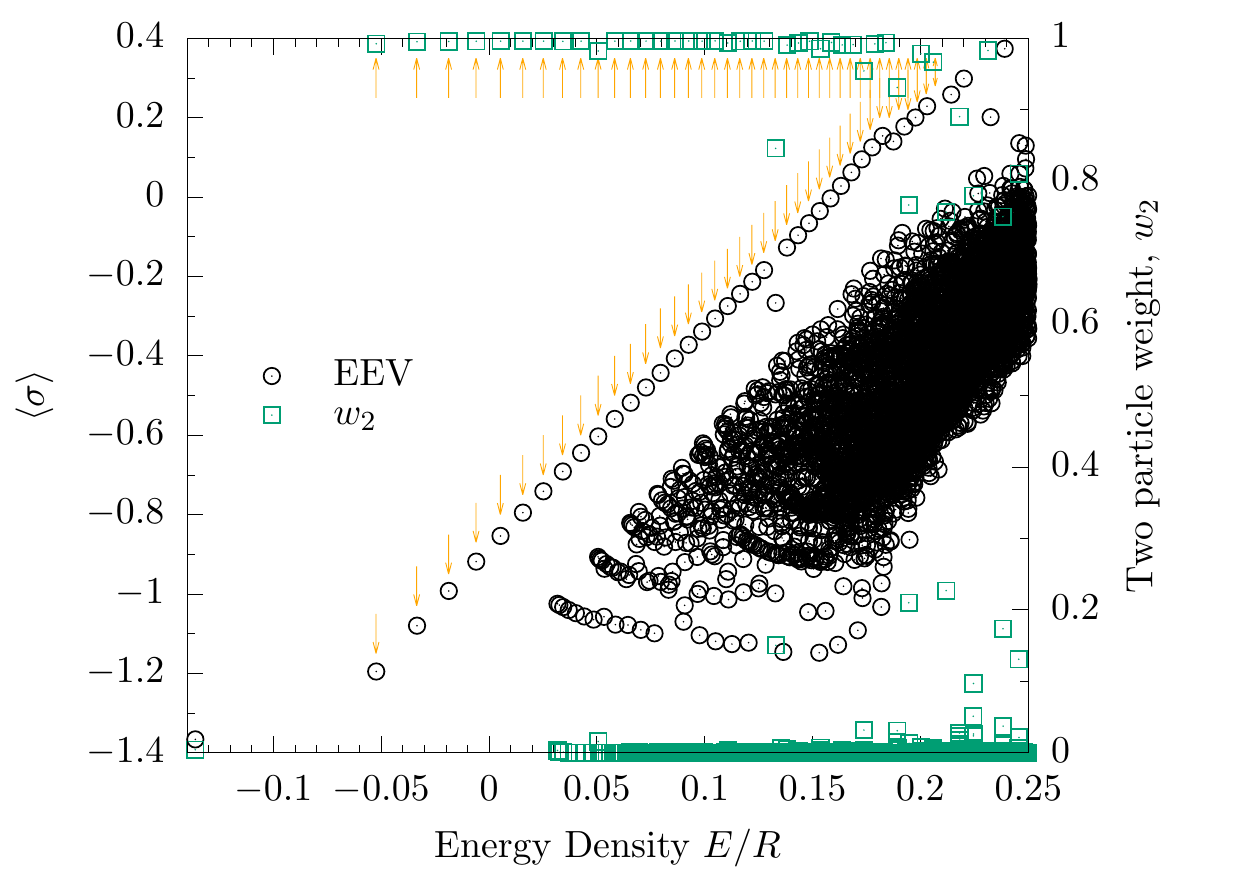}
\caption{The two particle weight $w_2$~\eqref{weight} and the eigenstate expectation values of the local magnetization operator $\la\s\ra$ in the field theory~\eqref{FT} with $m=1$, $g=0.1$ on the ring of size $R=35$. Eigenstates are constructed with the TSM for an energy cutoff of $E_\Lambda = 10.5$ and a maximum of ten particles in the basis states. Arrows (upper and lower) are drawn at the energies of the first forty confined states, $M_n$, as computed in Appendix~\ref{App:Semiclassical}. Higher particle weights, as well as histograms of their distributions, are provided in Appendix~\ref{App:Weights}.}
\label{Fig:TwoParticleWeight}
\end{figure}

To ascertain whether this is correct, we can use the information directly accessible to us from the TSM procedure. In the first case, we can check whether the rare states are (majority) two particle in nature by defining the particle weights. To do so, we recall that we construct states in terms of free fermion basis states 
\begin{align}
| E_m \ra = \sum_{N = 0,2,4,\ldots} \sum_j c^m_{N,j} | \{ p_j \}_N \ra, \label{eigenstates}
\end{align}
where $c^m_{N,j}$ are superposition coefficients and $ | \{ p_j \}_N \ra$ are $N$-fermion Fock states with fermions carrying momenta $\{p_j\}_N \equiv \{p_{j1},\ldots,p_{jN}\}$. Particle weights $w_N$ telling us the $N$-fermion fraction of state are: 
\begin{align}
w_N = \sum_{j} |c_{N,j}|^2. \label{weight}
\end{align}
We present a plot of the EEV spectrum for $R=35$, Fig.~\ref{Fig:EEVs}(b), with superimposed two particle weights in Fig.~\ref{Fig:TwoParticleWeight}. We see that those EEVs corresponding to rare states are also those which have $w_2 \approx 1$ and so we conclude the rare states are majority two-particle in nature, consistent with Eq.~\eqref{mesonState}.

To further strengthen our evidence that rare states are meson-like, we (semi)analytically compute the energy of the meson states. This can proceed in a number of manners; we consider in particular the energies computed via a semiclassical approximation (the salient points of which are summarized in Appendix~\ref{App:Semiclassical})~\cite{fonseca2003ising,rutkevich2005largen,fonseca2006ising}. One achieves consistent results by means of a Bethe-Salpeter analysis, which also allows one to obtain the wave function in Eq.~\eqref{mesonState}, see also Appendix~\ref{App:Wfn}. Computed results for the energies of meson states are also shown in Fig.~\ref{Fig:TwoParticleWeight} via the arrows, which are drawn at the (semiclassical) energies of the first forty meson states. They clearly coincide with the nonthermal EEVs and two particle dominated states. We thus conclude that the rare states are described by Eq.~\eqref{mesonState}, meson-like confined pairs of domain walls~\cite{fonseca2003ising,rutkevich2005largen,fonseca2006ising}.

As previously noted, the rare states exist far beyond the multiparticle continuum (occurring in Fig.~\ref{Fig:TwoParticleWeight} at $E/R \sim 0.03$), which is surprising when drawing analogies with mesons in quantum chromodynamics (QCD)~\cite{sulejmanpasic2017confinement}: in QCD we usually think of high energy mesons as splitting into multiple lower-energy mesons. Indeed, in the case considered here such processes \textit{are kinematically allowed}, so it is worth spending some time to understand why the meson states~\eqref{mesonState} exist above the continuum and have EEVs that remain well separated from the thermal result (cf. Fig.~\ref{Fig:fss}).
To reemphasize: with the TSM we have constructed well-converged \textit{eigenstates of the Hamiltonian~\eqref{FT}} that are still meson-like above the threshold of the multiparticle continuum.\footnote{We note that in the context of magnetic systems, one might conclude that the mesons are not present in the spectrum as one does not see signatures in dynamical spin-spin correlation functions: but from Fig.~\ref{Fig:EEVs} one could conclude that they are indeed there, though their signal is washed out by the thermal signal from the large number of states in the multiparticle continuum.}

\subsection{Stability of meson excitations above the multimeson threshold}

To gain some insight into why the rare states persist above the multimeson threshold, we take the mesons defined in Eq.~\eqref{mesonState},
\begin{align}
b\dg_{n} = \sum_{\nu = \text{NS,RM}} \sum_{p_\nu} \Psi_{n,\nu}(p_\nu) a\dg_{p_\nu} a\dg_{-p_\nu}, \label{meson}
\end{align}
which are \textit{approximate quasi-particles} of the problem---they do not have infinite lifetime because exact eigenstates of~\eqref{FT} contain a finite amount of $N\ge4$ particle dressing (see Appendix~\ref{App:Weights})---and discuss their hybridization with states with higher fermion number. 

We first recap known results for the zero temperature lifetime of meson excitations, before discussing the correction to the meson energies due to hybridization with higher fermion number states. We will see that the simple meson, Eq.~\eqref{meson}, has very long lifetime at zero temperature and only very weakly hybridizes with states containing four (or more) fermions. This is consistent with the idea that with a small amount of dressing the meson excitations~\eqref{meson} become absolutely stable. 

\subsubsection{Meson lifetime}
\label{sec:lifetime}

Let us consider the results of Rutkevich~\cite{rutkevich2005largen} for the zero temperature meson lifetime. Firstly, the lowest energy meson excitations (below the two-meson threshold, $E \lesssim 4m$) are absolutely stable, with no decay channels. Above the two meson threshold energy, the $n$th meson state has decay width $\Gamma_n$, which can be estimated to leading order using Fermi's golden rule 
\begin{align}
\Gamma_n = 2\pi \sum_{E_{\rm out}} \left| \la E_{\rm out} | V | \psi_n \ra \right|^2 \delta( M_n  - E_{\rm out} ).
\end{align}
Here $M_n$ is the meson mass (energy of the zero momentum meson state), $|\psi_n\ra$ is the meson wave function (cf. Eq.~\eqref{mesonState} and Appendix~\ref{App:Wfn}), $|E_{\rm out}\ra$ are eigenstates of the full Hamiltonian with energy $E_{\rm out}$ (measured relative to the ground state energy), and $V$ is the interaction term
\begin{align}
V = g \int \rd x\, \s(x). 
\end{align}
Under a semiclassical analysis, the decay width $\Gamma_n$ \textit{in the infinite volume} has been computed by Rutkevich~\cite{rutkevich2005largen}. This leads to a lifetime, above the two-meson threshold energy, for the meson states~\eqref{mesonState} that varies as 
\begin{align}
\tau \propto g^{-3},
\end{align}
which is very large at sufficiently small $g$. We note that this formula was derived for large meson number.  More generally one faces a challenge in computing the lifetime using Fermi's golden rule.  The matrix elements of the spin operator generically have IR divergences, which reflect the fact that the matrix elements represent both connected and disconnected diagrams.  To make sense of a lifetime computation in a fully quantum setting, one must be able to sensibly separate out only the connected part \cite{Zamolodchikov2011}.   Alternatively, one can extract the life time of meson excitations from finite-volume TSMs as shown in Ref.~\cite{pozsgay2006characterization}. Results can also be compared to computations from form factor perturbation theory, see also Refs.~\cite{delfino1996nonintegrable,delfino1998nonintegrable,delfino2009particle}.  We will do neither here -- we delay a detailed study of the lifetime (as well as the meson Green's function) to a later work.

\subsubsection{Corrections to the meson energy: energy dependence}
We instead will content ourselves here by considering the correction to the energy of the mesons~\eqref{meson} due to hybridization with zero and four fermion states. From this quantity we will be able to deduce that the mesons as constructed from two-particle states (i.e. Eq.~\eqref{meson}) are only weakly coupled to sectors of the unperturbed theory with different particle number.  

To second order in perturbation theory in $g$, the energy correction for the $j$th meson is 
\begin{align}
\Delta E_{2j}=& \frac{g^2}{M_j} \sum_{\nu = \text{RM,NS}} \left\vert \sum_{p_{\bar\nu}} \Psi_j(p_{\bar\nu}) \la \nu | \s(0) | \{ -p_{\bar \nu}, p_{\bar\nu}\}\ra  \right\vert^2 \nonumber \\
              & + g^2 \sum_{\nu = \text{NS,RM}} \sum_{p_{\nu1} < p_{\nu2} < \ldots}  \frac{\delta_{0,\sum_{i} p_{\nu i}} \left\vert f^{4,\nu}_{j,\{p_\nu\}_4} \right\vert^2}{M_j -  \sum_{i=1}^4 \omega_{p_{\nu i}}(R)},
                \label{encorr}
\end{align} 
where
\be
f^{4,\nu}_{j,\{p_\nu\}_4} = \sum_{p_{\bar\nu}} \Psi_j(p_{\bar\nu}) \la \{ p_{\nu} \}_4 | \s(0) | \{ -p_{\bar \nu}, p_{\bar\nu}\}\ra. 
\ee
The matrix elements of the spin operator $\s(0)$ required above are known, see Refs.~\cite{berg1979construction,fonseca2003ising,bugrij2000correlation,bugrij2001form,james2018nonperturbative}. 

\begin{figure}
\begin{tabular}{l}
(a) \\
\includegraphics[width=0.45\textwidth]{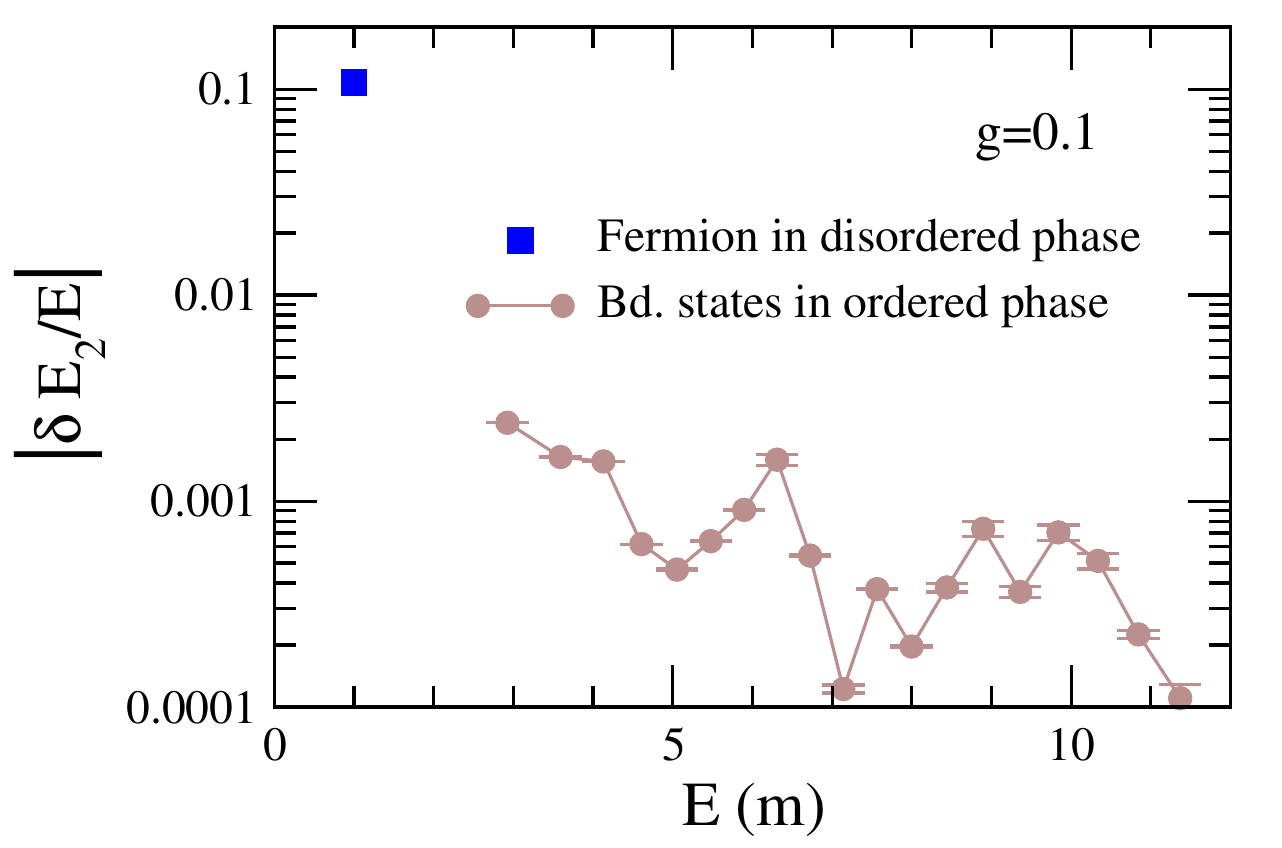}\\ 
(b) \\
\includegraphics[width=0.45\textwidth]{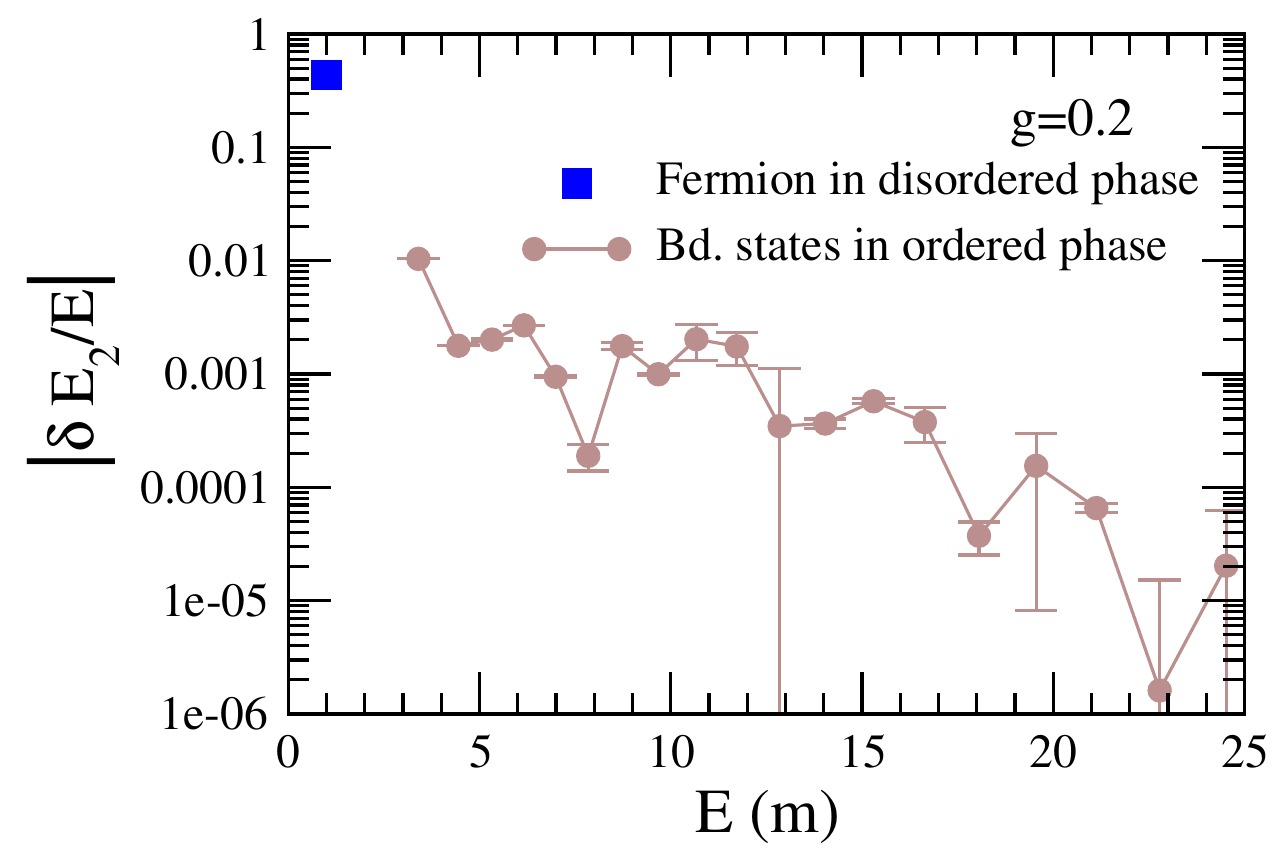}\\ 
\end{tabular}
\caption{The second order correction to the energy (measured relative to the ground state) of the first 19 meson masses due to mixing with zero and four domain wall fermion states. This should be compared to the correction to the mass of the spin flip excitation in the disordered phase ($m<0$), shown in blue, which is between two and four orders of magnitude larger. In the two cases we consider (a) $g=0.1$ (reproduced from \cite{james2018nonthermal}); (b) $g=0.2$. Error bars denote uncertainty in the extrapolation to infinite volume.}
\label{fig:encorr}
\end{figure}

The correction to the energy, $\Delta E_{2j}$ will take the form
\begin{align}
\Delta E_{2j} = \alpha R + \delta E_{2j},
\label{encorrR}
\end{align}
where the term scaling with volume, $\alpha R$, will be identical to the correction to the vacuum (ground) state due to hybridization with higher fermion number states.  This follows from the fact that the energy of a meson excitation will differ only by $O(1)$ compared to the ground state. Thus $\delta E_{2j}$ is the correction to the meson mass relative to the ground state energy. We compute $\Delta E_{2j}$ as a function of $R$ by numerically evaluating~\eqref{encorr} and fitting the result to Eq.~\eqref{encorrR}.  We present the results of this in Fig.~\ref{fig:encorr} for two values of the field, $g=0.1, 0.2$.

Some insights into the non-monotonic nature of $\delta E_{2j}$ with bound state number can be gained by looking at the wave functions of successive bound states.  We do so in Fig.~\ref{fig:wavefun} using the analytic forms of the wave function developed in Appendix~\ref{App:Wfn}.  The bound state wave functions have considerable structure and it is this, and its relative positioning relative the $E=4m$ threshold, that leads to the non-monotonic behavior of the energy corrections coming from hybridization with four particle states.

\begin{figure}
\includegraphics[width=0.45\textwidth]{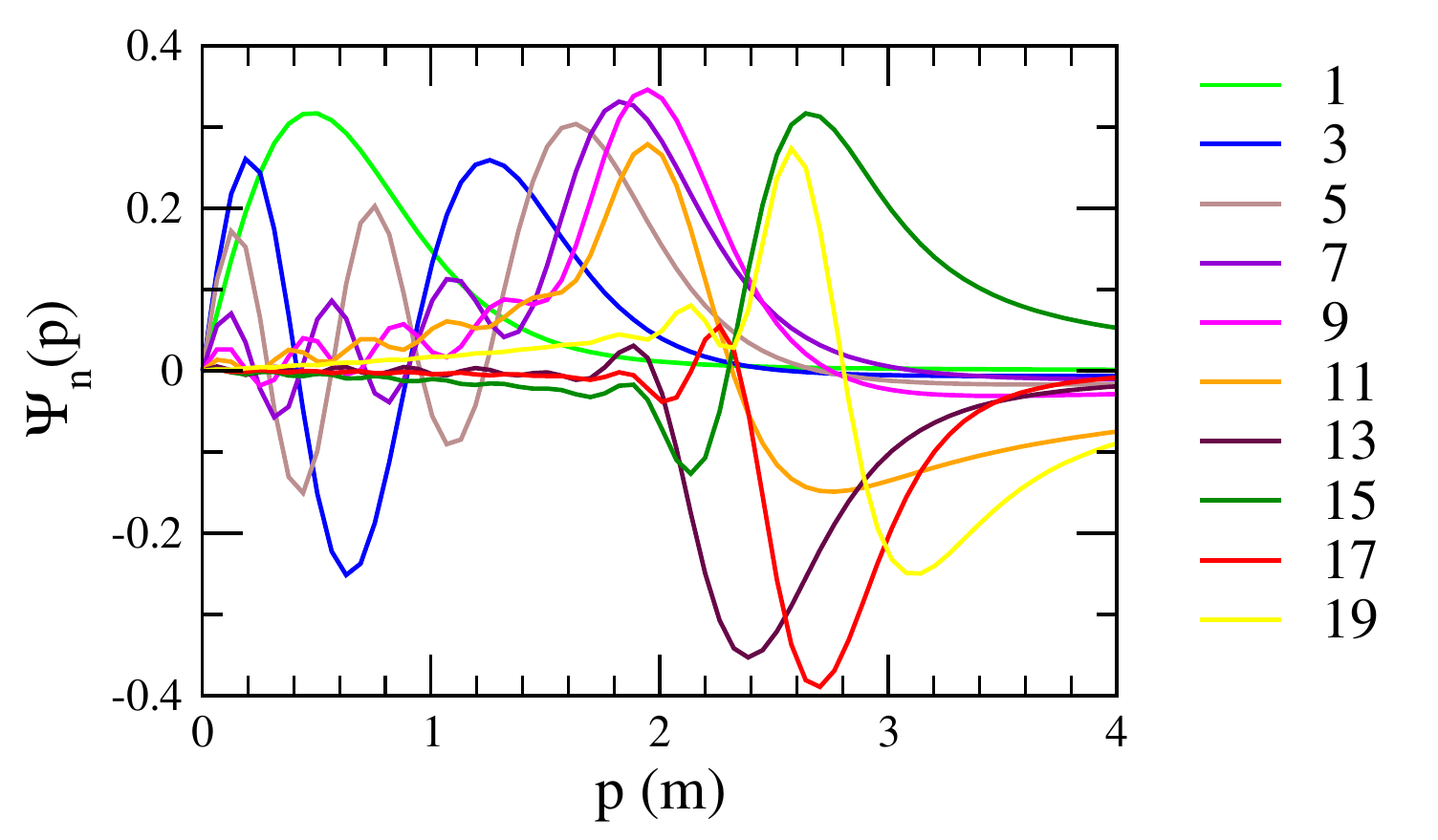}
\caption{The wave functions in momentum space for the first 10 odd numbered bound states for $g=0.1$. We see that
the maxima in the wave functions do not occur at momenta that are increasing in a uniform fashion.}
\label{fig:wavefun}
\end{figure}

We compare the correction to the meson energy to the analogous calculation in the disordered (paramagnetic) phase of the model, $m<0$, where single particle excitations are spin flips (i.e., we compute the correction to the energy of a single Ramond sector fermion). The calculation is similar to that above, with the explicit form for the correction being
\begin{align} 
\Delta E_{2,\text{sf}} &= \left( \frac{g^2 \bar\s^2}{m^2}\right) R  + \delta E_\text{sf}, \nonumber  \\
\delta E_\text{sf} &=  \frac{g^2 \bar s^2}{2mR} \sum_{q\in\text{NS}} \frac{1}{\omega_q(R)^2 [ m - 2\omega_q(R)]^2} \frac{\tanh^2 \theta_q}{\tanh^4(\theta_q/2)}. \nonumber
\end{align}
Here $\theta_q$ is a rapidity variable defined through the relation $|m|R\sinh\theta_q = 2\pi q$. The result for the correction to the spin flip mass, $\delta E_\text{sf}$, is also shown in Fig.~\ref{fig:encorr}. There we see that the correction to the spin flip energy is between two and four orders of magnitude larger than the corresponding corrections to the meson energies. Combined with the results of Sec.~\ref{sec:lifetime}, we see that the mesons~\eqref{meson} are almost unaffected by hybridization with states of higher fermion number.

\subsubsection{Corrections to the meson energy: field dependence}
\begin{figure}
\includegraphics[width=0.45\textwidth]{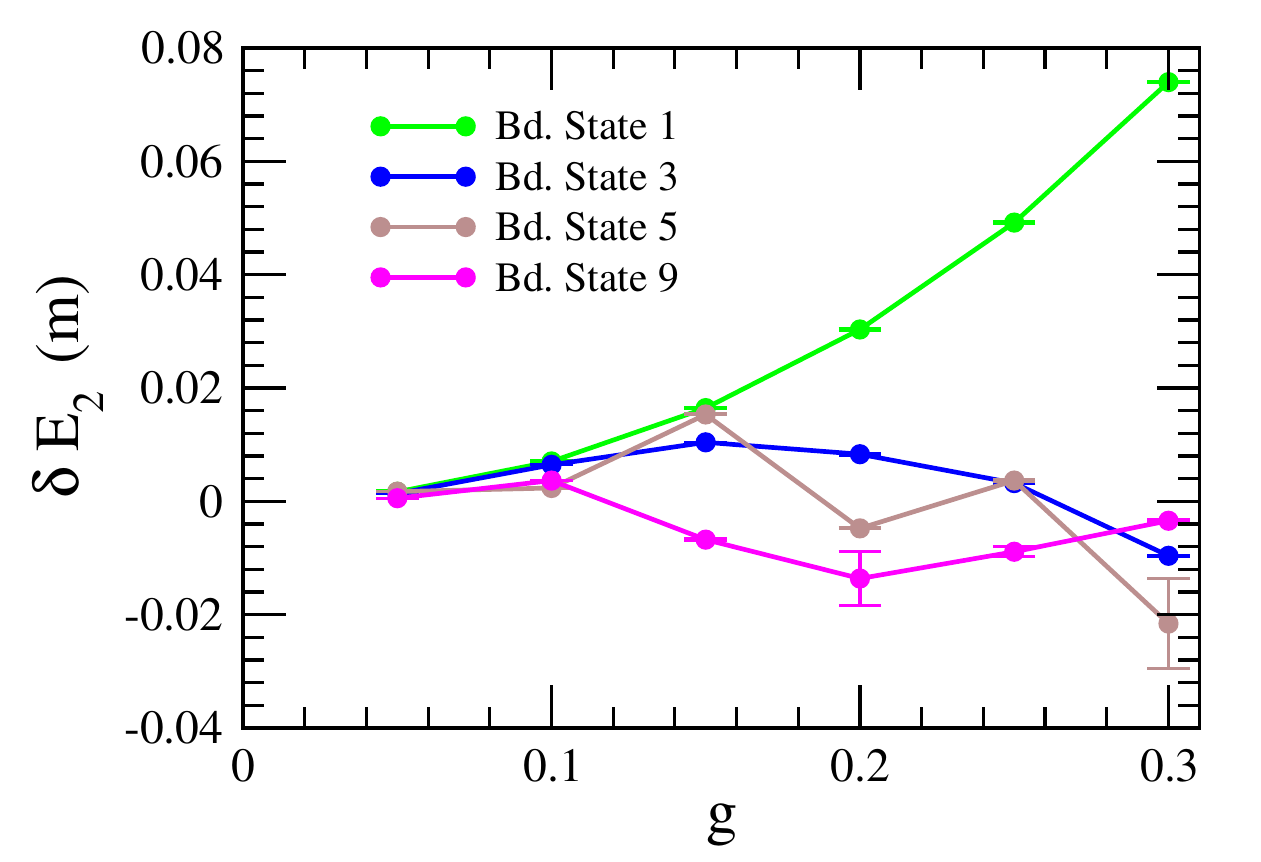}
\caption{The second order correction to the energy (measured relative to the ground state) as a function of longitudinal field strength $g$ for four mesons (the first, third, fifth and ninth ordered by energy) arising from mixing with zero and four domain wall fermion states. Error bars denote uncertainties in extrapolating to the infinite volume.}
\label{fig:encorrvarh}
\end{figure}

\begin{figure}
\includegraphics[width=0.45\textwidth]{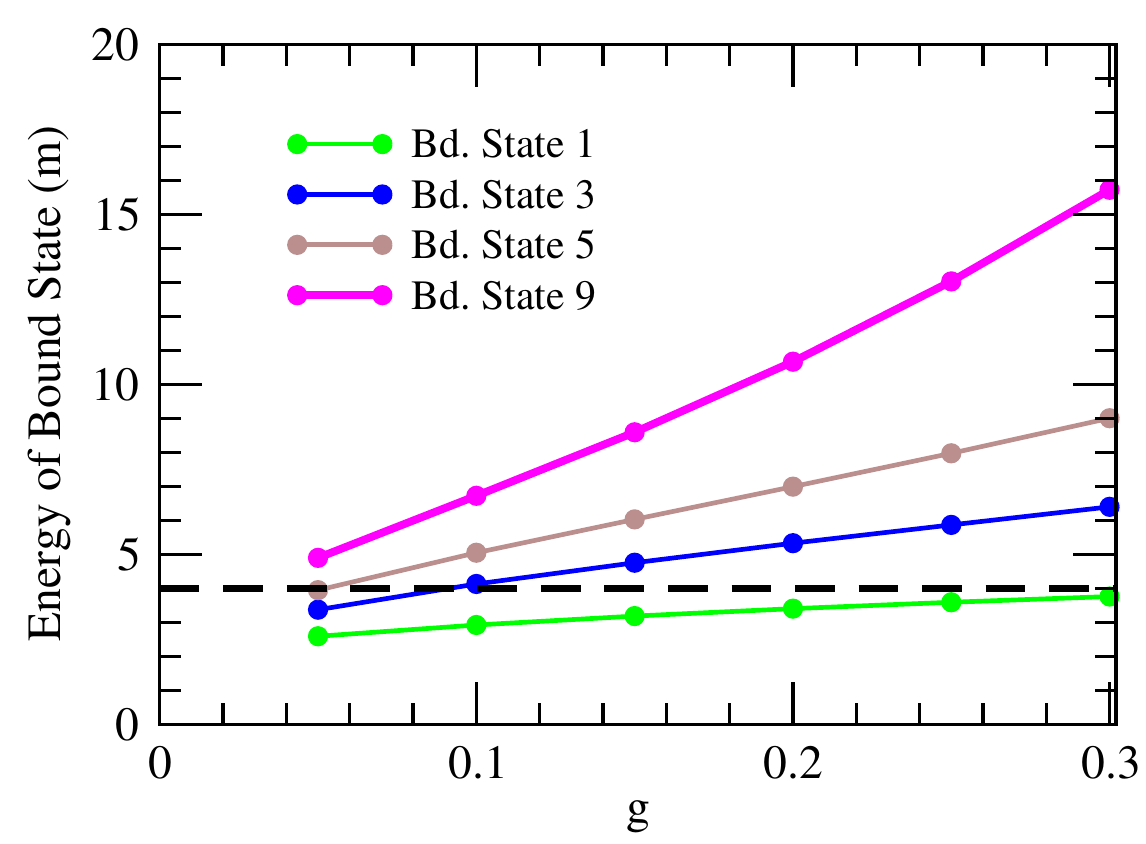}
\caption{The energies as a function of longitudinal field strength $g$ for four mesons (the first, third, fifth and ninth ordered by energy) whose
energy corrections are given in Fig.~\ref{fig:encorrvarh}.  The dashed line marks the $4m$ threshold.  Once the energy of the meson crosses this threshold, we no longer expect the energy of hybridization with the vacuum and four particle states to increase monotonically with increasing $g$.}
\label{fig:energies}
\end{figure}

We can also consider how the hybridization corrections to the meson energies evolve as a function of field strength $g$. Whilst naively the correction $\Delta E_{2j}$ is a $g^2$ diagram, it also features the meson wave function which evolves with $g$. In Fig.~\ref{fig:encorrvarh} we show the evolution of $\delta E_2$ as a function of $g$ for four meson states: the first, third, fifth and ninth as ordered by energy. One sees that $\delta E_2$ grows with $g$ provided the bound state energy is below the four domain-wall threshold (i.e. 4m). We show when this threshold is crossed, as a function of g, for the four different mesons in Fig.~\ref{fig:energies}.  For the lowest meson (the green line in Figs.~\ref{fig:encorrvarh} and \ref{fig:energies}), we see that for $0\leq g \leq 0.3$ this meson's energy never exceeds $4m$, and correspondingly the energy correction increases monotonically.  However for the fifth meson (the brown line in these two figures), its energy exceeds $4m$ once $g \gtrsim 0.06$.  We further see that while initially $\delta E_2$ increases with increasing $g$ for this meson, it begins to behave non-monotonically (with a decreasing trend) once $g$ exceeds $0.15$.


\subsubsection{Summary of insights}
In the preceding three sections we have shown that two fermion excitations, which are an approximation to the true meson excitations, have a long life time and very small  energy corrections due to hybridization. True meson states, as constructed via truncated spectrum methods in Fig.~\ref{Fig:EEVs}, are obtained by dressing the two fermion excitations with small contributions from four, six and more fermion states. The already long lifetime of the approximate meson states should increase upon inclusion of $N>2$ fermion states in the approximation (because they become closer to a true eigenstates), but such a calculation is a significant extension of those presented above and beyond the scope of this work. Instead, one can infer this directly from the results of the TSM in Fig.~\ref{Fig:EEVs}, where the meson states are eigenstates and thus possess infinite lifetimes.

\subsection{Approximate $U(1)$ symmetry at low energy}

In the previous section, we have seen that single meson excitations~\eqref{meson} only very weakly hybridize with the multi-meson continuum. This is in spite of the fact that the spin operator $\s(x)$ in principle couples states with widely differing numbers of fermions (and hence mesons). One might then suppose that there is an approximate low energy $U(1)$ symmetry, with the Hamiltonian being almost block diagonal in the space of fermion number. Here we present results for the EEV spectrum that show such an approximate block diagonalization works well within the low energy eigenstates. We leave a detailed study of this approximate symmetry to future work.

\begin{figure}
\includegraphics[width=0.45\textwidth]{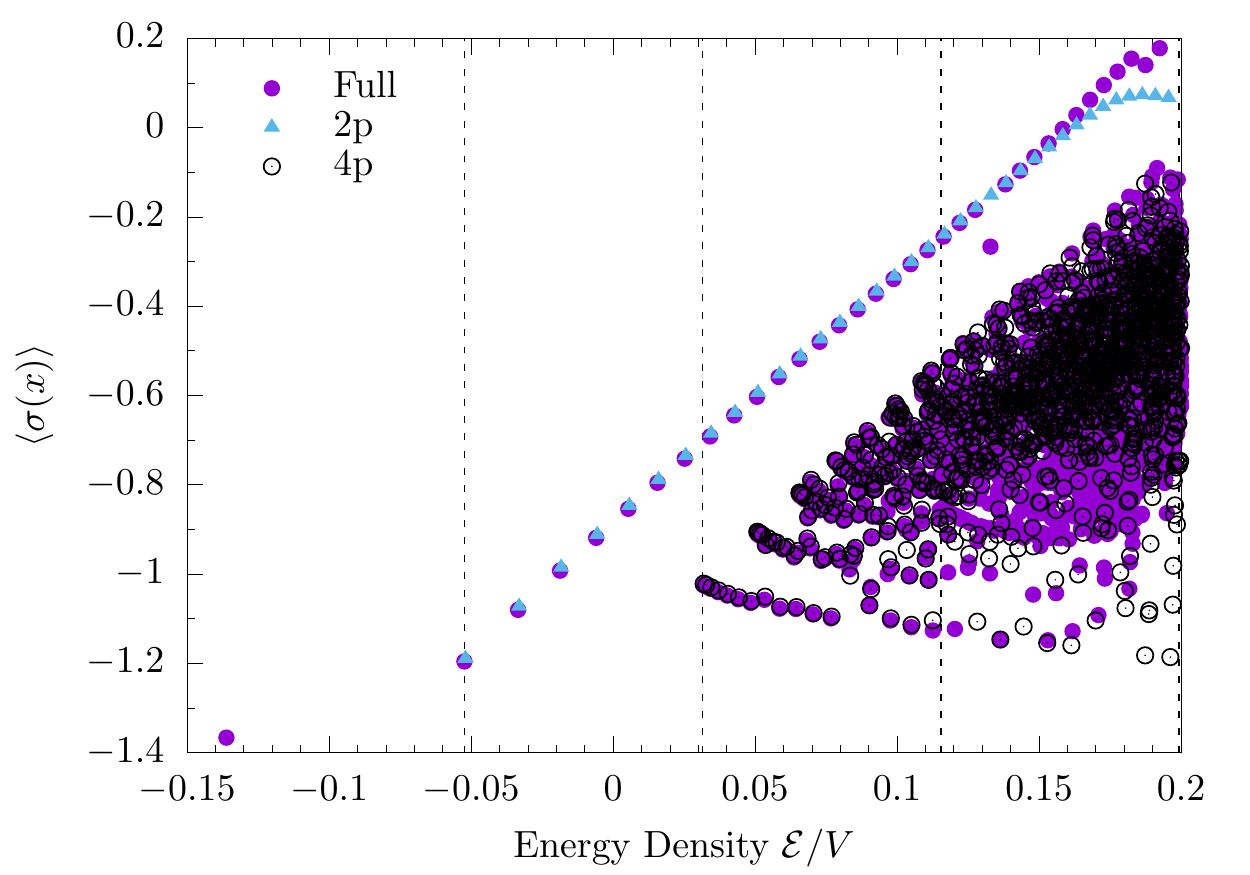}
\caption{We compare the results of a full TSM procedure with $E_\Lambda = 10.5$ (for a system with $g=0.1$, $m=1$ of size $R=35$) to an approximate block diagonalization by fermion number. Two particle states (2p) are considered up to cutoff $E_\Lambda = 200$; four particle states (4p) are considered up to cutoff $E_\Lambda = 16$. Dashed lines show the threshold energies for eigenstates containing $1-3$ mesons (i.e. up to 6 fermions).
}
\label{approxU1}
\end{figure}

In Fig.~\ref{approxU1}, we show the EEV spectrum from the full TSM treatment with energy cutoff $E_\Lambda$ (cf. Fig.~\ref{Fig:EEVs}(b)) and contrast it to the EEV spectrum computed from the Hamiltonian with fermion number conservation imposed \textit{by hand} on the interaction vertex (i.e. on the the spin operator). At low energies $E/R \lesssim0.1$ (below the six particle threshold), we see excellent agreement between the full and block diagonal treatments.

This agreement between \textit{by-hand} U(1) symmetric results and full results can be understood once one looks at the particle weights of the eigenstates. In Fig.~\ref{Fig:TwoParticleWeight} we presented the two particle weight, $w_2$, and saw that the meson states have $w_2 \approx 1$. Study of the higher particle weights (see Appendix~\ref{App:Weights}) similarly shows that the distribution of two and four particle weights is \textit{bimodal} in low-energy eigenstates, with either $(w_2 \approx 1,w_4 \approx 0)$ or $(w_2 \approx 0,w_4 \approx 1)$.

\section{Nonequilibrium Dynamics}
\label{Sec:Dynamics}

In the previous sections we have shown that rare states exist within the spectrum of the perturbed Ising field theory~\eqref{FT}. These states yield EEVs that do not conform with the MCE. One question that is natural to ask is whether these states can ever been seen? They exist above a large continuum of states that behave thermally, and hence one may imagine any response from the rare states is `washed out' and they cannot be observed. In this section, we will show that features of these rare states can be brought to bear on nonequilibrium dynamics following a quantum quench. 

We first show that at long times after a quench, one can arrive at a stationary state (the DE) that gives expectation values inconsistent with the appropriate MCE. Having ascertained that rare states show up in the long time limit, we then ask whether at finite times the presence of rare states influences nonequilibrium dynamics. We show that, indeed, they do. 

We note that the real-time dynamics following a quantum quench of both the fermion mass $m$ and the longitudinal field $g$ in the Ising field theory have been recently studied by~~\textcite{rakovsky2016hamiltonian} A similar quench in the closely-related lattice model has also recently received attention~\cite{kormos2016realtime}; it was shown that confinement leads to a suppression of the light cone spreading of correlations and a back-and-forth motion of domain walls. Similar behavior is also seen in a two dimensional model exhibiting confinement~\cite{james2018nonthermal}.

\subsection{Equilibration after a sudden quantum quench}
\label{Sec:DEandMCE}

Let us begin by defining our nonequilibrium protocol.
We study dynamics that are induced at time $t=0$ by an instantaneous change of the longitudinal field within the Hamiltonian~\eqref{FT}
\begin{align}
H_i \equiv H(m,g_i) \to H_f \equiv H(m,g_f),
\end{align}
restricting our analysis to quenches starting from eigenstates $|E_m\ra$ of the initial Hamiltonian. The expectation value of the local spin operator $\s(0)$ in the long-time limit is probed through the DE~\cite{kollar2008relaxation,rigol2008thermalization,rigol2009breakdown}
\begin{align}
\lim_{t\to\infty} \frac{1}{t} \int_0^t \rd t' &\la E_m|e^{iH_f t'} \s(0) e^{-iH_f t'} |E_m\ra\nonumber \\
& = \sum_n |\tilde c_{m,n}|^2 \la \tilde E_n| \s(0) |\tilde E_n\ra, \label{DE}
\end{align}
where $|\tilde E_n\ra$ are the eigenstates of $H_f$ and $\tilde c_{m,n} = \la E_m | \tilde E_n\ra$ are the overlap coefficients.  The DE predictions are compared to the MCE at the appropriate energy density constructed by averaging over an energy window $\Delta E$. Agreement between the DE and MCE signals thermalization, whilst disagreement suggests an absence of thermalization. The convergence of the DE and MCE are discussed in the Appendix~\ref{App:ConvTCSA}. 

Herein we consider quenches in which both the initial and final longitudinal fields are positive -- this avoids problems with convergence for sign changing quenches (see also Ref.~\cite{rakovsky2016hamiltonian}), which can be understood in terms of projecting onto the `false vacuum' in finite size systems.  

\begin{figure}[t]
\includegraphics[width=0.45\textwidth]{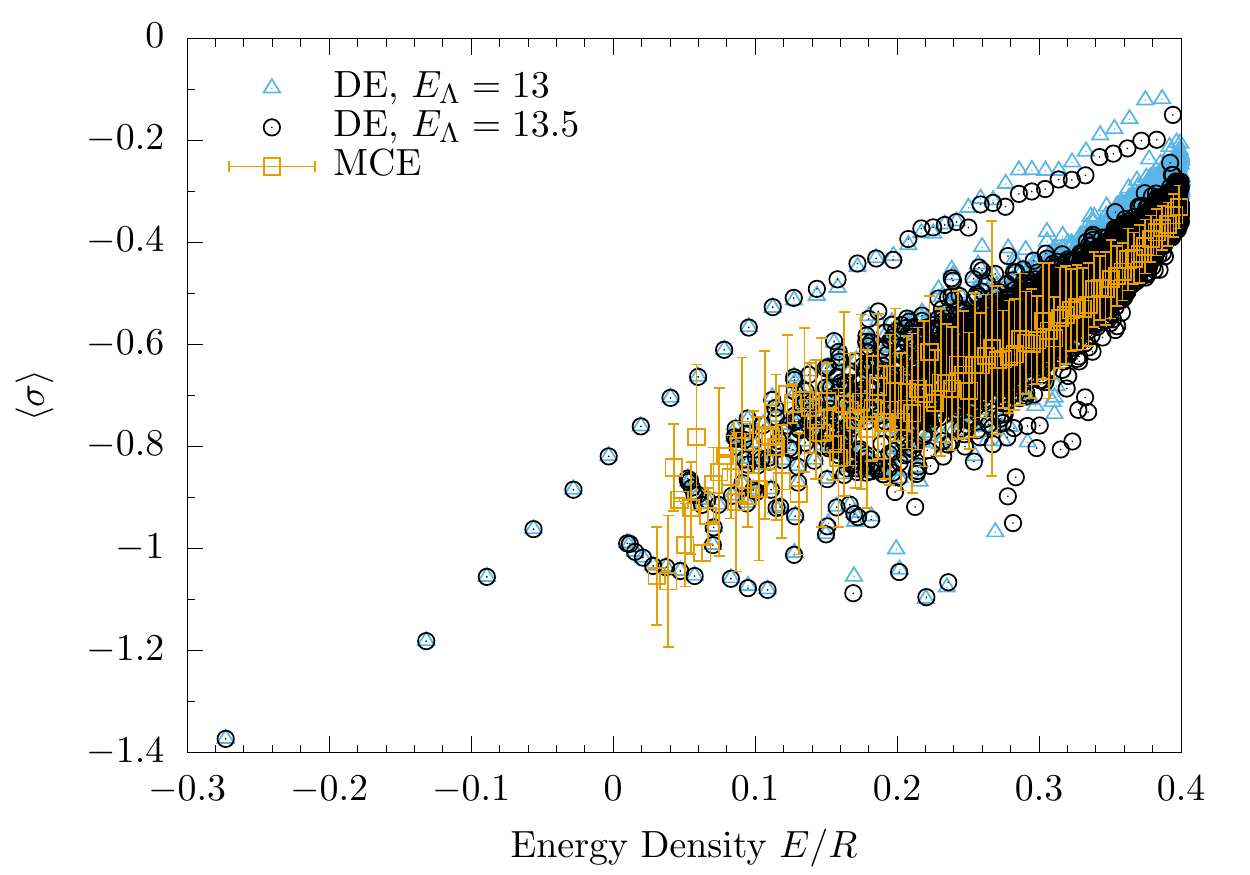}
\caption{The diagonal ensemble (DE) result for long-time expectation values and the microcanonical ensemble (MCE) prediction following the sudden quench $(m=1,g=0.1) \to (m=1,g=0.2)$ in the field theory~\eqref{FT}. Each data point corresponds to starting from a different low energy eigenstate of the initial Hamiltonian, $m=1,g=0.1$. In all cases, results are on a system of size $R=25$ and with energy cutoff $E_\Lambda$. We note that $E_\Lambda = 13$ ($E_\Lambda = 13.5$) corresponds to $23,238$ ($32,149$) computational basis states. The MCE is computed by averaging an energy window of size $\Delta E = 0.1$, and error bars denote the standard deviation of data averaged over.}
\label{Fig:DEMCE}
\end{figure}

We present DE and MCE results for the local spin operator $\s(0)$ at long-times after the quench $g = 0.1 \to 0.2$ in Fig.~\ref{Fig:DEMCE}. We use a system of size $R=25$ and fermion mass $m=1$, cf. Figs.~\ref{Fig:EEVs}. Each plotted point of the DE is constructed starting from a different eigenstate of the initial Hamiltonian; we see that the vast majority of initial states have DE and MCE results that agree (within one standard deviation, represented by the error bars on the MCE). However, much like in the equilibrium spectrum (see Figs.~\ref{Fig:EEVs}, and recall the finite size scaling in Fig.~\ref{Fig:fss}) we observe a well-separated band of states above the thermal continuum. We also include the DE ensemble for a smaller value of the TSM cutoff to highlight that these well-separated states are \textit{well converged}.

Figure~\ref{Fig:DEMCE} leads us to conclude that the rare states in the equilibrium spectrum indeed influence the nonequilibrium dynamics, leading to states that do not thermalize following a quantum quench. This is despite the fact that the Hamiltonian governing the time evolution is \textit{nonintegrable} and hence generically is expected to lead to thermalization. This is yet more support for ETH: if ETH is absent, thermalization can be avoided even in nonintegrable models. 

\subsection{Nonequilibrium real-time dynamics}

In the long time limit we see that there are states that do not thermalize. Now we want to understand whether such states have unusual behavior (or anything noticeably different from thermal states) in their short-to-intermediate time dynamics. This time window is of primary interest for experiments on cold atomic gases, and is also accessible in lattice models using numerical methods such as TEBD. Let us first discuss how we compute the real-time dynamics, before presenting results. 

\subsubsection{Time evolution by the Chebyshev Expansion}
\label{Sec:Cheb}

To compute the time evolution of states within TSMs, we use the recently developed formalism of~~\textcite{rakovsky2016hamiltonian} and expand the time evolution operator in terms of Chebyshev polynomials $T_n$. The expansion reads 
\begin{align}
e^{-iH_ft} = J_0\left(\tilde t\right) \mathbb{1} + 2 \sum_{n=1}^\infty \left(-i\right)^n J_n\left(\tilde t\right)T_n(\tilde H_f),
\label{Eq:EvolutionOp}
\end{align}
where we rescale both the time $\tilde t = E_{\rm max}t$ and the Hamiltonian $\tilde H_f = H_f/E_{\rm max}$ by the maximal eigenvalue of the Hamiltonian $H_f$, such that the eigenvalues of $\tilde H_f$ lie in the interval $[-1,1]$. $J_n(x)$ are the Bessel functions
\begin{align}
J_n(x) = \sum_{l=0}^\infty \frac{(-1)^l}{l! (n+l)!} \left( \frac{x}{2}\right)^{2l+n}, 
\end{align}
and the Chebyshev Polynomials are defined through the recursion relation
\begin{align}
T_{n+1}(x) = 2x T_n(x) - T_{n-1}(x), \label{Eq:recursion} \\
{\rm with}\quad T_0(x) = 1,\quad T_1(x) = x. \nonumber
\end{align}

With Eq.~\eqref{Eq:EvolutionOp} at hand, the procedure for computing the time evolution is straightforward. The initial state $|\Psi_i\ra$ is known in the free fermion basis, as is the final Hamiltonian, so, one generates the Chebyshev vectors $|t_n\ra = T_n(\tilde H_f)|\Psi_i\ra$ through the recursion relation~\eqref{Eq:recursion} and then sums them, Eq.~\eqref{Eq:EvolutionOp}, to obtain the time-evolved state. Observables, such as the local magnetization $\la \sigma(0)\ra$, can then be computed. Herein we refer to this approach as TSM+CHEB.

The Chebyshev expansion of the time evolution operator, Eq.~\eqref{Eq:EvolutionOp}, contains an infinite number of terms and so it is necessary to truncate the expansion. This truncation introduces a time scale after which the reported time evolution cannot be trusted. Qualitatively we see that this time scale increases (approximately) linearly with the number of terms kept in the sum. We discuss further convergence of TSM+CHEB in Appendix~\ref{App:ConvCheb}.

\begin{figure}[t]
\begin{tabular}{l}
(a) \\
\includegraphics[width=0.45\textwidth]{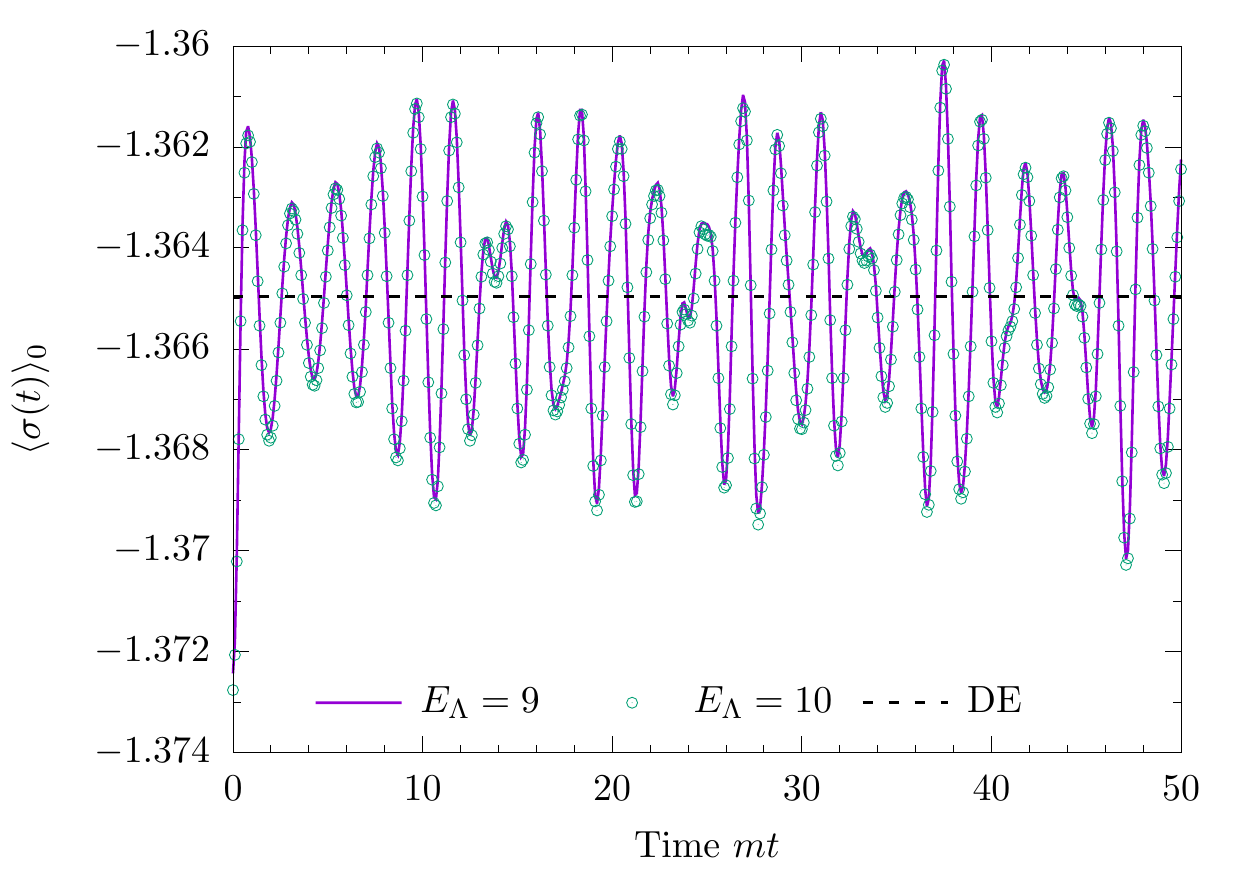}\\
(b) \\
\includegraphics[width=0.45\textwidth]{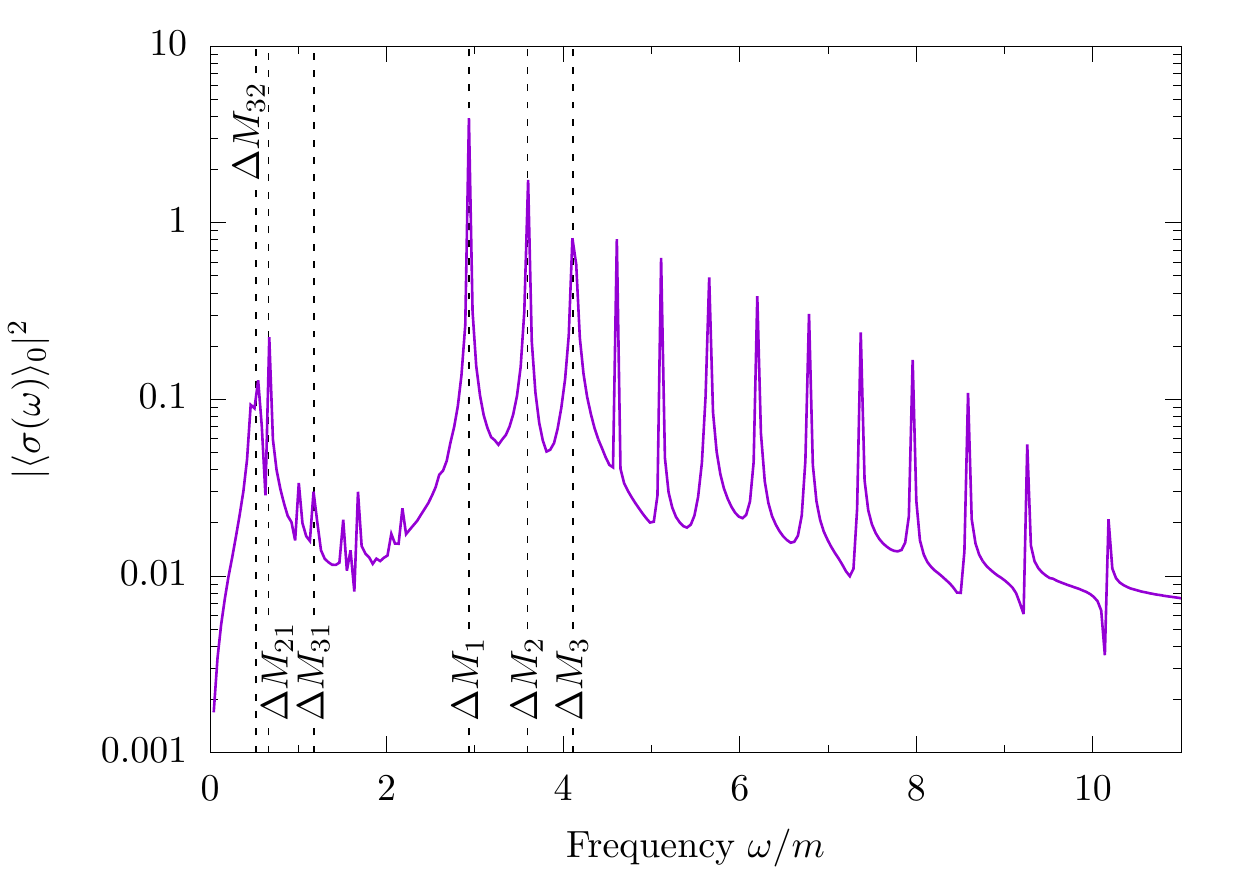}
\end{tabular}
\caption{(a) Time evolution of the local magnetization $\la \s(0)\ra$ following the quench $(m=1,g=0.2)\to(m=1,g=0.1)$ in~\eqref{FT} with $R=25$. Time evolution is computed via TSM+CHEB with cutoff $E_{\Lambda} = 9,\ 10$ and the expansion evaluated to order 2000. The magnetization is oscillating about its diagonal ensemble (DE) value, computed via the TSM, which also coincides with the microcanonical ensemble prediction. (b) The power spectrum of the time evolution. We note a few prominent frequencies in terms of the {\it post-quench} confined state energies: $\Delta M_{ij} = M_i-M_j$ and $\Delta M_j = M_j$.}
\label{Fig:GSquench}
\end{figure}

\begin{figure}[t]
\begin{tabular}{l}
(a) \\
\includegraphics[width=0.45\textwidth]{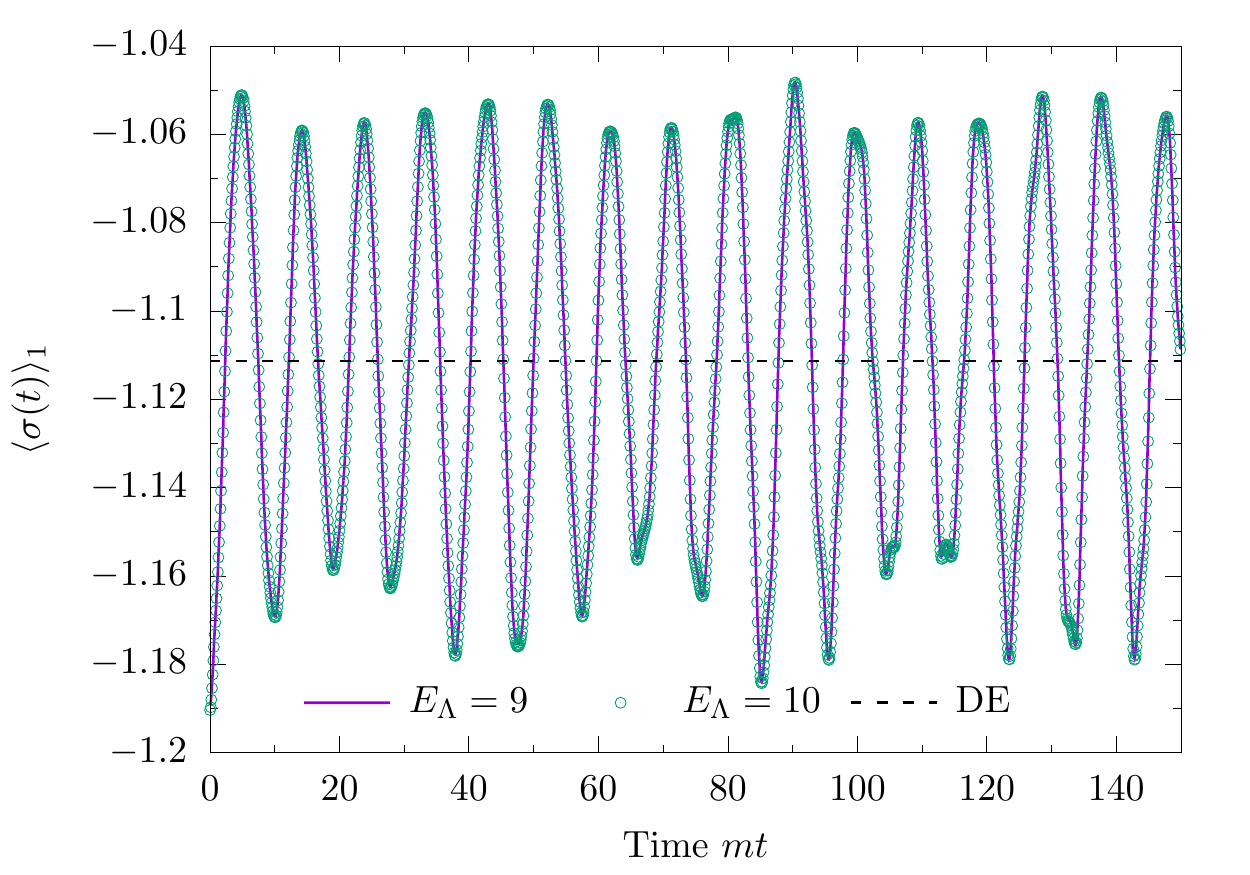}\\
(b) \\
\includegraphics[width=0.45\textwidth]{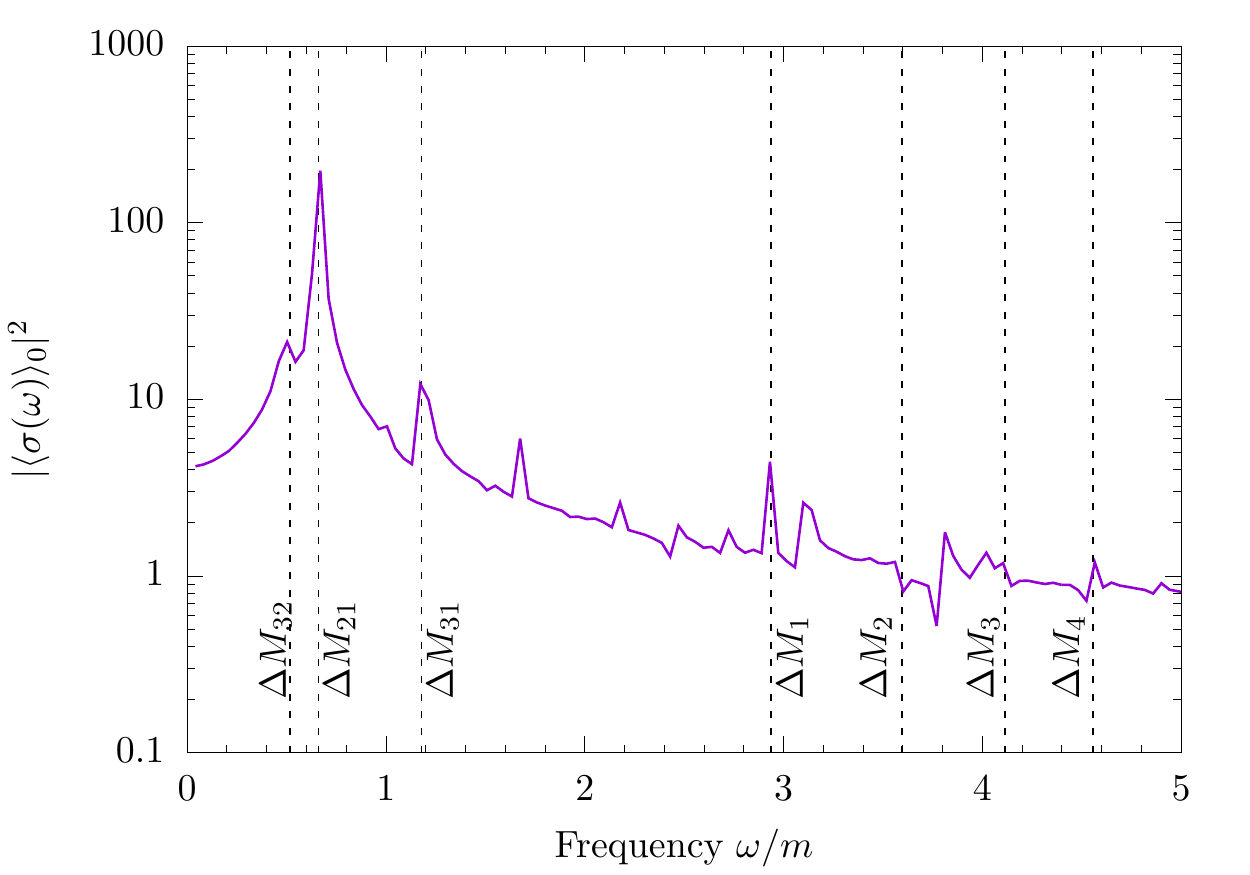}
\end{tabular}
\caption{(a) Time evolution of the local magnetization $\la \s(0)\ra$ when starting from the first excited state of~\eqref{FT} for the same quench described in Fig.~\ref{Fig:GSquench}. (b) The associated power spectrum. We note a few prominent frequencies in terms of the post-quench confined state energies:  $\Delta M_{ij} = M_i-M_j$ and $\Delta M_j = M_j$.}
\label{Fig:Firstquench}
\end{figure}

\subsubsection{Quenches from the ground and first excited states}

Let us first develop some intuition by studying quenches from the very lowest states. We present the time evolution of the local spin operator $\s(0)$ following the quench $H(m=1,g=0.2)\to H(1,0.1)$, when starting from the ground state of the initial $H$, in Fig.~\ref{Fig:GSquench}. Results are well converged  already for $E_\Lambda \sim 10$ and the expectation value oscillates about its DE prediction.\footnote{Interesting, we note that this particular quench has essentially no finite size effects, see the appendix~\ref{App:ConvCheb} for further information. A similar lack of finite-size effects was seen in quenches of the lattice Ising chain perturbed by a longitudinal field~\cite{kormos2016realtime}. $~$} These oscillations occur at many frequencies, as can be extracted through the power spectrum (Fourier transform) shown in the lower panel. These frequencies coincide with the post-quench meson energies (and their differences), giving an effective route for `meson spectroscopy' (for further details of how we compute these energies, see Appendix~\ref{App:Semiclassical}). 

Analogous behavior is seen for the quench in which we start from the first excited state, Fig.~\ref{Fig:Firstquench}. The power spectrum is now dominated by the frequency $\omega = M_2 - M_1$, implying the initial state projects heaving onto the first and second meson excitations. This is not entirely surprising, as we know in both the initial and final Hamiltonians the meson states are mostly two-particle in nature (see Sec.~\ref{Sec:Nature}). 

In both the above cases, Figs.~\ref{Fig:GSquench}--\ref{Fig:Firstquench}, oscillations of the expectation value are centered on the DE prediction. For the time-scales that we can reach within the TSM+CHEB procedure, it is not clear that these oscillations are decaying. This is an important question to address, but remains beyond the reach of existing approaches. 

\subsubsection{Quenches from a rare state and proximate thermal state}

Having looked at quenches starting from the very lowest states, let us now turn our attention to states with larger energy densities. In particular, we will compare the short time dynamics of a rare state to that of thermalizing states of similar energy. We choose the rare state by picking one of the initial states where the DE and MCE disagree; it is \textit{a priori} unclear whether we should see qualitatively different dynamics for rare and thermal states. 

To address this question, we consider the time evolution following the quench $H(m=1,g=0.1)\to H(1,0.2)$ (as in Sec.~\ref{Sec:DEandMCE}), and consider eigenstates of the initial Hamiltonian, $|\Psi_n\ra$, whose energy in the final basis are very close (computed via $E_n = \la \Psi_n | H(1,0.2) | \Psi_n\ra$). This in turn implies that the MCE prediction for the long-time expectation values are very close. The precise states that we consider are summarized in Table~\ref{Table:States}, with the $n=100$ initial eigenstate being a rare state (the DE and MCE disagree outside one standard deviation). All other states are thermal, in the sense that the DE agrees with the MCE prediction (within one standard deviation), cf. Fig.~\ref{Fig:DEMCE}(a).  

\begin{table}[t]
\caption{The energy $E_n = \la \Psi_n | H_f |\Psi_n\ra$ and diagonal ensemble result for the local spin operator $\la \s(0)\ra_{\rm DE}$ for the $n$th eigenstate of the initial Hamiltonian~\eqref{FT} $H_i = H(1,0.1)$, constructed with energy cutoff $E_\Lambda$ in the TSM procedure. The state is time evolved according to the final Hamiltonian $H_f = H(1,0.2)$. The $n=100$ state is a rare state, whilst the others are broadly consistent with the microcanonical ensemble, see Figs.~\ref{Fig:DEMCE}~and~\ref{Fig:TimeEvoRareThermal}. The microcanonical result is $\la \s(0)\ra_{\rm MCE} = -0.693097 \pm 0.129259$, with the uncertainty showing the standard deviation of values averaged over in the energy window of width $\Delta E = 0.1$.}
\vskip 10pt
\begin{tabular}{ccc}
\hline\hline
\hspace{0.75cm}$n$\hspace{0.75cm} & \hspace{0.75cm}$E_n$\hspace{0.75cm} & \hspace{0.75cm}$\la \s(0)\ra_{\rm DE}$\hspace{0.75cm} \\
\hline
100 & 5.19928 & -0.32158 \\
185 & 5.21496 & -0.550455 \\
333 & 5.15805 & -0.657668 \\
352 & 5.19533 & -0.718688 \\
502 & 5.19786 & -0.83252 \\
\hline 
\hline
\end{tabular}
\label{Table:States}
\end{table}

\begin{figure}
\includegraphics[width=0.45\textwidth]{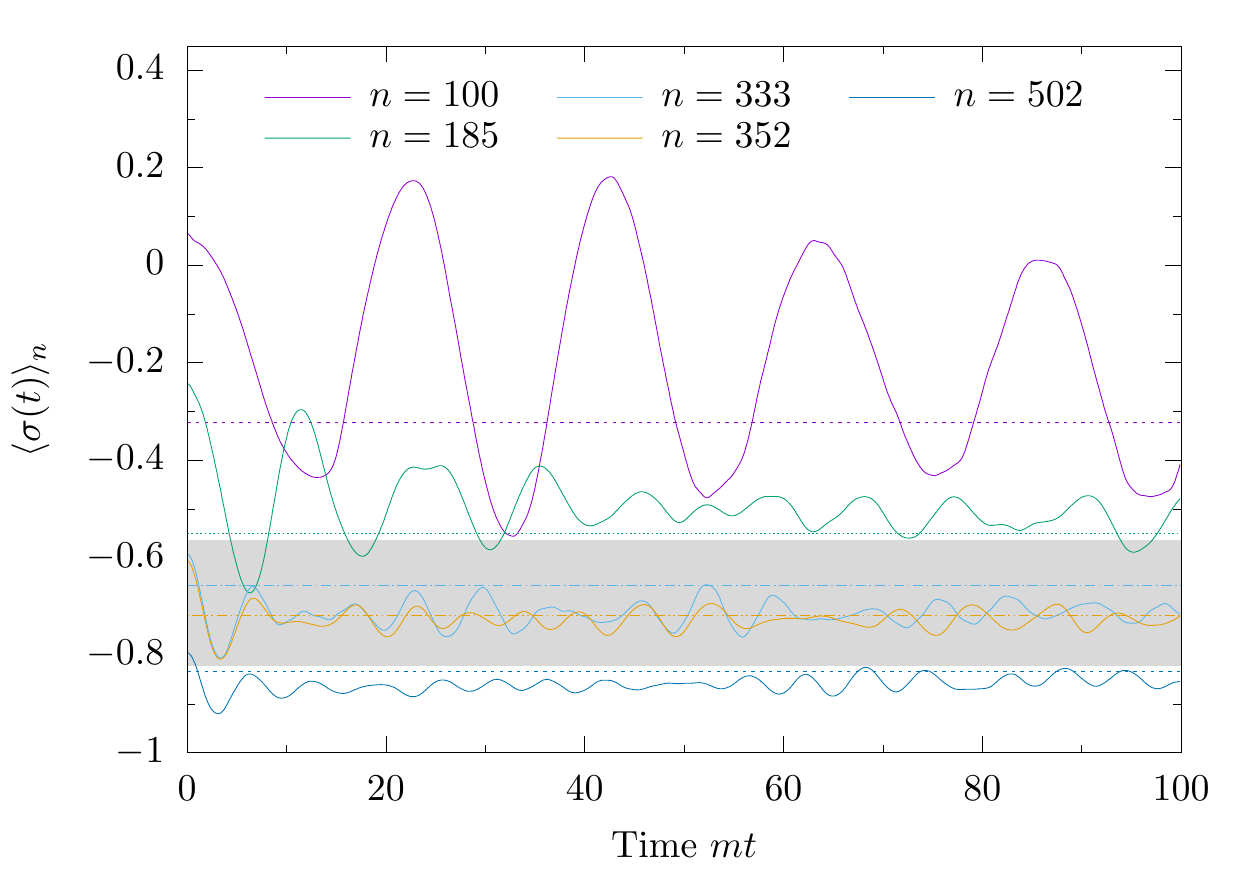}
\caption{Time evolution of the local spin operator $\s(0)$ following a quench $H_i = H(1,0.1) \to H_f = H(1,0.2)$ computed via TSM+CHEB. Each data set starts from a different eigenstate $n$ of the initial Hamiltonian. Dashed lines denote the diagonal ensemble prediction for the long time limit, whilst the shaded region shows the thermal result from the microcanonical ensemble (with the uncertainty representing the standard deviation of the data averaged over).}
\label{Fig:TimeEvoRareThermal}
\end{figure}

Having chosen our initial states with similar energies, we compute their time evolution using the TSM+CHEB explained in Sec.~\ref{Sec:Cheb}. The time evolution of the local magnetization $\s(0)$ is shown in the solid lines of Fig.~\ref{Fig:TimeEvoRareThermal}, with dashed lines of the same color corresponding to the DE prediction. The shaded region shows the MCE prediction, with the vertical extent denoting the standard deviation of the data averaged over. As expected, thermalizing states have time evolution compatible with relaxation to the MCE and DE predictions (some of the thermalizing states selected, e.g. $n=185$ and $n=502$, are on the edges of what we call thermal). On the other hand, the rare $n=100$ state is oscillating about a value consistent with the DE result, which is far from the MCE prediction.  

\begin{figure}
\begin{tabular}{l}
(a) \\
\includegraphics[width=0.45\textwidth]{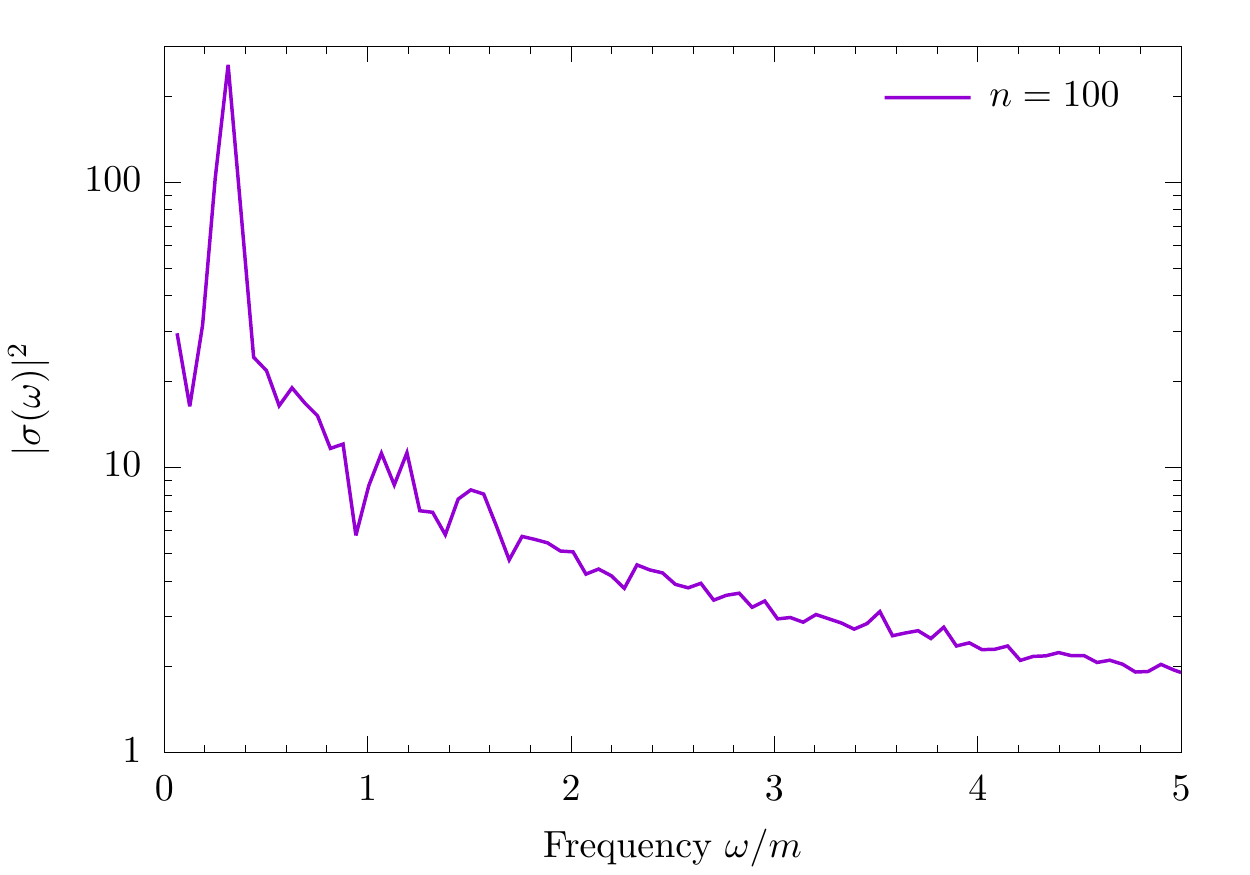}\\
(b) \\
\includegraphics[width=0.45\textwidth]{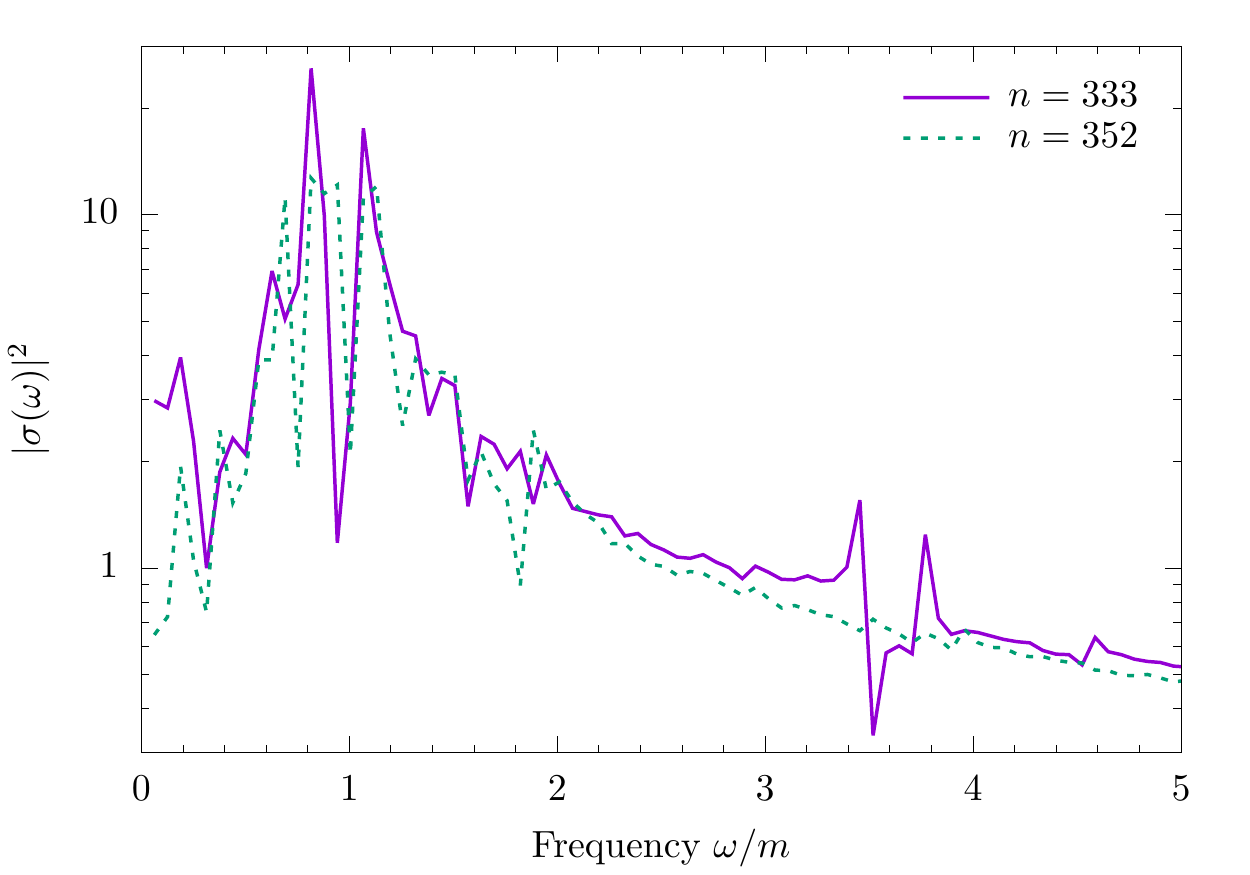}
\end{tabular}
\caption{Power spectrum $|\s(\omega)|^2$ for: (a) the $n=100$ rare state; (b) the $n=333,\,352$ thermalizing states as computed from data in Fig.~\ref{Fig:TimeEvoRareThermal}.}
\label{Fig:PowerRareThermal}
\end{figure}

In examining Fig.~\ref{Fig:TimeEvoRareThermal}, we do see some apparent differences between the time evolution of the rare states and close-in-energy thermal states. The rare state, $n=100$, exhibits large amplitude low frequency oscillations, whilst the thermal states have many more oscillation frequencies, highlighted in the power spectra of Fig.~\ref{Fig:PowerRareThermal}. The relaxation towards the DE/MCE is rapid for the thermal states, with only small amplitude oscillations beyond times $t\sim30$. In contrast, the rare state still has large amplitude oscillations at the longest accessible times, and it is not clear that these decay within the time frame $mt \sim 100$. 

These results suggest that one can diagnose rare states even in the short time dynamics by looking for states which exhibit large (and slow decaying) oscillations in observables. Indeed, this is reminiscent of the behavior observed in lattice simulations with confinement~\cite{kormos2016realtime}, where the light cone spreading of correlations is suppressed (but not entirely removed, cf. Ref.~\cite{delfino2018correlation} and Fig. 3b of Ref.~\cite{kormos2016realtime}), leading to dominant `back and forth' oscillations of correlations. 

\section{Relation to the lattice problem}
\label{Sec:Lattice}

Up until now we have considered states in the perturbed Ising field theory~\eqref{FT}. An important question is whether any of the discussed behavior carries through to the lattice, or if it is an artifact of the scaling limit? To address this, we consider a lattice model whose scaling limit is described by~\eqref{FT}, and work away from this limit. Our lattice Hamiltonian reads 
\begin{align}
H = J \sum_l \s^z_{l} \s^z_{l+1} + h^x \sum_l \s^x_{l} + h^z \sum_l \s^z_{l},
\label{lattice}
\end{align}
where we set the lattice spacing $a$ to one, $\s^{x,z}_n$ are the $x,z$ Pauli matrices at position $n$, $J$ is the Ising exchange, and $h^{x,z}$ are magnetic fields in the $x,z$ directions, respectively. Equation~\eqref{lattice} is a good description of CoNb$_2$O$_6$\cite{coldea2010quantum,kjall2011bound,cabrera2014excitations,robinson2014quasiparticle}, a quasi-one-dimensional Ising ferromagnet in which linearly confined domain wall excitations have been observed via inelastic neutron scattering and THz spectroscopy~\cite{coldea2010quantum,wang2015spinon,wang2016from,morris2014hierarchy}. The scaling limit that reproduces~\eqref{FT} is~\cite{zamolodchikov1989integrals,fonseca2003ising}
\begin{gather}
J \to \infty, \quad a \to 0, \quad h^x \to 1,\quad h^z \ll 1,\nonumber \\
m = 2J\vert 1 - h^x\vert,\quad 2Ja=1. \nonumber
\end{gather}
The lattice model~\eqref{lattice} has been investigated extensively, with the quench dynamics receiving particular attention~\cite{banuls2011strong,kim2013ballistic,kim2014testing,kim2015slowest,zhang2015thermalization,kormos2016realtime,lin2017quasiparticle,mazza2018suppression}. $\!$~~\textcite{banuls2011strong} showed that certain initial states do not appear to thermalize, in spite of the lack of integrability. It is thought that this is linked to the presence of rare states in the spectrum~\cite{kormos2016realtime,mazza2018suppression}. On the other hand, Ref.~\cite{kim2014testing} studied whether for certain parameter regimes all eigenstates of~\eqref{lattice} obey ETH---finding certain sets of parameters where this did appear to be the case from small system exact diagonalization. Putting these results together, it appears the precise details of the parameter regime matter. This is perhaps not surprising, as the physics contained within this simple lattice model is rather rich~\cite{mccoy2014twodimensional}.

In the following subsection we will show that, indeed, rare states exist within the spectrum of this lattice model, at least for certain values of the parameters. 

\subsection{Eigenstate expectation values}

To begin our study of the lattice model, Eq.~\eqref{lattice}, we consider the EEV spectrum for low-energy states. To do so, we use DMRG to construct eigenstates in a finite (open) chain of $N=20-40$ sites. We first use the standard finite-size DMRG technique~\cite{schollwock2011densitymatrix} to find the ground state, and subsequently construct excited states using projector methods (see, e.g., Ref.~\cite{stoudenmire2012studying} for a detailed explanation of this approach). With the projector method, it is important to remember that excited states may not be found in order, but provided one constructs a sufficient number of states (and chooses the `weight parameter' of the procedure carefully) it is possible to correctly construct the low energy spectrum. 

In our DMRG simulations, the truncation error was $10^{-10}$ and we allowed up to 20 finite size sweeps (20 left and 20 right sweeps) for each state. As a check of our routines, we compared the spectrum computed for $N=14$ sites with that obtained from exact diagonalization of the Hamiltonian, finding excellent agreement. 

To draw analogies between the lattice model~\eqref{lattice} and the continuum theory~\eqref{FT}, in particular in order to compare to Fig.~\ref{Fig:EEVs}, we compute the average magnetization across the chain. This mimics a projection onto the zero (quasi-)momentum sector of the theory, which is studied in the field theory. We remind the reader that eigenstates of the field theory are translationally invariant and hence satisfy 
\begin{align}
\la E| \s(0) | E \ra = \frac{1}{R} \int_0^R \rd x \la E | \s(x) | E \ra,
\end{align}
where $|E\ra$ is any eigenstate. 

\begin{figure}
\begin{tabular}{l}
(a) $N=20$ \\
\includegraphics[width=0.45\textwidth]{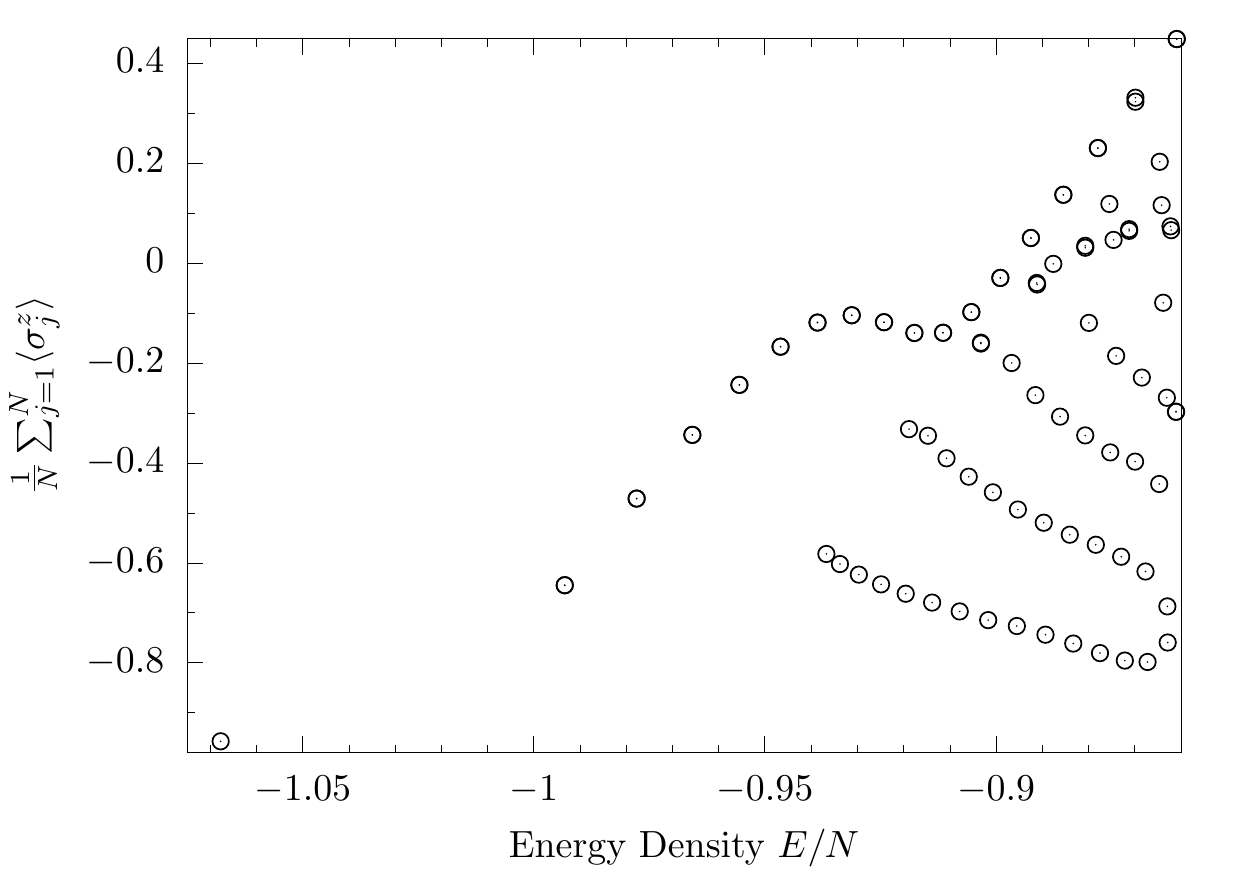} \\
(b) $N=30$ \\
\includegraphics[width=0.45\textwidth]{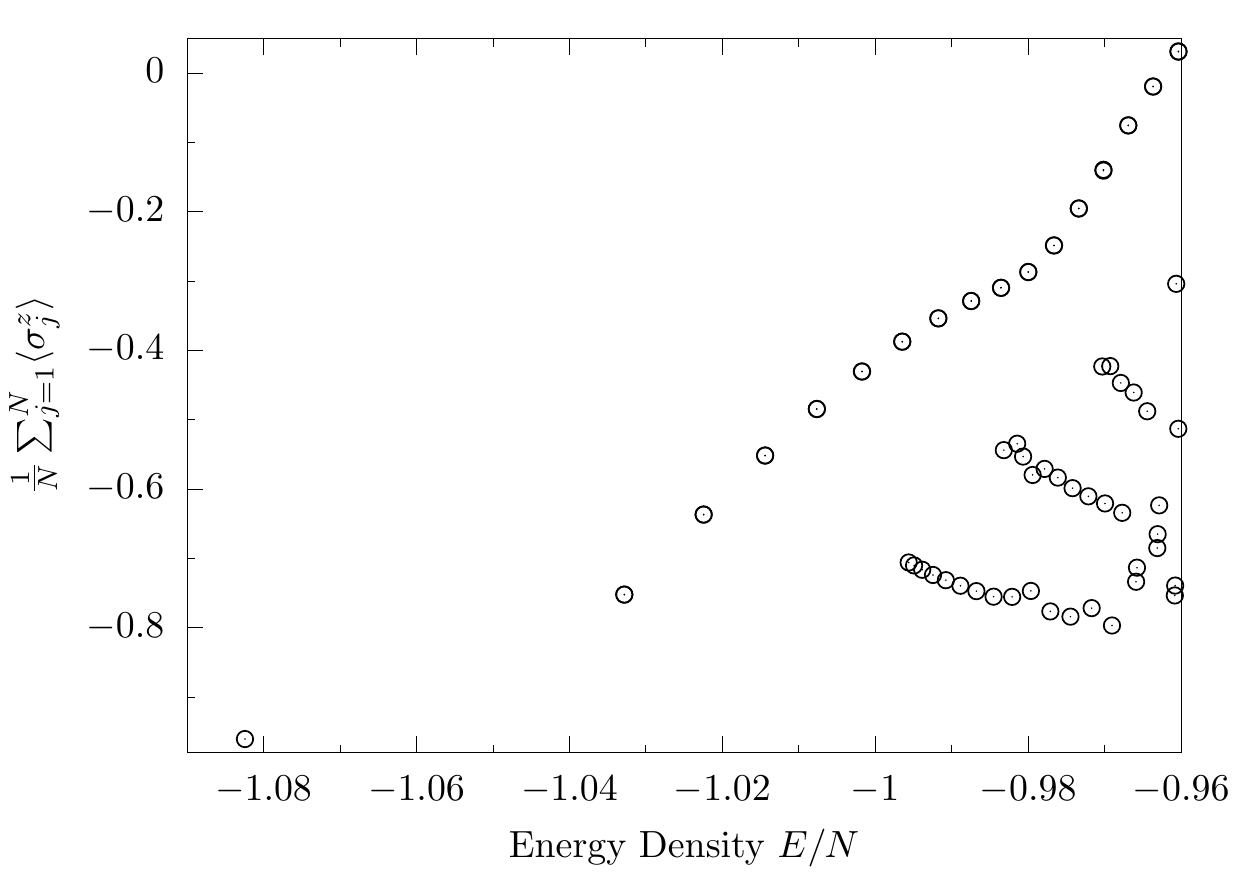} \\
(c) $N=40$ \\
\includegraphics[width=0.45\textwidth]{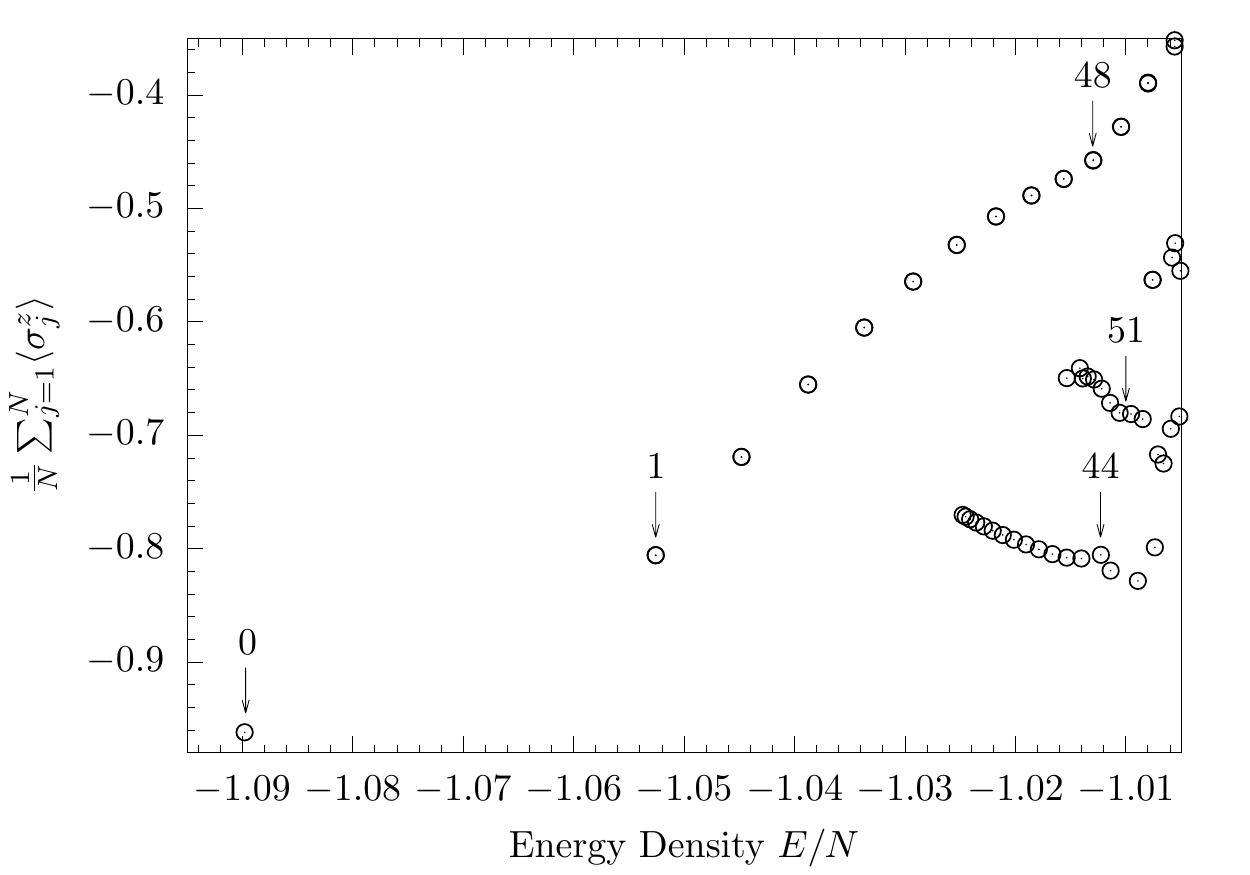}\\
\end{tabular}
\caption{The eigenstate expectation values for the chain-averaged magnetization $\frac{1}{N}\sum_{l=1}^N \langle \sigma_l^z \rangle$ in the lowest $\sim150$ states of the lattice model, Eq.~\eqref{lattice}, with $J=-1$, $h^x = -0.5$ and $h^z = 0.05$. Eigenstates were computed with DMRG for a chain of (a) $N=20$; (b) $N=30$; (c) $N=40$ sites. We encourage the reader to compare this to the analogous plot in the scaling limit, Fig.~\ref{Fig:EEVs}, and note the striking similarity. In plotting (a), (b), we exclude the false vacuum state. The labels in (c) indicate the states shown in Fig.~\ref{Fig:lattice_mag}.}
\label{Fig:LatticeExpectationValues}
\end{figure}

\begin{figure}
\includegraphics[width=0.45\textwidth]{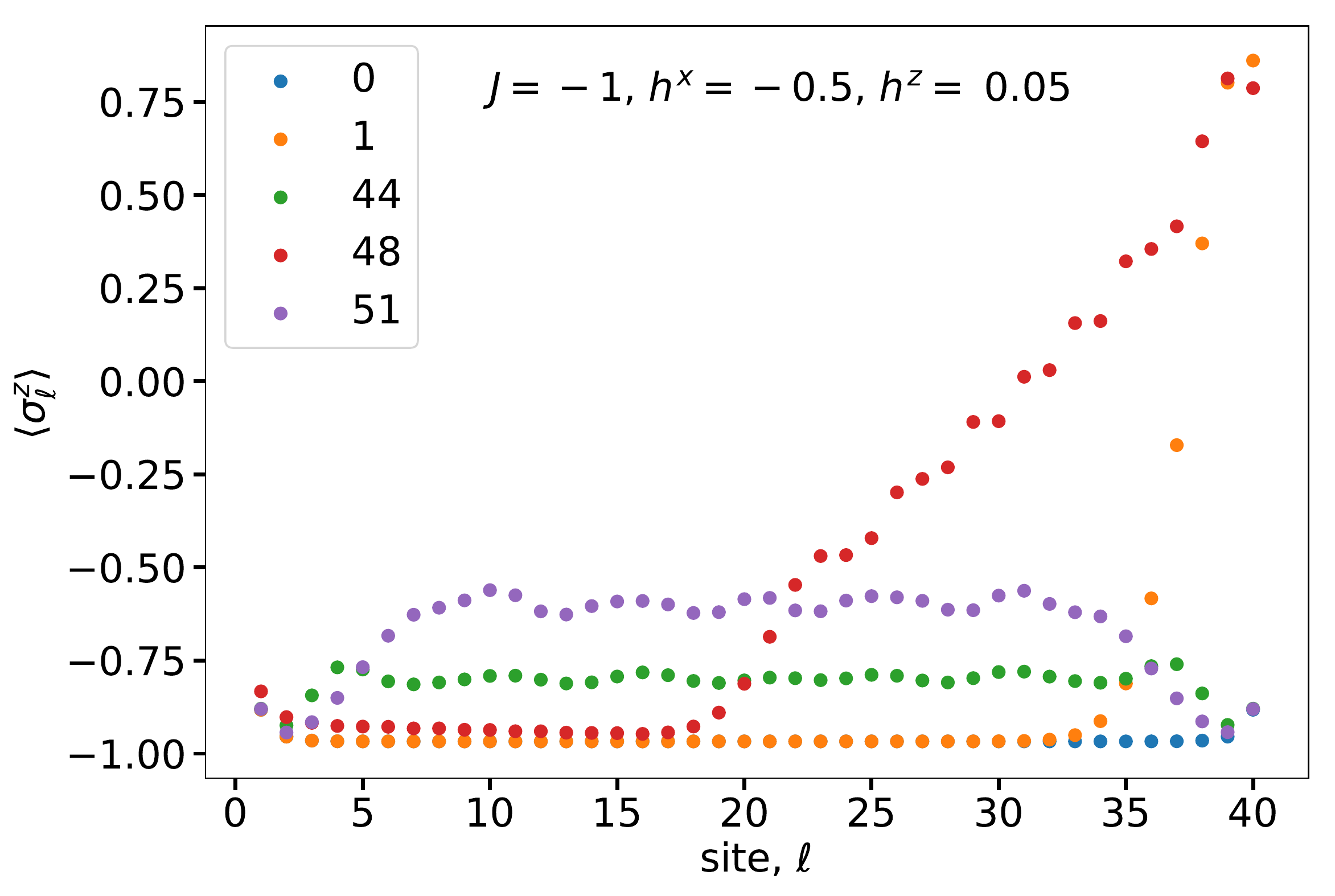}\\
\caption{The spin expectation value $\langle \sigma_\ell^z \rangle$, on each site $\ell =1,2,\cdots,40$, for the 0th (ground), 1st, 44th, 48th and 51st states in the spectrum as found using DMRG.
 These states are marked in Fig.~\ref{Fig:LatticeExpectationValues}(c).
 The 1st and 48th states are examples of mesons localized on the boundary.}
\label{Fig:lattice_mag}
\end{figure}

In Fig.~\ref{Fig:LatticeExpectationValues} we present data from our lattice simulations for~\eqref{lattice} with $J=-1$, $h^x = -0.5$ and $h^z = 0.05$ for three system sizes. This should be compared to analogous Figs.~\ref{Fig:EEVs} in the field theory; the similarity between lattice and field theory calculations is rather striking. We see a clear band of nonthermal states above a multiparticle `continuum' (this appears rather discrete due to the relatively small number of sites, $N=20-40$, but would broaden into a continuum in the large $N$ limit). It is worth noting that in the DMRG simulations on an open system, the meson excitations are localized in the vicinity of the boundaries due to the decreased energy cost of such excitations there. We plot the local magnetization $\la \s^z_l\ra$ in a number of representative states in Fig.~\ref{Fig:lattice_mag}; the corresponding states are labeled in Fig.~\ref{Fig:LatticeExpectationValues}(c).

It is also worth noting that we are only able to access the low-energy part of the spectrum. We will see in the following section that this part of the spectrum is well converged as a function of system size $N$, being representative of the thermodynamic limit -- implying that the low-energy meson states remain well-separated from the continuum in the infinite volume limit. We cannot, however, rule out that the meson states melt into the continuum with increasing energy (although, we note, there is no evidence of this for the states that we can construct). Speculatively, such behavior in the scaling limit could be caused by the presence of marginal or irrelevant operators (neglected in the field theory Hamiltonian~\eqref{FT}), which are difficult to incorporate into a truncated spectrum treatment. Further investigations of the lattice model, which is already known to exhibit anomalous dynamics~\cite{banuls2011strong,kormos2016realtime,lin2017quasiparticle,mazza2018suppression}, are undoubtedly warranted. 

\subsubsection{Comparison between the finite size scaling of the EEV spectrum on the lattice and in the continuum}
\label{sec:EEVRdep}

As in the continuum, we can ask to what extent the rare states on the lattice persist to large volumes. To do so, we consider the finite size scaling of the low-energy EEV spectrum, which we also compare to that in the continuum. In order to make a like-for-like comparison between results at different system sizes, in this part of the appendix we focus on the \textit{total} magnetization of the eigenstates relative to the ground state (i.e., the number of flipped spins in the eigenstate) as a function of energy (relative to the ground state) of the eigenstate. We focus on total magnetization to avoid obvious problems with scaling with volume of local magnetization: the states we construct have $O(1)$ flipped spins compared to the ground state and hence in the thermodynamic limit have the same magnetization \textit{density} as the ground state. 

\begin{figure}
\includegraphics[width=0.45\textwidth]{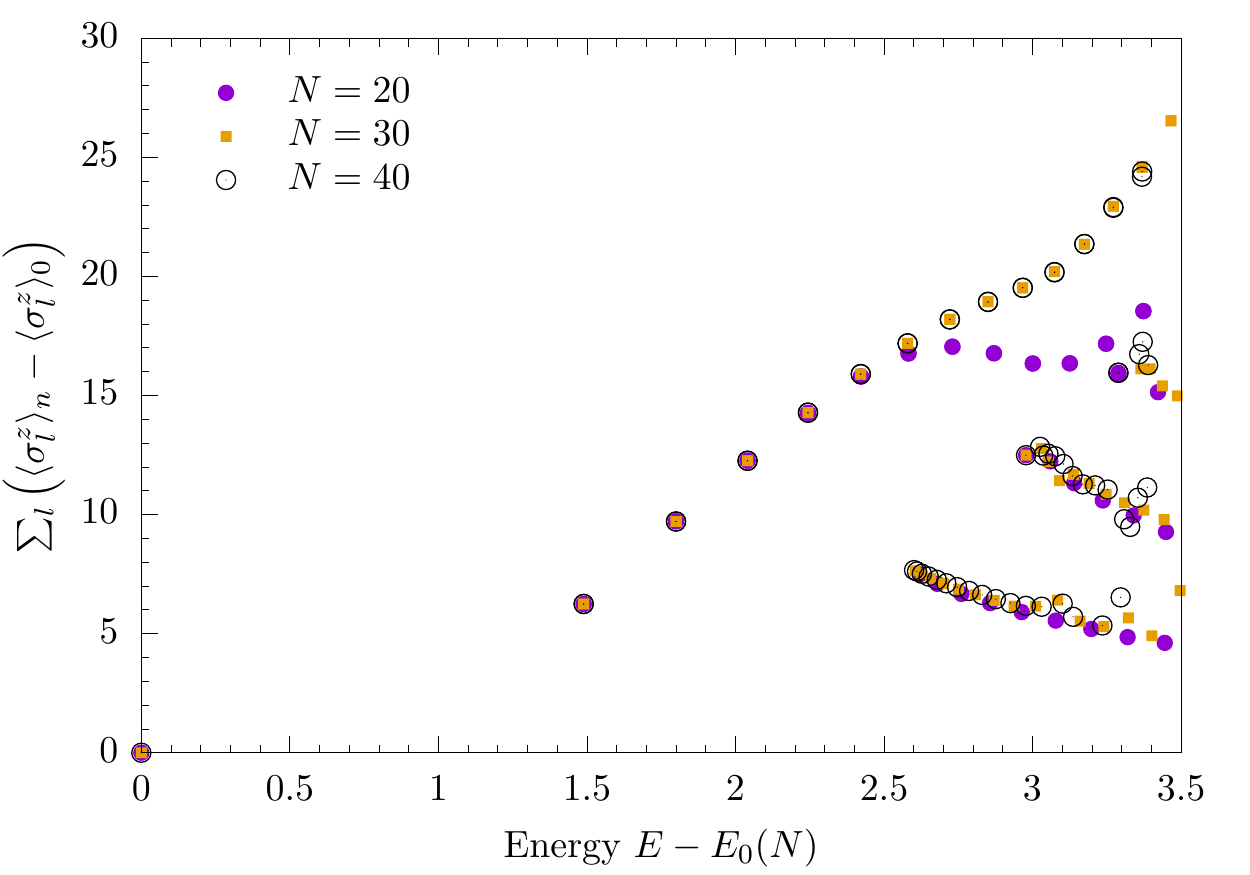}
\caption{The EEV spectrum obtained by DMRG on the finite lattice of $N$ sites for the Hamiltonian~\eqref{lattice} with $J=-1$, $h^x=-0.5$ and $h^z = 0.05$. Here we consider the \textit{total magnetization} of the eigenstates relative to the ground state, as a function of energy (relative to the ground state energy $E_0$). See Fig.~\ref{Fig:LatticeExpectationValues} for further details.}
\label{Fig:Rconvlatt}
\vspace{3mm}
\includegraphics[width=0.45\textwidth]{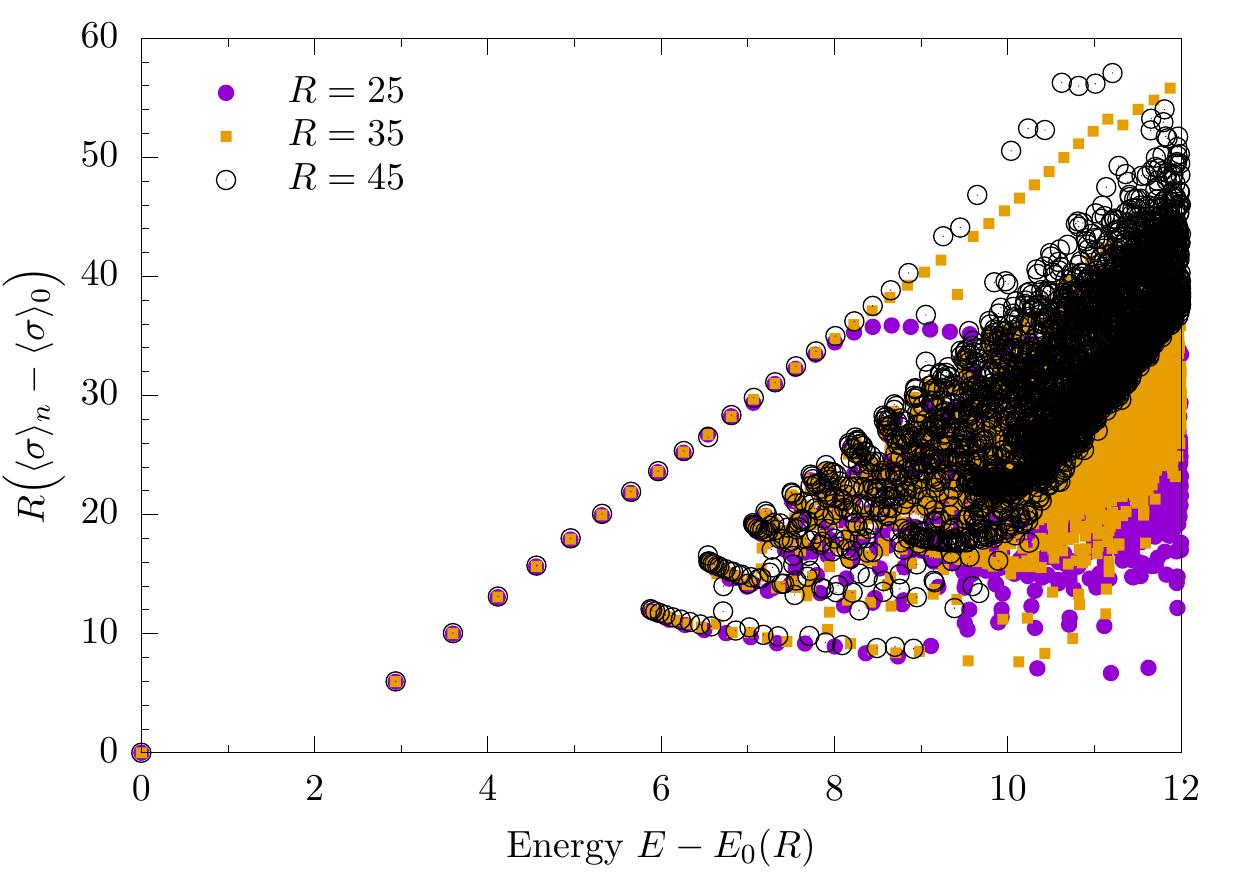}
\caption{The EEV spectrum for $m=1$, $g=0.1$ in systems of sizes $R=25-45$ (see Figs.~\ref{Fig:EEVs} for details). Here we plot the \textit{total} magnetization (relative to the ground state) as a function of energy (relative to the ground state). We see that the meson states are well converged as a function of system size $R$, while the thermal continuum is qualitatively converged (in the sense that the features are correct, but the density of state increases with increasing $R$). Note the striking similarity to Fig.~\ref{Fig:Rconvlatt}.}
\label{Fig:Rconv}
\end{figure}

With this in mind, the results of Fig.~\ref{Fig:LatticeExpectationValues} are translated into Fig.~\ref{Fig:Rconvlatt} (the corresponding continuum results, Fig.~\ref{Fig:EEVs}, become Fig.~\ref{Fig:Rconv}). We see that low energy EEV spectrum at different lattice sizes $N$ (system sizes $R$) is well converged: meson state EEVs are well converged and match for different values of $N$ ($R$). The thermal continuum is also qualitatively well converged at low energies, in the sense that the features are correct, but the density of states increases with increasing $N$ ($R$).  

In Figs.~\ref{Fig:Rconvlatt} and~\ref{Fig:Rconv}, it is nevertheless easy to see that the meson states in the low-energy part of the spectrum remain well-separated from the thermal continuum, and results are converged to the thermodynamic limit, being independent of $N$ and $R$. 

\subsection{Eigenstate entanglement properties}

\begin{figure}
  \begin{tabular}{l}
    (a) \\
    \includegraphics[width=0.45\textwidth]{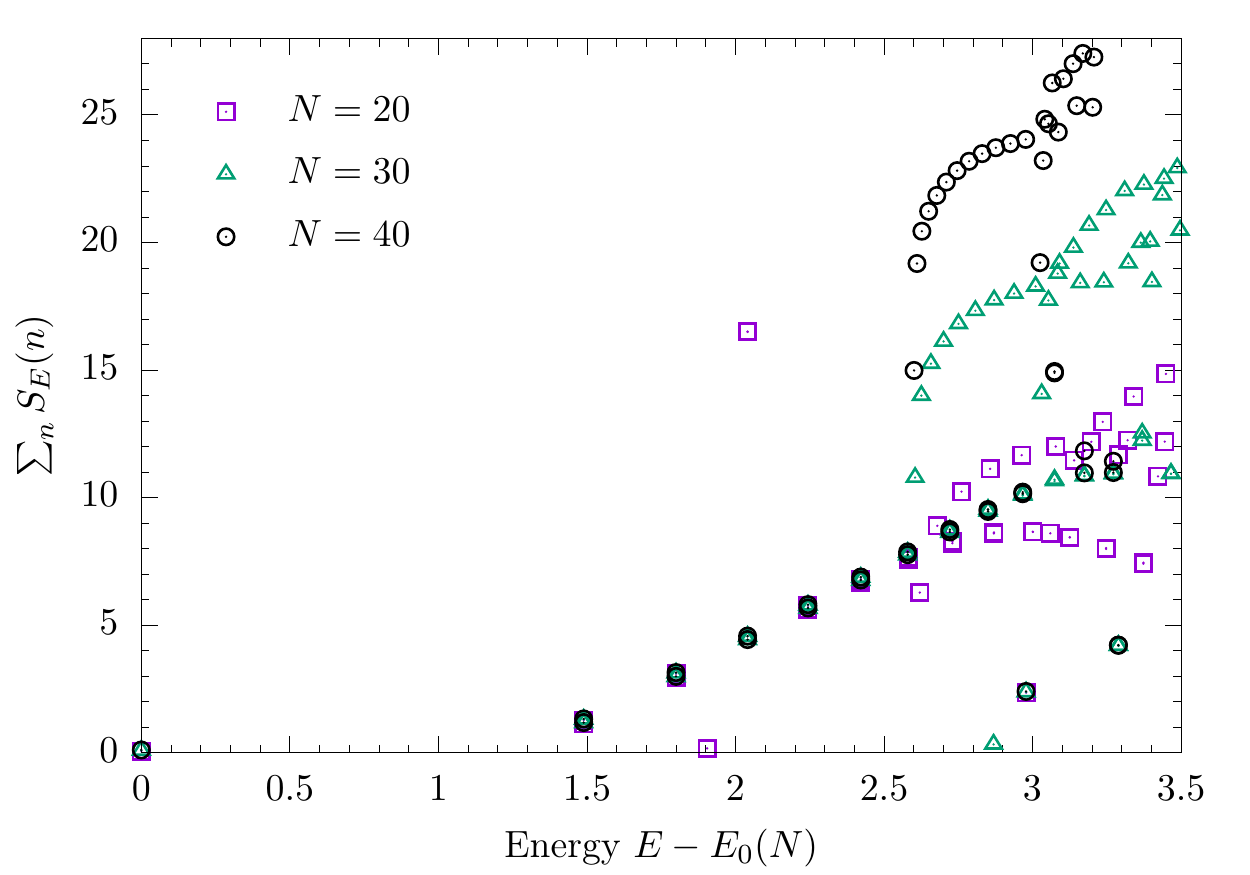} \\
    (b) \\
    \includegraphics[width=0.45\textwidth]{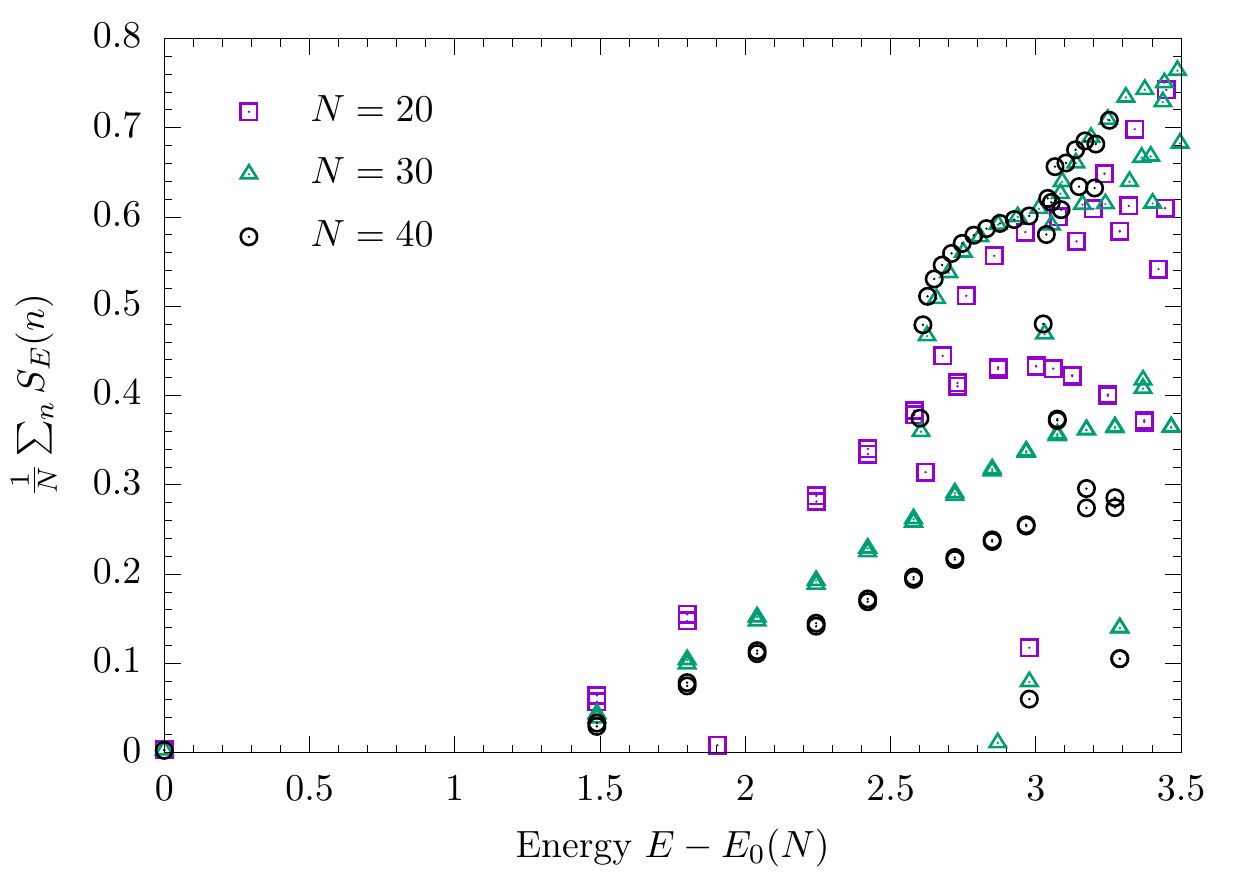}
  \end{tabular}
  \caption{(a) The entanglement entropy averaged over bipartitions of the system, $\sum_n S_E(n)$, in each eigenstate of energy $E$ relative to the ground state, for the lattice model ~\eqref{lattice} with $J=-1$, $h^x=-0.5$ and $h^z = 0.05$. Data is presented for three different system sizes $N$. We see that there are two classes of states: those that have $N$-independent entanglement (i.e. area law states) and those with volume law entanglement (linear in $N$). The presence of volume law entanglement states is shown more clearly in (b), where we scale the bipartition averaged entanglement by dividing through by the volume $N$. The nonthermal states have area law entanglement, while the thermal states have volume law.}
  \label{fig:SE}
\end{figure}

We finish our consideration of the lattice model with a study of the entanglement properties of the states constructed within DMRG. Much like with the EEV spectrum, we compute the entanglement entropy under a real space bipartition of the system, and average over all bipartitions to avoid probing real space structure of the states. That is, if $S_E(n)$ corresponds to the bipartition of the system on bond $n$, we plot $\sum_n S_E(n)$. Results are presented in Fig.~\ref{fig:SE}.

In our results, we see that there are two classes of states within the low-energy spectrum, which display different scaling with system size $N$. Beginning at the lowest energies and persisting through the spectrum, there are states whose entanglement is (essentially) $N$-independent -- so-called area law states. These are in one-to-one correspondence with the nonthermal states seen in the EEV spectrum. These are clear to see in the first panel, Fig.~\ref{fig:SE}(a). On the other hand, we have states whose entanglement varies with $N$, volume law states, which correspond to those states in the thermal continuum of the EEV spectrum. This is more clearly seen in Fig.~\ref{fig:SE}(b), where we have scaled the y-axis by the volume $N$. We thus see that the nonthermal meson states have properties, from the viewpoint of entanglement, that are very different from the thermal continuum. It seems reasonable to expect that the same dichotomy occurs in the field theory, where it is difficult to compute such entanglement measures. 

\section{Conclusions} 
\label{Sec:Conclusions}

ETH is key to understanding whether quantum systems thermalize. When ETH is valid, and matrix elements of operators in the eigenbasis satisfy Eq.~\eqref{eth}, expectation values of operators within an eigenstate are thermal, and dynamics following a quantum quench are expected to lead to thermalization. In this work, we have shown an explicit example of a nonintegrable model, Eq.~\eqref{FT} and its lattice regularization Eq.~\eqref{lattice}, where ETH is violated. This is signaled through the presence of rare states in the spectrum, with expectation values within these states being nonthermal. 

To show this, we used TSMs and explicitly construct the low-energy spectrum of the theory. With the eigenstates at hand, we established the nature of the rare states that violate ETH: they arise as a direct result of confinement, corresponding to the well-known ``meson'' states: single particle excitations that can have extensive energy ($E\propto R$). These excitations are formed from pairs of linearly confined ``domain wall fermions'' and remain kinematically stable far above the threshold of the multiparticle continuum. This surprising result can, however, be understood by means of perturbative calculations of the meson lifetime and energy corrections, combining both a standard Bethe-Salpeter analysis for the mesons with a perturbative treatment of the meson-to-four-fermion vertices. This reveals only a very weak hybridization of the meson with the multiparticle continuum. This implies that just slight dressing of the two-fermion excitation can render the excitation absolutely stable. A finite size scaling TSM analysis is consistent with the meson states remaining nonthermal in the infinite volume limit, see Figs.~\ref{Fig:fss} and~\ref{Fig:Rconv}.

With rare states established within the spectrum of the model through the study of EEVs, we turned our attention to understanding their influence on nonequilibrium dynamics. We first showed, through construction of the diagonal ensemble, that certain quantum quenches exhibit an absence of thermalization in the infinite time limit. This is in spite of the fact that our model is nonintegrable. Subsequent studies of the real-time nonequilibrium time evolution of observables revealed that rare states have EVs that show large amplitude, low frequency oscillations that appear not to decay on relatively long time scales $t\sim 100|m|^{-1}$. This should be contrasted to thermalizing states, which rapidly relax to their thermal values, showing only small fluctuations about their diagonal ensemble value. These difference may help diagnose rare states in experiments on, e.g., cold atoms. 

Finally, we addressed whether the violation of ETH in~\eqref{FT} is related to the scaling limit. To do so, we considered the lattice regularization of the model (e.g., the spin chain whose scaling limit gives the field theory) explicitly away from the scaling limit. Using DMRG we constructed the low-energy spectrum and present the EEV spectrum (at low energies). This has striking similarities to the EEV spectrum in the scaling limit, including the presence of a band of rare states above the multiparticle continuum. This finding suggests that confinement-induced rare states are not a remnant of the scaling limit, and supports previous results (see e.g. Ref.~\cite{kormos2016realtime}) that ascribed anomalous nonequilibrium dynamics to the presence of rare states in the spectrum. They may also be responsible for a lack of transport observed in time-evolution of the domain wall initial state~\cite{mazza2018suppression}. We do note, however, that we were unable to probe energies far into the spectrum and could not rule out that on the lattice such states melt into the continuum at finite energy densities.

The rare states in the field theory~\eqref{FT} are intimately related to the presence of a linearly confining potential for the domain wall excitations. We note that recent work on confined phases in holographic theories also suggests an absence of thermalization~\cite{myers2017holographic}, which may lead one to speculate that models with confinement in general exhibit a lack of thermalization. This is partially supported by recent results of the authors~\cite{james2018nonthermal} that show anomalous nonequilibrium dynamics in a two-dimensional model with confinement that are completely analogous to the one-dimensional problem~\cite{kormos2016realtime}. It would be interesting to test this conjecture in other theories, such as the Schwinger model~\cite{buyens2014matrix,buyens2016confinement,banuls2016chiral,buyens2017realtime,buyens2017finiterepresentation,banuls2017density,nandkishore2017many,akhtar2018symmetry} or the $q$-state Potts model~\cite{lencses2015confinement,rutkevich2017radiative}, with the view that this may have important consequences in the context of quantum chromodynamics (which has some parallels with the quantum magnetism discussed here~\cite{sulejmanpasic2017confinement}). 

We finish by noting that recent works suggest that other kinetic constraints, beyond those provided by confinement, can also lead to nonthermal behavior~\cite{ates2012thermalization,ji2013equilibration,vanhorssen2015dynamics,lan2018quantum,turner2017quantum,khemani2018signatures,ho2018periodic,lin2018exact,turner2018quantum,choi2018emergent}. This may have important implications for experiments on Rydberg gases~\cite{bernien2017probing}.

\acknowledgments{
We gratefully acknowledge useful conversations with Mari Carmen Ba\~nuls, Bruno Bertini, Mario Collura, Axel Cort\'es Cubero, Fabian Essler, Alessio Lerose, Dante Kennes, M\'arton Kormos, Andreas L\"auchli, Ian Mondragon-Shem, Marcos Rigol, and Gabor Tak\'acs. Parts of this work made use of ChainAMPS~\cite{james2013understanding,james2015quantum,james2018nonperturbative} and ALPS~\cite{albuquerque2007alps,bauer2011alps} routines.

Work at Brookhaven National Laboratory was supported by the U.S. Department of Energy, Office of Basic Energy Sciences, under Contract No. DE-SC0012704. A.J.A.J. acknowledges support from the Engineering and Physical Sciences Research Council, grant number EP/L010623/1. N.J.R. is supported by the EU's Horizon 2020 research and innovation programme under grant agreement No 745944.  

N.J.R., A.J.A.J. and R.M.K. acknowledge the Simons Collaboration Programme entitled ``The Nonperturbative Bootstrap'' as part of the ``Hamiltonian methods in strongly coupled Quantum Field Theory'' meeting at the IHES Universit\'e Paris-Saclay. N.J.R. and R.M.K. also acknowledge the support of the Erwin Schr\"odinger International Institute for Mathematics and Physics, University of Vienna, during the ``Quantum Paths'' program. N.J.R. would further like to acknowledge the Aspen Center for Physics (supported by NSF grant PHY-1066293), the Simons Center for Geometry and Physics at SUNY Stony Brook, and University College London for hospitality during parts of this work.  
}
\appendix

\section{Convergence of the TSM}
\label{App:ConvTCSA}

Truncated spectrum methods can be used to construct approximate eigenstates of a field theory in a nonperturbative manner~\cite{james2018nonperturbative}. The technique hinges on studying `known' theories that are perturbed by a relevant operator, where one uses the eigenstates of the `known' theory as an efficient computational basis. This basis is truncated by the introduction of an energy cutoff, which is motivated by properties of relevant operators: they strongly mix low energy degrees of freedom, while essentially leaving the high energy ones unperturbed. 

At a practical level, this can still remain a difficult problem. Even with the introduction of a cutoff, there is a rapid growth of the Hilbert space with increasing cutoff, limiting the truncation energies that one can reach. In the first part of this appendix, we examine the convergence of results with increasing energy cutoff at fixed system size $R$. Scaling with $R$ is discussed in the main body of the text.  We finish by examining the convergence of expectation values in the diagonal and microcanonical ensembles. 

\begin{figure}[t]
\includegraphics[width=0.45\textwidth]{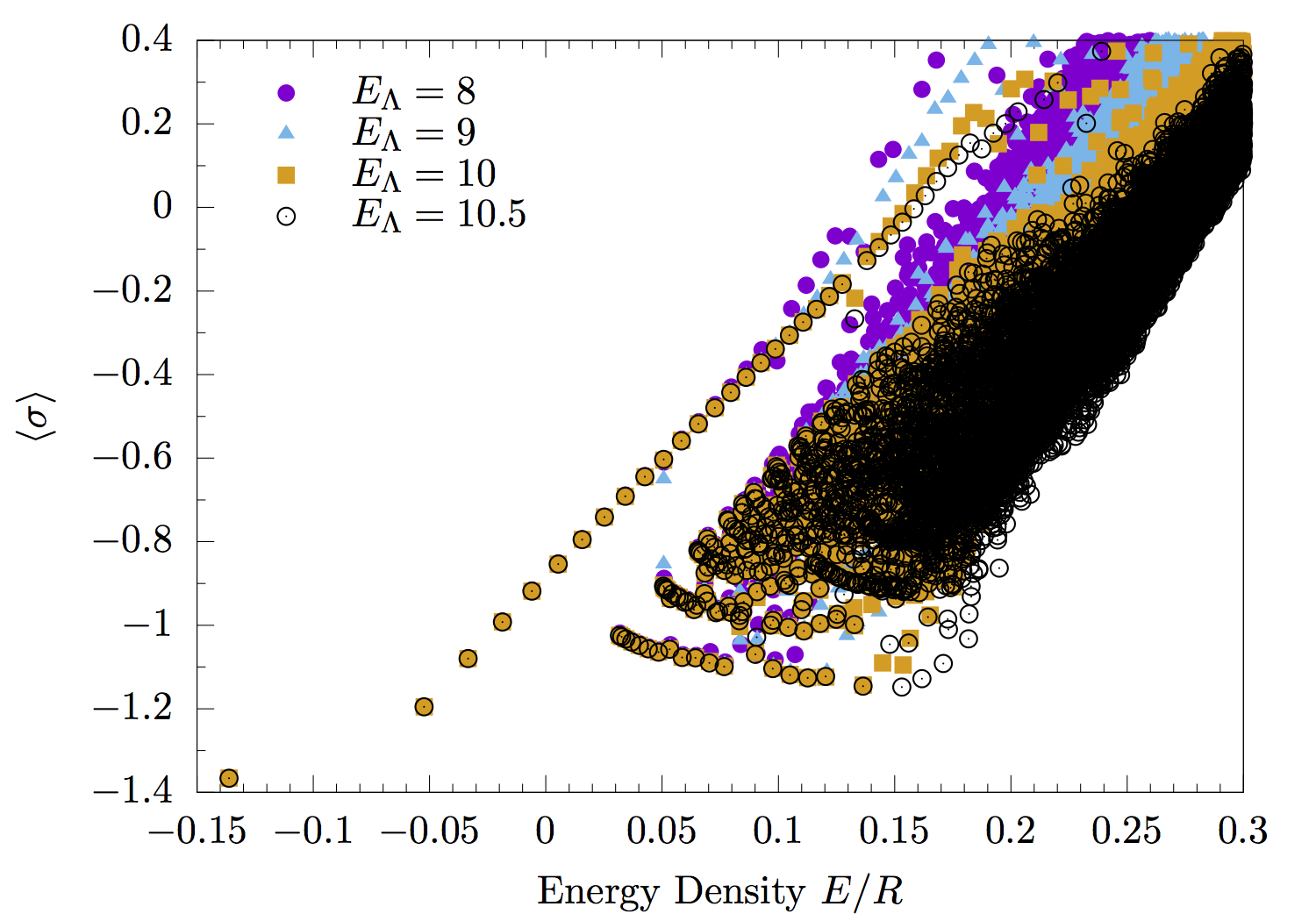} \\
\caption{The eigenstate expectation value spectrum of the local spin operator $\s(0)$ in~\eqref{FT} with $m=1$, $g=0.1$, $R=35$, for four values of the cutoff $E_{\Lambda}$.}
\label{Fig:EEV_conv}
\end{figure}

\subsection{With cutoff $E_\Lambda$}

We begin by discussing the convergence of TSM results with increasing energy cutoff $E_\Lambda$ at fixed system size $R$. Low energy eigenstates of the field theory~\eqref{FT} are constructed for $m=1$, $g=0.1$ on a ring of size $R=35$. In Fig.~\ref{Fig:EEV_conv} we show the EEVs of the local spin operator $\s(0)$ as a function of energy density of the eigenstate $E/R$ for four values of the cutoff $E_\Lambda = 8,\ 9,\ 10,\ 10.5$. We note that $E_\Lambda=10.5$ corresponds to $23500$ states in the truncated Hilbert space~\footnote{Throughout, we construct the free fermion basis with states containing at most ten fermions. In the case of $R=35$, $E_\Lambda = 10.5$ for $m=1$, the maximum number of fermions within states below the cutoff energy is eight.}.

As expected from the ideas underpinning TSMs, the energies of the low-energy eigenstates are well converged. With increasing energy density, convergence of the eigenstate energies decreases, with the highest constructed states showing behavior that is dominated by cutoff effects. We see that energy densities up to $E/R \sim 0.15$ are reasonably well converged by $E_\Lambda = 10.5$. Focusing on the very lowest states, we see that energies are very well-converged: to four or more decimal places. At reasonable energies densities $E/R \lesssim0.2$ we see that the behavior at all values of the cutoff is qualitatively converged -- while energies of states within the thermal continuum may evolve slightly, it does not significantly change the shape of the plot at low-energy densities (cutoff-dependent behavior is clearly seen on the right hand side of the plot).

When examining the expectation value of the spin operator in an eigenstate of energy density $E/R$, we see similar behavior to the convergence of the energy. Low energy EEVs are well converged, with convergence becoming poorer at higher energies, and clearly cutoff dependent behavior occurring towards the top of our energy scale. We note, however, that qualitative behavior does not change with increasing cutoff: the multiparticle continuum remains well separated from the band of meson states, with this gap being robust to increasing cutoff.

We note that, as can be seen clearly in the rare states on the right hand side of the plot, generally the expectation value of the spin operator converges more slowly with increasing energy cutoff than the energy of the eigenstate. This is evident from the EEVs moving downwards on the plot more than they move to the left.

\subsection{The diagonal ensemble}

The diagonal ensemble (DE) is assumed to describe the long-time limit of expectation values after a quantum quench. This can be motivated by considering off-diagonal matrix elements: in the long time limit these oscillate rapidly and average to zero. We we consider the DE result for the local spin operator $\s(0)$, Eq.~\eqref{DE}, 
\begin{align}
\lim_{t\to\infty} &\la E_m | e^{iH_f t} \s(0) e^{-iH_ft} |E_m\ra = \nonumber \\
&\sum_n |\tilde c_{m,n}|^2 \la \tilde E_n | \s(0) |\tilde E_n\ra. \label{DEspin}
\end{align}
Here $|E_m\ra$ are the \textit{prequench} eigenstates, $|\tilde E_n\ra$ are \textit{postquench} eigenstates (eigenvectors of the Hamiltonian performing the time evolution) and $c_{m,n} = \la E_m | \tilde E_n \ra$ are the overlap coefficients. 

From Eq.~\eqref{DEspin}, it is not \textit{a priori} obvious that we should expect accurate results when constructing the DE from the TSM: after all, it sums over \textit{all} eigenstates obtained in the procedure, and we know many of these will not be well converged. Thus it is important to check in what sense the DE from TSMs is correct.  

\begin{figure}[t]
\includegraphics[width=0.45\textwidth]{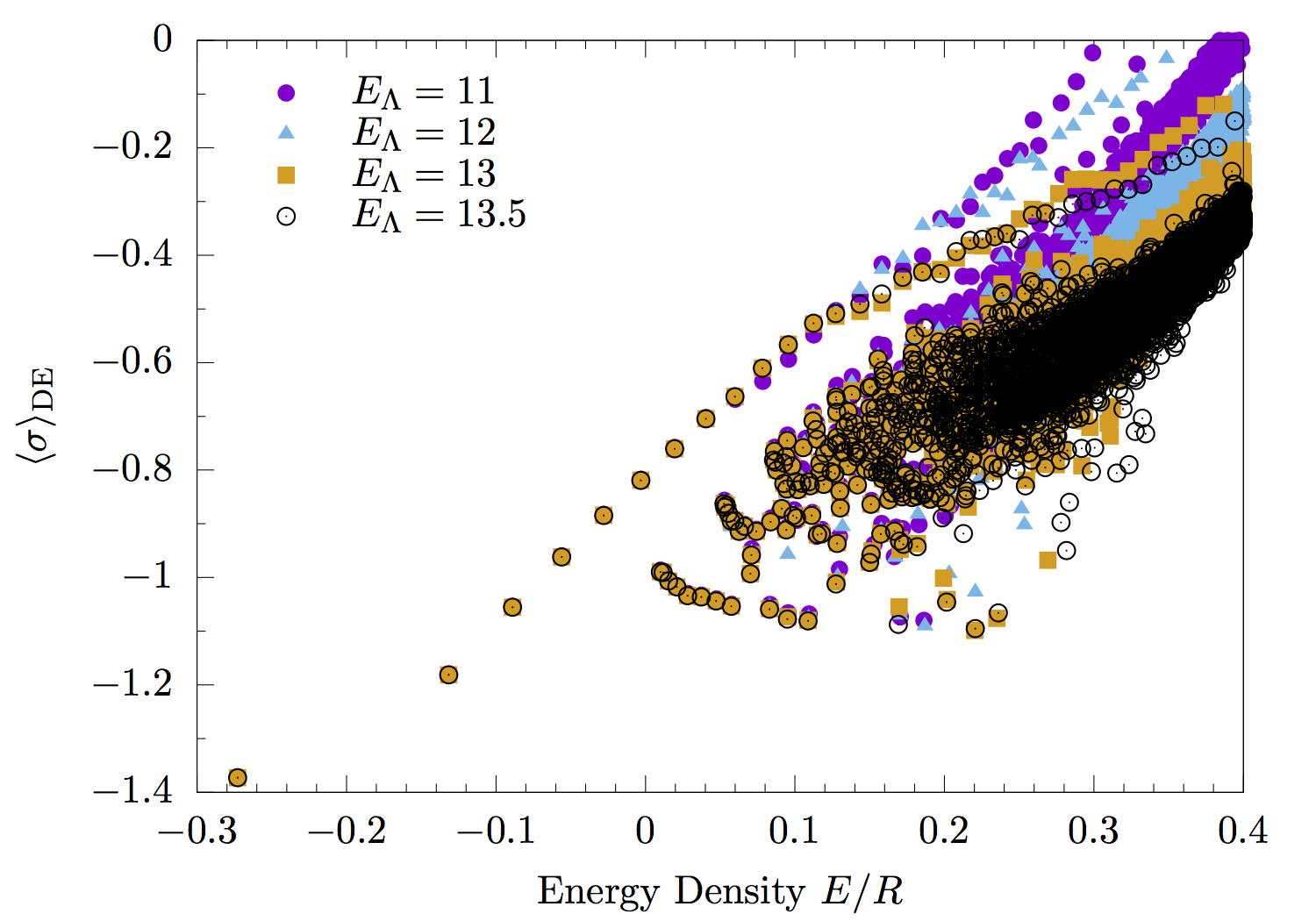}
\caption{The diagonal ensemble (DE) result for expectation values following a quench of the longitudinal field $g_i = 0.1 \to g_f = 0.2$ in the field theory~\eqref{FT} with $m=1$ and $R=25$ for four values of the energy cutoff. Free fermion basis states containing up to ten particles are considered.}
\label{Fig:DEconv}
\end{figure}

We present the DE result for $\s(0)$ following the quench $g_i \to g_f$ with both $g_i, g_f > 0$ in Fig.~\ref{Fig:DEconv}. In particular, our interest is in convergence of the result with increasing energy cutoff $E_\Lambda$ (we fix $g_i = 0.1$, $g_f=0.2$ and $R=25$). We see, once again, that the rare states are robustly separated from the multiparticle continuum, and there is qualitatively consistent behavior even at intermediate energy densities, for different values of the cutoff. At high energies, as with the EEV spectrum, results are dominated by cutoff effects. 

\subsection{The microcanonical ensemble}

\begin{figure}
\includegraphics[width=0.45\textwidth]{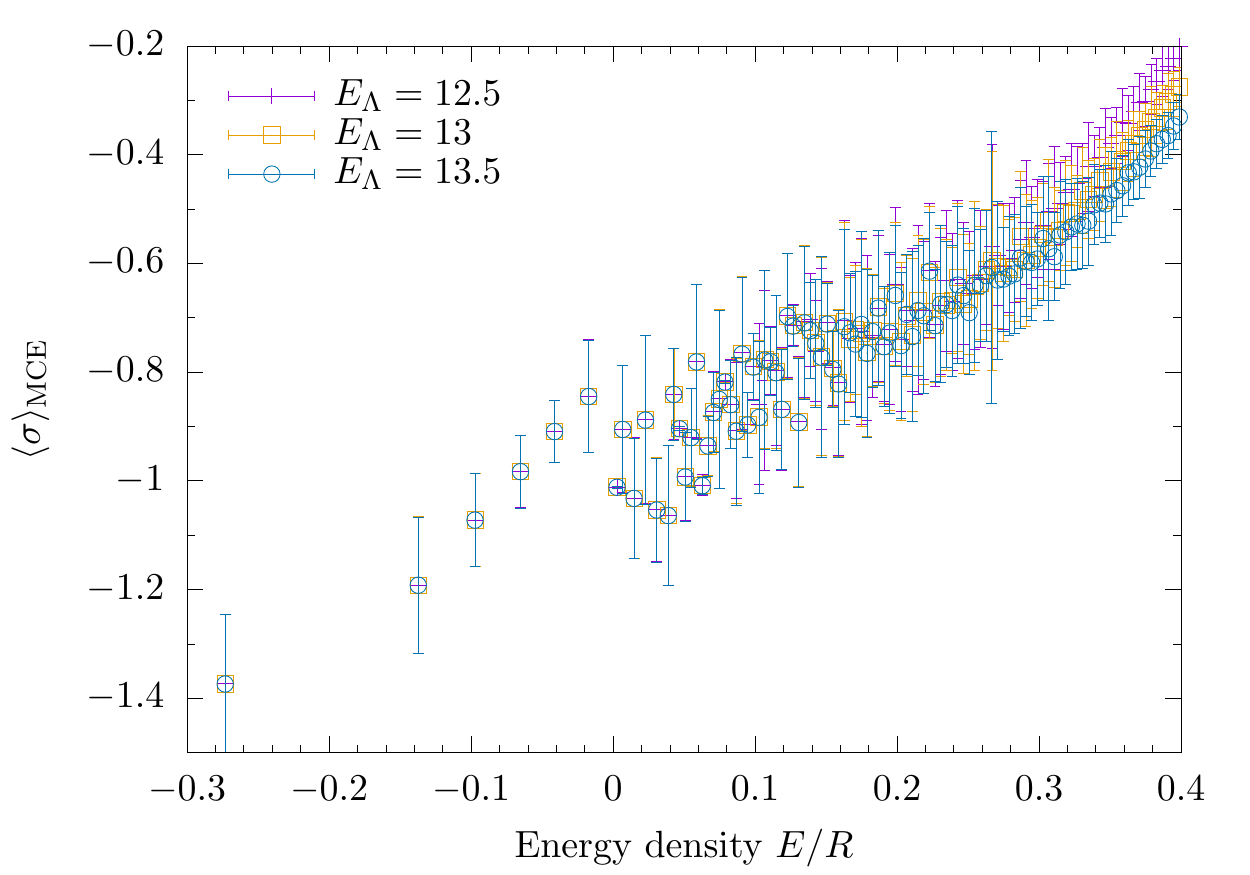}
\caption{The MCE prediction for $\la \s(0)\ra$ at a given energy density in the field theory~\eqref{FT} with $m=1$, $g=0.2$ and $R=25$. Error bars denote the standard deviation of the averaged data for the window $\Delta E = 0.1$.}
\label{Fig:MCEconv}
\end{figure}

With the convergence of the DE established, we now consider the other ensemble used in this work, the microcanonical ensemble (MCE). As defined in Eq.~\eqref{Eq:MCE}, this is constructed by averaging over all eigenstates within a given energy window $\Delta E$. The convergence of the MCE prediction for $\s(0)$ with increasing energy cutoff $E_\Lambda$ is shown in Fig.~\ref{Fig:MCEconv} for the field theory~\eqref{FT} with $m=1$, $g=0.2$ and $R=25$. We see that the MCE result (plus one standard deviation error bars) is well converged up to relatively high energy densities. This seems to reflect the fact that, even if individual EEVs are moving around a little with energy cut-off, their unweighted average is not significantly changing. Once again, at high energy results are dominated by cutoff effects, as would be expected from the EEV spectrum.  

\section{$N$-particle weights}
\label{App:Weights}

\begin{figure*}[p]
\begin{tabular}{lll}
(a) & & (b) \\
\includegraphics[height=0.33\textwidth]{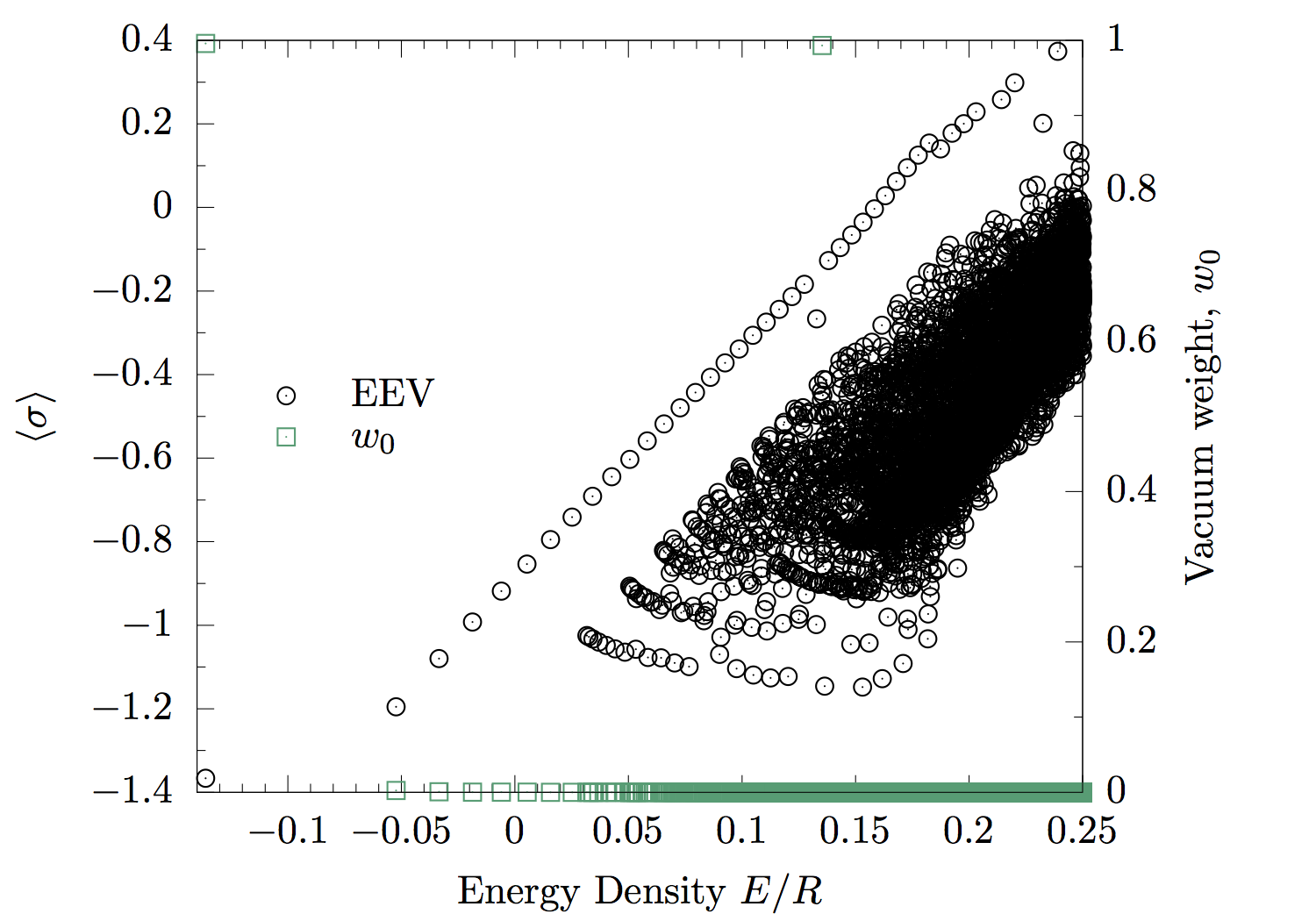} & & 
\includegraphics[height=0.33\textwidth]{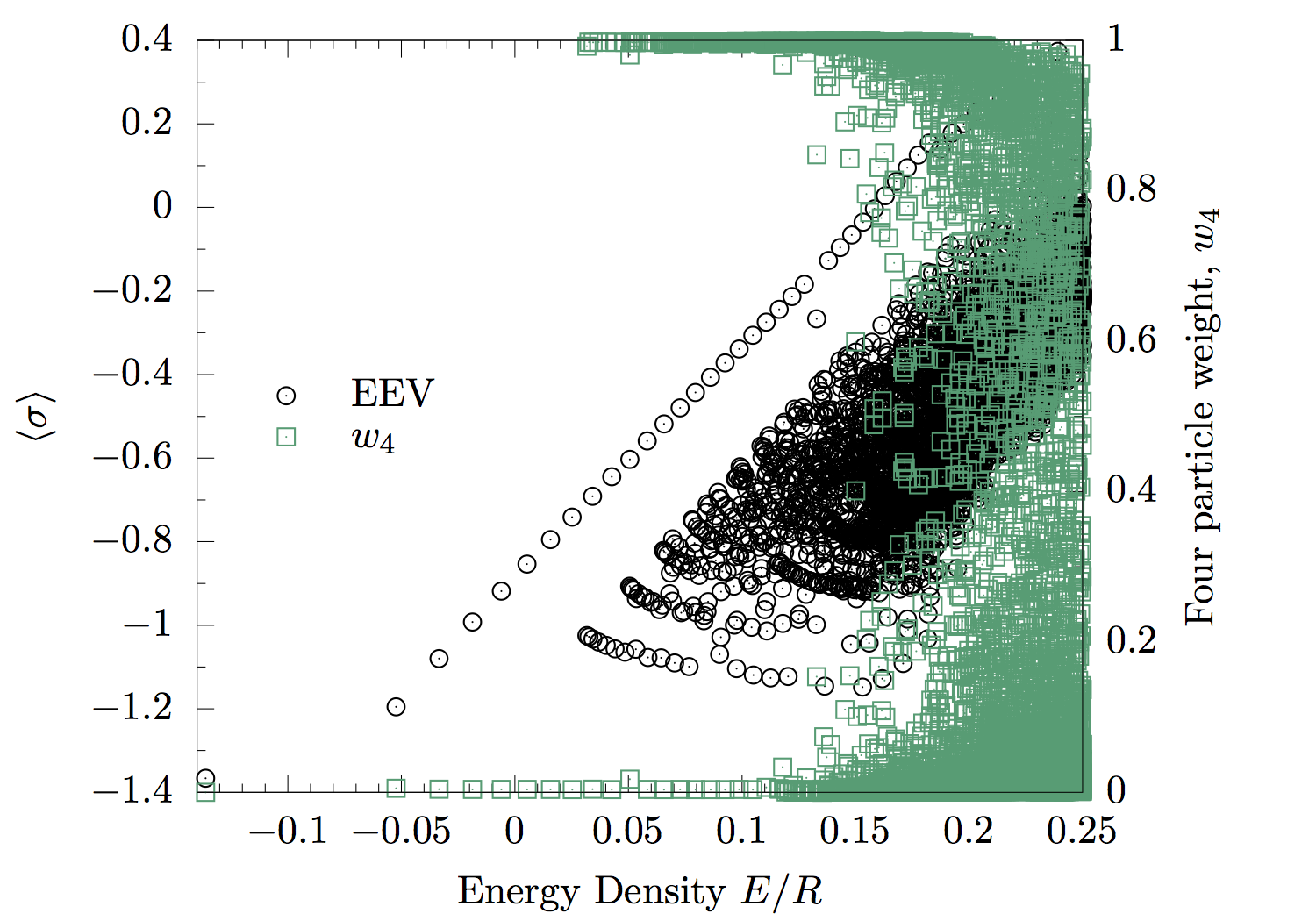} \\
(c) & & (d) \\ 
\includegraphics[height=0.33\textwidth]{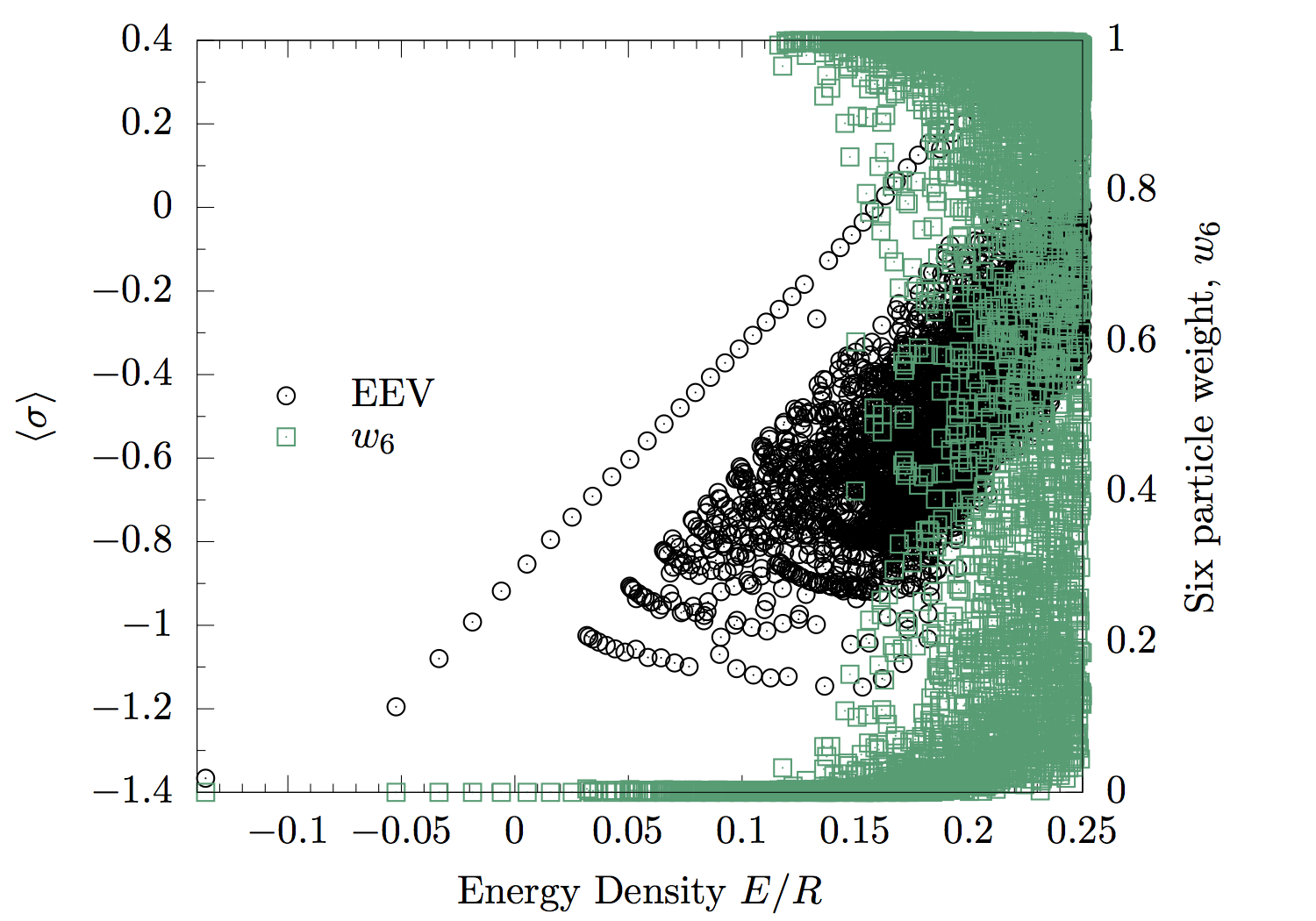} & & 
\includegraphics[height=0.33\textwidth]{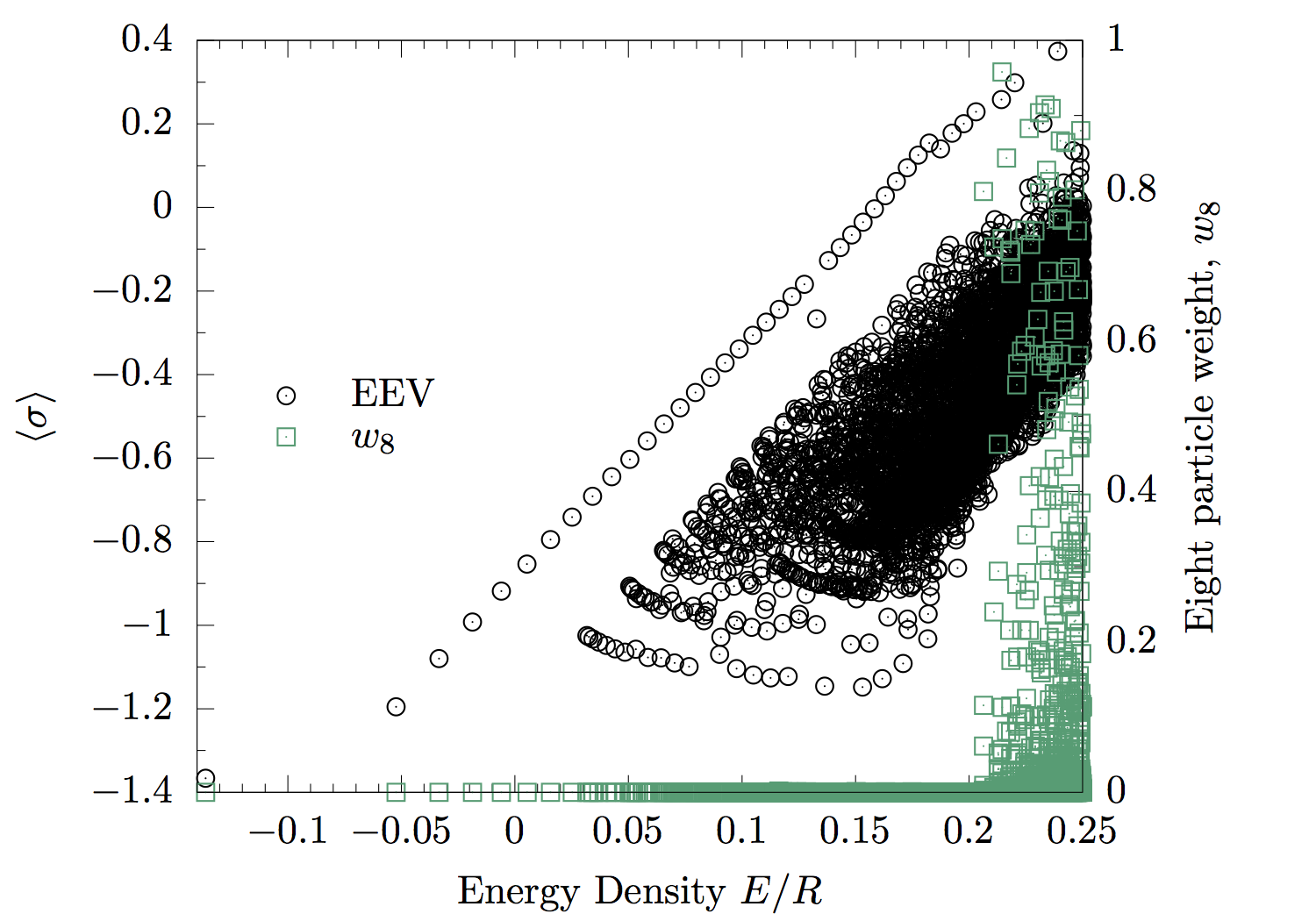} \\
(e) & & \\
\includegraphics[width=0.24\textwidth]{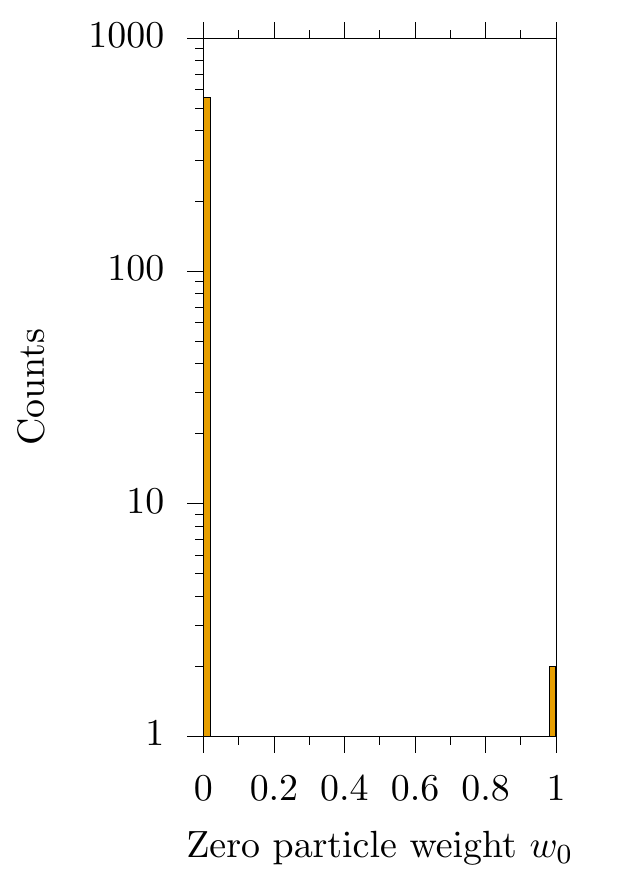} 
\includegraphics[width=0.24\textwidth]{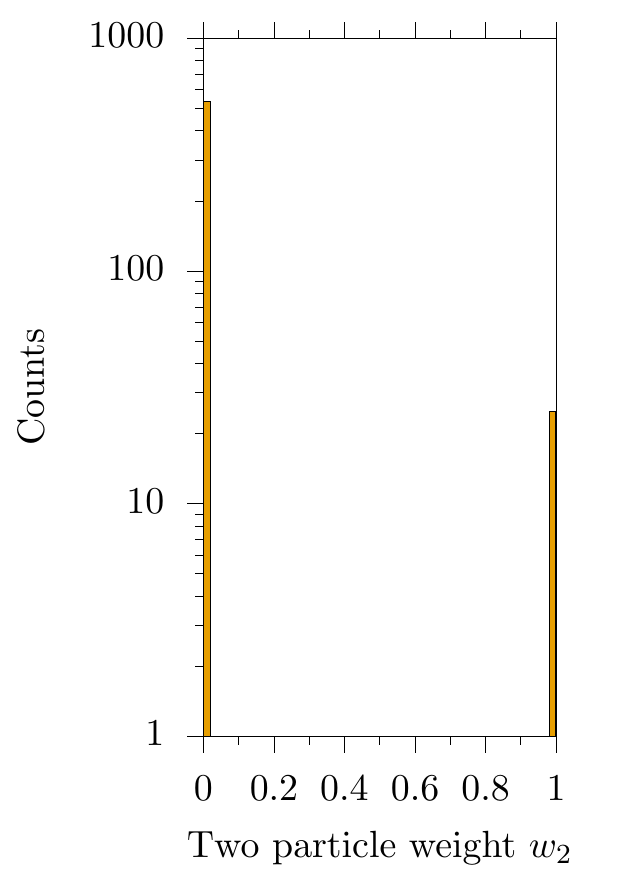} &&
\includegraphics[width=0.24\textwidth]{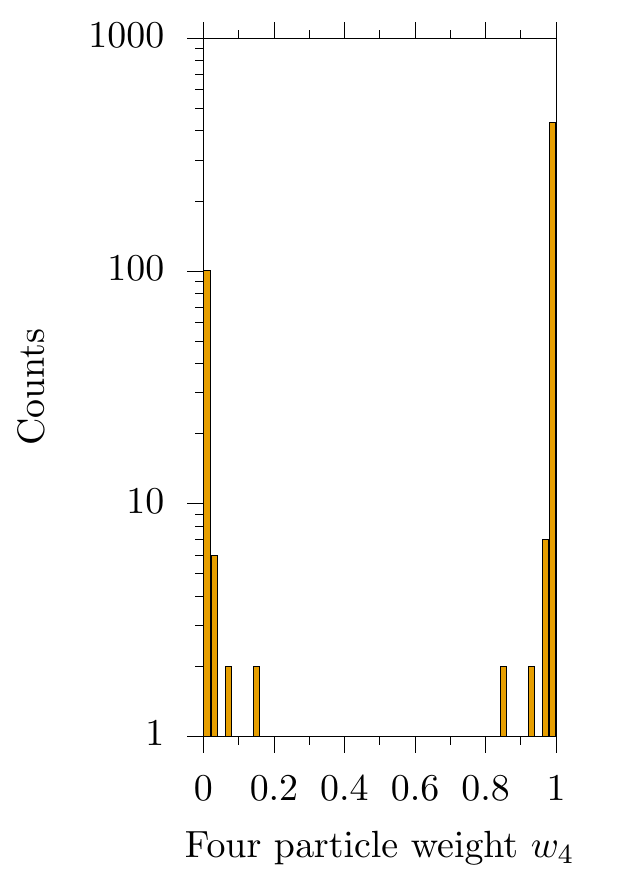} 
\includegraphics[width=0.24\textwidth]{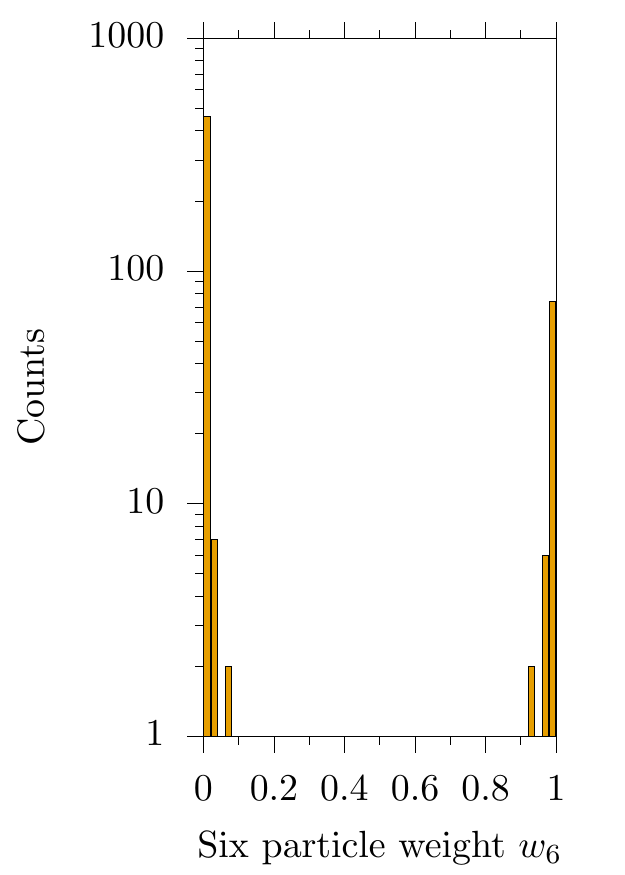}
\end{tabular}
\caption{The weights (a) $w_0$; (b) $w_4$; (c) $w_6$; (d) $w_8$. Eigenstates were constructed with the TSM on the length $R=35$ chain with $m=1$, $g=0.1$ and an energy cutoff $E_{\Lambda}=10.5$, with up to $10$ free fermions in the basis states. These plots illustrate the approximate low-energy U(1) symmetry discussed in the main text: below an energy density of $E/R \sim 0.15$, the weight for each state are either $w_N \approx 1$ or $w_N \approx 0$. This is emphasized in (e), where we show the distribution of the weights for eigenstates with energy density $E/R < 0.15$, where bins are of width $\Delta w_N = 0.02$. We see almost all states have weights close to zero or close to one, with the histograms being bimodal (note the logarithmic $y$-axis scale. The issue of where this approximate U(1) symmetry breaks is tough to address, with cutoff effects dominating results at higher energies.}
\label{Fig:Weights}
\end{figure*}

Within the ordered phase, $m>0$, the TSM provides us with (approximate) eigenstates $|E_m\ra$ of the field theory~\eqref{FT} of the form
\begin{align}
| E_m \ra = \sum_{N = 0,2,4,\ldots} \sum_j c^m_{N,j} | \{ p_j \})_N \ra, 
\end{align}
where $N$ labels the number of fermions within a basis state, $c^m_{N,j}$ are the superposition coefficients, and we defined the $N$-fermion Fock states 
\begin{align}
 | \{ p_j \}_{N} \ra = a\dg_{p_{j1}} \cdots a\dg_{p_{jN}} |0\ra. 
\end{align}
This leads to a natural definition of the $N$-particle weights: the sum of the absolute-value squared of the superposition coefficients restricted to a given $N$:
\begin{align}
w_N = \sum_{j} |c_{N,j}|^2. 
\end{align}
The quantity $w_N$ tells us what fraction of the state is described by the $N$ fermion basis states.

In the main body of the text, we used $w_2$ to support the assertion that the rare states are meson-like in nature, being built predominantly from linearly confined pairs of fermions. Here we present additional data for completeness for the $N=0,4,6,8$ particle weights within the constructed eigenstates, for the same eigenstates are presented in Fig.~\ref{Fig:TwoParticleWeight}. These weights are shown in Fig.~\ref{Fig:Weights}. 

At low energy densities $E/R \lesssim 0.15$, Fig.~\ref{Fig:Weights} supports the assertion in the main text that there is an approximate U(1) symmetry at low energies, with the states being predominantly ``$N$-particle'' in nature. To emphasize this we present a histogram of the zero, two, four and six particle weights, $w_0 - w_6$, restricted to lower energy densities $E/R < 0.15$ in Fig.~\ref{Fig:Weights}(e). We see that there is a bimodal distribution for both $w_2$ and $w_4$, with strong peaks at $w_N=0,1$. Detailed studies of how these results (and $w_6,w_8,\ldots$) evolve with increasing energy cutoff and system size are beyond the scope of this work and will be addressed in future studies.

\section{Semiclassical energies of the meson states}
\label{App:Semiclassical}
Here we provide semiclassical expressions for the energies of the meson states~\cite{fonseca2003ising,rutkevich2005largen,fonseca2006ising}. For weak longitudinal field $g$ and away from the two-particle threshold ($2m$) the energy of a meson state can be expressed as~\cite{fonseca2006ising}
\begin{align}
M_n = E_0 + 2m \cosh\theta_n, \quad n =1,2,\ldots, \label{En}
\end{align}
where $E_0$ is the ground state energy and $\theta_n$ is a rapidity that satisfies the nonlinear quantization condition
\begin{align}
\sinh2\theta_n - 2\theta_n = 2\pi \lambda \Big( n - \frac14\Big) - \lambda^2 S_1(\theta_n) - O(\lambda^3), \label{theta}
\end{align}
where $\lambda = 2 \bar \s g/m^2$ with $\bar \s = |m|^{1/8} \bar s$, and $\bar s = 2^{1/12} e^{-1/8} {\cal A}^{3/2}$ with ${\cal A} = 1.2824271291\ldots$ being Glashier's constant. We have also defined the function
\begin{align}
S_1(\theta) = - \frac{1}{\sinh2\theta}\bigg[ \frac{5}{24} \frac{1}{\sinh^2\theta} + \frac14 \frac{1}{\cosh^2\theta} - \frac{1}{12} - \frac{1}{6} \sinh^2\theta\bigg].\nonumber
\end{align}
Solutions for $m=1$, $g=0.1$ are shown in Fig.~\ref{Fig:TwoParticleWeight} (arrows are drawn at these energies) for the first forty mesons. 

\begin{widetext}

\section{The meson wave function at finite $R$}
\label{App:Wfn}

In this appendix, we consider states of the meson form 
\begin{align}
|\psi_n\ra = \sum_{v = {\rm NS}, {\rm RM}} \sum_{p_{v}} \Psi_{n,v}(p_v) a\dg_{p_v}a\dg_{-p_v} |v\ra, 
\label{twoQuarkState}
\end{align}
where $|v\ra$ is the vacuum state in the NS or RM sector, and $a\dg_{p_v}$ creates a fermion of momentum $p_v$ in the appropriate sector. Our aim is to constrain the form of the wave function such that these states are approximate eigenstates of the Hamiltonian~\eqref{FT}. To do so, we derive a Bethe-Salpeter equation for the wave function, see also~\cite{fonseca2006ising}.

\subsection{The Bethe-Salpeter Equation}

A Bethe-Salpeter equation for the meson wave function~\eqref{twoQuarkState} is obtained by restricting the Schr\"odinger equation to this manifold of states. To do this, we need to evaluate matrix elements of the form $\la v| a_{-p_v}a_{p_v} H |\psi\ra$. This gives rise to a restricted Schr\"odinger equation 
\begin{align}
E_n\Psi_{n,v}(p_v) = 2 \omega(p_v)\Psi_{n,v}(p_v)
 + \frac{gR}{2} \sum_{v',q_{v'}} \Psi_{n,v'}(q_{v'}) \la v | a_{-p_v}a_{p_v} \s(0) a\dg_{q_{v'}}a\dg_{-q_{v'}}|v'\ra.
\label{resSE}
\end{align}
Here $\omega(p) = \sqrt{p^2+m^2}$ is the dispersion relation for noninteracting fermions, and matrix elements of the spin operator in the finite volume are known~\cite{fonseca2003ising,bugrij2000correlation,bugrij2001form}. The spin operator only connects states in different sectors (NS and RM) of the Hilbert space, setting $v' = \bar v$ (where for $v=\text{NS},\text{RM}$, $\bar v = \text{RM},\text{NS}$). The matrix elements in the large but finite volume are given by 
\begin{align}
\sum_{q_{\bar v}}  \la v | a_{-p_v}a_{p_v} \s(0) a\dg_{q_{\bar v}}a\dg_{-q_{\bar v}}|\bar v\ra \Psi_{n, \bar v}(q_{\bar v}) = - \sum_{q_{\bar v}} \frac{\bar \s}{R^2} \frac{p_v q_{\bar v}}{\omega(p_v)^2\omega(q_{\bar v})^2} \left( \frac{\omega(p_v) + \omega(q_{\bar v})}{\omega(p_v)-\omega(q_{\bar v})}\right)^2 \Psi_{n,\bar v}(q_{\bar v}).
\label{ME}
\end{align}

The spatial wave function is obtained by Fourier transformation:
\begin{align}
\Psi_{n,\bar v}(q_{\bar v}) = \frac{1}{R} \int_0^R \rd x\, e^{i q_{\bar v} x} \Psi_{n,\bar v}(x). \nonumber
\end{align}
Fourier transforming the wave function in Eq.~\eqref{ME}, we can rewrite this equation as 
\begin{align}
 \sum_{q_{\bar v}}  \la v | a_{-p_v}a_{p_v} \s(0) a\dg_{q_{\bar v}}a\dg_{-q_{\bar v}}|\bar v\ra \Psi_{n,\bar v}(q_{\bar v})  = -\frac{\bar\s}{R^3} \int_0^R \rd x\, S(x,p_v) \Psi_{n,\bar v}(x),
\end{align}
where we define the Fourier transformed matrix elements
\begin{align}
S(x,p_v) = \sum_{q_{\bar v}} e^{i q_{\bar v} x} \frac{p_v q_{\bar v}}{\omega(p_v)^2\omega(q_{\bar v})^2} \left( \frac{\omega(p_v) + \omega(q_{\bar v})}{\omega(q_v)-\omega(q_{\bar v})}\right)^2. 
\label{Sxp}
\end{align}
A subsequent Fourier transform of Eq.~\eqref{Sxp}, where we expand in powers of $p_v$ and $q_{\bar v}$, leads to 
\begin{align}
\sum_{p_v} e^{-ip_v x'} S(x,p_v) = 
\sum_{p_v,q_{\bar v}} e^{i (q_{\bar v} x - p_v x')} \bigg[ 
\frac{4}{(p_v-q_{\bar v})^2} - \frac{4}{(p_v+q_{\bar v})^2} + \frac{2p_vq_{\bar v}}{m^4} 
+ O(p_vq_{\bar v}^2) + O(q_{\bar v}p_v^2) \bigg].
\label{sums}
\end{align}
Here we note that divergences that arise from vanishing denominators are avoided by $p_v$ and $q_{\bar v}$ being in different sectors of the Hilbert space (and hence the smallest difference being $\pm \pi/R$).

To continue, the sums on the right hand side of Eq.~\eqref{sums} are evaluated:
\begin{align}
\sum_{p_{\rm NS}} \frac{e^{-ip_{\rm NS}x}}{p_{\rm NS}^2} = \frac{R}{2} \left( \frac{R}{2} - |x| \right),
\qquad\sum_{p_{\rm RM}} e^{i p_{\rm RM} (x-x')} = \frac{R}{2\pi} \int \rd p\, e^{ip(x-x')} = R \delta(x-x'). \nonumber
\end{align}
In turn Eq.~\eqref{sums} becomes 
\begin{align}
\sum_{p_v} e^{-ip_v x'} S(x,p_v)  = 2R^2 \left(\frac{R}{2} - |x|\right)\Big[ \delta(x-x') - \delta(x+x'-R) \Big] 
+ \frac{2R^2}{m^4} \delta'(x)\delta'(x') + \ldots \nonumber
\end{align}
The restricted Schr\"odinger equation~\eqref{resSE} can now be recast into a differential equation for the real space wave function
\begin{align}
E_n \Psi_{n,v}(x)  = \left( 2m - \frac{1}{m} \p_x^2 - \frac{1}{4m^3} \p_x^4 - \frac{1}{8m^5}\p_x^6 \right) \Psi_{n,v}(x) - 2g \bar \s \left( \frac{R}{2} - |x| \right) \Psi_{n,\bar v}(x) + \frac{g\bar \s}{m^4} \delta'(x) \Psi_{n,\bar v}'(0).
\label{rse}
\end{align}
Here we keep only the leading (small momentum) terms in the expansion of the dispersion relation $\omega(p)$. Furthermore, we will herein assume that $\Psi_{n,\bar v}(x) = \Psi_{n,v}(x) \equiv \Psi_n(x)$, which will ultimately be justified in an \textit{a posteriori} manner. 

This assumption simplifies Eq.~\eqref{rse}: 
\begin{align}
\bigg(E_n +  g\bar\s R - 2m \bigg) \Psi_n(x)
= - \left( \frac{1}{m} \p_x^2 + \frac{1}{4m^3} \p_x^4 + \frac{1}{8m^5}\p_x^6 -  g \bar \s |x|  \right) \Psi_n(x) 
 + \frac{g\bar \s}{m^4}\delta'(x) \Psi'_n(0).  \nonumber
\end{align}
Performing a change of variables, $y = xmt$ with 
\begin{align}
t = \left( \frac{2g\bar\s}{m^2}\right)^{\frac13}, \label{deft}
\end{align}
the restricted Schr\"odinger equation is 
\begin{align}
\epsilon_n \Psi_n(y) = \Big(|y| -  \p_y^2 - \frac{t^2}{4} \p_y^4 - \frac{t^4}{8} \p_y^6 \Big) \Psi_n(y) - t^4 \delta'(y) \Psi'_n(0),
\label{BSEq} 
\end{align}
with $\epsilon_n m t^2 = (E_n - 2m + g\bar\s R)$. We call Eq.~\eqref{BSEq} the Bethe-Salpeter equation.  

\subsection{Solution of the Bethe-Salpeter Equation}

Now let us solve Eq.~\eqref{BSEq}. We consider the general equation
\begin{align}
\epsilon_n\Psi_n(y) = \Big( |y| - \p_y^2 - \mu t^2 \p_y^4 - \nu t^4 \p_y^6\Big) \Psi_n(y) + \rho t^4 \delta'(y)\Psi'_n(0),
\label{genBS}
\end{align}
where Eq.~\eqref{BSEq} is recovered by setting $\mu= 1/4$, $\nu = 1/8$ and $\rho=1/2$. To begin we construct solutions of 
\begin{align}
0 = \Big(y - \p_y^2 - \mu t^2 \p_y^4 - \nu t^4 \p_y^6\Big) F(y) .
\label{BSgenred}
\end{align}
These can be written in terms of solutions $A(y)$ of Airy's equation
\begin{align}
(y - \p_y^2)A(y) = 0. \label{airyeq}
\end{align}
Neglecting terms higher order in $t$ than $t^4$, the solutions of Eq.~\eqref{BSgenred} are $F_A(y) = A(y) + t^2 F_2(y) + t^4 F_4(y)$, where
\begin{align}
F_2(y) &= - \frac{4\mu}{5} y A(y) - \frac{\mu}{5} y^2 A'(y), \nn
F_4(y) &= - \left(2\mu^2 - \frac{9}{7}\nu\right) y^2 A(y) + \frac{\mu^2}{50} y^5 A(y)
 + \left(\frac{8\mu^2}{5} - \frac{10\nu}{7} \right)A'(y) + \left(\frac{14\mu^2}{35} - \frac{\nu}{7} \right)y^3 A'(y). \nonumber
\end{align}
To show that these are solutions, it is useful to remember that $A''(y) = y A(y)$, allowing one to write all terms as functions of only $A(y)$ and $A'(y)$. 

Let us now consider the full equation~\eqref{genBS}, and make the ansatz that the solution has the form 
\begin{align}
\Psi_n(y) = {\rm sgn}(y) F_n(|y|-\epsilon_n), 
\label{propsol}
\end{align}
with $F_n(y)$ a solution of Eq.~\eqref{BSgenred}. This will be a solution provided the function $F$ satisfies the following boundary conditions: 
\begin{align}
&(i)~~~~F_n(-\epsilon_n) = O(t^2), \nn
&(ii)~~~ \mu F_n(-\epsilon_n) + vt^2 F''_n(-\epsilon_n) = O(t^4), \label{BCs}  \\
&(iii)~~ F_n(-\epsilon_n) + \mu t^2 F''_n(-\epsilon_n) + v t^4 F^{(4)}_n(-\epsilon_n)
 - \frac{\rho}{2} t^4 F'_n(-\epsilon_n) = O(t^6). \nonumber
\end{align}
To ensure these are satisfied, we look for a solution of the form $F_n(y) = F_{{\rm Ai}}(y) + \alpha_n(\epsilon_n) F_{{\rm Bi}}(y)$, where ${\rm Ai}(y)$, ${\rm Bi}(y)$ are the linearly independent solutions to Airy's equation~\eqref{airyeq}. Furthermore, we assume that one can write an expansion in powers of $t$ for $\alpha_n(\epsilon_n)$: $\alpha_n(\epsilon_n) = \alpha_{0,n}(\epsilon_n) + t^2 \alpha_{2,n}(\epsilon_n) + t^4 \alpha_{4,n}(\epsilon_n) + O(t^6)$. The terms $\alpha_{i,n}(\epsilon_n)$ are fixed by imposing the boundary conditions~\eqref{BCs}, leading to
\begin{align}
\alpha_{0,n}(\epsilon_n) &= - \frac{{\rm Ai}(-\epsilon_n)}{{\rm Bi}(-\epsilon_n)}, \qquad \alpha_{2,n}(\epsilon_n) = \frac{\mu\epsilon_n^2}{5} \frac{ {\rm Ai}(-\epsilon_n)}{{\rm Bi}(-\epsilon_n)} \bigg( \frac{{\rm Ai}'(-\epsilon_n)}{{\rm Ai}(-\epsilon_n)} - \frac{{\rm Bi}'(-\epsilon_n)}{{\rm Bi}(-\epsilon_n)}\bigg), \nn
\alpha_{4,n}(\epsilon_n) &= \frac{{\rm Ai}(-\epsilon_n)}{{\rm Bi}(-\epsilon_n)}\bigg( \frac{{\rm Ai}'(-\epsilon_n)}{{\rm Ai}(-\epsilon_n)} - \frac{{\rm Bi}'(-\epsilon_n)}{{\rm Bi}(-\epsilon_n)}\bigg) \bigg[  \frac{2\mu^2}{5} - \frac{4\nu}{7} + \frac{\rho}{2} +  \bigg( \frac{84\mu^2}{350}  - \frac{\nu}{7}\bigg) \epsilon_n^3 + \frac{{\rm Bi}'(-\epsilon_n)}{{\rm Bi}(-\epsilon_n)} \frac{\mu^2 \epsilon_n^4}{25} \bigg]. \nonumber
\end{align}

To complete the solution of Eq.~\eqref{genBS}, we restrict our attention to normalizable solutions $\Psi_n(y)$. Using that $\lim_{y\to\infty} {\rm Bi}(y) = \infty$, this forces us to find $\epsilon_n$ such that $\alpha_n(\epsilon_n)=0$. Combining this condition with the above, we arrive at
\begin{align}
\epsilon_n = - z_n + \delta_{2,n} t^2 + \delta_{4,n} t^4 + O(t^6), \label{BSenergies}
\end{align}
where ${\rm Ai}(z_n) = 0$, and 
\begin{align}
\delta_{2,n} = - \frac{\mu}{5} z_n^2, \qquad \delta_{4,n} = \bigg( \frac{84\mu^2}{350} - \frac{2\mu^2}{25} - \frac{\nu}{7}\bigg) z_n^3 - \bigg( \frac{2\mu^2}{5} - \frac{4\nu}{7} + \frac{\rho}{2}\bigg). \nonumber
\end{align}

\subsection{Explicit expressions in original units}

To recover normalizable solutions of~\eqref{BSEq}, we set $\mu = 1/4$, $\nu = 1/8$ and $\rho=1/2$. The meson energies, $E_n$, are a power series in $t = (2g \bar \s/ m)^{1/3}$ 
\begin{align}
E_n - E_0 = 2m \Big( 1 + a_{2,n} t^2 +  a_{4,n} t^4 + a_{6,n} t^6 + O(t^8) \Big),  \label{En2}
\end{align}
with $E_0 =-g\bar \s R$  the energy of the ground state, and dimensionless parameters
\begin{align}
a_{2,n} = - \frac{z_n}{2}, \qquad 
a_{4,n} = -\frac{z_n^2}{40}, \qquad
a_{6,n} = -\frac{57}{280} - \frac{11z_n^3}{2800}. \nonumber
\end{align}
Equation~\eqref{En2} agrees with previous calculations by other authors~\cite{fonseca2003ising,fonseca2006ising}. The $n$th meson wave function is 
\begin{align}
\Psi_n(x) = \frac{1}{\sqrt{{\cal N}_n}} {\rm sgn}(x) {\cal F}_n\left( mt |x| - \frac{1}{mt^2}\Big( E_n - E_0 - 2m \Big) \right), \nonumber
\end{align}
where ${\cal N}_n$ sets the normalization, and ${\cal F}_n(y) = G_{0,n}(y) + G_{2,n}(y) t^2 + G_{4,n}(y) t^4 + O(t^6)$, with 
\begin{align}
G_{0,n}(y) &= {\rm Ai}(y) - \frac{{\rm Ai}(-\epsilon_n)}{{\rm Bi}(-\epsilon_n)} {\rm Bi}(y), \nn
G_{2,n}(y) &= \frac{1}{20} \bigg\{ - y \Big[ 4{\rm Ai}(y) + y {\rm Ai}'(y) \Big] + \frac{{\rm Ai}(-\epsilon_n)}{{\rm Bi}(-\epsilon_n)} y \Big[ 4 {\rm Bi}(y) + y {\rm Bi}'(y)\Big] + \epsilon_n^2 \frac{{\rm Ai}(-\epsilon_n)}{{\rm Bi}(-\epsilon_n)} \bigg[ \frac{{\rm Ai}'(-\epsilon_n)}{{\rm Ai}(-\epsilon_n)}  - \frac{{\rm Bi}'(-\epsilon_n)}{{\rm Bi}(-\epsilon_n)} \bigg] {\rm Bi}(y) \bigg\}, \nonumber
\end{align}
and
\begin{align}
G_{4,n}(y) &= \frac{1}{28} y^2 {\rm Ai}(y) + \frac{1}{800} y^5 {\rm Ai}(y) + \frac{1}{140} \big( y^3 - 11\big)  {\rm Ai}'(y)\nn
& - \frac{{\rm Ai}(-\epsilon_n)}{{\rm Bi}(-\epsilon_n)} \bigg[ \frac{1}{28} y^2 {\rm Bi}(y) + \frac{1}{800} y^5 {\rm Bi}(y) + \frac{1}{140}\big( y^3 - 11\big) {\rm Bi}'(y)\bigg] \nn
& - \frac{\epsilon_n y }{400} \frac{{\rm Ai}(-\epsilon_n)}{{\rm Bi}(-\epsilon_n)} \bigg[ \frac{{\rm Ai}'(-\epsilon_n)}{{\rm Ai}(-\epsilon_n)} - \frac{{\rm Bi}'(-\epsilon_n)}{{\rm Bi}(-\epsilon_n)} \bigg] \Big[ 4 {\rm Bi}(y) + y {\rm Bi}'(y)\Big] \nn
& +  \frac{{\rm Ai}(-\epsilon_n)}{{\rm Bi}(-\epsilon_n)} \bigg[ \frac{{\rm Ai}'(-\epsilon_n)}{{\rm Ai}(-\epsilon_n)} - \frac{{\rm Bi}'(-\epsilon_n)}{{\rm Bi}(-\epsilon_n)} \bigg] \bigg[ \frac{127}{280} - \frac{\epsilon_n^3}{350}  + \frac{{\rm Bi}'(-\epsilon_n)}{{\rm Bi}(-\epsilon_n)} \frac{\epsilon_n^4}{25} \bigg] {\rm Bi}(y). \nonumber
\end{align}

\end{widetext}

\section{Convergence of the TSM+CHEB}
\label{App:ConvCheb}

In this final appendix, we consider the convergence of the TSM+CHEB with expansion order and energy cutoff. We then examine results for the nonequilibrium dynamics at different system sizes. 

\subsection{With expansion order $N_{\rm max}$}

\begin{figure}
\begin{tabular}{l}
(a) \\
\includegraphics[width=0.45\textwidth]{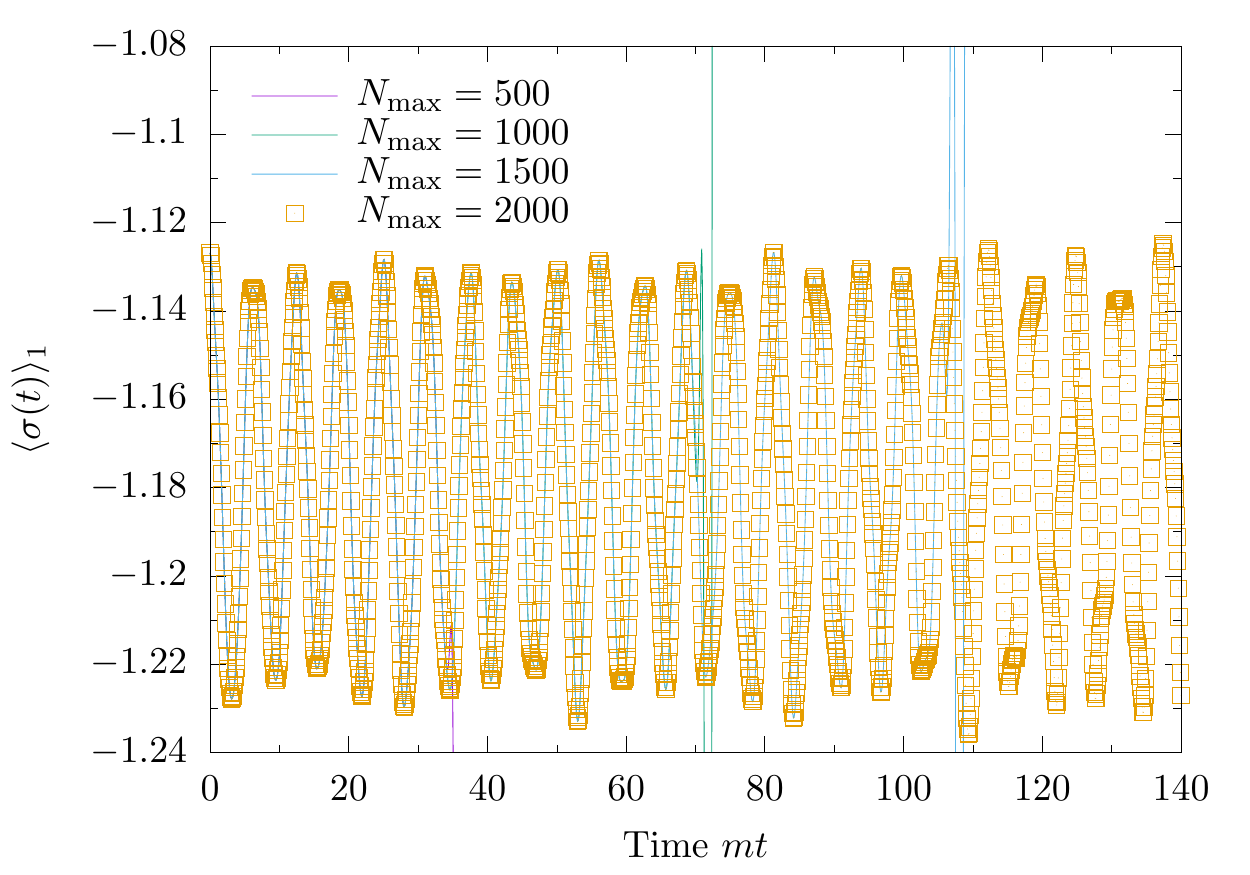} \\
(b) \\
\includegraphics[width=0.45\textwidth]{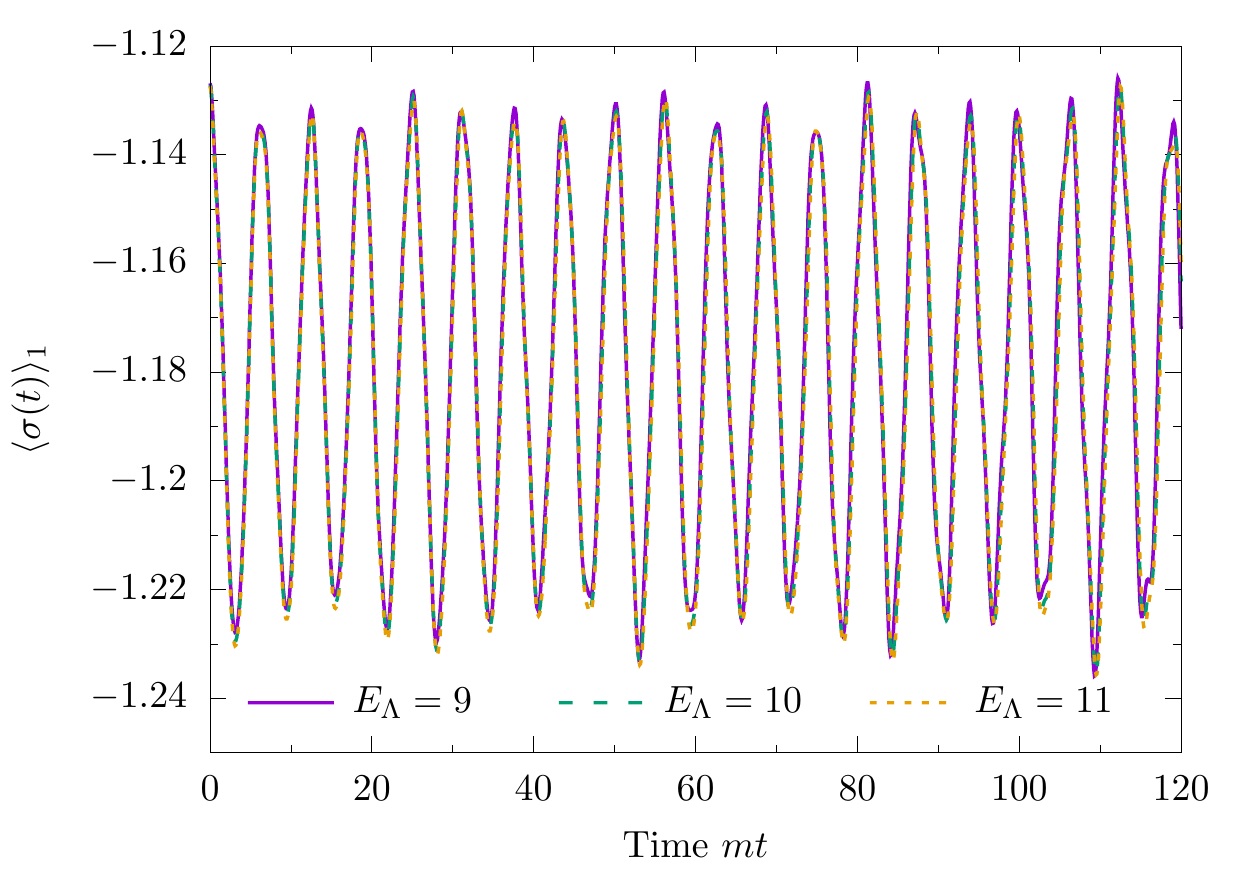}
\end{tabular}
  \caption{(a) Time evolution via the $N_{\rm max}$-order Chebyshev expansion for the quantum quench $(m,g) = (1,0.1) \to (1,0.2)$ of~\eqref{FT} for $R=25$. We start from the first excited state of the initial Hamiltonian, computed via the TSM with an energy cutoff $E_{\Lambda} = 9$. The breakdown time $t_\text{br}$ is seen by the divergence of the result from higher order predictions, see the purple line at $mt\sim35$, the green line at $mt\sim70$ and the cyan line at $mt\sim110$. (b) Time evolution shown for three values of $E_{\Lambda}$ with $N_{\rm max}=2000$. We see that the energy cutoff plays only a small role in the convergence of the presented results.}
\label{Fig:Cheby}
\end{figure}

Let us first consider convergence of the TSM+CHEB with order of the expansion, $N_\text{max}$. The upper panel of Fig.~\ref{Fig:Cheby} presents the time evolution of $\s(0)$ following the quench $(m,g)=(1,0.1) \to (1,0.2)$ in a system of size $R=25$ with energy cutoff $E_\Lambda = 9$. This is computed following Sec.~\ref{Sec:Dynamics} for four values of $N_\text{max}$: we see that results are well converged up to a time $t_\text{br}$, where the finite order of expansion leads to a divergence in the result. This breakdown time $t_\text{br}$ increases approximately linearly with expansion order, and manifests as a divergence of the result from its converged value. 

\subsection{With energy cutoff $E_{\Lambda}$}

\begin{figure}
\begin{tabular}{l}
(a) \\
\includegraphics[width=0.45\textwidth]{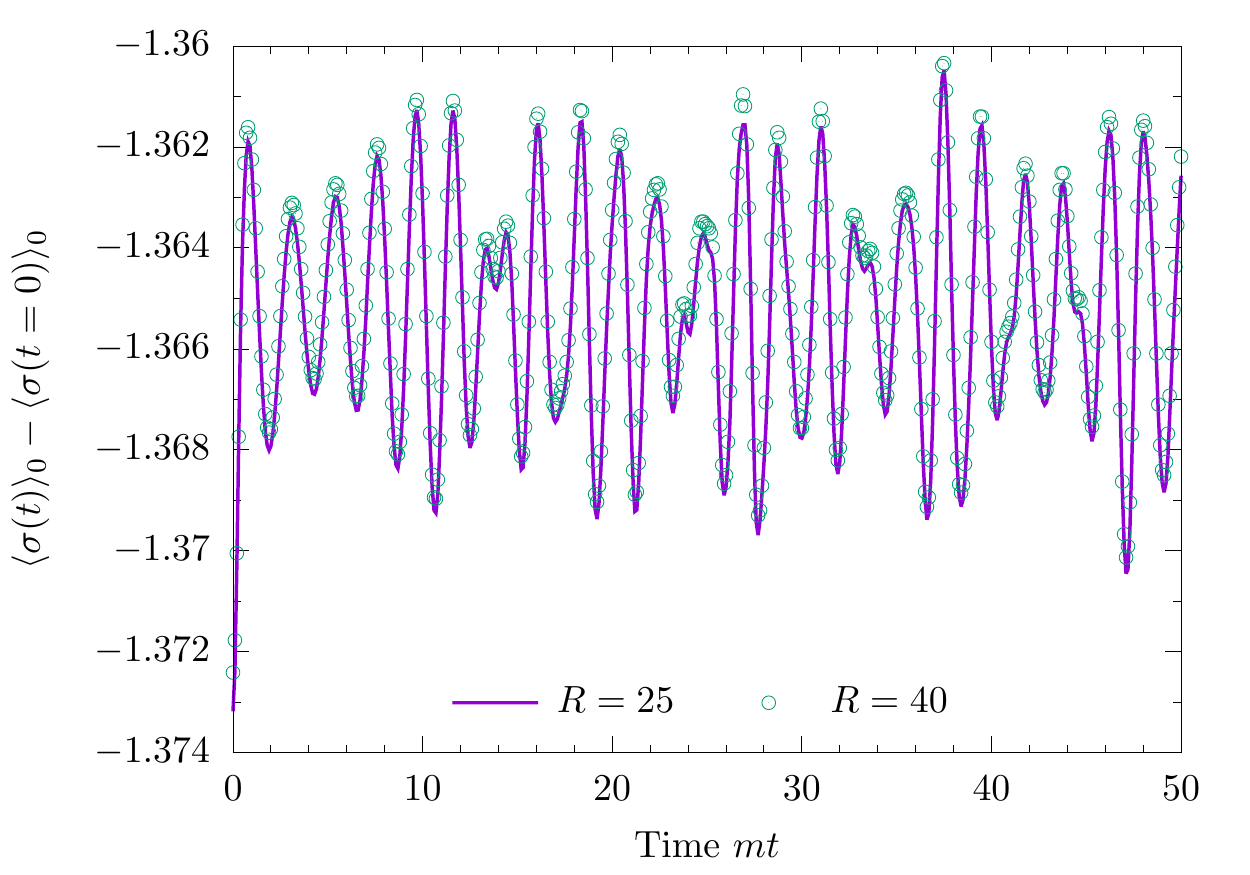} \\
(b) \\
\includegraphics[width=0.45\textwidth]{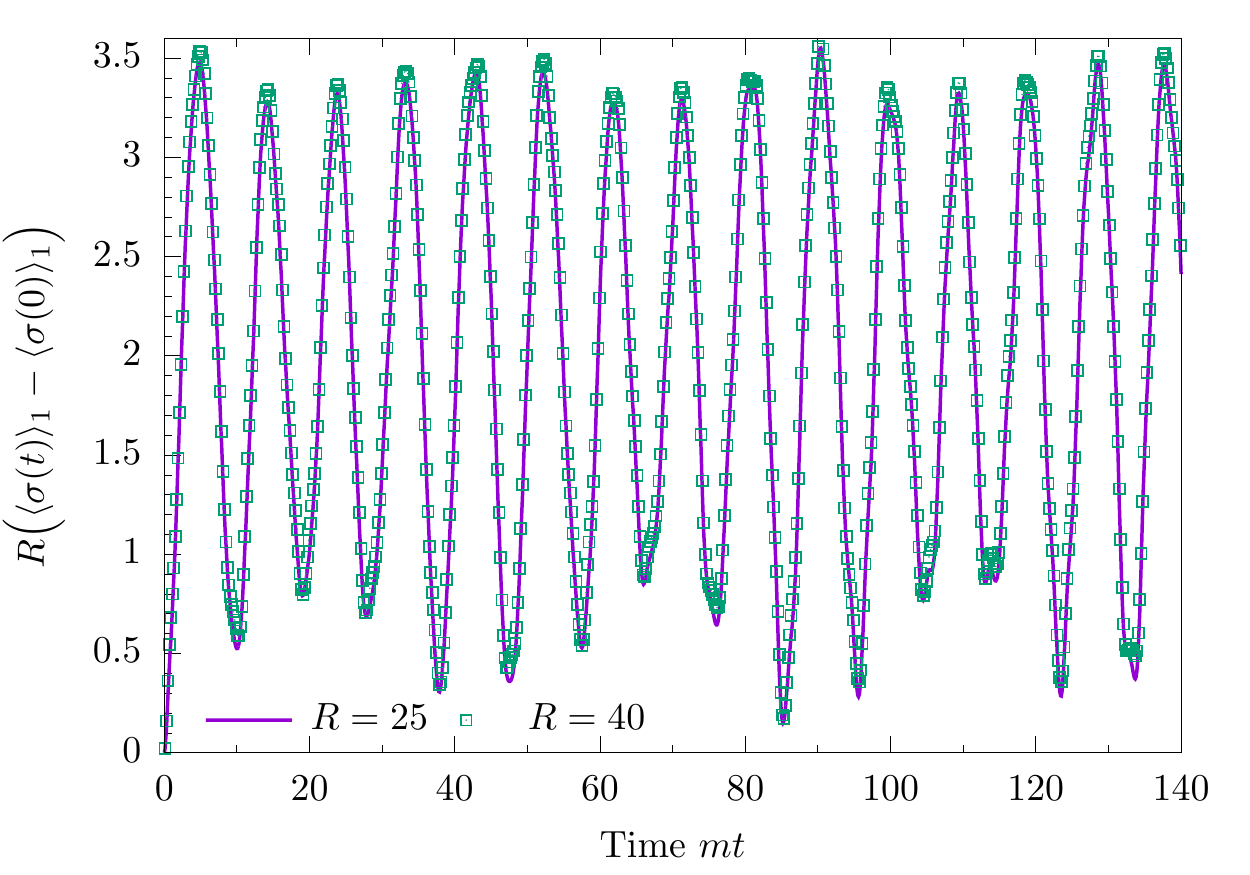}
\end{tabular}
\caption{Time evolution via the $N_{\rm max}=2000$-order Chebyshev expansion for the quantum quench $(m,g) = (1,0.1) \to (1,0.2)$ of~\eqref{FT} for $R=25,\ 40$. We start from the (a) ground state; (b) first excited state of the initial Hamiltonian, computed via the TSM with an energy cutoff $E_{\Lambda} = 9$. Good agreement of the results is seen until $t\sim145$ (not shown in (a)), where the finite order of the Chebyshev expansion causes the results for the large system size to breakdown.}
\label{Fig:TimeEvoRdep}
\end{figure}

Next we consider how the TSM+CHEB results change with energy cutoff $E_\Lambda$. In the right panel of Fig.~\ref{Fig:Cheby}(b) we see the time evolution of $\s(0)$ for the quench $(m,g)=(1,0.1)\to(1,0.2)$ at fixed order of expansion $N_\text{max}=2000$ on the system of size $R=25$. Results are present for three different cutoff energies $E_\Lambda$, showing that there is very good convergence in the energy cutoff. We note also that $t_\text{br}$ decreases with increasing $E_\Lambda$ (not shown).

\subsection{With volume $R$}

Finally we examine results of the TSM+CHEB at different system sizes $R$. Time evolution is induced by the quench $(m,g) = (1,0.2)\to(1,0.1)$ with fixed energy cutoff $E_\Lambda = 9$ and expansion order $N_\text{max} = 2000$. In Fig.~\ref{Fig:TimeEvoRdep}(a) we start from the ground state of the initial Hamiltonian and compare results for $R=25,\ 40$. The two data sets are almost identical and show no signs of finite size revivals. Similarly good agreement is observed up to $t_\text{br} = 145$, where the finite order of expansion is felt and the results become unreliable. The lack of finite size revival effects is similar to that observed in the lattice model~\eqref{lattice}, studied by~~\textcite{kormos2016realtime}

For quenches starting from higher states, we need to be careful with the observable computed. Starting from the first meson, we know that the number of flipped spins, compared to the ordered ground state, will be $O(1)$ and hence \textit{local} observables will vary as $1/R$. To counter this, we compute the \textit{total} magnetization by multiplying the local magnetization by $R$ (using translational invariance of the states). We faced similar issues with the finite size scaling analysis of observables in the meson states in the main text. This is shown in Fig.~\ref{Fig:TimeEvoRdep}(b), where we start from the first meson state. We see similar excellent agreement of the time-evolution of observables at different system sizes $R$ as in the left panel.

\twocolumngrid

\bibliography{RareStatesBib}

\begin{thebibliography}{146}%
\makeatletter
\providecommand \@ifxundefined [1]{%
 \@ifx{#1\undefined}
}%
\providecommand \@ifnum [1]{%
 \ifnum #1\expandafter \@firstoftwo
 \else \expandafter \@secondoftwo
 \fi
}%
\providecommand \@ifx [1]{%
 \ifx #1\expandafter \@firstoftwo
 \else \expandafter \@secondoftwo
 \fi
}%
\providecommand \natexlab [1]{#1}%
\providecommand \enquote  [1]{``#1''}%
\providecommand \bibnamefont  [1]{#1}%
\providecommand \bibfnamefont [1]{#1}%
\providecommand \citenamefont [1]{#1}%
\providecommand \href@noop [0]{\@secondoftwo}%
\providecommand \href [0]{\begingroup \@sanitize@url \@href}%
\providecommand \@href[1]{\@@startlink{#1}\@@href}%
\providecommand \@@href[1]{\endgroup#1\@@endlink}%
\providecommand \@sanitize@url [0]{\catcode `\\12\catcode `\$12\catcode
  `\&12\catcode `\#12\catcode `\^12\catcode `\_12\catcode `\%12\relax}%
\providecommand \@@startlink[1]{}%
\providecommand \@@endlink[0]{}%
\providecommand \url  [0]{\begingroup\@sanitize@url \@url }%
\providecommand \@url [1]{\endgroup\@href {#1}{\urlprefix }}%
\providecommand \urlprefix  [0]{URL }%
\providecommand \Eprint [0]{\href }%
\providecommand \doibase [0]{https://doi.org/}%
\providecommand \selectlanguage [0]{\@gobble}%
\providecommand \bibinfo  [0]{\@secondoftwo}%
\providecommand \bibfield  [0]{\@secondoftwo}%
\providecommand \translation [1]{[#1]}%
\providecommand \BibitemOpen [0]{}%
\providecommand \bibitemStop [0]{}%
\providecommand \bibitemNoStop [0]{.\EOS\space}%
\providecommand \EOS [0]{\spacefactor3000\relax}%
\providecommand \BibitemShut  [1]{\csname bibitem#1\endcsname}%
\let\auto@bib@innerbib\@empty
\bibitem [{\citenamefont {Polkovnikov}\ \emph {et~al.}(2011)\citenamefont
  {Polkovnikov}, \citenamefont {Sengupta}, \citenamefont {Silva},\ and\
  \citenamefont {Vengalattore}}]{polkovnikov2011nonequilibrium}%
  \BibitemOpen
  \bibfield  {author} {\bibinfo {author} {\bibfnamefont {A.}~\bibnamefont
  {Polkovnikov}}, \bibinfo {author} {\bibfnamefont {K.}~\bibnamefont
  {Sengupta}}, \bibinfo {author} {\bibfnamefont {A.}~\bibnamefont {Silva}},\
  and\ \bibinfo {author} {\bibfnamefont {M.}~\bibnamefont {Vengalattore}},\
  }\bibfield  {title} {\bibinfo {title} {\textit{Colloquium} : Nonequilibrium
  dynamics of closed interacting quantum systems},\ }\href
  {https://doi.org/10.1103/RevModPhys.83.863} {\bibfield  {journal} {\bibinfo
  {journal} {Rev. Mod. Phys.}\ }\textbf {\bibinfo {volume} {83}},\ \bibinfo
  {pages} {863} (\bibinfo {year} {2011})}\BibitemShut {NoStop}%
\bibitem [{\citenamefont {Gogolin}\ and\ \citenamefont
  {Eisert}(2016)}]{gogolin2016equilibration}%
  \BibitemOpen
  \bibfield  {author} {\bibinfo {author} {\bibfnamefont {C.}~\bibnamefont
  {Gogolin}}\ and\ \bibinfo {author} {\bibfnamefont {J.}~\bibnamefont
  {Eisert}},\ }\bibfield  {title} {\bibinfo {title} {Equilibration,
  thermalisation, and the emergence of statistical mechanics in closed quantum
  systems},\ }\href {http://stacks.iop.org/0034-4885/79/i=5/a=056001}
  {\bibfield  {journal} {\bibinfo  {journal} {Rep. Prog. Phys.}\ }\textbf
  {\bibinfo {volume} {79}},\ \bibinfo {pages} {056001} (\bibinfo {year}
  {2016})}\BibitemShut {NoStop}%
\bibitem [{\citenamefont {D'Alessio}\ \emph {et~al.}(2016)\citenamefont
  {D'Alessio}, \citenamefont {Kafri}, \citenamefont {Polkovnikov},\ and\
  \citenamefont {Rigol}}]{dalessio2016quantum}%
  \BibitemOpen
  \bibfield  {author} {\bibinfo {author} {\bibfnamefont {L.}~\bibnamefont
  {D'Alessio}}, \bibinfo {author} {\bibfnamefont {Y.}~\bibnamefont {Kafri}},
  \bibinfo {author} {\bibfnamefont {A.}~\bibnamefont {Polkovnikov}},\ and\
  \bibinfo {author} {\bibfnamefont {M.}~\bibnamefont {Rigol}},\ }\bibfield
  {title} {\bibinfo {title} {From quantum chaos and eigenstate thermalization
  to statistical mechanics and thermodynamics},\ }\href
  {https://doi.org/10.1080/00018732.2016.1198134} {\bibfield  {journal}
  {\bibinfo  {journal} {Adv. Phys.}\ }\textbf {\bibinfo {volume} {65}},\
  \bibinfo {pages} {239} (\bibinfo {year} {2016})}\BibitemShut {NoStop}%
\bibitem [{\citenamefont {Essler}\ and\ \citenamefont
  {Fagotti}(2016)}]{essler2016quench}%
  \BibitemOpen
  \bibfield  {author} {\bibinfo {author} {\bibfnamefont {F.~H.~L.}\
  \bibnamefont {Essler}}\ and\ \bibinfo {author} {\bibfnamefont
  {M.}~\bibnamefont {Fagotti}},\ }\bibfield  {title} {\bibinfo {title} {Quench
  dynamics and relaxation in isolated integrable quantum spin chains},\ }\href
  {http://stacks.iop.org/1742-5468/2016/i=6/a=064002} {\bibfield  {journal}
  {\bibinfo  {journal} {J. Stat. Mech.}\ }\textbf {\bibinfo {volume} {2016}},\
  \bibinfo {pages} {064002} (\bibinfo {year} {2016})}\BibitemShut {NoStop}%
\bibitem [{\citenamefont {Calabrese}\ and\ \citenamefont
  {Cardy}(2016)}]{calabrese2016quantum}%
  \BibitemOpen
  \bibfield  {author} {\bibinfo {author} {\bibfnamefont {P.}~\bibnamefont
  {Calabrese}}\ and\ \bibinfo {author} {\bibfnamefont {J.}~\bibnamefont
  {Cardy}},\ }\bibfield  {title} {\bibinfo {title} {Quantum quenches in 1+1
  dimensional conformal field theories},\ }\href
  {http://stacks.iop.org/1742-5468/2016/i=6/a=064003} {\bibfield  {journal}
  {\bibinfo  {journal} {J. Stat. Mech.}\ }\textbf {\bibinfo {volume} {2016}},\
  \bibinfo {pages} {064003} (\bibinfo {year} {2016})}\BibitemShut {NoStop}%
\bibitem [{\citenamefont {Cazalilla}\ and\ \citenamefont
  {Chung}(2016)}]{cazalilla2016quantum}%
  \BibitemOpen
  \bibfield  {author} {\bibinfo {author} {\bibfnamefont {M.~A.}\ \bibnamefont
  {Cazalilla}}\ and\ \bibinfo {author} {\bibfnamefont {M.-C.}\ \bibnamefont
  {Chung}},\ }\bibfield  {title} {\bibinfo {title} {Quantum quenches in the
  luttinger model and its close relatives},\ }\href
  {http://stacks.iop.org/1742-5468/2016/i=6/a=064004} {\bibfield  {journal}
  {\bibinfo  {journal} {J. Stat. Mech.}\ }\textbf {\bibinfo {volume} {2016}},\
  \bibinfo {pages} {064004} (\bibinfo {year} {2016})}\BibitemShut {NoStop}%
\bibitem [{\citenamefont {Bernard}\ and\ \citenamefont
  {Doyon}(2016)}]{bernard2016conformal}%
  \BibitemOpen
  \bibfield  {author} {\bibinfo {author} {\bibfnamefont {D.}~\bibnamefont
  {Bernard}}\ and\ \bibinfo {author} {\bibfnamefont {B.}~\bibnamefont
  {Doyon}},\ }\bibfield  {title} {\bibinfo {title} {Conformal field theory out
  of equilibrium: a review},\ }\href
  {http://stacks.iop.org/1742-5468/2016/i=6/a=064005} {\bibfield  {journal}
  {\bibinfo  {journal} {J. Stat. Mech.}\ }\textbf {\bibinfo {volume} {2016}},\
  \bibinfo {pages} {064005} (\bibinfo {year} {2016})}\BibitemShut {NoStop}%
\bibitem [{\citenamefont {Caux}(2016)}]{caux2016quench}%
  \BibitemOpen
  \bibfield  {author} {\bibinfo {author} {\bibfnamefont {J.-S.}\ \bibnamefont
  {Caux}},\ }\bibfield  {title} {\bibinfo {title} {The {Q}uench {A}ction},\
  }\href {http://stacks.iop.org/1742-5468/2016/i=6/a=064006} {\bibfield
  {journal} {\bibinfo  {journal} {J. Stat. Mech.}\ }\textbf {\bibinfo {volume}
  {2016}},\ \bibinfo {pages} {064006} (\bibinfo {year} {2016})}\BibitemShut
  {NoStop}%
\bibitem [{\citenamefont {Vidmar}\ and\ \citenamefont
  {Rigol}(2016)}]{vidmar2016generalized}%
  \BibitemOpen
  \bibfield  {author} {\bibinfo {author} {\bibfnamefont {L.}~\bibnamefont
  {Vidmar}}\ and\ \bibinfo {author} {\bibfnamefont {M.}~\bibnamefont {Rigol}},\
  }\bibfield  {title} {\bibinfo {title} {Generalized gibbs ensemble in
  integrable lattice models},\ }\href
  {http://stacks.iop.org/1742-5468/2016/i=6/a=064007} {\bibfield  {journal}
  {\bibinfo  {journal} {J. Stat. Mech.}\ }\textbf {\bibinfo {volume} {2016}},\
  \bibinfo {pages} {064007} (\bibinfo {year} {2016})}\BibitemShut {NoStop}%
\bibitem [{\citenamefont {{Langen}}\ \emph {et~al.}(2016)\citenamefont
  {{Langen}}, \citenamefont {{Gasenzer}},\ and\ \citenamefont
  {{Schmiedmayer}}}]{langen2016prethermalization}%
  \BibitemOpen
  \bibfield  {author} {\bibinfo {author} {\bibfnamefont {T.}~\bibnamefont
  {{Langen}}}, \bibinfo {author} {\bibfnamefont {T.}~\bibnamefont
  {{Gasenzer}}},\ and\ \bibinfo {author} {\bibfnamefont {J.}~\bibnamefont
  {{Schmiedmayer}}},\ }\bibfield  {title} {\bibinfo {title} {Prethermalization
  and universal dynamics in near-integrable quantum systems},\ }\href
  {http://stacks.iop.org/1742-5468/2016/i=6/a=064009} {\bibfield  {journal}
  {\bibinfo  {journal} {J. Stat. Mech.}\ }\textbf {\bibinfo {volume} {2016}},\
  \bibinfo {pages} {064009} (\bibinfo {year} {2016})}\BibitemShut {NoStop}%
\bibitem [{\citenamefont {Ilievski}\ \emph {et~al.}(2016)\citenamefont
  {Ilievski}, \citenamefont {Medenjak}, \citenamefont {Prosen},\ and\
  \citenamefont {Zadnik}}]{ilievski2016quasilocal}%
  \BibitemOpen
  \bibfield  {author} {\bibinfo {author} {\bibfnamefont {E.}~\bibnamefont
  {Ilievski}}, \bibinfo {author} {\bibfnamefont {M.}~\bibnamefont {Medenjak}},
  \bibinfo {author} {\bibfnamefont {T.}~\bibnamefont {Prosen}},\ and\ \bibinfo
  {author} {\bibfnamefont {L.}~\bibnamefont {Zadnik}},\ }\bibfield  {title}
  {\bibinfo {title} {Quasilocal charges in integrable lattice systems},\ }\href
  {http://stacks.iop.org/1742-5468/2016/i=6/a=064008} {\bibfield  {journal}
  {\bibinfo  {journal} {J. Stat. Mech.}\ }\textbf {\bibinfo {volume} {2016}},\
  \bibinfo {pages} {064008} (\bibinfo {year} {2016})}\BibitemShut {NoStop}%
\bibitem [{\citenamefont {Vasseur}\ and\ \citenamefont
  {Moore}(2016)}]{vasseur2016nonequilibrium}%
  \BibitemOpen
  \bibfield  {author} {\bibinfo {author} {\bibfnamefont {R.}~\bibnamefont
  {Vasseur}}\ and\ \bibinfo {author} {\bibfnamefont {J.~E.}\ \bibnamefont
  {Moore}},\ }\bibfield  {title} {\bibinfo {title} {Nonequilibrium quantum
  dynamics and transport: from integrability to many-body localization},\
  }\href {http://stacks.iop.org/1742-5468/2016/i=6/a=064010} {\bibfield
  {journal} {\bibinfo  {journal} {J. Stat. Mech.}\ }\textbf {\bibinfo {volume}
  {2016}},\ \bibinfo {pages} {064010} (\bibinfo {year} {2016})}\BibitemShut
  {NoStop}%
\bibitem [{\citenamefont {{De Luca}}\ and\ \citenamefont
  {{Mussardo}}(2016)}]{deluca2016equilibration}%
  \BibitemOpen
  \bibfield  {author} {\bibinfo {author} {\bibfnamefont {A.}~\bibnamefont {{De
  Luca}}}\ and\ \bibinfo {author} {\bibfnamefont {G.}~\bibnamefont
  {{Mussardo}}},\ }\bibfield  {title} {\bibinfo {title} {Equilibration
  properties of classical integrable field theories},\ }\href
  {http://stacks.iop.org/1742-5468/2016/i=6/a=064011} {\bibfield  {journal}
  {\bibinfo  {journal} {J. Stat. Mech.}\ }\textbf {\bibinfo {volume} {2016}},\
  \bibinfo {pages} {064011} (\bibinfo {year} {2016})}\BibitemShut {NoStop}%
\bibitem [{\citenamefont {Kinoshita}\ \emph {et~al.}(2006)\citenamefont
  {Kinoshita}, \citenamefont {Wenger},\ and\ \citenamefont
  {Weiss}}]{weiss2006quantum}%
  \BibitemOpen
  \bibfield  {author} {\bibinfo {author} {\bibfnamefont {T.}~\bibnamefont
  {Kinoshita}}, \bibinfo {author} {\bibfnamefont {T.}~\bibnamefont {Wenger}},\
  and\ \bibinfo {author} {\bibfnamefont {D.~S.}\ \bibnamefont {Weiss}},\
  }\bibfield  {title} {\bibinfo {title} {A quantum {N}ewton's cradle},\ }\href
  {http://dx.doi.org/10.1038/nature04693} {\bibfield  {journal} {\bibinfo
  {journal} {Nature}\ }\textbf {\bibinfo {volume} {440}},\ \bibinfo {pages}
  {900} (\bibinfo {year} {2006})}\BibitemShut {NoStop}%
\bibitem [{\citenamefont {Lewenstein}\ \emph {et~al.}(2007)\citenamefont
  {Lewenstein}, \citenamefont {Sanpera}, \citenamefont {Ahufinger},
  \citenamefont {Damski}, \citenamefont {Sen(De)},\ and\ \citenamefont
  {Sen}}]{lewenstein2007ultracold}%
  \BibitemOpen
  \bibfield  {author} {\bibinfo {author} {\bibfnamefont {M.}~\bibnamefont
  {Lewenstein}}, \bibinfo {author} {\bibfnamefont {A.}~\bibnamefont {Sanpera}},
  \bibinfo {author} {\bibfnamefont {V.}~\bibnamefont {Ahufinger}}, \bibinfo
  {author} {\bibfnamefont {B.}~\bibnamefont {Damski}}, \bibinfo {author}
  {\bibfnamefont {A.}~\bibnamefont {Sen(De)}},\ and\ \bibinfo {author}
  {\bibfnamefont {U.}~\bibnamefont {Sen}},\ }\bibfield  {title} {\bibinfo
  {title} {Ultracold atomic gases in optical lattices: mimicking condensed
  matter physics and beyond},\ }\href
  {https://doi.org/10.1080/00018730701223200} {\bibfield  {journal} {\bibinfo
  {journal} {Adv. Phys.}\ }\textbf {\bibinfo {volume} {56}},\ \bibinfo {pages}
  {243} (\bibinfo {year} {2007})}\BibitemShut {NoStop}%
\bibitem [{\citenamefont {Deutsch}(1991)}]{deutsch1991quantum}%
  \BibitemOpen
  \bibfield  {author} {\bibinfo {author} {\bibfnamefont {J.~M.}\ \bibnamefont
  {Deutsch}},\ }\bibfield  {title} {\bibinfo {title} {Quantum statistical
  mechanics in a closed system},\ }\href
  {https://doi.org/10.1103/PhysRevA.43.2046} {\bibfield  {journal} {\bibinfo
  {journal} {Phys. Rev. A}\ }\textbf {\bibinfo {volume} {43}},\ \bibinfo
  {pages} {2046} (\bibinfo {year} {1991})}\BibitemShut {NoStop}%
\bibitem [{\citenamefont {Srednicki}(1994)}]{srednicki1994chaos}%
  \BibitemOpen
  \bibfield  {author} {\bibinfo {author} {\bibfnamefont {M.}~\bibnamefont
  {Srednicki}},\ }\bibfield  {title} {\bibinfo {title} {Chaos and quantum
  thermalization},\ }\href {https://doi.org/10.1103/PhysRevE.50.888} {\bibfield
   {journal} {\bibinfo  {journal} {Phys. Rev. E}\ }\textbf {\bibinfo {volume}
  {50}},\ \bibinfo {pages} {888} (\bibinfo {year} {1994})}\BibitemShut
  {NoStop}%
\bibitem [{\citenamefont {Rigol}\ \emph {et~al.}(2008)\citenamefont {Rigol},
  \citenamefont {Dunjko},\ and\ \citenamefont
  {Olshanii}}]{rigol2008thermalization}%
  \BibitemOpen
  \bibfield  {author} {\bibinfo {author} {\bibfnamefont {M.}~\bibnamefont
  {Rigol}}, \bibinfo {author} {\bibfnamefont {V.}~\bibnamefont {Dunjko}},\ and\
  \bibinfo {author} {\bibfnamefont {M.}~\bibnamefont {Olshanii}},\ }\bibfield
  {title} {\bibinfo {title} {Thermalization and its mechanism for generic
  isolated quantum systems},\ }\href {http://dx.doi.org/10.1038/nature06838}
  {\bibfield  {journal} {\bibinfo  {journal} {Nature}\ }\textbf {\bibinfo
  {volume} {452}},\ \bibinfo {pages} {854} (\bibinfo {year}
  {2008})}\BibitemShut {NoStop}%
\bibitem [{\citenamefont {Rigol}(2009)}]{rigol2009breakdown}%
  \BibitemOpen
  \bibfield  {author} {\bibinfo {author} {\bibfnamefont {M.}~\bibnamefont
  {Rigol}},\ }\bibfield  {title} {\bibinfo {title} {{Breakdown of
  Thermalization in Finite One-Dimensional Systems}},\ }\href
  {https://doi.org/10.1103/PhysRevLett.103.100403} {\bibfield  {journal}
  {\bibinfo  {journal} {Phys. Rev. Lett.}\ }\textbf {\bibinfo {volume} {103}},\
  \bibinfo {pages} {100403} (\bibinfo {year} {2009})}\BibitemShut {NoStop}%
\bibitem [{\citenamefont {Rigol}\ and\ \citenamefont
  {Srednicki}(2012)}]{rigol2012alternatives}%
  \BibitemOpen
  \bibfield  {author} {\bibinfo {author} {\bibfnamefont {M.}~\bibnamefont
  {Rigol}}\ and\ \bibinfo {author} {\bibfnamefont {M.}~\bibnamefont
  {Srednicki}},\ }\bibfield  {title} {\bibinfo {title} {{Alternatives to
  Eigenstate Thermalization}},\ }\href
  {https://doi.org/10.1103/PhysRevLett.108.110601} {\bibfield  {journal}
  {\bibinfo  {journal} {Phys. Rev. Lett.}\ }\textbf {\bibinfo {volume} {108}},\
  \bibinfo {pages} {110601} (\bibinfo {year} {2012})}\BibitemShut {NoStop}%
\bibitem [{\citenamefont {Reimann}(2015)}]{reimann2015eigenstate}%
  \BibitemOpen
  \bibfield  {author} {\bibinfo {author} {\bibfnamefont {P.}~\bibnamefont
  {Reimann}},\ }\bibfield  {title} {\bibinfo {title} {Eigenstate
  thermalization: {D}eutsch's approach and beyond},\ }\href
  {http://stacks.iop.org/1367-2630/17/i=5/a=055025} {\bibfield  {journal}
  {\bibinfo  {journal} {New J. Phys.}\ }\textbf {\bibinfo {volume} {17}},\
  \bibinfo {pages} {055025} (\bibinfo {year} {2015})}\BibitemShut {NoStop}%
\bibitem [{\citenamefont {Deutsch}(2018)}]{deutsch2018eigenstate}%
  \BibitemOpen
  \bibfield  {author} {\bibinfo {author} {\bibfnamefont {J.~M.}\ \bibnamefont
  {Deutsch}},\ }\bibfield  {title} {\bibinfo {title} {Eigenstate thermalization
  hypothesis},\ }\href {http://stacks.iop.org/0034-4885/81/i=8/a=082001}
  {\bibfield  {journal} {\bibinfo  {journal} {Rep. Prog. Phys.}\ }\textbf
  {\bibinfo {volume} {81}},\ \bibinfo {pages} {082001} (\bibinfo {year}
  {2018})}\BibitemShut {NoStop}%
\bibitem [{Note1()}]{Note1}%
  \BibitemOpen
  \bibinfo {note} {For neighboring energy eigenvalues $E$ and $E'$ ($E'>E$)
  this function satisfies $A(E')-A(E) \propto e^{-R/\protect \mathaccentV
  {tilde}07ER}$, with $R$ being the system size and $\protect \mathaccentV
  {tilde}07ER$ being some (possibly $E$-dependent) constant.}\BibitemShut
  {Stop}%
\bibitem [{\citenamefont {Mondaini}\ and\ \citenamefont
  {Rigol}(2017)}]{mondaini2017eigenstate}%
  \BibitemOpen
  \bibfield  {author} {\bibinfo {author} {\bibfnamefont {R.}~\bibnamefont
  {Mondaini}}\ and\ \bibinfo {author} {\bibfnamefont {M.}~\bibnamefont
  {Rigol}},\ }\bibfield  {title} {\bibinfo {title} {{Eigenstate thermalization
  in the two-dimensional transverse field Ising model. II. Off-diagonal matrix
  elements of observables}},\ }\href
  {https://doi.org/10.1103/PhysRevE.96.012157} {\bibfield  {journal} {\bibinfo
  {journal} {Phys. Rev. E}\ }\textbf {\bibinfo {volume} {96}},\ \bibinfo
  {pages} {012157} (\bibinfo {year} {2017})}\BibitemShut {NoStop}%
\bibitem [{\citenamefont {Biroli}\ \emph {et~al.}(2010)\citenamefont {Biroli},
  \citenamefont {Kollath},\ and\ \citenamefont {L\"auchli}}]{biroli2010effect}%
  \BibitemOpen
  \bibfield  {author} {\bibinfo {author} {\bibfnamefont {G.}~\bibnamefont
  {Biroli}}, \bibinfo {author} {\bibfnamefont {C.}~\bibnamefont {Kollath}},\
  and\ \bibinfo {author} {\bibfnamefont {A.~M.}\ \bibnamefont {L\"auchli}},\
  }\bibfield  {title} {\bibinfo {title} {{Effect of Rare Fluctuations on the
  Thermalization of Isolated Quantum Systems}},\ }\href
  {https://doi.org/10.1103/PhysRevLett.105.250401} {\bibfield  {journal}
  {\bibinfo  {journal} {Phys. Rev. Lett.}\ }\textbf {\bibinfo {volume} {105}},\
  \bibinfo {pages} {250401} (\bibinfo {year} {2010})}\BibitemShut {NoStop}%
\bibitem [{\citenamefont {Santos}\ and\ \citenamefont
  {Rigol}(2010{\natexlab{a}})}]{santos2010localization}%
  \BibitemOpen
  \bibfield  {author} {\bibinfo {author} {\bibfnamefont {L.~F.}\ \bibnamefont
  {Santos}}\ and\ \bibinfo {author} {\bibfnamefont {M.}~\bibnamefont {Rigol}},\
  }\bibfield  {title} {\bibinfo {title} {Localization and the effects of
  symmetries in the thermalization properties of one-dimensional quantum
  systems},\ }\href {https://doi.org/10.1103/PhysRevE.82.031130} {\bibfield
  {journal} {\bibinfo  {journal} {Phys. Rev. E}\ }\textbf {\bibinfo {volume}
  {82}},\ \bibinfo {pages} {031130} (\bibinfo {year}
  {2010}{\natexlab{a}})}\BibitemShut {NoStop}%
\bibitem [{\citenamefont {Santos}\ and\ \citenamefont
  {Rigol}(2010{\natexlab{b}})}]{santos2010onset}%
  \BibitemOpen
  \bibfield  {author} {\bibinfo {author} {\bibfnamefont {L.~F.}\ \bibnamefont
  {Santos}}\ and\ \bibinfo {author} {\bibfnamefont {M.}~\bibnamefont {Rigol}},\
  }\bibfield  {title} {\bibinfo {title} {Onset of quantum chaos in
  one-dimensional bosonic and fermionic systems and its relation to
  thermalization},\ }\href {https://doi.org/10.1103/PhysRevE.81.036206}
  {\bibfield  {journal} {\bibinfo  {journal} {Phys. Rev. E}\ }\textbf {\bibinfo
  {volume} {81}},\ \bibinfo {pages} {036206} (\bibinfo {year}
  {2010}{\natexlab{b}})}\BibitemShut {NoStop}%
\bibitem [{\citenamefont {Richter}\ \emph {et~al.}(2018)\citenamefont
  {Richter}, \citenamefont {Jin}, \citenamefont {De~Raedt}, \citenamefont
  {Michielsen}, \citenamefont {Gemmer},\ and\ \citenamefont
  {Steinigeweg}}]{richter2018realtime}%
  \BibitemOpen
  \bibfield  {author} {\bibinfo {author} {\bibfnamefont {J.}~\bibnamefont
  {Richter}}, \bibinfo {author} {\bibfnamefont {F.}~\bibnamefont {Jin}},
  \bibinfo {author} {\bibfnamefont {H.}~\bibnamefont {De~Raedt}}, \bibinfo
  {author} {\bibfnamefont {K.}~\bibnamefont {Michielsen}}, \bibinfo {author}
  {\bibfnamefont {J.}~\bibnamefont {Gemmer}},\ and\ \bibinfo {author}
  {\bibfnamefont {R.}~\bibnamefont {Steinigeweg}},\ }\bibfield  {title}
  {\bibinfo {title} {Real-time dynamics of typical and untypical states in
  nonintegrable systems},\ }\href {https://doi.org/10.1103/PhysRevB.97.174430}
  {\bibfield  {journal} {\bibinfo  {journal} {Phys. Rev. B}\ }\textbf {\bibinfo
  {volume} {97}},\ \bibinfo {pages} {174430} (\bibinfo {year}
  {2018})}\BibitemShut {NoStop}%
\bibitem [{\citenamefont {Ikeda}\ \emph {et~al.}(2013)\citenamefont {Ikeda},
  \citenamefont {Watanabe},\ and\ \citenamefont {Ueda}}]{ikeda2013finitesize}%
  \BibitemOpen
  \bibfield  {author} {\bibinfo {author} {\bibfnamefont {T.~N.}\ \bibnamefont
  {Ikeda}}, \bibinfo {author} {\bibfnamefont {Y.}~\bibnamefont {Watanabe}},\
  and\ \bibinfo {author} {\bibfnamefont {M.}~\bibnamefont {Ueda}},\ }\bibfield
  {title} {\bibinfo {title} {Finite-size scaling analysis of the eigenstate
  thermalization hypothesis in a one-dimensional interacting {B}ose gas},\
  }\href {https://doi.org/10.1103/PhysRevE.87.012125} {\bibfield  {journal}
  {\bibinfo  {journal} {Phys. Rev. E}\ }\textbf {\bibinfo {volume} {87}},\
  \bibinfo {pages} {012125} (\bibinfo {year} {2013})}\BibitemShut {NoStop}%
\bibitem [{\citenamefont {Alba}(2015)}]{alba2015eigenstate}%
  \BibitemOpen
  \bibfield  {author} {\bibinfo {author} {\bibfnamefont {V.}~\bibnamefont
  {Alba}},\ }\bibfield  {title} {\bibinfo {title} {Eigenstate thermalization
  hypothesis and integrability in quantum spin chains},\ }\href
  {https://doi.org/10.1103/PhysRevB.91.155123} {\bibfield  {journal} {\bibinfo
  {journal} {Phys. Rev. B}\ }\textbf {\bibinfo {volume} {91}},\ \bibinfo
  {pages} {155123} (\bibinfo {year} {2015})}\BibitemShut {NoStop}%
\bibitem [{\citenamefont {{Khlebnikov}}\ and\ \citenamefont
  {{Kruczenski}}(2013)}]{khlebnikov2013thermalization}%
  \BibitemOpen
  \bibfield  {author} {\bibinfo {author} {\bibfnamefont {S.}~\bibnamefont
  {{Khlebnikov}}}\ and\ \bibinfo {author} {\bibfnamefont {M.}~\bibnamefont
  {{Kruczenski}}},\ }\bibfield  {title} {\bibinfo {title} {{Thermalization of
  isolated quantum systems}},\ }\href@noop {} {\bibfield  {journal} {\bibinfo
  {journal} {ArXiv e-prints}\ } (\bibinfo {year} {2013})},\ \Eprint
  {https://arxiv.org/abs/1312.4612} {arXiv:1312.4612 [cond-mat.stat-mech]}
  \BibitemShut {NoStop}%
\bibitem [{\citenamefont {Beugeling}\ \emph {et~al.}(2014)\citenamefont
  {Beugeling}, \citenamefont {Moessner},\ and\ \citenamefont
  {Haque}}]{beugeling2014finitesize}%
  \BibitemOpen
  \bibfield  {author} {\bibinfo {author} {\bibfnamefont {W.}~\bibnamefont
  {Beugeling}}, \bibinfo {author} {\bibfnamefont {R.}~\bibnamefont
  {Moessner}},\ and\ \bibinfo {author} {\bibfnamefont {M.}~\bibnamefont
  {Haque}},\ }\bibfield  {title} {\bibinfo {title} {Finite-size scaling of
  eigenstate thermalization},\ }\href
  {https://doi.org/10.1103/PhysRevE.89.042112} {\bibfield  {journal} {\bibinfo
  {journal} {Phys. Rev. E}\ }\textbf {\bibinfo {volume} {89}},\ \bibinfo
  {pages} {042112} (\bibinfo {year} {2014})}\BibitemShut {NoStop}%
\bibitem [{\citenamefont {Sorg}\ \emph {et~al.}(2014)\citenamefont {Sorg},
  \citenamefont {Vidmar}, \citenamefont {Pollet},\ and\ \citenamefont
  {Heidrich-Meisner}}]{sorg2014relaxation}%
  \BibitemOpen
  \bibfield  {author} {\bibinfo {author} {\bibfnamefont {S.}~\bibnamefont
  {Sorg}}, \bibinfo {author} {\bibfnamefont {L.}~\bibnamefont {Vidmar}},
  \bibinfo {author} {\bibfnamefont {L.}~\bibnamefont {Pollet}},\ and\ \bibinfo
  {author} {\bibfnamefont {F.}~\bibnamefont {Heidrich-Meisner}},\ }\bibfield
  {title} {\bibinfo {title} {{Relaxation and thermalization in the
  one-dimensional {B}ose-{H}ubbard model: A case study for the interaction
  quantum quench from the atomic limit}},\ }\href
  {https://doi.org/10.1103/PhysRevA.90.033606} {\bibfield  {journal} {\bibinfo
  {journal} {Phys. Rev. A}\ }\textbf {\bibinfo {volume} {90}},\ \bibinfo
  {pages} {033606} (\bibinfo {year} {2014})}\BibitemShut {NoStop}%
\bibitem [{\citenamefont {Khodja}\ \emph {et~al.}(2015)\citenamefont {Khodja},
  \citenamefont {Steinigeweg},\ and\ \citenamefont
  {Gemmer}}]{khodja2015relevance}%
  \BibitemOpen
  \bibfield  {author} {\bibinfo {author} {\bibfnamefont {A.}~\bibnamefont
  {Khodja}}, \bibinfo {author} {\bibfnamefont {R.}~\bibnamefont
  {Steinigeweg}},\ and\ \bibinfo {author} {\bibfnamefont {J.}~\bibnamefont
  {Gemmer}},\ }\bibfield  {title} {\bibinfo {title} {Relevance of the
  eigenstate thermalization hypothesis for thermal relaxation},\ }\href
  {https://doi.org/10.1103/PhysRevE.91.012120} {\bibfield  {journal} {\bibinfo
  {journal} {Phys. Rev. E}\ }\textbf {\bibinfo {volume} {91}},\ \bibinfo
  {pages} {012120} (\bibinfo {year} {2015})}\BibitemShut {NoStop}%
\bibitem [{\citenamefont {Cosme}\ and\ \citenamefont
  {Fialko}(2015)}]{cosme2015relaxation}%
  \BibitemOpen
  \bibfield  {author} {\bibinfo {author} {\bibfnamefont {J.~G.}\ \bibnamefont
  {Cosme}}\ and\ \bibinfo {author} {\bibfnamefont {O.}~\bibnamefont {Fialko}},\
  }\bibfield  {title} {\bibinfo {title} {{Relaxation dynamics of ultracold
  bosons in a double-well potential: Thermalization and prethermalization in a
  nearly integrable model}},\ }\href
  {https://doi.org/10.1103/PhysRevA.92.033607} {\bibfield  {journal} {\bibinfo
  {journal} {Phys. Rev. A}\ }\textbf {\bibinfo {volume} {92}},\ \bibinfo
  {pages} {033607} (\bibinfo {year} {2015})}\BibitemShut {NoStop}%
\bibitem [{\citenamefont {Mondaini}\ \emph {et~al.}(2016)\citenamefont
  {Mondaini}, \citenamefont {Fratus}, \citenamefont {Srednicki},\ and\
  \citenamefont {Rigol}}]{mondaini2016eigenstate}%
  \BibitemOpen
  \bibfield  {author} {\bibinfo {author} {\bibfnamefont {R.}~\bibnamefont
  {Mondaini}}, \bibinfo {author} {\bibfnamefont {K.~R.}\ \bibnamefont
  {Fratus}}, \bibinfo {author} {\bibfnamefont {M.}~\bibnamefont {Srednicki}},\
  and\ \bibinfo {author} {\bibfnamefont {M.}~\bibnamefont {Rigol}},\ }\bibfield
   {title} {\bibinfo {title} {Eigenstate thermalization in the two-dimensional
  transverse field {I}sing model},\ }\href
  {https://doi.org/10.1103/PhysRevE.93.032104} {\bibfield  {journal} {\bibinfo
  {journal} {Phys. Rev. E}\ }\textbf {\bibinfo {volume} {93}},\ \bibinfo
  {pages} {032104} (\bibinfo {year} {2016})}\BibitemShut {NoStop}%
\bibitem [{\citenamefont {Lan}\ and\ \citenamefont
  {Powell}(2017)}]{lan2017eigenstate}%
  \BibitemOpen
  \bibfield  {author} {\bibinfo {author} {\bibfnamefont {Z.}~\bibnamefont
  {Lan}}\ and\ \bibinfo {author} {\bibfnamefont {S.}~\bibnamefont {Powell}},\
  }\bibfield  {title} {\bibinfo {title} {Eigenstate thermalization hypothesis
  in quantum dimer models},\ }\href
  {https://doi.org/10.1103/PhysRevB.96.115140} {\bibfield  {journal} {\bibinfo
  {journal} {Phys. Rev. B}\ }\textbf {\bibinfo {volume} {96}},\ \bibinfo
  {pages} {115140} (\bibinfo {year} {2017})}\BibitemShut {NoStop}%
\bibitem [{\citenamefont {Essler}\ and\ \citenamefont
  {Konik}(2005)}]{essler2005applications}%
  \BibitemOpen
  \bibfield  {author} {\bibinfo {author} {\bibfnamefont {F.~H.~L.}\
  \bibnamefont {Essler}}\ and\ \bibinfo {author} {\bibfnamefont {R.~M.}\
  \bibnamefont {Konik}},\ }\bibfield  {title} {\bibinfo {title} {{Applications
  of Massive Integrable Quantum Field Theories to Problems in Condensed Matter
  Physics}},\ }in\ \href@noop {} {\emph {\bibinfo {booktitle} {From Fields to
  Strings: Circumnavigating Theoretical Physics}}},\ \bibinfo {editor} {edited
  by\ \bibinfo {editor} {\bibfnamefont {M.}~\bibnamefont {Shifman}}, \bibinfo
  {editor} {\bibfnamefont {A.}~\bibnamefont {Vainshtein}},\ and\ \bibinfo
  {editor} {\bibfnamefont {J.}~\bibnamefont {Wheater}}}\ (\bibinfo  {publisher}
  {World Scientific},\ \bibinfo {address} {Singapore},\ \bibinfo {year}
  {2005})\ \Eprint {https://arxiv.org/abs/arXiv:0412421 (2004)} {arXiv:0412421
  (2004)} \BibitemShut {NoStop}%
\bibitem [{\citenamefont {Kim}\ \emph {et~al.}(2014)\citenamefont {Kim},
  \citenamefont {Ikeda},\ and\ \citenamefont {Huse}}]{kim2014testing}%
  \BibitemOpen
  \bibfield  {author} {\bibinfo {author} {\bibfnamefont {H.}~\bibnamefont
  {Kim}}, \bibinfo {author} {\bibfnamefont {T.~N.}\ \bibnamefont {Ikeda}},\
  and\ \bibinfo {author} {\bibfnamefont {D.~A.}\ \bibnamefont {Huse}},\
  }\bibfield  {title} {\bibinfo {title} {Testing whether all eigenstates obey
  the eigenstate thermalization hypothesis},\ }\href
  {https://doi.org/10.1103/PhysRevE.90.052105} {\bibfield  {journal} {\bibinfo
  {journal} {Phys. Rev. E}\ }\textbf {\bibinfo {volume} {90}},\ \bibinfo
  {pages} {052105} (\bibinfo {year} {2014})}\BibitemShut {NoStop}%
\bibitem [{\citenamefont {Moudgalya}\ \emph
  {et~al.}(2018{\natexlab{a}})\citenamefont {Moudgalya}, \citenamefont
  {Rachel}, \citenamefont {Bernevig},\ and\ \citenamefont
  {Regnault}}]{moudgalya2018exact}%
  \BibitemOpen
  \bibfield  {author} {\bibinfo {author} {\bibfnamefont {S.}~\bibnamefont
  {Moudgalya}}, \bibinfo {author} {\bibfnamefont {S.}~\bibnamefont {Rachel}},
  \bibinfo {author} {\bibfnamefont {B.~A.}\ \bibnamefont {Bernevig}},\ and\
  \bibinfo {author} {\bibfnamefont {N.}~\bibnamefont {Regnault}},\ }\bibfield
  {title} {\bibinfo {title} {Exact excited states of nonintegrable models},\
  }\href {https://doi.org/10.1103/PhysRevB.98.235155} {\bibfield  {journal}
  {\bibinfo  {journal} {Phys. Rev. B}\ }\textbf {\bibinfo {volume} {98}},\
  \bibinfo {pages} {235155} (\bibinfo {year} {2018}{\natexlab{a}})}\BibitemShut
  {NoStop}%
\bibitem [{\citenamefont {Moudgalya}\ \emph
  {et~al.}(2018{\natexlab{b}})\citenamefont {Moudgalya}, \citenamefont
  {Regnault},\ and\ \citenamefont {Bernevig}}]{moudgalya2018entanglement}%
  \BibitemOpen
  \bibfield  {author} {\bibinfo {author} {\bibfnamefont {S.}~\bibnamefont
  {Moudgalya}}, \bibinfo {author} {\bibfnamefont {N.}~\bibnamefont
  {Regnault}},\ and\ \bibinfo {author} {\bibfnamefont {B.~A.}\ \bibnamefont
  {Bernevig}},\ }\bibfield  {title} {\bibinfo {title} {{Entanglement of exact
  excited states of Affleck-Kennedy-Lieb-Tasaki models: Exact results,
  many-body scars, and violation of the strong eigenstate thermalization
  hypothesis}},\ }\href {https://doi.org/10.1103/PhysRevB.98.235156} {\bibfield
   {journal} {\bibinfo  {journal} {Phys. Rev. B}\ }\textbf {\bibinfo {volume}
  {98}},\ \bibinfo {pages} {235156} (\bibinfo {year}
  {2018}{\natexlab{b}})}\BibitemShut {NoStop}%
\bibitem [{\citenamefont {{Bernien}}\ \emph {et~al.}(2017)\citenamefont
  {{Bernien}}, \citenamefont {{Schwartz}}, \citenamefont {{Keesling}},
  \citenamefont {{Levine}}, \citenamefont {{Omran}}, \citenamefont {{Pichler}},
  \citenamefont {{Choi}}, \citenamefont {{Zibrov}}, \citenamefont {{Endres}},
  \citenamefont {{Greiner}}, \citenamefont {{Vuleti{\'c}}},\ and\ \citenamefont
  {{Lukin}}}]{bernien2017probing}%
  \BibitemOpen
  \bibfield  {author} {\bibinfo {author} {\bibfnamefont {H.}~\bibnamefont
  {{Bernien}}}, \bibinfo {author} {\bibfnamefont {S.}~\bibnamefont
  {{Schwartz}}}, \bibinfo {author} {\bibfnamefont {A.}~\bibnamefont
  {{Keesling}}}, \bibinfo {author} {\bibfnamefont {H.}~\bibnamefont
  {{Levine}}}, \bibinfo {author} {\bibfnamefont {A.}~\bibnamefont {{Omran}}},
  \bibinfo {author} {\bibfnamefont {H.}~\bibnamefont {{Pichler}}}, \bibinfo
  {author} {\bibfnamefont {S.}~\bibnamefont {{Choi}}}, \bibinfo {author}
  {\bibfnamefont {A.~S.}\ \bibnamefont {{Zibrov}}}, \bibinfo {author}
  {\bibfnamefont {M.}~\bibnamefont {{Endres}}}, \bibinfo {author}
  {\bibfnamefont {M.}~\bibnamefont {{Greiner}}}, \bibinfo {author}
  {\bibfnamefont {V.}~\bibnamefont {{Vuleti{\'c}}}},\ and\ \bibinfo {author}
  {\bibfnamefont {M.~D.}\ \bibnamefont {{Lukin}}},\ }\bibfield  {title}
  {\bibinfo {title} {{Probing many-body dynamics on a 51-atom quantum
  simulator}},\ }\href {https://doi.org/10.1038/nature24622} {\bibfield
  {journal} {\bibinfo  {journal} {\nat}\ }\textbf {\bibinfo {volume} {551}},\
  \bibinfo {pages} {579} (\bibinfo {year} {2017})}\BibitemShut {NoStop}%
\bibitem [{\citenamefont {Turner}\ \emph
  {et~al.}(2018{\natexlab{a}})\citenamefont {Turner}, \citenamefont
  {Michailidis}, \citenamefont {Abanin}, \citenamefont {Serbyn},\ and\
  \citenamefont {Papi{\'c}}}]{turner2017quantum}%
  \BibitemOpen
  \bibfield  {author} {\bibinfo {author} {\bibfnamefont {C.~J.}\ \bibnamefont
  {Turner}}, \bibinfo {author} {\bibfnamefont {A.~A.}\ \bibnamefont
  {Michailidis}}, \bibinfo {author} {\bibfnamefont {D.~A.}\ \bibnamefont
  {Abanin}}, \bibinfo {author} {\bibfnamefont {M.}~\bibnamefont {Serbyn}},\
  and\ \bibinfo {author} {\bibfnamefont {Z.}~\bibnamefont {Papi{\'c}}},\
  }\bibfield  {title} {\bibinfo {title} {Weak ergodicity breaking from quantum
  many-body scars},\ }\href {https://doi.org/10.1038/s41567-018-0137-5}
  {\bibfield  {journal} {\bibinfo  {journal} {Nat. Phys.}\ }\textbf {\bibinfo
  {volume} {14}},\ \bibinfo {pages} {745} (\bibinfo {year}
  {2018}{\natexlab{a}})}\BibitemShut {NoStop}%
\bibitem [{\citenamefont {Khemani}\ \emph {et~al.}(2019)\citenamefont
  {Khemani}, \citenamefont {Laumann},\ and\ \citenamefont
  {Chandran}}]{khemani2018signatures}%
  \BibitemOpen
  \bibfield  {author} {\bibinfo {author} {\bibfnamefont {V.}~\bibnamefont
  {Khemani}}, \bibinfo {author} {\bibfnamefont {C.~R.}\ \bibnamefont
  {Laumann}},\ and\ \bibinfo {author} {\bibfnamefont {A.}~\bibnamefont
  {Chandran}},\ }\bibfield  {title} {\bibinfo {title} {{Signatures of
  integrability in the dynamics of Rydberg-blockaded chains}},\ }\href
  {https://doi.org/10.1103/PhysRevB.99.161101} {\bibfield  {journal} {\bibinfo
  {journal} {Phys. Rev. B}\ }\textbf {\bibinfo {volume} {99}},\ \bibinfo
  {pages} {161101} (\bibinfo {year} {2019})}\BibitemShut {NoStop}%
\bibitem [{\citenamefont {Ho}\ \emph {et~al.}(2019)\citenamefont {Ho},
  \citenamefont {Choi}, \citenamefont {Pichler},\ and\ \citenamefont
  {Lukin}}]{ho2018periodic}%
  \BibitemOpen
  \bibfield  {author} {\bibinfo {author} {\bibfnamefont {W.~W.}\ \bibnamefont
  {Ho}}, \bibinfo {author} {\bibfnamefont {S.}~\bibnamefont {Choi}}, \bibinfo
  {author} {\bibfnamefont {H.}~\bibnamefont {Pichler}},\ and\ \bibinfo {author}
  {\bibfnamefont {M.~D.}\ \bibnamefont {Lukin}},\ }\bibfield  {title} {\bibinfo
  {title} {{Periodic Orbits, Entanglement, and Quantum Many-Body Scars in
  Constrained Models: Matrix Product State Approach}},\ }\href
  {https://doi.org/10.1103/PhysRevLett.122.040603} {\bibfield  {journal}
  {\bibinfo  {journal} {Phys. Rev. Lett.}\ }\textbf {\bibinfo {volume} {122}},\
  \bibinfo {pages} {040603} (\bibinfo {year} {2019})}\BibitemShut {NoStop}%
\bibitem [{\citenamefont {{Lin}}\ and\ \citenamefont
  {{Motrunich}}(2018)}]{lin2018exact}%
  \BibitemOpen
  \bibfield  {author} {\bibinfo {author} {\bibfnamefont {C.-J.}\ \bibnamefont
  {{Lin}}}\ and\ \bibinfo {author} {\bibfnamefont {O.~I.}\ \bibnamefont
  {{Motrunich}}},\ }\bibfield  {title} {\bibinfo {title} {{Exact Strong-ETH
  Violating Eigenstates in the Rydberg-blockaded Atom Chain}},\ }\href@noop {}
  {\bibfield  {journal} {\bibinfo  {journal} {arXiv e-prints}\ ,\ \bibinfo
  {eid} {arXiv:1810.00888}} (\bibinfo {year} {2018})},\ \Eprint
  {https://arxiv.org/abs/1810.00888} {arXiv:1810.00888 [cond-mat.quant-gas]}
  \BibitemShut {NoStop}%
\bibitem [{\citenamefont {Turner}\ \emph
  {et~al.}(2018{\natexlab{b}})\citenamefont {Turner}, \citenamefont
  {Michailidis}, \citenamefont {Abanin}, \citenamefont {Serbyn},\ and\
  \citenamefont {Papi{\'c}}}]{turner2018quantum}%
  \BibitemOpen
  \bibfield  {author} {\bibinfo {author} {\bibfnamefont {C.~J.}\ \bibnamefont
  {Turner}}, \bibinfo {author} {\bibfnamefont {A.~A.}\ \bibnamefont
  {Michailidis}}, \bibinfo {author} {\bibfnamefont {D.~A.}\ \bibnamefont
  {Abanin}}, \bibinfo {author} {\bibfnamefont {M.}~\bibnamefont {Serbyn}},\
  and\ \bibinfo {author} {\bibfnamefont {Z.}~\bibnamefont {Papi{\'c}}},\
  }\bibfield  {title} {\bibinfo {title} {{Quantum scarred eigenstates in a
  Rydberg atom chain: Entanglement, breakdown of thermalization, and stability
  to perturbations}},\ }\href {https://doi.org/10.1103/PhysRevB.98.155134}
  {\bibfield  {journal} {\bibinfo  {journal} {Phys. Rev. B}\ }\textbf {\bibinfo
  {volume} {98}},\ \bibinfo {pages} {155134} (\bibinfo {year}
  {2018}{\natexlab{b}})}\BibitemShut {NoStop}%
\bibitem [{\citenamefont {{Choi}}\ \emph {et~al.}(2018)\citenamefont {{Choi}},
  \citenamefont {{Turner}}, \citenamefont {{Pichler}}, \citenamefont {{Ho}},
  \citenamefont {{Michailidis}}, \citenamefont {{Papi{\'c}}}, \citenamefont
  {{Serbyn}}, \citenamefont {{Lukin}},\ and\ \citenamefont
  {{Abanin}}}]{choi2018emergent}%
  \BibitemOpen
  \bibfield  {author} {\bibinfo {author} {\bibfnamefont {S.}~\bibnamefont
  {{Choi}}}, \bibinfo {author} {\bibfnamefont {C.~J.}\ \bibnamefont
  {{Turner}}}, \bibinfo {author} {\bibfnamefont {H.}~\bibnamefont {{Pichler}}},
  \bibinfo {author} {\bibfnamefont {W.~W.}\ \bibnamefont {{Ho}}}, \bibinfo
  {author} {\bibfnamefont {A.~A.}\ \bibnamefont {{Michailidis}}}, \bibinfo
  {author} {\bibfnamefont {Z.}~\bibnamefont {{Papi{\'c}}}}, \bibinfo {author}
  {\bibfnamefont {M.}~\bibnamefont {{Serbyn}}}, \bibinfo {author}
  {\bibfnamefont {M.~D.}\ \bibnamefont {{Lukin}}},\ and\ \bibinfo {author}
  {\bibfnamefont {D.~A.}\ \bibnamefont {{Abanin}}},\ }\bibfield  {title}
  {\bibinfo {title} {{Emergent SU(2) dynamics and perfect quantum many-body
  scars}},\ }\href@noop {} {\bibfield  {journal} {\bibinfo  {journal} {arXiv
  e-prints}\ ,\ \bibinfo {eid} {arXiv:1812.05561}} (\bibinfo {year} {2018})},\
  \Eprint {https://arxiv.org/abs/1812.05561} {arXiv:1812.05561 [quant-ph]}
  \BibitemShut {NoStop}%
\bibitem [{\citenamefont {Ates}\ \emph {et~al.}(2012)\citenamefont {Ates},
  \citenamefont {Garrahan},\ and\ \citenamefont
  {Lesanovsky}}]{ates2012thermalization}%
  \BibitemOpen
  \bibfield  {author} {\bibinfo {author} {\bibfnamefont {C.}~\bibnamefont
  {Ates}}, \bibinfo {author} {\bibfnamefont {J.~P.}\ \bibnamefont {Garrahan}},\
  and\ \bibinfo {author} {\bibfnamefont {I.}~\bibnamefont {Lesanovsky}},\
  }\bibfield  {title} {\bibinfo {title} {{Thermalization of a Strongly
  Interacting Closed Spin System: From Coherent Many-Body Dynamics to a
  Fokker-Planck Equation}},\ }\href
  {https://doi.org/10.1103/PhysRevLett.108.110603} {\bibfield  {journal}
  {\bibinfo  {journal} {Phys. Rev. Lett.}\ }\textbf {\bibinfo {volume} {108}},\
  \bibinfo {pages} {110603} (\bibinfo {year} {2012})}\BibitemShut {NoStop}%
\bibitem [{\citenamefont {Ji}\ \emph {et~al.}(2013)\citenamefont {Ji},
  \citenamefont {Ates}, \citenamefont {Garrahan},\ and\ \citenamefont
  {Lesanovsky}}]{ji2013equilibration}%
  \BibitemOpen
  \bibfield  {author} {\bibinfo {author} {\bibfnamefont {S.}~\bibnamefont
  {Ji}}, \bibinfo {author} {\bibfnamefont {C.}~\bibnamefont {Ates}}, \bibinfo
  {author} {\bibfnamefont {J.~P.}\ \bibnamefont {Garrahan}},\ and\ \bibinfo
  {author} {\bibfnamefont {I.}~\bibnamefont {Lesanovsky}},\ }\bibfield  {title}
  {\bibinfo {title} {Equilibration of quantum hard rods in one dimension},\
  }\href {https://doi.org/10.1088/1742-5468/2013/02/p02005} {\bibfield
  {journal} {\bibinfo  {journal} {J. Stat. Mech.}\ }\textbf {\bibinfo {volume}
  {2013}},\ \bibinfo {pages} {P02005} (\bibinfo {year} {2013})}\BibitemShut
  {NoStop}%
\bibitem [{\citenamefont {van Horssen}\ \emph {et~al.}(2015)\citenamefont {van
  Horssen}, \citenamefont {Levi},\ and\ \citenamefont
  {Garrahan}}]{vanhorssen2015dynamics}%
  \BibitemOpen
  \bibfield  {author} {\bibinfo {author} {\bibfnamefont {M.}~\bibnamefont {van
  Horssen}}, \bibinfo {author} {\bibfnamefont {E.}~\bibnamefont {Levi}},\ and\
  \bibinfo {author} {\bibfnamefont {J.~P.}\ \bibnamefont {Garrahan}},\
  }\bibfield  {title} {\bibinfo {title} {Dynamics of many-body localization in
  a translation-invariant quantum glass model},\ }\href
  {https://doi.org/10.1103/PhysRevB.92.100305} {\bibfield  {journal} {\bibinfo
  {journal} {Phys. Rev. B}\ }\textbf {\bibinfo {volume} {92}},\ \bibinfo
  {pages} {100305} (\bibinfo {year} {2015})}\BibitemShut {NoStop}%
\bibitem [{\citenamefont {Lan}\ \emph {et~al.}(2018)\citenamefont {Lan},
  \citenamefont {van Horssen}, \citenamefont {Powell},\ and\ \citenamefont
  {Garrahan}}]{lan2018quantum}%
  \BibitemOpen
  \bibfield  {author} {\bibinfo {author} {\bibfnamefont {Z.}~\bibnamefont
  {Lan}}, \bibinfo {author} {\bibfnamefont {M.}~\bibnamefont {van Horssen}},
  \bibinfo {author} {\bibfnamefont {S.}~\bibnamefont {Powell}},\ and\ \bibinfo
  {author} {\bibfnamefont {J.~P.}\ \bibnamefont {Garrahan}},\ }\bibfield
  {title} {\bibinfo {title} {{Quantum Slow Relaxation and Metastability due to
  Dynamical Constraints}},\ }\href
  {https://doi.org/10.1103/PhysRevLett.121.040603} {\bibfield  {journal}
  {\bibinfo  {journal} {Phys. Rev. Lett.}\ }\textbf {\bibinfo {volume} {121}},\
  \bibinfo {pages} {040603} (\bibinfo {year} {2018})}\BibitemShut {NoStop}%
\bibitem [{\citenamefont {{Liu}}\ \emph {et~al.}(2018)\citenamefont {{Liu}},
  \citenamefont {{Lundgren}}, \citenamefont {{Titum}}, \citenamefont
  {{Pagano}}, \citenamefont {{Zhang}}, \citenamefont {{Monroe}},\ and\
  \citenamefont {{Gorshkov}}}]{liu2018confined}%
  \BibitemOpen
  \bibfield  {author} {\bibinfo {author} {\bibfnamefont {F.}~\bibnamefont
  {{Liu}}}, \bibinfo {author} {\bibfnamefont {R.}~\bibnamefont {{Lundgren}}},
  \bibinfo {author} {\bibfnamefont {P.}~\bibnamefont {{Titum}}}, \bibinfo
  {author} {\bibfnamefont {G.}~\bibnamefont {{Pagano}}}, \bibinfo {author}
  {\bibfnamefont {J.}~\bibnamefont {{Zhang}}}, \bibinfo {author} {\bibfnamefont
  {C.}~\bibnamefont {{Monroe}}},\ and\ \bibinfo {author} {\bibfnamefont
  {A.~V.}\ \bibnamefont {{Gorshkov}}},\ }\bibfield  {title} {\bibinfo {title}
  {{Confined Dynamics in Long-Range Interacting Quantum Spin Chains}},\
  }\href@noop {} {\bibfield  {journal} {\bibinfo  {journal} {arXiv e-prints}\
  ,\ \bibinfo {eid} {arXiv:1810.02365}} (\bibinfo {year} {2018})},\ \Eprint
  {https://arxiv.org/abs/1810.02365} {arXiv:1810.02365 [cond-mat.quant-gas]}
  \BibitemShut {NoStop}%
\bibitem [{\citenamefont {Lerose}\ \emph {et~al.}(2019)\citenamefont {Lerose},
  \citenamefont {\ifmmode \check{Z}\else
  \v{Z}\fi{}unkovi\ifmmode~\check{c}\else \v{c}\fi{}}, \citenamefont {Silva},\
  and\ \citenamefont {Gambassi}}]{lerose2019quasilocalized}%
  \BibitemOpen
  \bibfield  {author} {\bibinfo {author} {\bibfnamefont {A.}~\bibnamefont
  {Lerose}}, \bibinfo {author} {\bibfnamefont {B.}~\bibnamefont {\ifmmode
  \check{Z}\else \v{Z}\fi{}unkovi\ifmmode~\check{c}\else \v{c}\fi{}}}, \bibinfo
  {author} {\bibfnamefont {A.}~\bibnamefont {Silva}},\ and\ \bibinfo {author}
  {\bibfnamefont {A.}~\bibnamefont {Gambassi}},\ }\bibfield  {title} {\bibinfo
  {title} {{Quasilocalized excitations induced by long-range interactions in
  translationally invariant quantum spin chains}},\ }\href
  {https://doi.org/10.1103/PhysRevB.99.121112} {\bibfield  {journal} {\bibinfo
  {journal} {Phys. Rev. B}\ }\textbf {\bibinfo {volume} {99}},\ \bibinfo
  {pages} {121112} (\bibinfo {year} {2019})}\BibitemShut {NoStop}%
\bibitem [{\citenamefont {Steinigeweg}\ \emph {et~al.}(2014)\citenamefont
  {Steinigeweg}, \citenamefont {Khodja}, \citenamefont {Niemeyer},
  \citenamefont {Gogolin},\ and\ \citenamefont
  {Gemmer}}]{steinigeweg2014pushing}%
  \BibitemOpen
  \bibfield  {author} {\bibinfo {author} {\bibfnamefont {R.}~\bibnamefont
  {Steinigeweg}}, \bibinfo {author} {\bibfnamefont {A.}~\bibnamefont {Khodja}},
  \bibinfo {author} {\bibfnamefont {H.}~\bibnamefont {Niemeyer}}, \bibinfo
  {author} {\bibfnamefont {C.}~\bibnamefont {Gogolin}},\ and\ \bibinfo {author}
  {\bibfnamefont {J.}~\bibnamefont {Gemmer}},\ }\bibfield  {title} {\bibinfo
  {title} {{Pushing the Limits of the Eigenstate Thermalization Hypothesis
  towards Mesoscopic Quantum Systems}},\ }\href
  {https://doi.org/10.1103/PhysRevLett.112.130403} {\bibfield  {journal}
  {\bibinfo  {journal} {Phys. Rev. Lett.}\ }\textbf {\bibinfo {volume} {112}},\
  \bibinfo {pages} {130403} (\bibinfo {year} {2014})}\BibitemShut {NoStop}%
\bibitem [{\citenamefont {Bartsch}\ and\ \citenamefont
  {Gemmer}(2017)}]{bartsch2017necessity}%
  \BibitemOpen
  \bibfield  {author} {\bibinfo {author} {\bibfnamefont {C.}~\bibnamefont
  {Bartsch}}\ and\ \bibinfo {author} {\bibfnamefont {J.}~\bibnamefont
  {Gemmer}},\ }\bibfield  {title} {\bibinfo {title} {Necessity of eigenstate
  thermalisation for equilibration towards unique expectation values when
  starting from generic initial states},\ }\href
  {http://stacks.iop.org/0295-5075/118/i=1/a=10006} {\bibfield  {journal}
  {\bibinfo  {journal} {EPL (Europhysics Letters)}\ }\textbf {\bibinfo {volume}
  {118}},\ \bibinfo {pages} {10006} (\bibinfo {year} {2017})}\BibitemShut
  {NoStop}%
\bibitem [{\citenamefont {Kollath}\ \emph {et~al.}(2007)\citenamefont
  {Kollath}, \citenamefont {L\"auchli},\ and\ \citenamefont
  {Altman}}]{kollath2007quench}%
  \BibitemOpen
  \bibfield  {author} {\bibinfo {author} {\bibfnamefont {C.}~\bibnamefont
  {Kollath}}, \bibinfo {author} {\bibfnamefont {A.~M.}\ \bibnamefont
  {L\"auchli}},\ and\ \bibinfo {author} {\bibfnamefont {E.}~\bibnamefont
  {Altman}},\ }\bibfield  {title} {\bibinfo {title} {{Quench Dynamics and
  Nonequilibrium Phase Diagram of the {B}ose-{H}ubbard Model}},\ }\href
  {https://doi.org/10.1103/PhysRevLett.98.180601} {\bibfield  {journal}
  {\bibinfo  {journal} {Phys. Rev. Lett.}\ }\textbf {\bibinfo {volume} {98}},\
  \bibinfo {pages} {180601} (\bibinfo {year} {2007})}\BibitemShut {NoStop}%
\bibitem [{\citenamefont {Manmana}\ \emph {et~al.}(2007)\citenamefont
  {Manmana}, \citenamefont {Wessel}, \citenamefont {Noack},\ and\ \citenamefont
  {Muramatsu}}]{manmana2007strongly}%
  \BibitemOpen
  \bibfield  {author} {\bibinfo {author} {\bibfnamefont {S.~R.}\ \bibnamefont
  {Manmana}}, \bibinfo {author} {\bibfnamefont {S.}~\bibnamefont {Wessel}},
  \bibinfo {author} {\bibfnamefont {R.~M.}\ \bibnamefont {Noack}},\ and\
  \bibinfo {author} {\bibfnamefont {A.}~\bibnamefont {Muramatsu}},\ }\bibfield
  {title} {\bibinfo {title} {{Strongly Correlated Fermions after a Quantum
  Quench}},\ }\href {https://doi.org/10.1103/PhysRevLett.98.210405} {\bibfield
  {journal} {\bibinfo  {journal} {Phys. Rev. Lett.}\ }\textbf {\bibinfo
  {volume} {98}},\ \bibinfo {pages} {210405} (\bibinfo {year}
  {2007})}\BibitemShut {NoStop}%
\bibitem [{\citenamefont {Trotzky}\ \emph {et~al.}(2012)\citenamefont
  {Trotzky}, \citenamefont {Chen}, \citenamefont {Flesch}, \citenamefont
  {McCulloch}, \citenamefont {Schollwock}, \citenamefont {Eisert},\ and\
  \citenamefont {Bloch}}]{trotzky2012probing}%
  \BibitemOpen
  \bibfield  {author} {\bibinfo {author} {\bibfnamefont {S.}~\bibnamefont
  {Trotzky}}, \bibinfo {author} {\bibfnamefont {Y.-A.}\ \bibnamefont {Chen}},
  \bibinfo {author} {\bibfnamefont {A.}~\bibnamefont {Flesch}}, \bibinfo
  {author} {\bibfnamefont {I.~P.}\ \bibnamefont {McCulloch}}, \bibinfo {author}
  {\bibfnamefont {U.}~\bibnamefont {Schollwock}}, \bibinfo {author}
  {\bibfnamefont {J.}~\bibnamefont {Eisert}},\ and\ \bibinfo {author}
  {\bibfnamefont {I.}~\bibnamefont {Bloch}},\ }\bibfield  {title} {\bibinfo
  {title} {Probing the relaxation towards equilibrium in an isolated strongly
  correlated one-dimensional {B}ose gas},\ }\href
  {http://dx.doi.org/10.1038/nphys2232} {\bibfield  {journal} {\bibinfo
  {journal} {Nat. Phys.}\ }\textbf {\bibinfo {volume} {8}},\ \bibinfo {pages}
  {325} (\bibinfo {year} {2012})}\BibitemShut {NoStop}%
\bibitem [{\citenamefont {James}\ and\ \citenamefont
  {Konik}(2015)}]{james2015quantum}%
  \BibitemOpen
  \bibfield  {author} {\bibinfo {author} {\bibfnamefont {A.~J.~A.}\
  \bibnamefont {James}}\ and\ \bibinfo {author} {\bibfnamefont {R.~M.}\
  \bibnamefont {Konik}},\ }\bibfield  {title} {\bibinfo {title} {Quantum
  quenches in two spatial dimensions using chain array matrix product states},\
  }\href {https://doi.org/10.1103/PhysRevB.92.161111} {\bibfield  {journal}
  {\bibinfo  {journal} {Phys. Rev. B}\ }\textbf {\bibinfo {volume} {92}},\
  \bibinfo {pages} {161111} (\bibinfo {year} {2015})}\BibitemShut {NoStop}%
\bibitem [{\citenamefont {James}\ \emph {et~al.}(2018)\citenamefont {James},
  \citenamefont {Konik}, \citenamefont {Lecheminant}, \citenamefont
  {Robinson},\ and\ \citenamefont {Tsvelik}}]{james2018nonperturbative}%
  \BibitemOpen
  \bibfield  {author} {\bibinfo {author} {\bibfnamefont {A.~J.~A.}\
  \bibnamefont {James}}, \bibinfo {author} {\bibfnamefont {R.~M.}\ \bibnamefont
  {Konik}}, \bibinfo {author} {\bibfnamefont {P.}~\bibnamefont {Lecheminant}},
  \bibinfo {author} {\bibfnamefont {N.~J.}\ \bibnamefont {Robinson}},\ and\
  \bibinfo {author} {\bibfnamefont {A.~M.}\ \bibnamefont {Tsvelik}},\
  }\bibfield  {title} {\bibinfo {title} {{Non-perturbative methodologies for
  low-dimensional strongly-correlated systems: From non-Abelian bosonization to
  truncated spectrum methods}},\ }\href
  {https://doi.org/10.1088/1361-6633/aa91ea} {\bibfield  {journal} {\bibinfo
  {journal} {Rep. Prog. Phys.}\ }\textbf {\bibinfo {volume} {81}},\ \bibinfo
  {pages} {046002} (\bibinfo {year} {2018})}\BibitemShut {NoStop}%
\bibitem [{\citenamefont {Brandino}\ \emph {et~al.}(2015)\citenamefont
  {Brandino}, \citenamefont {Caux},\ and\ \citenamefont
  {Konik}}]{brandino2015glimmers}%
  \BibitemOpen
  \bibfield  {author} {\bibinfo {author} {\bibfnamefont {G.~P.}\ \bibnamefont
  {Brandino}}, \bibinfo {author} {\bibfnamefont {J.-S.}\ \bibnamefont {Caux}},\
  and\ \bibinfo {author} {\bibfnamefont {R.~M.}\ \bibnamefont {Konik}},\
  }\bibfield  {title} {\bibinfo {title} {{Glimmers of a Quantum {KAM} Theorem:
  Insights from Quantum Quenches in One-Dimensional {B}ose Gases}},\ }\href
  {https://doi.org/10.1103/PhysRevX.5.041043} {\bibfield  {journal} {\bibinfo
  {journal} {Phys. Rev. X}\ }\textbf {\bibinfo {volume} {5}},\ \bibinfo {pages}
  {041043} (\bibinfo {year} {2015})}\BibitemShut {NoStop}%
\bibitem [{\citenamefont {Moeckel}\ and\ \citenamefont
  {Kehrein}(2008)}]{moeckel2008interaction}%
  \BibitemOpen
  \bibfield  {author} {\bibinfo {author} {\bibfnamefont {M.}~\bibnamefont
  {Moeckel}}\ and\ \bibinfo {author} {\bibfnamefont {S.}~\bibnamefont
  {Kehrein}},\ }\bibfield  {title} {\bibinfo {title} {{Interaction Quench in
  the Hubbard Model}},\ }\href {https://doi.org/10.1103/PhysRevLett.100.175702}
  {\bibfield  {journal} {\bibinfo  {journal} {Phys. Rev. Lett.}\ }\textbf
  {\bibinfo {volume} {100}},\ \bibinfo {pages} {175702} (\bibinfo {year}
  {2008})}\BibitemShut {NoStop}%
\bibitem [{\citenamefont {Moeckel}\ and\ \citenamefont
  {Kehrein}(2009)}]{moeckel2009realtime}%
  \BibitemOpen
  \bibfield  {author} {\bibinfo {author} {\bibfnamefont {M.}~\bibnamefont
  {Moeckel}}\ and\ \bibinfo {author} {\bibfnamefont {S.}~\bibnamefont
  {Kehrein}},\ }\bibfield  {title} {\bibinfo {title} {{Real-time evolution for
  weak interaction quenches in quantum systems}},\ }\href
  {https://doi.org/http://doi.org/10.1016/j.aop.2009.03.009} {\bibfield
  {journal} {\bibinfo  {journal} {Ann. Phys. (N.Y.)}\ }\textbf {\bibinfo
  {volume} {324}},\ \bibinfo {pages} {2146 } (\bibinfo {year}
  {2009})}\BibitemShut {NoStop}%
\bibitem [{\citenamefont {Eckstein}\ \emph {et~al.}(2009)\citenamefont
  {Eckstein}, \citenamefont {Hackl}, \citenamefont {Kehrein}, \citenamefont
  {Kollar}, \citenamefont {Moeckel}, \citenamefont {Werner},\ and\
  \citenamefont {Wolf}}]{eckstein2009new}%
  \BibitemOpen
  \bibfield  {author} {\bibinfo {author} {\bibfnamefont {M.}~\bibnamefont
  {Eckstein}}, \bibinfo {author} {\bibfnamefont {A.}~\bibnamefont {Hackl}},
  \bibinfo {author} {\bibfnamefont {S.}~\bibnamefont {Kehrein}}, \bibinfo
  {author} {\bibfnamefont {M.}~\bibnamefont {Kollar}}, \bibinfo {author}
  {\bibfnamefont {M.}~\bibnamefont {Moeckel}}, \bibinfo {author} {\bibfnamefont
  {P.}~\bibnamefont {Werner}},\ and\ \bibinfo {author} {\bibfnamefont
  {F.}~\bibnamefont {Wolf}},\ }\bibfield  {title} {\bibinfo {title} {{New
  theoretical approaches for correlated systems in nonequilibrium}},\ }\href
  {https://doi.org/10.1140/epjst/e2010-01219-x} {\bibfield  {journal} {\bibinfo
   {journal} {Eur. Phys. J. Special Topics}\ }\textbf {\bibinfo {volume}
  {180}},\ \bibinfo {pages} {217} (\bibinfo {year} {2009})}\BibitemShut
  {NoStop}%
\bibitem [{\citenamefont {Essler}\ \emph {et~al.}(2014)\citenamefont {Essler},
  \citenamefont {Kehrein}, \citenamefont {Manmana},\ and\ \citenamefont
  {Robinson}}]{essler2014quench}%
  \BibitemOpen
  \bibfield  {author} {\bibinfo {author} {\bibfnamefont {F.~H.~L.}\
  \bibnamefont {Essler}}, \bibinfo {author} {\bibfnamefont {S.}~\bibnamefont
  {Kehrein}}, \bibinfo {author} {\bibfnamefont {S.~R.}\ \bibnamefont
  {Manmana}},\ and\ \bibinfo {author} {\bibfnamefont {N.~J.}\ \bibnamefont
  {Robinson}},\ }\bibfield  {title} {\bibinfo {title} {Quench dynamics in a
  model with tuneable integrability breaking},\ }\href
  {https://doi.org/10.1103/PhysRevB.89.165104} {\bibfield  {journal} {\bibinfo
  {journal} {Phys. Rev. B}\ }\textbf {\bibinfo {volume} {89}},\ \bibinfo
  {pages} {165104} (\bibinfo {year} {2014})}\BibitemShut {NoStop}%
\bibitem [{\citenamefont {Nessi}\ and\ \citenamefont
  {Iucci}(2014)}]{nessi2014equations}%
  \BibitemOpen
  \bibfield  {author} {\bibinfo {author} {\bibfnamefont {N.}~\bibnamefont
  {Nessi}}\ and\ \bibinfo {author} {\bibfnamefont {A.}~\bibnamefont {Iucci}},\
  }\bibfield  {title} {\bibinfo {title} {{Equations of Motion for the
  Out-of-Equilibrium Dynamics of Isolated Quantum Systems from the Projection
  Operator Technique}},\ }\href
  {http://stacks.iop.org/1742-6596/568/i=1/a=012013} {\bibfield  {journal}
  {\bibinfo  {journal} {J. Phys.: Conf. Ser.}\ }\textbf {\bibinfo {volume}
  {568}},\ \bibinfo {pages} {012013} (\bibinfo {year} {2014})}\BibitemShut
  {NoStop}%
\bibitem [{\citenamefont {{Nessi}}\ and\ \citenamefont
  {{Iucci}}(2015)}]{nessi2015glasslike}%
  \BibitemOpen
  \bibfield  {author} {\bibinfo {author} {\bibfnamefont {N.}~\bibnamefont
  {{Nessi}}}\ and\ \bibinfo {author} {\bibfnamefont {A.}~\bibnamefont
  {{Iucci}}},\ }\bibfield  {title} {\bibinfo {title} {{Glass-like Behavior in a
  System of One Dimensional Fermions after a Quantum Quench}},\ }\href@noop {}
  {\bibfield  {journal} {\bibinfo  {journal} {ArXiv e-prints}\ } (\bibinfo
  {year} {2015})},\ \Eprint {https://arxiv.org/abs/1503.02507}
  {arXiv:1503.02507 [cond-mat.quant-gas]} \BibitemShut {NoStop}%
\bibitem [{\citenamefont {Bertini}\ \emph {et~al.}(2015)\citenamefont
  {Bertini}, \citenamefont {Essler}, \citenamefont {Groha},\ and\ \citenamefont
  {Robinson}}]{bertini2015prethermalization}%
  \BibitemOpen
  \bibfield  {author} {\bibinfo {author} {\bibfnamefont {B.}~\bibnamefont
  {Bertini}}, \bibinfo {author} {\bibfnamefont {F.~H.~L.}\ \bibnamefont
  {Essler}}, \bibinfo {author} {\bibfnamefont {S.}~\bibnamefont {Groha}},\ and\
  \bibinfo {author} {\bibfnamefont {N.~J.}\ \bibnamefont {Robinson}},\
  }\bibfield  {title} {\bibinfo {title} {{Prethermalization and Thermalization
  in Models with Weak Integrability Breaking}},\ }\href
  {https://doi.org/10.1103/PhysRevLett.115.180601} {\bibfield  {journal}
  {\bibinfo  {journal} {Phys. Rev. Lett.}\ }\textbf {\bibinfo {volume} {115}},\
  \bibinfo {pages} {180601} (\bibinfo {year} {2015})}\BibitemShut {NoStop}%
\bibitem [{\citenamefont {Bertini}\ \emph {et~al.}(2016)\citenamefont
  {Bertini}, \citenamefont {Essler}, \citenamefont {Groha},\ and\ \citenamefont
  {Robinson}}]{bertini2016thermalization}%
  \BibitemOpen
  \bibfield  {author} {\bibinfo {author} {\bibfnamefont {B.}~\bibnamefont
  {Bertini}}, \bibinfo {author} {\bibfnamefont {F.~H.~L.}\ \bibnamefont
  {Essler}}, \bibinfo {author} {\bibfnamefont {S.}~\bibnamefont {Groha}},\ and\
  \bibinfo {author} {\bibfnamefont {N.~J.}\ \bibnamefont {Robinson}},\
  }\bibfield  {title} {\bibinfo {title} {Thermalization and light cones in a
  model with weak integrability breaking},\ }\href
  {https://doi.org/10.1103/PhysRevB.94.245117} {\bibfield  {journal} {\bibinfo
  {journal} {Phys. Rev. B}\ }\textbf {\bibinfo {volume} {94}},\ \bibinfo
  {pages} {245117} (\bibinfo {year} {2016})}\BibitemShut {NoStop}%
\bibitem [{\citenamefont {F\"urst}\ \emph {et~al.}(2012)\citenamefont
  {F\"urst}, \citenamefont {Mendl},\ and\ \citenamefont
  {Spohn}}]{furst2012matrixvalued}%
  \BibitemOpen
  \bibfield  {author} {\bibinfo {author} {\bibfnamefont {M.~L.~R.}\
  \bibnamefont {F\"urst}}, \bibinfo {author} {\bibfnamefont {C.~B.}\
  \bibnamefont {Mendl}},\ and\ \bibinfo {author} {\bibfnamefont
  {H.}~\bibnamefont {Spohn}},\ }\bibfield  {title} {\bibinfo {title}
  {{Matrix-valued Boltzmann equation for the Hubbard chain}},\ }\href
  {https://doi.org/10.1103/PhysRevE.86.031122} {\bibfield  {journal} {\bibinfo
  {journal} {Phys. Rev. E}\ }\textbf {\bibinfo {volume} {86}},\ \bibinfo
  {pages} {031122} (\bibinfo {year} {2012})}\BibitemShut {NoStop}%
\bibitem [{\citenamefont {F\"urst}\ \emph {et~al.}(2013)\citenamefont
  {F\"urst}, \citenamefont {Mendl},\ and\ \citenamefont
  {Spohn}}]{furst2013matrixvalued}%
  \BibitemOpen
  \bibfield  {author} {\bibinfo {author} {\bibfnamefont {M.~L.~R.}\
  \bibnamefont {F\"urst}}, \bibinfo {author} {\bibfnamefont {C.~B.}\
  \bibnamefont {Mendl}},\ and\ \bibinfo {author} {\bibfnamefont
  {H.}~\bibnamefont {Spohn}},\ }\bibfield  {title} {\bibinfo {title}
  {{Matrix-valued Boltzmann equation for the nonintegrable Hubbard chain}},\
  }\href {https://doi.org/10.1103/PhysRevE.88.012108} {\bibfield  {journal}
  {\bibinfo  {journal} {Phys. Rev. E}\ }\textbf {\bibinfo {volume} {88}},\
  \bibinfo {pages} {012108} (\bibinfo {year} {2013})}\BibitemShut {NoStop}%
\bibitem [{\citenamefont {Biebl}\ and\ \citenamefont
  {Kehrein}(2017)}]{biebl2017thermalization}%
  \BibitemOpen
  \bibfield  {author} {\bibinfo {author} {\bibfnamefont {F.~R.~A.}\
  \bibnamefont {Biebl}}\ and\ \bibinfo {author} {\bibfnamefont
  {S.}~\bibnamefont {Kehrein}},\ }\bibfield  {title} {\bibinfo {title}
  {{Thermalization rates in the one-dimensional Hubbard model with
  next-to-nearest neighbor hopping}},\ }\href
  {https://doi.org/10.1103/PhysRevB.95.104304} {\bibfield  {journal} {\bibinfo
  {journal} {Phys. Rev. B}\ }\textbf {\bibinfo {volume} {95}},\ \bibinfo
  {pages} {104304} (\bibinfo {year} {2017})}\BibitemShut {NoStop}%
\bibitem [{\citenamefont {Rakovszky}\ \emph {et~al.}(2016)\citenamefont
  {Rakovszky}, \citenamefont {Mesty{\'a}n}, \citenamefont {Collura},
  \citenamefont {Kormos},\ and\ \citenamefont
  {Tak{\'a}cs}}]{rakovsky2016hamiltonian}%
  \BibitemOpen
  \bibfield  {author} {\bibinfo {author} {\bibfnamefont {T.}~\bibnamefont
  {Rakovszky}}, \bibinfo {author} {\bibfnamefont {M.}~\bibnamefont
  {Mesty{\'a}n}}, \bibinfo {author} {\bibfnamefont {M.}~\bibnamefont
  {Collura}}, \bibinfo {author} {\bibfnamefont {M.}~\bibnamefont {Kormos}},\
  and\ \bibinfo {author} {\bibfnamefont {G.}~\bibnamefont {Tak{\'a}cs}},\
  }\bibfield  {title} {\bibinfo {title} {{Hamiltonian truncation approach to
  quenches in the Ising field theory}},\ }\href
  {https://doi.org/http://dx.doi.org/10.1016/j.nuclphysb.2016.08.024}
  {\bibfield  {journal} {\bibinfo  {journal} {Nucl. Phys. B}\ }\textbf
  {\bibinfo {volume} {911}},\ \bibinfo {pages} {805 } (\bibinfo {year}
  {2016})}\BibitemShut {NoStop}%
\bibitem [{\citenamefont {{Basko}}\ \emph {et~al.}(2006)\citenamefont
  {{Basko}}, \citenamefont {{Aleiner}},\ and\ \citenamefont
  {{Altshuler}}}]{basko2006metal}%
  \BibitemOpen
  \bibfield  {author} {\bibinfo {author} {\bibfnamefont {D.~M.}\ \bibnamefont
  {{Basko}}}, \bibinfo {author} {\bibfnamefont {I.~L.}\ \bibnamefont
  {{Aleiner}}},\ and\ \bibinfo {author} {\bibfnamefont {B.~L.}\ \bibnamefont
  {{Altshuler}}},\ }\bibfield  {title} {\bibinfo {title} {{Metal insulator
  transition in a weakly interacting many-electron system with localized
  single-particle states}},\ }\href {https://doi.org/10.1016/j.aop.2005.11.014}
  {\bibfield  {journal} {\bibinfo  {journal} {Ann. Phys. (N.Y.)}\ }\textbf
  {\bibinfo {volume} {321}},\ \bibinfo {pages} {1126} (\bibinfo {year}
  {2006})}\BibitemShut {NoStop}%
\bibitem [{\citenamefont {Imbrie}\ \emph {et~al.}(2017)\citenamefont {Imbrie},
  \citenamefont {Ros},\ and\ \citenamefont {Scardicchio}}]{imbrie2017local}%
  \BibitemOpen
  \bibfield  {author} {\bibinfo {author} {\bibfnamefont {J.~Z.}\ \bibnamefont
  {Imbrie}}, \bibinfo {author} {\bibfnamefont {V.}~\bibnamefont {Ros}},\ and\
  \bibinfo {author} {\bibfnamefont {A.}~\bibnamefont {Scardicchio}},\
  }\bibfield  {title} {\bibinfo {title} {Local integrals of motion in many-body
  localized systems},\ }\href {https://doi.org/10.1002/andp.201600278}
  {\bibfield  {journal} {\bibinfo  {journal} {Ann. Phys. (Berl.)}\ }\textbf
  {\bibinfo {volume} {529}},\ \bibinfo {pages} {1600278} (\bibinfo {year}
  {2017})}\BibitemShut {NoStop}%
\bibitem [{\citenamefont {Parameswaran}\ \emph {et~al.}(2017)\citenamefont
  {Parameswaran}, \citenamefont {Potter},\ and\ \citenamefont
  {Vasseur}}]{parameswaran2017eigenstate}%
  \BibitemOpen
  \bibfield  {author} {\bibinfo {author} {\bibfnamefont {S.~A.}\ \bibnamefont
  {Parameswaran}}, \bibinfo {author} {\bibfnamefont {A.~C.}\ \bibnamefont
  {Potter}},\ and\ \bibinfo {author} {\bibfnamefont {R.}~\bibnamefont
  {Vasseur}},\ }\bibfield  {title} {\bibinfo {title} {Eigenstate phase
  transitions and the emergence of universal dynamics in highly excited
  states},\ }\href {https://doi.org/10.1002/andp.201600302} {\bibfield
  {journal} {\bibinfo  {journal} {Ann. Phys. (Berl.)}\ }\textbf {\bibinfo
  {volume} {529}},\ \bibinfo {pages} {1600302} (\bibinfo {year}
  {2017})}\BibitemShut {NoStop}%
\bibitem [{\citenamefont {Agarwal}\ \emph {et~al.}(2017)\citenamefont
  {Agarwal}, \citenamefont {Altman}, \citenamefont {Demler}, \citenamefont
  {Gopalakrishnan}, \citenamefont {Huse},\ and\ \citenamefont
  {Knap}}]{agarwal2017rareregion}%
  \BibitemOpen
  \bibfield  {author} {\bibinfo {author} {\bibfnamefont {K.}~\bibnamefont
  {Agarwal}}, \bibinfo {author} {\bibfnamefont {E.}~\bibnamefont {Altman}},
  \bibinfo {author} {\bibfnamefont {E.}~\bibnamefont {Demler}}, \bibinfo
  {author} {\bibfnamefont {S.}~\bibnamefont {Gopalakrishnan}}, \bibinfo
  {author} {\bibfnamefont {D.~A.}\ \bibnamefont {Huse}},\ and\ \bibinfo
  {author} {\bibfnamefont {M.}~\bibnamefont {Knap}},\ }\bibfield  {title}
  {\bibinfo {title} {Rare-region effects and dynamics near the many-body
  localization transition},\ }\href {https://doi.org/10.1002/andp.201600326}
  {\bibfield  {journal} {\bibinfo  {journal} {Ann. Phys. (Berl.)}\ }\textbf
  {\bibinfo {volume} {529}},\ \bibinfo {pages} {1600326} (\bibinfo {year}
  {2017})}\BibitemShut {NoStop}%
\bibitem [{\citenamefont {Deng}\ \emph {et~al.}(2017)\citenamefont {Deng},
  \citenamefont {Ganeshan}, \citenamefont {Li}, \citenamefont {Modak},
  \citenamefont {Mukerjee},\ and\ \citenamefont {Pixley}}]{deng2017manybody}%
  \BibitemOpen
  \bibfield  {author} {\bibinfo {author} {\bibfnamefont {D.-L.}\ \bibnamefont
  {Deng}}, \bibinfo {author} {\bibfnamefont {S.}~\bibnamefont {Ganeshan}},
  \bibinfo {author} {\bibfnamefont {X.}~\bibnamefont {Li}}, \bibinfo {author}
  {\bibfnamefont {R.}~\bibnamefont {Modak}}, \bibinfo {author} {\bibfnamefont
  {S.}~\bibnamefont {Mukerjee}},\ and\ \bibinfo {author} {\bibfnamefont
  {J.~H.}\ \bibnamefont {Pixley}},\ }\bibfield  {title} {\bibinfo {title}
  {Many-body localization in incommensurate models with a mobility edge},\
  }\href {https://doi.org/10.1002/andp.201600399} {\bibfield  {journal}
  {\bibinfo  {journal} {Ann. Phys. (Berl.)}\ }\textbf {\bibinfo {volume}
  {529}},\ \bibinfo {pages} {1600399} (\bibinfo {year} {2017})}\BibitemShut
  {NoStop}%
\bibitem [{\citenamefont {{Abanin}}\ and\ \citenamefont
  {{Papi{\'c}}}(2017)}]{abanin2017recent}%
  \BibitemOpen
  \bibfield  {author} {\bibinfo {author} {\bibfnamefont {D.~A.}\ \bibnamefont
  {{Abanin}}}\ and\ \bibinfo {author} {\bibfnamefont {Z.}~\bibnamefont
  {{Papi{\'c}}}},\ }\bibfield  {title} {\bibinfo {title} {{Recent progress in
  many-body localization}},\ }\href {https://doi.org/10.1002/andp.201700169}
  {\bibfield  {journal} {\bibinfo  {journal} {Ann. Phys. (Berl.)}\ }\textbf
  {\bibinfo {volume} {529}},\ \bibinfo {pages} {1700169} (\bibinfo {year}
  {2017})}\BibitemShut {NoStop}%
\bibitem [{\citenamefont {Alet}\ and\ \citenamefont
  {Laflorencie}(2018)}]{alet2018manybody}%
  \BibitemOpen
  \bibfield  {author} {\bibinfo {author} {\bibfnamefont {F.}~\bibnamefont
  {Alet}}\ and\ \bibinfo {author} {\bibfnamefont {N.}~\bibnamefont
  {Laflorencie}},\ }\bibfield  {title} {\bibinfo {title} {Many-body
  localization: An introduction and selected topics},\ }\bibfield  {journal}
  {\bibinfo  {journal} {C. R. Phys.}\ }\href
  {https://doi.org/https://doi.org/10.1016/j.crhy.2018.03.003}
  {https://doi.org/10.1016/j.crhy.2018.03.003} (\bibinfo {year}
  {2018})\BibitemShut {NoStop}%
\bibitem [{\citenamefont {Khemani}\ \emph {et~al.}(2016)\citenamefont
  {Khemani}, \citenamefont {Pollmann},\ and\ \citenamefont
  {Sondhi}}]{khemani2016obtaining}%
  \BibitemOpen
  \bibfield  {author} {\bibinfo {author} {\bibfnamefont {V.}~\bibnamefont
  {Khemani}}, \bibinfo {author} {\bibfnamefont {F.}~\bibnamefont {Pollmann}},\
  and\ \bibinfo {author} {\bibfnamefont {S.~L.}\ \bibnamefont {Sondhi}},\
  }\bibfield  {title} {\bibinfo {title} {{Obtaining Highly Excited Eigenstates
  of Many-Body Localized Hamiltonians by the Density Matrix Renormalization
  Group Approach}},\ }\href {https://doi.org/10.1103/PhysRevLett.116.247204}
  {\bibfield  {journal} {\bibinfo  {journal} {Phys. Rev. Lett.}\ }\textbf
  {\bibinfo {volume} {116}},\ \bibinfo {pages} {247204} (\bibinfo {year}
  {2016})}\BibitemShut {NoStop}%
\bibitem [{\citenamefont {Yu}\ \emph {et~al.}(2017)\citenamefont {Yu},
  \citenamefont {Pekker},\ and\ \citenamefont {Clark}}]{yu2017finding}%
  \BibitemOpen
  \bibfield  {author} {\bibinfo {author} {\bibfnamefont {X.}~\bibnamefont
  {Yu}}, \bibinfo {author} {\bibfnamefont {D.}~\bibnamefont {Pekker}},\ and\
  \bibinfo {author} {\bibfnamefont {B.~K.}\ \bibnamefont {Clark}},\ }\bibfield
  {title} {\bibinfo {title} {{Finding Matrix Product State Representations of
  Highly Excited Eigenstates of Many-Body Localized Hamiltonians}},\ }\href
  {https://doi.org/10.1103/PhysRevLett.118.017201} {\bibfield  {journal}
  {\bibinfo  {journal} {Phys. Rev. Lett.}\ }\textbf {\bibinfo {volume} {118}},\
  \bibinfo {pages} {017201} (\bibinfo {year} {2017})}\BibitemShut {NoStop}%
\bibitem [{\citenamefont {{Imbrie}}(2016)}]{imbrie2016manybody}%
  \BibitemOpen
  \bibfield  {author} {\bibinfo {author} {\bibfnamefont {J.~Z.}\ \bibnamefont
  {{Imbrie}}},\ }\bibfield  {title} {\bibinfo {title} {{On Many-Body
  Localization for Quantum Spin Chains}},\ }\href
  {https://doi.org/10.1007/s10955-016-1508-x} {\bibfield  {journal} {\bibinfo
  {journal} {J. Stat. Phys.}\ }\textbf {\bibinfo {volume} {163}},\ \bibinfo
  {pages} {998} (\bibinfo {year} {2016})}\BibitemShut {NoStop}%
\bibitem [{\citenamefont {McCoy}\ and\ \citenamefont
  {Wu}(1978)}]{mccoy1978twodimension}%
  \BibitemOpen
  \bibfield  {author} {\bibinfo {author} {\bibfnamefont {B.~M.}\ \bibnamefont
  {McCoy}}\ and\ \bibinfo {author} {\bibfnamefont {T.~T.}\ \bibnamefont {Wu}},\
  }\bibfield  {title} {\bibinfo {title} {Two-dimensional {I}sing field theory
  in a magnetic field: Breakup of the cut in the two-point function},\ }\href
  {https://doi.org/10.1103/PhysRevD.18.1259} {\bibfield  {journal} {\bibinfo
  {journal} {Phys. Rev. D}\ }\textbf {\bibinfo {volume} {18}},\ \bibinfo
  {pages} {1259} (\bibinfo {year} {1978})}\BibitemShut {NoStop}%
\bibitem [{\citenamefont {Sachdev}(2011)}]{sachdev2011quantum}%
  \BibitemOpen
  \bibfield  {author} {\bibinfo {author} {\bibfnamefont {S.}~\bibnamefont
  {Sachdev}},\ }\href {https://books.google.com/books?id=F3IkpxwpqSgC} {\emph
  {\bibinfo {title} {Quantum Phase Transitions}}}\ (\bibinfo  {publisher}
  {Cambridge University Press},\ \bibinfo {year} {2011})\BibitemShut {NoStop}%
\bibitem [{\citenamefont {Onsager}(1944)}]{onsager1944crystal}%
  \BibitemOpen
  \bibfield  {author} {\bibinfo {author} {\bibfnamefont {L.}~\bibnamefont
  {Onsager}},\ }\bibfield  {title} {\bibinfo {title} {Crystal statistics. {I}.
  {A} two-dimensional model with an order-disorder transition},\ }\href
  {https://doi.org/10.1103/PhysRev.65.117} {\bibfield  {journal} {\bibinfo
  {journal} {Phys. Rev.}\ }\textbf {\bibinfo {volume} {65}},\ \bibinfo {pages}
  {117} (\bibinfo {year} {1944})}\BibitemShut {NoStop}%
\bibitem [{\citenamefont {Zamolodchikov}(1989)}]{zamolodchikov1989integrals}%
  \BibitemOpen
  \bibfield  {author} {\bibinfo {author} {\bibfnamefont {A.~B.}\ \bibnamefont
  {Zamolodchikov}},\ }\bibfield  {title} {\bibinfo {title} {Integrals of motion
  and {S}-matrix of the (scaled) ${T}= {T}_c$ {I}sing model with magnetic
  field},\ }\href {https://doi.org/10.1142/S0217751X8900176X} {\bibfield
  {journal} {\bibinfo  {journal} {Int. J. Mod. Phys. A}\ }\textbf {\bibinfo
  {volume} {04}},\ \bibinfo {pages} {4235} (\bibinfo {year}
  {1989})}\BibitemShut {NoStop}%
\bibitem [{\citenamefont {Delfino}\ \emph {et~al.}(1996)\citenamefont
  {Delfino}, \citenamefont {Mussardo},\ and\ \citenamefont
  {Simonetti}}]{delfino1996nonintegrable}%
  \BibitemOpen
  \bibfield  {author} {\bibinfo {author} {\bibfnamefont {G.}~\bibnamefont
  {Delfino}}, \bibinfo {author} {\bibfnamefont {G.}~\bibnamefont {Mussardo}},\
  and\ \bibinfo {author} {\bibfnamefont {P.}~\bibnamefont {Simonetti}},\
  }\bibfield  {title} {\bibinfo {title} {Non-integrable quantum field theories
  as perturbations of certain integrable models},\ }\href
  {https://doi.org/https://doi.org/10.1016/0550-3213(96)00265-9} {\bibfield
  {journal} {\bibinfo  {journal} {Nucl. Phys. B}\ }\textbf {\bibinfo {volume}
  {473}},\ \bibinfo {pages} {469 } (\bibinfo {year} {1996})}\BibitemShut
  {NoStop}%
\bibitem [{\citenamefont {Delfino}\ and\ \citenamefont
  {Mussardo}(1998)}]{delfino1998nonintegrable}%
  \BibitemOpen
  \bibfield  {author} {\bibinfo {author} {\bibfnamefont {G.}~\bibnamefont
  {Delfino}}\ and\ \bibinfo {author} {\bibfnamefont {G.}~\bibnamefont
  {Mussardo}},\ }\bibfield  {title} {\bibinfo {title} {Non-integrable aspects
  of the multi-frequency sine-gordon model},\ }\href
  {https://doi.org/https://doi.org/10.1016/S0550-3213(98)00063-7} {\bibfield
  {journal} {\bibinfo  {journal} {Nucl. Phys. B}\ }\textbf {\bibinfo {volume}
  {516}},\ \bibinfo {pages} {675 } (\bibinfo {year} {1998})}\BibitemShut
  {NoStop}%
\bibitem [{\citenamefont {Sulejmanpasic}\ \emph {et~al.}(2017)\citenamefont
  {Sulejmanpasic}, \citenamefont {Shao}, \citenamefont {Sandvik},\ and\
  \citenamefont {\"Unsal}}]{sulejmanpasic2017confinement}%
  \BibitemOpen
  \bibfield  {author} {\bibinfo {author} {\bibfnamefont {T.}~\bibnamefont
  {Sulejmanpasic}}, \bibinfo {author} {\bibfnamefont {H.}~\bibnamefont {Shao}},
  \bibinfo {author} {\bibfnamefont {A.~W.}\ \bibnamefont {Sandvik}},\ and\
  \bibinfo {author} {\bibfnamefont {M.}~\bibnamefont {\"Unsal}},\ }\bibfield
  {title} {\bibinfo {title} {{Confinement in the Bulk, Deconfinement on the
  Wall: Infrared Equivalence between Compactified QCD and Quantum Magnets}},\
  }\href {https://doi.org/10.1103/PhysRevLett.119.091601} {\bibfield  {journal}
  {\bibinfo  {journal} {Phys. Rev. Lett.}\ }\textbf {\bibinfo {volume} {119}},\
  \bibinfo {pages} {091601} (\bibinfo {year} {2017})}\BibitemShut {NoStop}%
\bibitem [{\citenamefont {Coldea}\ \emph {et~al.}(2010)\citenamefont {Coldea},
  \citenamefont {Tennant}, \citenamefont {Wheeler}, \citenamefont {Wawrzynska},
  \citenamefont {Prabhakaran}, \citenamefont {Telling}, \citenamefont
  {Habicht}, \citenamefont {Smeibidl},\ and\ \citenamefont
  {Kiefer}}]{coldea2010quantum}%
  \BibitemOpen
  \bibfield  {author} {\bibinfo {author} {\bibfnamefont {R.}~\bibnamefont
  {Coldea}}, \bibinfo {author} {\bibfnamefont {D.~A.}\ \bibnamefont {Tennant}},
  \bibinfo {author} {\bibfnamefont {E.~M.}\ \bibnamefont {Wheeler}}, \bibinfo
  {author} {\bibfnamefont {E.}~\bibnamefont {Wawrzynska}}, \bibinfo {author}
  {\bibfnamefont {D.}~\bibnamefont {Prabhakaran}}, \bibinfo {author}
  {\bibfnamefont {M.}~\bibnamefont {Telling}}, \bibinfo {author} {\bibfnamefont
  {K.}~\bibnamefont {Habicht}}, \bibinfo {author} {\bibfnamefont
  {P.}~\bibnamefont {Smeibidl}},\ and\ \bibinfo {author} {\bibfnamefont
  {K.}~\bibnamefont {Kiefer}},\ }\bibfield  {title} {\bibinfo {title} {{Quantum
  Criticality in an {I}sing Chain: Experimental Evidence for Emergent {E8}
  Symmetry}},\ }\href {https://doi.org/10.1126/science.1180085} {\bibfield
  {journal} {\bibinfo  {journal} {Science}\ }\textbf {\bibinfo {volume}
  {327}},\ \bibinfo {pages} {177} (\bibinfo {year} {2010})}\BibitemShut
  {NoStop}%
\bibitem [{\citenamefont {Morris}\ \emph {et~al.}(2014)\citenamefont {Morris},
  \citenamefont {Vald\'es~Aguilar}, \citenamefont {Ghosh}, \citenamefont
  {Koohpayeh}, \citenamefont {Krizan}, \citenamefont {Cava}, \citenamefont
  {Tchernyshyov}, \citenamefont {McQueen},\ and\ \citenamefont
  {Armitage}}]{morris2014hierarchy}%
  \BibitemOpen
  \bibfield  {author} {\bibinfo {author} {\bibfnamefont {C.~M.}\ \bibnamefont
  {Morris}}, \bibinfo {author} {\bibfnamefont {R.}~\bibnamefont
  {Vald\'es~Aguilar}}, \bibinfo {author} {\bibfnamefont {A.}~\bibnamefont
  {Ghosh}}, \bibinfo {author} {\bibfnamefont {S.~M.}\ \bibnamefont
  {Koohpayeh}}, \bibinfo {author} {\bibfnamefont {J.}~\bibnamefont {Krizan}},
  \bibinfo {author} {\bibfnamefont {R.~J.}\ \bibnamefont {Cava}}, \bibinfo
  {author} {\bibfnamefont {O.}~\bibnamefont {Tchernyshyov}}, \bibinfo {author}
  {\bibfnamefont {T.~M.}\ \bibnamefont {McQueen}},\ and\ \bibinfo {author}
  {\bibfnamefont {N.~P.}\ \bibnamefont {Armitage}},\ }\bibfield  {title}
  {\bibinfo {title} {{Hierarchy of Bound States in the One-Dimensional
  Ferromagnetic {I}sing Chain {C}o{N}b$_{2}${O}$_{6}$ Investigated by
  High-Resolution Time-Domain Terahertz Spectroscopy}},\ }\href
  {https://doi.org/10.1103/PhysRevLett.112.137403} {\bibfield  {journal}
  {\bibinfo  {journal} {Phys. Rev. Lett.}\ }\textbf {\bibinfo {volume} {112}},\
  \bibinfo {pages} {137403} (\bibinfo {year} {2014})}\BibitemShut {NoStop}%
\bibitem [{\citenamefont {Wang}\ \emph {et~al.}(2015)\citenamefont {Wang},
  \citenamefont {Schmidt}, \citenamefont {Bera}, \citenamefont {Islam},
  \citenamefont {Lake}, \citenamefont {Loidl},\ and\ \citenamefont
  {Deisenhofer}}]{wang2015spinon}%
  \BibitemOpen
  \bibfield  {author} {\bibinfo {author} {\bibfnamefont {Z.}~\bibnamefont
  {Wang}}, \bibinfo {author} {\bibfnamefont {M.}~\bibnamefont {Schmidt}},
  \bibinfo {author} {\bibfnamefont {A.~K.}\ \bibnamefont {Bera}}, \bibinfo
  {author} {\bibfnamefont {A.~T. M.~N.}\ \bibnamefont {Islam}}, \bibinfo
  {author} {\bibfnamefont {B.}~\bibnamefont {Lake}}, \bibinfo {author}
  {\bibfnamefont {A.}~\bibnamefont {Loidl}},\ and\ \bibinfo {author}
  {\bibfnamefont {J.}~\bibnamefont {Deisenhofer}},\ }\bibfield  {title}
  {\bibinfo {title} {{Spinon confinement in the one-dimensional Ising-like
  antiferromagnet {SrCo}$_{2}${V}$_{2}${O}$_{8}$}},\ }\href
  {https://doi.org/10.1103/PhysRevB.91.140404} {\bibfield  {journal} {\bibinfo
  {journal} {Phys. Rev. B}\ }\textbf {\bibinfo {volume} {91}},\ \bibinfo
  {pages} {140404} (\bibinfo {year} {2015})}\BibitemShut {NoStop}%
\bibitem [{\citenamefont {Wang}\ \emph {et~al.}(2016)\citenamefont {Wang},
  \citenamefont {Wu}, \citenamefont {Xu}, \citenamefont {Yang}, \citenamefont
  {Wu}, \citenamefont {Bera}, \citenamefont {Islam}, \citenamefont {Lake},
  \citenamefont {Kamenskyi}, \citenamefont {Gogoi}, \citenamefont {Engelkamp},
  \citenamefont {Wang}, \citenamefont {Deisenhofer},\ and\ \citenamefont
  {Loidl}}]{wang2016from}%
  \BibitemOpen
  \bibfield  {author} {\bibinfo {author} {\bibfnamefont {Z.}~\bibnamefont
  {Wang}}, \bibinfo {author} {\bibfnamefont {J.}~\bibnamefont {Wu}}, \bibinfo
  {author} {\bibfnamefont {S.}~\bibnamefont {Xu}}, \bibinfo {author}
  {\bibfnamefont {W.}~\bibnamefont {Yang}}, \bibinfo {author} {\bibfnamefont
  {C.}~\bibnamefont {Wu}}, \bibinfo {author} {\bibfnamefont {A.~K.}\
  \bibnamefont {Bera}}, \bibinfo {author} {\bibfnamefont {A.~T. M.~N.}\
  \bibnamefont {Islam}}, \bibinfo {author} {\bibfnamefont {B.}~\bibnamefont
  {Lake}}, \bibinfo {author} {\bibfnamefont {D.}~\bibnamefont {Kamenskyi}},
  \bibinfo {author} {\bibfnamefont {P.}~\bibnamefont {Gogoi}}, \bibinfo
  {author} {\bibfnamefont {H.}~\bibnamefont {Engelkamp}}, \bibinfo {author}
  {\bibfnamefont {N.}~\bibnamefont {Wang}}, \bibinfo {author} {\bibfnamefont
  {J.}~\bibnamefont {Deisenhofer}},\ and\ \bibinfo {author} {\bibfnamefont
  {A.}~\bibnamefont {Loidl}},\ }\bibfield  {title} {\bibinfo {title} {{From
  confined spinons to emergent fermions: Observation of elementary magnetic
  excitations in a transverse-field Ising chain}},\ }\href
  {https://doi.org/10.1103/PhysRevB.94.125130} {\bibfield  {journal} {\bibinfo
  {journal} {Phys. Rev. B}\ }\textbf {\bibinfo {volume} {94}},\ \bibinfo
  {pages} {125130} (\bibinfo {year} {2016})}\BibitemShut {NoStop}%
\bibitem [{\citenamefont {James}\ \emph {et~al.}(2019)\citenamefont {James},
  \citenamefont {Konik},\ and\ \citenamefont {Robinson}}]{james2018nonthermal}%
  \BibitemOpen
  \bibfield  {author} {\bibinfo {author} {\bibfnamefont {A.~J.~A.}\
  \bibnamefont {James}}, \bibinfo {author} {\bibfnamefont {R.~M.}\ \bibnamefont
  {Konik}},\ and\ \bibinfo {author} {\bibfnamefont {N.~J.}\ \bibnamefont
  {Robinson}},\ }\bibfield  {title} {\bibinfo {title} {{Nonthermal States
  Arising from Confinement in One and Two Dimensions}},\ }\href
  {https://doi.org/10.1103/PhysRevLett.122.130603} {\bibfield  {journal}
  {\bibinfo  {journal} {Phys. Rev. Lett.}\ }\textbf {\bibinfo {volume} {122}},\
  \bibinfo {pages} {130603} (\bibinfo {year} {2019})}\BibitemShut {NoStop}%
\bibitem [{\citenamefont {Yurov}\ and\ \citenamefont
  {Zamolodchikov}(1990)}]{yurov1990truncated}%
  \BibitemOpen
  \bibfield  {author} {\bibinfo {author} {\bibfnamefont {V.~P.}\ \bibnamefont
  {Yurov}}\ and\ \bibinfo {author} {\bibfnamefont {A.~B.}\ \bibnamefont
  {Zamolodchikov}},\ }\bibfield  {title} {\bibinfo {title} {Truncated conformal
  space approach to scaling {L}ee-{Y}ang model},\ }\href
  {https://doi.org/10.1142/S0217751X9000218X} {\bibfield  {journal} {\bibinfo
  {journal} {Int. J. Mod. Phys. A}\ }\textbf {\bibinfo {volume} {05}},\
  \bibinfo {pages} {3221} (\bibinfo {year} {1990})}\BibitemShut {NoStop}%
\bibitem [{\citenamefont {Di~Francesco}\ \emph {et~al.}(1997)\citenamefont
  {Di~Francesco}, \citenamefont {Mathieu},\ and\ \citenamefont
  {S{\'e}n{\'e}chal}}]{BigYellowBook}%
  \BibitemOpen
  \bibfield  {author} {\bibinfo {author} {\bibfnamefont {P.}~\bibnamefont
  {Di~Francesco}}, \bibinfo {author} {\bibfnamefont {P.}~\bibnamefont
  {Mathieu}},\ and\ \bibinfo {author} {\bibfnamefont {D.}~\bibnamefont
  {S{\'e}n{\'e}chal}},\ }\href {https://books.google.com/books?id=keUrdME5rhIC}
  {\emph {\bibinfo {title} {Conformal Field Theory}}},\ Graduate Texts in
  Contemporary Physics\ (\bibinfo  {publisher} {Springer},\ \bibinfo {year}
  {1997})\BibitemShut {NoStop}%
\bibitem [{\citenamefont {Yurov}\ and\ \citenamefont
  {Zamolodchikov}(1991)}]{yurov1991truncated}%
  \BibitemOpen
  \bibfield  {author} {\bibinfo {author} {\bibfnamefont {V.~P.}\ \bibnamefont
  {Yurov}}\ and\ \bibinfo {author} {\bibfnamefont {A.~B.}\ \bibnamefont
  {Zamolodchikov}},\ }\bibfield  {title} {\bibinfo {title}
  {Truncated-fermionic-space approach to the critical 2{D} {I}sing model with
  magnetic field},\ }\href {https://doi.org/10.1142/S0217751X91002161}
  {\bibfield  {journal} {\bibinfo  {journal} {Int. J. Mod. Phys. A}\ }\textbf
  {\bibinfo {volume} {06}},\ \bibinfo {pages} {4557} (\bibinfo {year}
  {1991})}\BibitemShut {NoStop}%
\bibitem [{\citenamefont {Konik}\ and\ \citenamefont
  {Adamov}(2007)}]{konik2007numerical}%
  \BibitemOpen
  \bibfield  {author} {\bibinfo {author} {\bibfnamefont {R.~M.}\ \bibnamefont
  {Konik}}\ and\ \bibinfo {author} {\bibfnamefont {Y.}~\bibnamefont {Adamov}},\
  }\bibfield  {title} {\bibinfo {title} {{Numerical Renormalization Group for
  Continuum One-Dimensional Systems}},\ }\href
  {https://doi.org/10.1103/PhysRevLett.98.147205} {\bibfield  {journal}
  {\bibinfo  {journal} {Phys. Rev. Lett.}\ }\textbf {\bibinfo {volume} {98}},\
  \bibinfo {pages} {147205} (\bibinfo {year} {2007})}\BibitemShut {NoStop}%
\bibitem [{\citenamefont {Konik}\ and\ \citenamefont
  {Adamov}(2009)}]{konik2009renormalization}%
  \BibitemOpen
  \bibfield  {author} {\bibinfo {author} {\bibfnamefont {R.~M.}\ \bibnamefont
  {Konik}}\ and\ \bibinfo {author} {\bibfnamefont {Y.}~\bibnamefont {Adamov}},\
  }\bibfield  {title} {\bibinfo {title} {{Renormalization Group for Treating
  2{D} Coupled Arrays of Continuum 1{D} Systems}},\ }\href
  {https://doi.org/10.1103/PhysRevLett.102.097203} {\bibfield  {journal}
  {\bibinfo  {journal} {Phys. Rev. Lett.}\ }\textbf {\bibinfo {volume} {102}},\
  \bibinfo {pages} {097203} (\bibinfo {year} {2009})}\BibitemShut {NoStop}%
\bibitem [{\citenamefont {Brandino}\ \emph {et~al.}(2010)\citenamefont
  {Brandino}, \citenamefont {Konik},\ and\ \citenamefont
  {Mussardo}}]{brandino2010energy}%
  \BibitemOpen
  \bibfield  {author} {\bibinfo {author} {\bibfnamefont {G.~P.}\ \bibnamefont
  {Brandino}}, \bibinfo {author} {\bibfnamefont {R.~M.}\ \bibnamefont
  {Konik}},\ and\ \bibinfo {author} {\bibfnamefont {G.}~\bibnamefont
  {Mussardo}},\ }\bibfield  {title} {\bibinfo {title} {Energy level
  distribution of perturbed conformal field theories},\ }\href
  {http://stacks.iop.org/1742-5468/2010/i=07/a=P07013} {\bibfield  {journal}
  {\bibinfo  {journal} {J. Stat. Mech.}\ }\textbf {\bibinfo {volume} {2010}},\
  \bibinfo {pages} {P07013} (\bibinfo {year} {2010})}\BibitemShut {NoStop}%
\bibitem [{\citenamefont {H\'ods\'agi}\ \emph {et~al.}(2018)\citenamefont
  {H\'ods\'agi}, \citenamefont {Kormos},\ and\ \citenamefont
  {Tak\'acs}}]{hodsagi2018quench}%
  \BibitemOpen
  \bibfield  {author} {\bibinfo {author} {\bibfnamefont {K.}~\bibnamefont
  {H\'ods\'agi}}, \bibinfo {author} {\bibfnamefont {M.}~\bibnamefont
  {Kormos}},\ and\ \bibinfo {author} {\bibfnamefont {G.}~\bibnamefont
  {Tak\'acs}},\ }\bibfield  {title} {\bibinfo {title} {{Quench dynamics of the
  Ising field theory in a magnetic field}},\ }\href
  {https://doi.org/10.21468/SciPostPhys.5.3.027} {\bibfield  {journal}
  {\bibinfo  {journal} {SciPost Phys.}\ }\textbf {\bibinfo {volume} {5}},\
  \bibinfo {pages} {27} (\bibinfo {year} {2018})}\BibitemShut {NoStop}%
\bibitem [{\citenamefont {Berg}\ \emph {et~al.}(1979)\citenamefont {Berg},
  \citenamefont {Karowski},\ and\ \citenamefont
  {Weisz}}]{berg1979construction}%
  \BibitemOpen
  \bibfield  {author} {\bibinfo {author} {\bibfnamefont {B.}~\bibnamefont
  {Berg}}, \bibinfo {author} {\bibfnamefont {M.}~\bibnamefont {Karowski}},\
  and\ \bibinfo {author} {\bibfnamefont {P.}~\bibnamefont {Weisz}},\ }\bibfield
   {title} {\bibinfo {title} {Construction of {G}reen's functions from an exact
  {$S$} matrix},\ }\href {https://doi.org/10.1103/PhysRevD.19.2477} {\bibfield
  {journal} {\bibinfo  {journal} {Phys. Rev. D}\ }\textbf {\bibinfo {volume}
  {19}},\ \bibinfo {pages} {2477} (\bibinfo {year} {1979})}\BibitemShut
  {NoStop}%
\bibitem [{\citenamefont {{Fonseca}}\ and\ \citenamefont
  {{Zamolodchikov}}(2001)}]{fonseca2001ising}%
  \BibitemOpen
  \bibfield  {author} {\bibinfo {author} {\bibfnamefont {P.}~\bibnamefont
  {{Fonseca}}}\ and\ \bibinfo {author} {\bibfnamefont {A.}~\bibnamefont
  {{Zamolodchikov}}},\ }\bibfield  {title} {\bibinfo {title} {{Ising field
  theory in a magnetic field: analytic properties of the free energy}},\
  }\href@noop {} {\bibfield  {journal} {\bibinfo  {journal} {ArXiv e-prints}\ }
  (\bibinfo {year} {2001})},\ \Eprint {https://arxiv.org/abs/hep-th/0112167}
  {hep-th/0112167} \BibitemShut {NoStop}%
\bibitem [{\citenamefont {{Bugrij}}(2000)}]{bugrij2000correlation}%
  \BibitemOpen
  \bibfield  {author} {\bibinfo {author} {\bibfnamefont {A.~I.}\ \bibnamefont
  {{Bugrij}}},\ }\bibfield  {title} {\bibinfo {title} {{The correlation
  function in two dimensional {I}sing model on the finite size lattice. {I}}},\
  }\href@noop {} {\bibfield  {journal} {\bibinfo  {journal} {ArXiv e-prints}\ }
  (\bibinfo {year} {2000})},\ \Eprint {https://arxiv.org/abs/hep-th/0011104}
  {hep-th/0011104} \BibitemShut {NoStop}%
\bibitem [{\citenamefont {{Bugrij}}(2001)}]{bugrij2001form}%
  \BibitemOpen
  \bibfield  {author} {\bibinfo {author} {\bibfnamefont {A.~I.}\ \bibnamefont
  {{Bugrij}}},\ }\bibfield  {title} {\bibinfo {title} {{Form factor
  representation of the correlation function of the two dimensional {I}sing
  model on a cylinder}},\ }\href@noop {} {\bibfield  {journal} {\bibinfo
  {journal} {ArXiv e-prints}\ } (\bibinfo {year} {2001})},\ \Eprint
  {https://arxiv.org/abs/hep-th/0107117} {hep-th/0107117} \BibitemShut
  {NoStop}%
\bibitem [{Note2()}]{Note2}%
  \BibitemOpen
  \bibinfo {note} {We say that the EEVs and MCE agree when the EEVs fall within
  one standard deviation of the MCE. This spread of the EEVs reflects the
  second term in Eq.~\protect \textup {\hbox {\mathsurround \z@ \protect
  \normalfont (\ignorespaces \ref {eth}\unskip \@@italiccorr )}}, which
  features the random variable $R_{\alpha ,\beta }$, which has mean zero and
  unit variance. We see that the standard deviation decreases with increasing
  system size $R$, reflecting the extensivity of the thermodynamic entropy
  $S(E)$.}\BibitemShut {Stop}%
\bibitem [{\citenamefont {Fonseca}\ and\ \citenamefont
  {Zamolodchikov}(2003)}]{fonseca2003ising}%
  \BibitemOpen
  \bibfield  {author} {\bibinfo {author} {\bibfnamefont {P.}~\bibnamefont
  {Fonseca}}\ and\ \bibinfo {author} {\bibfnamefont {A.}~\bibnamefont
  {Zamolodchikov}},\ }\bibfield  {title} {\bibinfo {title} {{Ising Field Theory
  in a Magnetic Field: Analytic Properties of the Free Energy}},\ }\href
  {https://doi.org/10.1023/A:1022147532606} {\bibfield  {journal} {\bibinfo
  {journal} {J. Stat. Phys.}\ }\textbf {\bibinfo {volume} {110}},\ \bibinfo
  {pages} {527} (\bibinfo {year} {2003})}\BibitemShut {NoStop}%
\bibitem [{\citenamefont {Rutkevich}(2005)}]{rutkevich2005largen}%
  \BibitemOpen
  \bibfield  {author} {\bibinfo {author} {\bibfnamefont {S.~B.}\ \bibnamefont
  {Rutkevich}},\ }\bibfield  {title} {\bibinfo {title} {{Large-$n$ Excitations
  in the Ferromagnetic Ising Field Theory in a Weak Magnetic Field: Mass
  Spectrum and Decay Widths}},\ }\href
  {https://doi.org/10.1103/PhysRevLett.95.250601} {\bibfield  {journal}
  {\bibinfo  {journal} {Phys. Rev. Lett.}\ }\textbf {\bibinfo {volume} {95}},\
  \bibinfo {pages} {250601} (\bibinfo {year} {2005})}\BibitemShut {NoStop}%
\bibitem [{\citenamefont {{Fonseca}}\ and\ \citenamefont
  {{Zamolodchikov}}(2006)}]{fonseca2006ising}%
  \BibitemOpen
  \bibfield  {author} {\bibinfo {author} {\bibfnamefont {P.}~\bibnamefont
  {{Fonseca}}}\ and\ \bibinfo {author} {\bibfnamefont {A.}~\bibnamefont
  {{Zamolodchikov}}},\ }\bibfield  {title} {\bibinfo {title} {{Ising
  Spectroscopy I: Mesons at $T < T_c$}},\ }\href@noop {} {\bibfield  {journal}
  {\bibinfo  {journal} {ArXiv e-prints}\ } (\bibinfo {year} {2006})},\ \Eprint
  {https://arxiv.org/abs/hep-th/0612304} {hep-th/0612304} \BibitemShut
  {NoStop}%
\bibitem [{Note3()}]{Note3}%
  \BibitemOpen
  \bibinfo {note} {We note that in the context of magnetic systems, one might
  conclude that the mesons are not present in the spectrum as one does not see
  signatures in dynamical spin-spin correlation functions: but from Fig.~\ref
  {Fig:EEVs} one could conclude that they are indeed there, though their signal
  is washed out by the thermal signal from the large number of states in the
  multiparticle continuum.}\BibitemShut {Stop}%
\bibitem [{\citenamefont {Zamolodchikov}\ and\ \citenamefont
  {Ziyatdinov}(2011)}]{Zamolodchikov2011}%
  \BibitemOpen
  \bibfield  {author} {\bibinfo {author} {\bibfnamefont {A.}~\bibnamefont
  {Zamolodchikov}}\ and\ \bibinfo {author} {\bibfnamefont {I.}~\bibnamefont
  {Ziyatdinov}},\ }\bibfield  {title} {\bibinfo {title} {{Inelastic scattering
  and elastic amplitude in Ising field theory in a weak magnetic field at
  $T>T_c$. Perturbative analysis}},\ }\href
  {https://doi.org/https://doi.org/10.1016/j.nuclphysb.2011.04.005} {\bibfield
  {journal} {\bibinfo  {journal} {Nucl. Phys. B}\ }\textbf {\bibinfo {volume}
  {849}},\ \bibinfo {pages} {654 } (\bibinfo {year} {2011})}\BibitemShut
  {NoStop}%
\bibitem [{\citenamefont {Pozsgay}\ and\ \citenamefont
  {Tak\'acs}(2006)}]{pozsgay2006characterization}%
  \BibitemOpen
  \bibfield  {author} {\bibinfo {author} {\bibfnamefont {B.}~\bibnamefont
  {Pozsgay}}\ and\ \bibinfo {author} {\bibfnamefont {G.}~\bibnamefont
  {Tak\'acs}},\ }\bibfield  {title} {\bibinfo {title} {Characterization of
  resonances using finite size effects},\ }\href
  {http://www.sciencedirect.com/science/article/pii/S055032130600397X}
  {\bibfield  {journal} {\bibinfo  {journal} {Nucl. Phys. B}\ }\textbf
  {\bibinfo {volume} {748}},\ \bibinfo {pages} {485 } (\bibinfo {year}
  {2006})}\BibitemShut {NoStop}%
\bibitem [{\citenamefont {Delfino}(2009)}]{delfino2009particle}%
  \BibitemOpen
  \bibfield  {author} {\bibinfo {author} {\bibfnamefont {G.}~\bibnamefont
  {Delfino}},\ }\bibinfo {title} {{Particle Decay in Ising Field Theory with
  Magnetic Field}},\ in\ \href {https://doi.org/10.1007/978-90-481-2810-5_15}
  {\emph {\bibinfo {booktitle} {{New Trends in Mathematical Physics: Selected
  contributions of the XVth International Congress on Mathematical
  Physics}}}},\ \bibinfo {editor} {edited by\ \bibinfo {editor} {\bibfnamefont
  {V.}~\bibnamefont {Sidoravi{\v{c}}ius}}}\ (\bibinfo  {publisher} {Springer
  Netherlands},\ \bibinfo {address} {Dordrecht},\ \bibinfo {year} {2009})\ pp.\
  \bibinfo {pages} {173--185}\BibitemShut {NoStop}%
\bibitem [{\citenamefont {Kormos}\ \emph {et~al.}(2017)\citenamefont {Kormos},
  \citenamefont {Collura}, \citenamefont {Tak{\'a}cs},\ and\ \citenamefont
  {Calabrese}}]{kormos2016realtime}%
  \BibitemOpen
  \bibfield  {author} {\bibinfo {author} {\bibfnamefont {M.}~\bibnamefont
  {Kormos}}, \bibinfo {author} {\bibfnamefont {M.}~\bibnamefont {Collura}},
  \bibinfo {author} {\bibfnamefont {G.}~\bibnamefont {Tak{\'a}cs}},\ and\
  \bibinfo {author} {\bibfnamefont {P.}~\bibnamefont {Calabrese}},\ }\bibfield
  {title} {\bibinfo {title} {{Real-time confinement following a quantum quench
  to a non-integrable model}},\ }\href {http://dx.doi.org/10.1038/nphys3934}
  {\bibfield  {journal} {\bibinfo  {journal} {Nat. Phys.}\ }\textbf {\bibinfo
  {volume} {13}},\ \bibinfo {pages} {246} (\bibinfo {year} {2017})}\BibitemShut
  {NoStop}%
\bibitem [{\citenamefont {Kollar}\ and\ \citenamefont
  {Eckstein}(2008)}]{kollar2008relaxation}%
  \BibitemOpen
  \bibfield  {author} {\bibinfo {author} {\bibfnamefont {M.}~\bibnamefont
  {Kollar}}\ and\ \bibinfo {author} {\bibfnamefont {M.}~\bibnamefont
  {Eckstein}},\ }\bibfield  {title} {\bibinfo {title} {Relaxation of a
  one-dimensional {M}ott insulator after an interaction quench},\ }\href
  {https://doi.org/10.1103/PhysRevA.78.013626} {\bibfield  {journal} {\bibinfo
  {journal} {Phys. Rev. A}\ }\textbf {\bibinfo {volume} {78}},\ \bibinfo
  {pages} {013626} (\bibinfo {year} {2008})}\BibitemShut {NoStop}%
\bibitem [{Note4()}]{Note4}%
  \BibitemOpen
  \bibinfo {note} {Interesting, we note that this particular quench has
  essentially no finite size effects, see the appendix~\ref {App:ConvCheb} for
  further information. A similar lack of finite-size effects was seen in
  quenches of the lattice Ising chain perturbed by a longitudinal field~\cite
  {kormos2016realtime}. $~$}\BibitemShut {NoStop}%
\bibitem [{\citenamefont {Delfino}(2018)}]{delfino2018correlation}%
  \BibitemOpen
  \bibfield  {author} {\bibinfo {author} {\bibfnamefont {G.}~\bibnamefont
  {Delfino}},\ }\bibfield  {title} {\bibinfo {title} {{Correlation spreading
  and properties of the quantum state in quench dynamics}},\ }\href
  {https://doi.org/10.1103/PhysRevE.97.062138} {\bibfield  {journal} {\bibinfo
  {journal} {Phys. Rev. E}\ }\textbf {\bibinfo {volume} {97}},\ \bibinfo
  {pages} {062138} (\bibinfo {year} {2018})}\BibitemShut {NoStop}%
\bibitem [{\citenamefont {Kj\"all}\ \emph {et~al.}(2011)\citenamefont
  {Kj\"all}, \citenamefont {Pollmann},\ and\ \citenamefont
  {Moore}}]{kjall2011bound}%
  \BibitemOpen
  \bibfield  {author} {\bibinfo {author} {\bibfnamefont {J.~A.}\ \bibnamefont
  {Kj\"all}}, \bibinfo {author} {\bibfnamefont {F.}~\bibnamefont {Pollmann}},\
  and\ \bibinfo {author} {\bibfnamefont {J.~E.}\ \bibnamefont {Moore}},\
  }\bibfield  {title} {\bibinfo {title} {{Bound states and ${E}_{8}$ symmetry
  effects in perturbed quantum Ising chains}},\ }\href
  {https://doi.org/10.1103/PhysRevB.83.020407} {\bibfield  {journal} {\bibinfo
  {journal} {Phys. Rev. B}\ }\textbf {\bibinfo {volume} {83}},\ \bibinfo
  {pages} {020407} (\bibinfo {year} {2011})}\BibitemShut {NoStop}%
\bibitem [{\citenamefont {Cabrera}\ \emph {et~al.}(2014)\citenamefont
  {Cabrera}, \citenamefont {Thompson}, \citenamefont {Coldea}, \citenamefont
  {Prabhakaran}, \citenamefont {Bewley}, \citenamefont {Guidi}, \citenamefont
  {Rodriguez-Rivera},\ and\ \citenamefont {Stock}}]{cabrera2014excitations}%
  \BibitemOpen
  \bibfield  {author} {\bibinfo {author} {\bibfnamefont {I.}~\bibnamefont
  {Cabrera}}, \bibinfo {author} {\bibfnamefont {J.~D.}\ \bibnamefont
  {Thompson}}, \bibinfo {author} {\bibfnamefont {R.}~\bibnamefont {Coldea}},
  \bibinfo {author} {\bibfnamefont {D.}~\bibnamefont {Prabhakaran}}, \bibinfo
  {author} {\bibfnamefont {R.~I.}\ \bibnamefont {Bewley}}, \bibinfo {author}
  {\bibfnamefont {T.}~\bibnamefont {Guidi}}, \bibinfo {author} {\bibfnamefont
  {J.~A.}\ \bibnamefont {Rodriguez-Rivera}},\ and\ \bibinfo {author}
  {\bibfnamefont {C.}~\bibnamefont {Stock}},\ }\bibfield  {title} {\bibinfo
  {title} {{Excitations in the quantum paramagnetic phase of the
  quasi-one-dimensional Ising magnet ${\mathrm{CoNb}}_{2}$${\mathrm{O}}_{6}$ in
  a transverse field: Geometric frustration and quantum renormalization
  effects}},\ }\href {https://doi.org/10.1103/PhysRevB.90.014418} {\bibfield
  {journal} {\bibinfo  {journal} {Phys. Rev. B}\ }\textbf {\bibinfo {volume}
  {90}},\ \bibinfo {pages} {014418} (\bibinfo {year} {2014})}\BibitemShut
  {NoStop}%
\bibitem [{\citenamefont {Robinson}\ \emph {et~al.}(2014)\citenamefont
  {Robinson}, \citenamefont {Essler}, \citenamefont {Cabrera},\ and\
  \citenamefont {Coldea}}]{robinson2014quasiparticle}%
  \BibitemOpen
  \bibfield  {author} {\bibinfo {author} {\bibfnamefont {N.~J.}\ \bibnamefont
  {Robinson}}, \bibinfo {author} {\bibfnamefont {F.~H.~L.}\ \bibnamefont
  {Essler}}, \bibinfo {author} {\bibfnamefont {I.}~\bibnamefont {Cabrera}},\
  and\ \bibinfo {author} {\bibfnamefont {R.}~\bibnamefont {Coldea}},\
  }\bibfield  {title} {\bibinfo {title} {{Quasiparticle breakdown in the
  quasi-one-dimensional Ising ferromagnet
  ${\mathrm{CoNb}}_{2}{\mathrm{O}}_{6}$}},\ }\href
  {https://doi.org/10.1103/PhysRevB.90.174406} {\bibfield  {journal} {\bibinfo
  {journal} {Phys. Rev. B}\ }\textbf {\bibinfo {volume} {90}},\ \bibinfo
  {pages} {174406} (\bibinfo {year} {2014})}\BibitemShut {NoStop}%
\bibitem [{\citenamefont {Ba\~nuls}\ \emph {et~al.}(2011)\citenamefont
  {Ba\~nuls}, \citenamefont {Cirac},\ and\ \citenamefont
  {Hastings}}]{banuls2011strong}%
  \BibitemOpen
  \bibfield  {author} {\bibinfo {author} {\bibfnamefont {M.~C.}\ \bibnamefont
  {Ba\~nuls}}, \bibinfo {author} {\bibfnamefont {J.~I.}\ \bibnamefont
  {Cirac}},\ and\ \bibinfo {author} {\bibfnamefont {M.~B.}\ \bibnamefont
  {Hastings}},\ }\bibfield  {title} {\bibinfo {title} {{Strong and Weak
  Thermalization of Infinite Nonintegrable Quantum Systems}},\ }\href
  {https://doi.org/10.1103/PhysRevLett.106.050405} {\bibfield  {journal}
  {\bibinfo  {journal} {Phys. Rev. Lett.}\ }\textbf {\bibinfo {volume} {106}},\
  \bibinfo {pages} {050405} (\bibinfo {year} {2011})}\BibitemShut {NoStop}%
\bibitem [{\citenamefont {Kim}\ and\ \citenamefont
  {Huse}(2013)}]{kim2013ballistic}%
  \BibitemOpen
  \bibfield  {author} {\bibinfo {author} {\bibfnamefont {H.}~\bibnamefont
  {Kim}}\ and\ \bibinfo {author} {\bibfnamefont {D.~A.}\ \bibnamefont {Huse}},\
  }\bibfield  {title} {\bibinfo {title} {{Ballistic Spreading of Entanglement
  in a Diffusive Nonintegrable System}},\ }\href
  {https://doi.org/10.1103/PhysRevLett.111.127205} {\bibfield  {journal}
  {\bibinfo  {journal} {Phys. Rev. Lett.}\ }\textbf {\bibinfo {volume} {111}},\
  \bibinfo {pages} {127205} (\bibinfo {year} {2013})}\BibitemShut {NoStop}%
\bibitem [{\citenamefont {Kim}\ \emph {et~al.}(2015)\citenamefont {Kim},
  \citenamefont {Ba\~nuls}, \citenamefont {Cirac}, \citenamefont {Hastings},\
  and\ \citenamefont {Huse}}]{kim2015slowest}%
  \BibitemOpen
  \bibfield  {author} {\bibinfo {author} {\bibfnamefont {H.}~\bibnamefont
  {Kim}}, \bibinfo {author} {\bibfnamefont {M.~C.}\ \bibnamefont {Ba\~nuls}},
  \bibinfo {author} {\bibfnamefont {J.~I.}\ \bibnamefont {Cirac}}, \bibinfo
  {author} {\bibfnamefont {M.~B.}\ \bibnamefont {Hastings}},\ and\ \bibinfo
  {author} {\bibfnamefont {D.~A.}\ \bibnamefont {Huse}},\ }\bibfield  {title}
  {\bibinfo {title} {{Slowest local operators in quantum spin chains}},\ }\href
  {https://doi.org/10.1103/PhysRevE.92.012128} {\bibfield  {journal} {\bibinfo
  {journal} {Phys. Rev. E}\ }\textbf {\bibinfo {volume} {92}},\ \bibinfo
  {pages} {012128} (\bibinfo {year} {2015})}\BibitemShut {NoStop}%
\bibitem [{\citenamefont {Zhang}\ \emph {et~al.}(2015)\citenamefont {Zhang},
  \citenamefont {Kim},\ and\ \citenamefont {Huse}}]{zhang2015thermalization}%
  \BibitemOpen
  \bibfield  {author} {\bibinfo {author} {\bibfnamefont {L.}~\bibnamefont
  {Zhang}}, \bibinfo {author} {\bibfnamefont {H.}~\bibnamefont {Kim}},\ and\
  \bibinfo {author} {\bibfnamefont {D.~A.}\ \bibnamefont {Huse}},\ }\bibfield
  {title} {\bibinfo {title} {{Thermalization of entanglement}},\ }\href
  {https://doi.org/10.1103/PhysRevE.91.062128} {\bibfield  {journal} {\bibinfo
  {journal} {Phys. Rev. E}\ }\textbf {\bibinfo {volume} {91}},\ \bibinfo
  {pages} {062128} (\bibinfo {year} {2015})}\BibitemShut {NoStop}%
\bibitem [{\citenamefont {Lin}\ and\ \citenamefont
  {Motrunich}(2017)}]{lin2017quasiparticle}%
  \BibitemOpen
  \bibfield  {author} {\bibinfo {author} {\bibfnamefont {C.-J.}\ \bibnamefont
  {Lin}}\ and\ \bibinfo {author} {\bibfnamefont {O.~I.}\ \bibnamefont
  {Motrunich}},\ }\bibfield  {title} {\bibinfo {title} {{Quasiparticle
  explanation of the weak-thermalization regime under quench in a nonintegrable
  quantum spin chain}},\ }\href {https://doi.org/10.1103/PhysRevA.95.023621}
  {\bibfield  {journal} {\bibinfo  {journal} {Phys. Rev. A}\ }\textbf {\bibinfo
  {volume} {95}},\ \bibinfo {pages} {023621} (\bibinfo {year}
  {2017})}\BibitemShut {NoStop}%
\bibitem [{\citenamefont {{Mazza}}\ \emph {et~al.}(2018)\citenamefont
  {{Mazza}}, \citenamefont {{Perfetto}}, \citenamefont {{Lerose}},
  \citenamefont {{Collura}},\ and\ \citenamefont
  {{Gambassi}}}]{mazza2018suppression}%
  \BibitemOpen
  \bibfield  {author} {\bibinfo {author} {\bibfnamefont {P.~P.}\ \bibnamefont
  {{Mazza}}}, \bibinfo {author} {\bibfnamefont {G.}~\bibnamefont {{Perfetto}}},
  \bibinfo {author} {\bibfnamefont {A.}~\bibnamefont {{Lerose}}}, \bibinfo
  {author} {\bibfnamefont {M.}~\bibnamefont {{Collura}}},\ and\ \bibinfo
  {author} {\bibfnamefont {A.}~\bibnamefont {{Gambassi}}},\ }\bibfield  {title}
  {\bibinfo {title} {{Suppression of transport in non-disordered quantum spin
  chains due to confined excitations}},\ }\href@noop {} {\bibfield  {journal}
  {\bibinfo  {journal} {ArXiv e-prints}\ } (\bibinfo {year} {2018})},\ \Eprint
  {https://arxiv.org/abs/1806.09674} {arXiv:1806.09674 [cond-mat.stat-mech]}
  \BibitemShut {NoStop}%
\bibitem [{\citenamefont {McCoy}\ and\ \citenamefont
  {Wu}(2014)}]{mccoy2014twodimensional}%
  \BibitemOpen
  \bibfield  {author} {\bibinfo {author} {\bibfnamefont {B.~M.}\ \bibnamefont
  {McCoy}}\ and\ \bibinfo {author} {\bibfnamefont {T.~T.}\ \bibnamefont {Wu}},\
  }\href {https://books.google.com/books?id=nB\_-AgAAQBAJ} {\emph {\bibinfo
  {title} {{The Two-Dimensional Ising Model: Second Edition}}}},\ Dover books
  on physics\ (\bibinfo  {publisher} {Dover Publications},\ \bibinfo {year}
  {2014})\BibitemShut {NoStop}%
\bibitem [{\citenamefont {{U.
  Schollw\"ock}}(2011)}]{schollwock2011densitymatrix}%
  \BibitemOpen
  \bibfield  {author} {\bibinfo {author} {\bibnamefont {{U. Schollw\"ock}}},\
  }\bibfield  {title} {\bibinfo {title} {The density-matrix renormalization
  group in the age of matrix product states},\ }\href
  {https://doi.org/10.1016/j.aop.2010.09.012} {\bibfield  {journal} {\bibinfo
  {journal} {Ann. Phys. (N.Y.)}\ }\textbf {\bibinfo {volume} {326}},\ \bibinfo
  {pages} {96 } (\bibinfo {year} {2011})}\BibitemShut {NoStop}%
\bibitem [{\citenamefont {Stoudenmire}\ and\ \citenamefont
  {White}(2012)}]{stoudenmire2012studying}%
  \BibitemOpen
  \bibfield  {author} {\bibinfo {author} {\bibfnamefont {E.~M.}\ \bibnamefont
  {Stoudenmire}}\ and\ \bibinfo {author} {\bibfnamefont {S.~R.}\ \bibnamefont
  {White}},\ }\bibfield  {title} {\bibinfo {title} {{Studying Two-Dimensional
  Systems with the Density Matrix Renormalization Group}},\ }\href
  {https://doi.org/10.1146/annurev-conmatphys-020911-125018} {\bibfield
  {journal} {\bibinfo  {journal} {Annu. Rev. Condens. Matt. Phys.}\ }\textbf
  {\bibinfo {volume} {3}},\ \bibinfo {pages} {111} (\bibinfo {year}
  {2012})}\BibitemShut {NoStop}%
\bibitem [{\citenamefont {Myers}\ \emph {et~al.}(2017)\citenamefont {Myers},
  \citenamefont {Rozali},\ and\ \citenamefont {Way}}]{myers2017holographic}%
  \BibitemOpen
  \bibfield  {author} {\bibinfo {author} {\bibfnamefont {R.~C.}\ \bibnamefont
  {Myers}}, \bibinfo {author} {\bibfnamefont {M.}~\bibnamefont {Rozali}},\ and\
  \bibinfo {author} {\bibfnamefont {B.}~\bibnamefont {Way}},\ }\bibfield
  {title} {\bibinfo {title} {Holographic quenches in a confined phase},\ }\href
  {http://stacks.iop.org/1751-8121/50/i=49/a=494002} {\bibfield  {journal}
  {\bibinfo  {journal} {J. Phys. A}\ }\textbf {\bibinfo {volume} {50}},\
  \bibinfo {pages} {494002} (\bibinfo {year} {2017})}\BibitemShut {NoStop}%
\bibitem [{\citenamefont {Buyens}\ \emph {et~al.}(2014)\citenamefont {Buyens},
  \citenamefont {Haegeman}, \citenamefont {Van~Acoleyen}, \citenamefont
  {Verschelde},\ and\ \citenamefont {Verstraete}}]{buyens2014matrix}%
  \BibitemOpen
  \bibfield  {author} {\bibinfo {author} {\bibfnamefont {B.}~\bibnamefont
  {Buyens}}, \bibinfo {author} {\bibfnamefont {J.}~\bibnamefont {Haegeman}},
  \bibinfo {author} {\bibfnamefont {K.}~\bibnamefont {Van~Acoleyen}}, \bibinfo
  {author} {\bibfnamefont {H.}~\bibnamefont {Verschelde}},\ and\ \bibinfo
  {author} {\bibfnamefont {F.}~\bibnamefont {Verstraete}},\ }\bibfield  {title}
  {\bibinfo {title} {{Matrix Product States for Gauge Field Theories}},\ }\href
  {https://doi.org/10.1103/PhysRevLett.113.091601} {\bibfield  {journal}
  {\bibinfo  {journal} {Phys. Rev. Lett.}\ }\textbf {\bibinfo {volume} {113}},\
  \bibinfo {pages} {091601} (\bibinfo {year} {2014})}\BibitemShut {NoStop}%
\bibitem [{\citenamefont {Buyens}\ \emph {et~al.}(2016)\citenamefont {Buyens},
  \citenamefont {Haegeman}, \citenamefont {Verschelde}, \citenamefont
  {Verstraete},\ and\ \citenamefont {Van~Acoleyen}}]{buyens2016confinement}%
  \BibitemOpen
  \bibfield  {author} {\bibinfo {author} {\bibfnamefont {B.}~\bibnamefont
  {Buyens}}, \bibinfo {author} {\bibfnamefont {J.}~\bibnamefont {Haegeman}},
  \bibinfo {author} {\bibfnamefont {H.}~\bibnamefont {Verschelde}}, \bibinfo
  {author} {\bibfnamefont {F.}~\bibnamefont {Verstraete}},\ and\ \bibinfo
  {author} {\bibfnamefont {K.}~\bibnamefont {Van~Acoleyen}},\ }\bibfield
  {title} {\bibinfo {title} {{Confinement and String Breaking for
  ${\mathrm{QED}}_{2}$ in the Hamiltonian Picture}},\ }\href
  {https://doi.org/10.1103/PhysRevX.6.041040} {\bibfield  {journal} {\bibinfo
  {journal} {Phys. Rev. X}\ }\textbf {\bibinfo {volume} {6}},\ \bibinfo {pages}
  {041040} (\bibinfo {year} {2016})}\BibitemShut {NoStop}%
\bibitem [{\citenamefont {Ba\~nuls}\ \emph {et~al.}(2016)\citenamefont
  {Ba\~nuls}, \citenamefont {Cichy}, \citenamefont {Jansen},\ and\
  \citenamefont {Saito}}]{banuls2016chiral}%
  \BibitemOpen
  \bibfield  {author} {\bibinfo {author} {\bibfnamefont {M.~C.}\ \bibnamefont
  {Ba\~nuls}}, \bibinfo {author} {\bibfnamefont {K.}~\bibnamefont {Cichy}},
  \bibinfo {author} {\bibfnamefont {K.}~\bibnamefont {Jansen}},\ and\ \bibinfo
  {author} {\bibfnamefont {H.}~\bibnamefont {Saito}},\ }\bibfield  {title}
  {\bibinfo {title} {Chiral condensate in the schwinger model with matrix
  product operators},\ }\href {https://doi.org/10.1103/PhysRevD.93.094512}
  {\bibfield  {journal} {\bibinfo  {journal} {Phys. Rev. D}\ }\textbf {\bibinfo
  {volume} {93}},\ \bibinfo {pages} {094512} (\bibinfo {year}
  {2016})}\BibitemShut {NoStop}%
\bibitem [{\citenamefont {Buyens}\ \emph
  {et~al.}(2017{\natexlab{a}})\citenamefont {Buyens}, \citenamefont {Haegeman},
  \citenamefont {Hebenstreit}, \citenamefont {Verstraete},\ and\ \citenamefont
  {Van~Acoleyen}}]{buyens2017realtime}%
  \BibitemOpen
  \bibfield  {author} {\bibinfo {author} {\bibfnamefont {B.}~\bibnamefont
  {Buyens}}, \bibinfo {author} {\bibfnamefont {J.}~\bibnamefont {Haegeman}},
  \bibinfo {author} {\bibfnamefont {F.}~\bibnamefont {Hebenstreit}}, \bibinfo
  {author} {\bibfnamefont {F.}~\bibnamefont {Verstraete}},\ and\ \bibinfo
  {author} {\bibfnamefont {K.}~\bibnamefont {Van~Acoleyen}},\ }\bibfield
  {title} {\bibinfo {title} {{Real-time simulation of the Schwinger effect with
  matrix product states}},\ }\href {https://doi.org/10.1103/PhysRevD.96.114501}
  {\bibfield  {journal} {\bibinfo  {journal} {Phys. Rev. D}\ }\textbf {\bibinfo
  {volume} {96}},\ \bibinfo {pages} {114501} (\bibinfo {year}
  {2017}{\natexlab{a}})}\BibitemShut {NoStop}%
\bibitem [{\citenamefont {Buyens}\ \emph
  {et~al.}(2017{\natexlab{b}})\citenamefont {Buyens}, \citenamefont
  {Montangero}, \citenamefont {Haegeman}, \citenamefont {Verstraete},\ and\
  \citenamefont {Van~Acoleyen}}]{buyens2017finiterepresentation}%
  \BibitemOpen
  \bibfield  {author} {\bibinfo {author} {\bibfnamefont {B.}~\bibnamefont
  {Buyens}}, \bibinfo {author} {\bibfnamefont {S.}~\bibnamefont {Montangero}},
  \bibinfo {author} {\bibfnamefont {J.}~\bibnamefont {Haegeman}}, \bibinfo
  {author} {\bibfnamefont {F.}~\bibnamefont {Verstraete}},\ and\ \bibinfo
  {author} {\bibfnamefont {K.}~\bibnamefont {Van~Acoleyen}},\ }\bibfield
  {title} {\bibinfo {title} {Finite-representation approximation of lattice
  gauge theories at the continuum limit with tensor networks},\ }\href
  {https://doi.org/10.1103/PhysRevD.95.094509} {\bibfield  {journal} {\bibinfo
  {journal} {Phys. Rev. D}\ }\textbf {\bibinfo {volume} {95}},\ \bibinfo
  {pages} {094509} (\bibinfo {year} {2017}{\natexlab{b}})}\BibitemShut
  {NoStop}%
\bibitem [{\citenamefont {Ba\~nuls}\ \emph {et~al.}(2017)\citenamefont
  {Ba\~nuls}, \citenamefont {Cichy}, \citenamefont {Cirac}, \citenamefont
  {Jansen},\ and\ \citenamefont {K\"uhn}}]{banuls2017density}%
  \BibitemOpen
  \bibfield  {author} {\bibinfo {author} {\bibfnamefont {M.~C.}\ \bibnamefont
  {Ba\~nuls}}, \bibinfo {author} {\bibfnamefont {K.}~\bibnamefont {Cichy}},
  \bibinfo {author} {\bibfnamefont {J.~I.}\ \bibnamefont {Cirac}}, \bibinfo
  {author} {\bibfnamefont {K.}~\bibnamefont {Jansen}},\ and\ \bibinfo {author}
  {\bibfnamefont {S.}~\bibnamefont {K\"uhn}},\ }\bibfield  {title} {\bibinfo
  {title} {{Density Induced Phase Transitions in the Schwinger Model: A Study
  with Matrix Product States}},\ }\href
  {https://doi.org/10.1103/PhysRevLett.118.071601} {\bibfield  {journal}
  {\bibinfo  {journal} {Phys. Rev. Lett.}\ }\textbf {\bibinfo {volume} {118}},\
  \bibinfo {pages} {071601} (\bibinfo {year} {2017})}\BibitemShut {NoStop}%
\bibitem [{\citenamefont {Nandkishore}\ and\ \citenamefont
  {Sondhi}(2017)}]{nandkishore2017many}%
  \BibitemOpen
  \bibfield  {author} {\bibinfo {author} {\bibfnamefont {R.~M.}\ \bibnamefont
  {Nandkishore}}\ and\ \bibinfo {author} {\bibfnamefont {S.~L.}\ \bibnamefont
  {Sondhi}},\ }\bibfield  {title} {\bibinfo {title} {{Many-Body Localization
  with Long-Range Interactions}},\ }\href
  {https://doi.org/10.1103/PhysRevX.7.041021} {\bibfield  {journal} {\bibinfo
  {journal} {Phys. Rev. X}\ }\textbf {\bibinfo {volume} {7}},\ \bibinfo {pages}
  {041021} (\bibinfo {year} {2017})}\BibitemShut {NoStop}%
\bibitem [{\citenamefont {Akhtar}\ \emph {et~al.}(2018)\citenamefont {Akhtar},
  \citenamefont {Nandkishore},\ and\ \citenamefont
  {Sondhi}}]{akhtar2018symmetry}%
  \BibitemOpen
  \bibfield  {author} {\bibinfo {author} {\bibfnamefont {A.~A.}\ \bibnamefont
  {Akhtar}}, \bibinfo {author} {\bibfnamefont {R.~M.}\ \bibnamefont
  {Nandkishore}},\ and\ \bibinfo {author} {\bibfnamefont {S.~L.}\ \bibnamefont
  {Sondhi}},\ }\bibfield  {title} {\bibinfo {title} {{Symmetry breaking and
  localization in a random Schwinger model with commensuration}},\ }\href
  {https://doi.org/10.1103/PhysRevB.98.115109} {\bibfield  {journal} {\bibinfo
  {journal} {Phys. Rev. B}\ }\textbf {\bibinfo {volume} {98}},\ \bibinfo
  {pages} {115109} (\bibinfo {year} {2018})}\BibitemShut {NoStop}%
\bibitem [{\citenamefont {Lencs{\'e}s}\ and\ \citenamefont
  {Tak{\'a}cs}(2015)}]{lencses2015confinement}%
  \BibitemOpen
  \bibfield  {author} {\bibinfo {author} {\bibfnamefont {M.}~\bibnamefont
  {Lencs{\'e}s}}\ and\ \bibinfo {author} {\bibfnamefont {G.}~\bibnamefont
  {Tak{\'a}cs}},\ }\bibfield  {title} {\bibinfo {title} {{Confinement in the
  q-state Potts model: an RG-TCSA study}},\ }\href
  {https://doi.org/10.1007/JHEP09(2015)146} {\bibfield  {journal} {\bibinfo
  {journal} {JHEP}\ }\textbf {\bibinfo {volume} {2015}}\bibinfo  {number} {
  (9)},\ \bibinfo {pages} {146}}\BibitemShut {NoStop}%
\bibitem [{\citenamefont {Rutkevich}(2017)}]{rutkevich2017radiative}%
  \BibitemOpen
\bibfield  {number} {  }\bibfield  {author} {\bibinfo {author} {\bibfnamefont
  {S.~B.}\ \bibnamefont {Rutkevich}},\ }\bibfield  {title} {\bibinfo {title}
  {{Radiative corrections to the quark masses in the ferromagnetic Ising and
  Potts field theories}},\ }\href
  {https://doi.org/https://doi.org/10.1016/j.nuclphysb.2017.08.009} {\bibfield
  {journal} {\bibinfo  {journal} {Nucl. Phys. B}\ }\textbf {\bibinfo {volume}
  {923}},\ \bibinfo {pages} {508 } (\bibinfo {year} {2017})}\BibitemShut
  {NoStop}%
\bibitem [{\citenamefont {James}\ and\ \citenamefont
  {Konik}(2013)}]{james2013understanding}%
  \BibitemOpen
  \bibfield  {author} {\bibinfo {author} {\bibfnamefont {A.~J.~A.}\
  \bibnamefont {James}}\ and\ \bibinfo {author} {\bibfnamefont {R.~M.}\
  \bibnamefont {Konik}},\ }\bibfield  {title} {\bibinfo {title} {Understanding
  the entanglement entropy and spectra of 2{D} quantum systems through arrays
  of coupled 1{D} chains},\ }\href {https://doi.org/10.1103/PhysRevB.87.241103}
  {\bibfield  {journal} {\bibinfo  {journal} {Phys. Rev. B}\ }\textbf {\bibinfo
  {volume} {87}},\ \bibinfo {pages} {241103} (\bibinfo {year}
  {2013})}\BibitemShut {NoStop}%
\bibitem [{\citenamefont {Albuquerque}\ \emph {et~al.}(2007)\citenamefont
  {Albuquerque}, \citenamefont {Alet}, \citenamefont {Corboz}, \citenamefont
  {Dayal}, \citenamefont {Feiguin}, \citenamefont {Fuchs}, \citenamefont
  {Gamper}, \citenamefont {Gull}, \citenamefont {G\"urtler}, \citenamefont
  {Honecker}, \citenamefont {Igarashi}, \citenamefont {K\"orner}, \citenamefont
  {Kozhevnikov}, \citenamefont {L\"auchli}, \citenamefont {Manmana} \emph
  {et~al.}}]{albuquerque2007alps}%
  \BibitemOpen
  \bibfield  {author} {\bibinfo {author} {\bibfnamefont {A.~F.}\ \bibnamefont
  {Albuquerque}}, \bibinfo {author} {\bibfnamefont {F.}~\bibnamefont {Alet}},
  \bibinfo {author} {\bibfnamefont {P.}~\bibnamefont {Corboz}}, \bibinfo
  {author} {\bibfnamefont {P.}~\bibnamefont {Dayal}}, \bibinfo {author}
  {\bibfnamefont {A.}~\bibnamefont {Feiguin}}, \bibinfo {author} {\bibfnamefont
  {S.}~\bibnamefont {Fuchs}}, \bibinfo {author} {\bibfnamefont
  {L.}~\bibnamefont {Gamper}}, \bibinfo {author} {\bibfnamefont
  {E.}~\bibnamefont {Gull}}, \bibinfo {author} {\bibfnamefont {S.}~\bibnamefont
  {G\"urtler}}, \bibinfo {author} {\bibfnamefont {A.}~\bibnamefont {Honecker}},
  \bibinfo {author} {\bibfnamefont {R.}~\bibnamefont {Igarashi}}, \bibinfo
  {author} {\bibfnamefont {M.}~\bibnamefont {K\"orner}}, \bibinfo {author}
  {\bibfnamefont {A.}~\bibnamefont {Kozhevnikov}}, \bibinfo {author}
  {\bibfnamefont {A.}~\bibnamefont {L\"auchli}}, \bibinfo {author}
  {\bibfnamefont {S.~R.}\ \bibnamefont {Manmana}}, \emph {et~al.},\ }\bibfield
  {title} {\bibinfo {title} {{The ALPS project release 1.3: Open-source
  software for strongly correlated systems}},\ }\href
  {https://doi.org/http://doi.org/10.1016/j.jmmm.2006.10.304} {\bibfield
  {journal} {\bibinfo  {journal} {J. Magn. Magn. Mater.}\ }\textbf {\bibinfo
  {volume} {310}},\ \bibinfo {pages} {1187 } (\bibinfo {year} {2007})},\
  \bibinfo {note} {proceedings of the 17th International Conference on
  Magnetism}\BibitemShut {NoStop}%
\bibitem [{\citenamefont {Bauer}\ \emph {et~al.}(2011)\citenamefont {Bauer},
  \citenamefont {Carr}, \citenamefont {Evertz}, \citenamefont {Feiguin},
  \citenamefont {Freire}, \citenamefont {Fuchs}, \citenamefont {Gamper},
  \citenamefont {Gukelberger}, \citenamefont {Gull}, \citenamefont {Guertler},
  \citenamefont {Hehn}, \citenamefont {Igarashi}, \citenamefont {Isakov},
  \citenamefont {Koop}, \citenamefont {Ma} \emph {et~al.}}]{bauer2011alps}%
  \BibitemOpen
  \bibfield  {author} {\bibinfo {author} {\bibfnamefont {B.}~\bibnamefont
  {Bauer}}, \bibinfo {author} {\bibfnamefont {L.~D.}\ \bibnamefont {Carr}},
  \bibinfo {author} {\bibfnamefont {H.~G.}\ \bibnamefont {Evertz}}, \bibinfo
  {author} {\bibfnamefont {A.}~\bibnamefont {Feiguin}}, \bibinfo {author}
  {\bibfnamefont {J.}~\bibnamefont {Freire}}, \bibinfo {author} {\bibfnamefont
  {S.}~\bibnamefont {Fuchs}}, \bibinfo {author} {\bibfnamefont
  {L.}~\bibnamefont {Gamper}}, \bibinfo {author} {\bibfnamefont
  {J.}~\bibnamefont {Gukelberger}}, \bibinfo {author} {\bibfnamefont
  {E.}~\bibnamefont {Gull}}, \bibinfo {author} {\bibfnamefont {S.}~\bibnamefont
  {Guertler}}, \bibinfo {author} {\bibfnamefont {A.}~\bibnamefont {Hehn}},
  \bibinfo {author} {\bibfnamefont {R.}~\bibnamefont {Igarashi}}, \bibinfo
  {author} {\bibfnamefont {S.~V.}\ \bibnamefont {Isakov}}, \bibinfo {author}
  {\bibfnamefont {D.}~\bibnamefont {Koop}}, \bibinfo {author} {\bibfnamefont
  {P.~N.}\ \bibnamefont {Ma}}, \emph {et~al.},\ }\bibfield  {title} {\bibinfo
  {title} {{The ALPS project release 2.0: open source software for strongly
  correlated systems}},\ }\href
  {http://stacks.iop.org/1742-5468/2011/i=05/a=P05001} {\bibfield  {journal}
  {\bibinfo  {journal} {J. Stat. Mech.}\ }\textbf {\bibinfo {volume} {2011}},\
  \bibinfo {pages} {P05001} (\bibinfo {year} {2011})}\BibitemShut {NoStop}%
\bibitem [{Note5()}]{Note5}%
  \BibitemOpen
  \bibinfo {note} {Throughout, we construct the free fermion basis with states
  containing at most ten fermions. In the case of $R=35$, $E_\Lambda = 10.5$
  for $m=1$, the maximum number of fermions within states below the cutoff
  energy is eight.}\BibitemShut {Stop}%
\end{thebibliography}%

\end{document}